\pdfoutput=1
\documentclass[useAMS]{mn2e}
\usepackage[tight]{subfigure}
\usepackage{longtable}
\usepackage{amssymb}
\usepackage{lscape}
\usepackage{graphicx}

\title[]{The detailed nature of active central cluster galaxies}
\author[Loubser et al.]{S. I. Loubser$^{1}$\thanks{E-mail:
Ilani.Loubser@nwu.ac.za (SIL)}, I. K. Soechting$^{2}$\\
$^{1}$Centre for Space Research, North-West University, Potchefstroom 2520, South Africa\\
$^{2}$Astrophysics, Department of Physics, University of Oxford, Oxford, OX1 3RH, UK}
\begin{document}


\pagerange{\pageref{firstpage}--\pageref{lastpage}} \pubyear{2012}

\maketitle

\label{firstpage}

\begin{abstract}
We present detailed integral field unit (IFU) observations of the central few kiloparsecs of the ionised nebulae surrounding four active 
central cluster galaxies (CCGs) in cooling flow clusters (Abell 0496, 0780, 1644 and 2052). Our sample consists of CCGs with H$\alpha$ filaments, and 
have existing data from the X-ray regime available. Here, we present the detailed optical 
emission-line (and simultaneous absorption line) data over a broad wavelength range to probe the dominant ionisation processes, excitation sources, morphology and 
kinematics of the hot gas (as well as the morphology and kinematics of the stars). This, combined with the other multiwavelength data, will form a complete view of the 
different phases (hot and cold gas and stars) and how they interact in the processes of star formation and 
feedback detected in central galaxies in cooling flow clusters, as well as the influence of the host cluster. We derive the optical dust extinction maps of the four nebulae.
We also derive a range of different kinematic properties, given the small sample size. For Abell 0496 and 0780, we find that the stars and gas are kinematically decoupled, and in the case of Abell 1644 we find that these components are aligned. For Abell 2052, we find that the gaseous components show rotation even though no rotation is apparent in the stellar components. To the degree that our spatial resolution reveals, it appears that all the optical forbidden and hydrogen recombination lines originate in the same gas for all the galaxies. Based on optical diagnostic ratios ([OIII]$\lambda$5007/H$\beta$ against [NII]$\lambda$6584/H$\alpha$, [SII]$\lambda\lambda$6717,6731/H$\alpha$, and [OI]$\lambda$6300/H$\alpha$), all galaxies show extended LINER emission, but that at least one has significant Seyfert emission areas, and at least one other has significant HII like emission line ratios for many pixels. We also show that the hardness of the ionising continuum do not decrease with galactocentric distance within our field-of-view (with the exception of one galaxy that show a core separation) as the emission line ratios do not vary significantly with radius. This also indicates that the derived nebular properties are spatially homogeneous. We fit AGN and pAGB stars photoionisation models as well as shock excitation models to our derived diagnostic ratios. These fits, combined with information from multiwavelength studies reveal that AGN photoionisation is the most likely ionisation mechanism in at least two cases even though shocks and pAGB stars can not be conclusively eliminated.
\end{abstract}


\section{Introduction}

Decades ago, elliptical galaxies were thought to contain very little, if any, gas. Studies of galaxy formation, therefore, often focussed on the stellar properties, however we now know that a large fraction of the baryonic mass in massive galaxies is believe to be in diffuse form. Thus a complete view of galaxy formation and evolution necessarily incorporates both the stars and hot gas and an understanding of the processes by which these phases interact (McCarthy et al.\ 2010).

Cooling-flow clusters are common in the local Universe and massive central cluster galaxies (CCGs) are often found at the centres of these systems (Peres et al.\ 1998). If the central cluster density is high enough, intracluster gas can condense and form stars at the bottom of the potential well. Since the radiative cooling times for intracluster gas are short enough that gas can cool and settle to the cluster centre (Edge, Stewart $\&$ Fabian 1992), it has been suggested that the big envelopes of CCGs may arise from the gradual deposition of this cool gas. More recently, high spectral resolution \textit{XMM-Newton} observations showed that the X-ray gas in cluster centres does not cool significantly below a threshold temperature of $kT\sim1-2$ keV (Jord\'an et al.\ 2004, and references therein). This initially contradicted the model that these young stars are formed in cooling flows. However, it is possible that star formation is ongoing in cool-core clusters at a much reduced rate (Bildfell et al.\ 2008).   

Previous studies have reported several examples of ongoing star formation in CCGs, in particular those hosted by cooling-flow clusters (Cardiel, Gorgas $\&$ Arag\'{o}n-Salamanca 1998; Crawford et al.\ 1999; McNamara et al.\ 2006; Edwards et al.\ 2007; O'Dea et al.\ 2008; Bildfell et al.\ 2008; Pipino et al.\ 2009; Loubser et al.\ 2009). However, the origin of the gas fuelling this star formation is not yet known. Possible explanations include processes involving cooling flows or cold gas deposited during a merging event (Bildfell et al.\ 2008). These processes will leave different imprints in the dynamical properties, the detailed chemical abundances, and the star formation histories of these galaxies, which can be studied using high-quality spectroscopy (Loubser et al. 2008; 2009; 2012).

Observations that support this idea are blue- and ultraviolet-colour (UV-colour) excesses observed (indicative of star formation) in the central galaxy of Abell 1795 by McNamara et al.\ (1996) and molecular gas detected in 10 out of 32 central cluster galaxies by Salom\'{e} $\&$ Combes (2003). The observations by Cardiel et al.\ (1998) were consistent with an evolutionary sequence in which star formation bursts, triggered by radio sources, take place several times during the lifetime of the cooling flow in the centre of the cluster. However, McNamara $\&$ O'Connell (1992) found only colour anomalies with small amplitudes, implying star formation rates that account for at most a few percent of the material that is cooling and accreting onto the central galaxy. Cooling-flow models for CCG formation also imply the formation of larger numbers of new stars, for which there is no good observational evidence (Athanassoula, Garijo $\&$ Garc\'{\i}a-G\'{o}mez 2001). The central cluster galaxies often host radio-loud AGN which may account for the necessary heating to counteract radiative cooling (Von der Linden et al.\ 2007). 

In summary, cooling flow models predict more cooled gas than is observed (Bohringer et al. 2001). Thus, it is possible that the 
mass deposited into the molecular clouds is heated by one of several processes - hot young stellar populations, 
radio-loud AGN, X-rays or heat conduction from the intracluster medium itself, shocks and turbulent mixing 
layers, and cosmic rays. Therefore, only a small fraction of the cooled gas is detected (Crawford et al. 2005; Ferland 
et al. 2009). Thus, CCGs lie at the interface where it is crucial to understand the role of feedback and accretion in 
star formation. Within these cooling-flow CCGs, cool molecular clouds, warm ionized hydrogen, and the cooling 
intracluster medium are related. A complete view of the star formation process incorporates the stars with the gas 
and an understanding of the processes by which these phases interact, and therefore, requires information from 
several wavelength regimes. 

In conclusion, although CCGs are probably not completely formed in cooling flows, the flows play an important role in regulating the rate at which gas cools at the centres of groups and clusters. In the $\Lambda$CDM cosmology it is now understood that local massive haloes assemble late through the merging of smaller systems. In this picture, cooling flows seem to be the main fuel for galaxy mass growth at high redshift. This source is removed only at low redshifts in group or cluster environments, due to AGN feedback (De Lucia $\&$ Blaizot 2007; Voit et al.\ 2008). Indeed, if AGN feedback is not properly assumed in hydrodynamical simulations, an apparently bluer BCG is formed as a result of an accelerated late stellar birthrate even after the epoch of quiescent star formation (Romeo et al.\ 2008).

We proposed and obtained IFU observations of the central few kiloparsecs of the ionised nebulae in active CCGs in cooling-flow 
clusters. These observations will map the morphology, kinematics and ionisation state of the nebulae to gain an
understanding of their formation, heating and relationship to the cluster centre. We selected galaxies from the McDonald et al.\ (2010) study, which consisted of an H$\alpha$ survey of 23 cooling flow clusters. Amongst their conclusions, McDonald, Veilleux $\&$ Rupke (2012) find a stong correlation between the H$\alpha$ luminosity contained in filaments and the X-ray cooling flow rate of the cluster, suggesting that the warm, ionised gas is linked to the cooling flow. We chose objects with
confirmed extended H$\alpha$ emission, and with near-IR (2MASS), ultraviolet (GALEX), X-ray data (Chandra), 
and in some cases VLA 1.4 GHz fluxes, already available (McDonald et al. 2010). Detailed properties of the host 
clusters, which are reported to influence the activity in the central galaxy, such as central cooling times and the 
offset between the cluster X-ray peak and the central galaxy, have been derived. 

Farage et al.\ (2010) presented IFU observations of one nearby BCG showing LINER-like emission, and Brough et al.\ (2011) presented IFU data on four CCGs at z $\sim 0.1$ to calculate the dynamical masses of CCGs and measure their stellar angular momentum (although none of their four CCGs contained emission lines). The SAURON sample includes only one CCG (M87). Hatch, Crawford $\&$ Fabian (2007) and Edwards et al. (2009) analysed data comparable to the data presented here. Hatch et al.\ (2007) presents IFU observations of six emission-line nebulae in cool-core clusters (selected from the optical ROSAT follow-up by Crawford et al.\ 1999) with OASIS on the WHT (with a limited wavelength range centered around H$\alpha$). Edwards et al.\ (2009) presents IFU observations of nine CCGs in cooling and non-cooling clusters (also from Crawford et al.\ 1999) and within 50 kpc of the X-ray centre, with GMOS IFU and OASIS on WHT (most galaxies with wavelength ranges also limited to the H$\alpha$ region, but three also included H$\beta$). Emission line maps of morphology, kinematics, often line ratios, and the stellar continuum have been published for Abell 496 by Hatch et al.\ (2007), and for Abell 2052 by Edwards et al.\ (2009). These studies concluded that no proposed heating mechanism reproduces all the emission-line properties within their observed wavelength within a single source. Thus, a single dominant mechanism may not apply to all CCG nebulae, and there may be a mixture of heating mechanisms acting within a single nebula (Wilman, Edge $\&$ Swinbank 2006; Hatch et al.\ 2007). Thus to get more information which will enable the dominant mechanism(s) to be identified, we observed more line ratios over a longer wavelength range (around H$\alpha$ and H$\beta$ which also allowed extinction to be appropriately measured). We will be adding the detailed stellar population analysis, and place the derived information from the optical spectra in context with multiwavelength data over the full spectrum in a future paper, to explain the diverse nature of these galaxies. All the data for Abell 780 and 1644 is new, and all four of the extinction maps are new. Also, the construction of ionisation diagrams and analysis of ionisation processes with shock models and pAGB models is also new. Our exposure times (improved from previous studies) are shown in Table \ref{table:objects}. This data is also complimentary to the long-slit spectroscopy (on KECK and Magellan) along the H$\alpha$ filaments of the four objects (studied here) by McDonald et al.\ (2012).

We introduce the sample and detail of the data reductions in Sections \ref{Sample} and \ref{reduction}. We then derive the optical extinction as well as the line strengths in Section \ref{extinction}. We proceed to discuss the four individual cases in Section \ref{figures_NVSS}, and identify the mechanism producing the emission and the ionised gas in these four CCGs in Section \ref{ionisation}. We summarise the findings of this paper in \ref{summary}. We have used the following set of cosmological parameters: $\Omega_{m}$ = 0.3, $\Omega_{\Lambda}$ = 0.7, H$_{0}$ = 70km s$^{-1}$ Mpc$^{-1}$. 

\section{Sample}
\label{Sample}

\begin{figure*}
   \begin{center}
 \mbox{\subfigure[Abell 0496 -- MCG-02-12-039]{\includegraphics[width=4.2cm,height=4.2cm]{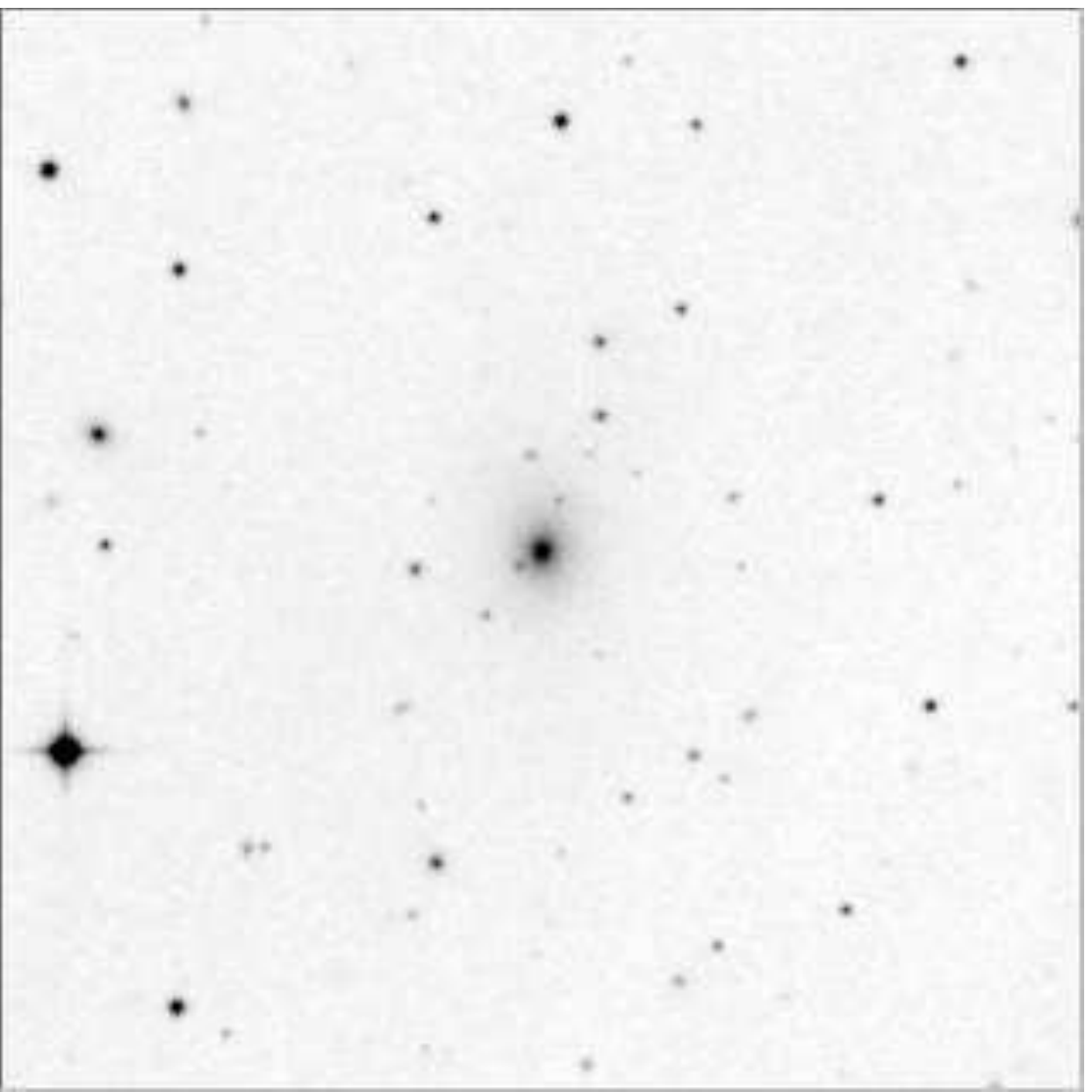}}\quad
         \subfigure[Abell 0780 -- PGC026269]{\includegraphics[width=4.2cm,height=4.2cm]{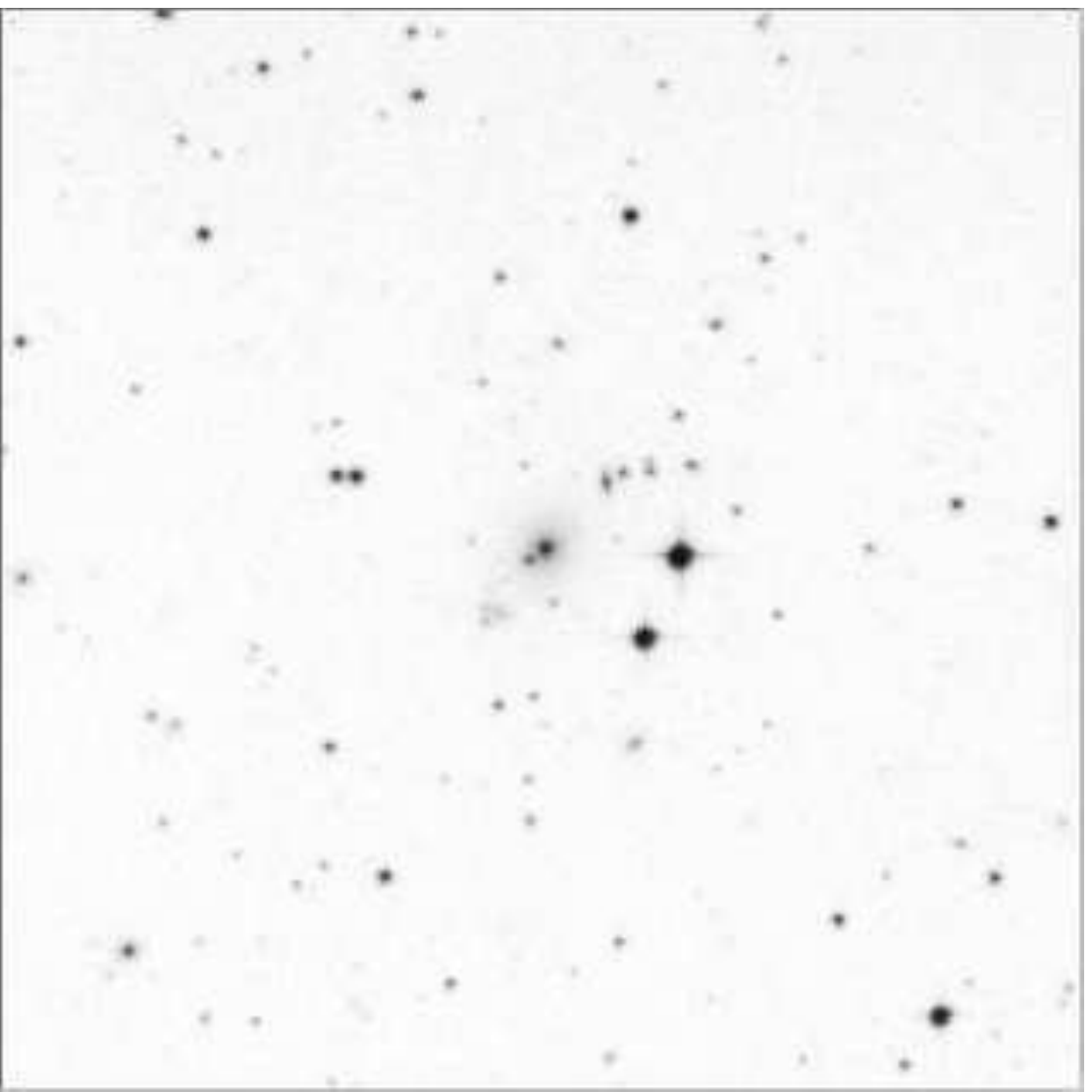}}\quad
         \subfigure[Abell 1644 -- PGC044257]{\includegraphics[width=4.2cm,height=4.2cm]{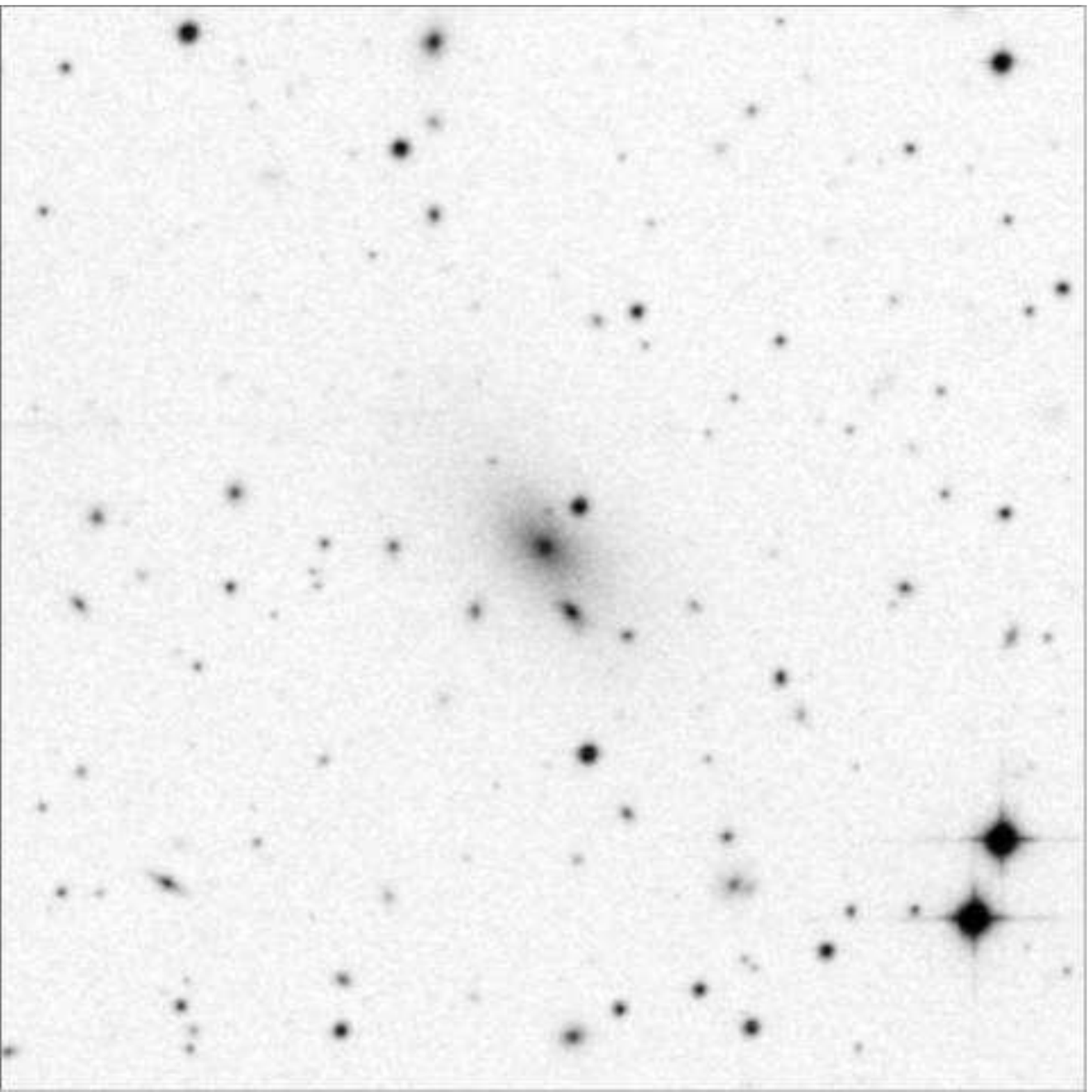}}\quad
\subfigure[Abell 2052 -- UGC09799]{\includegraphics[width=4.2cm,height=4.2cm]{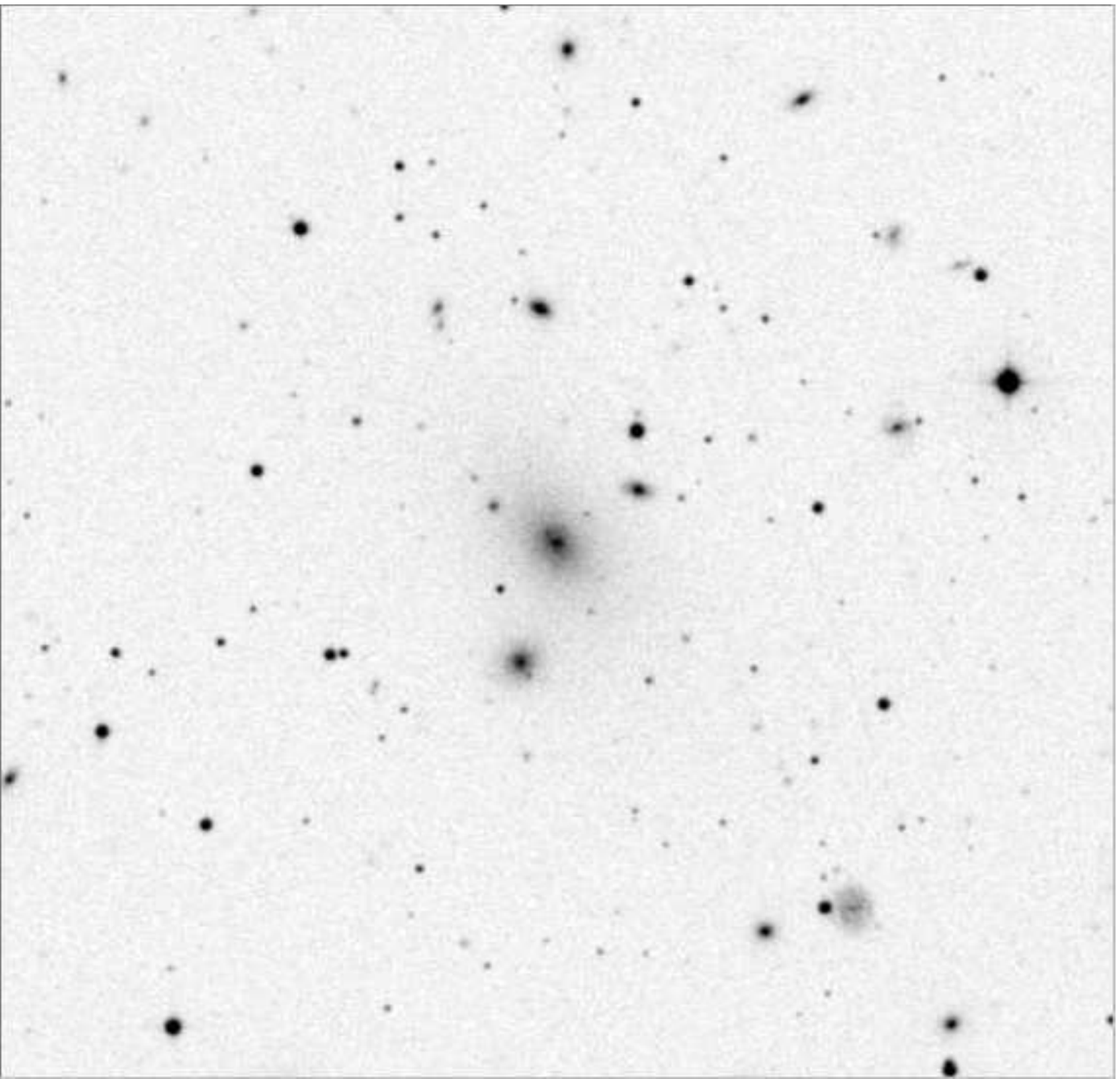}}}
   \mbox{\subfigure[Abell 0496 -- MCG-02-12-039]{\includegraphics[width=4.2cm,height=4.2cm]{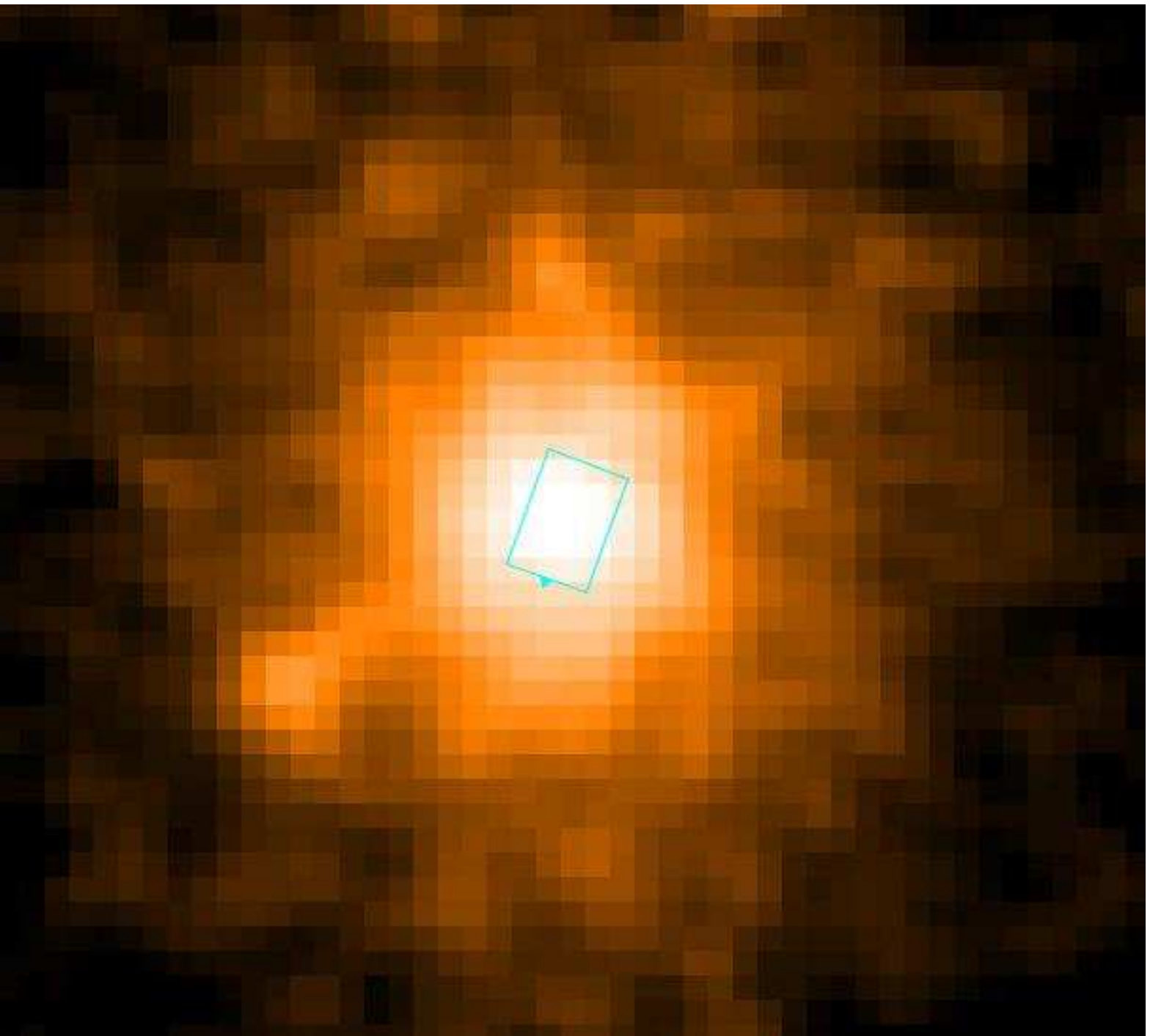}}\quad
         \subfigure[Abell 0780 -- PGC026269]{\includegraphics[width=4.2cm,height=4.2cm]{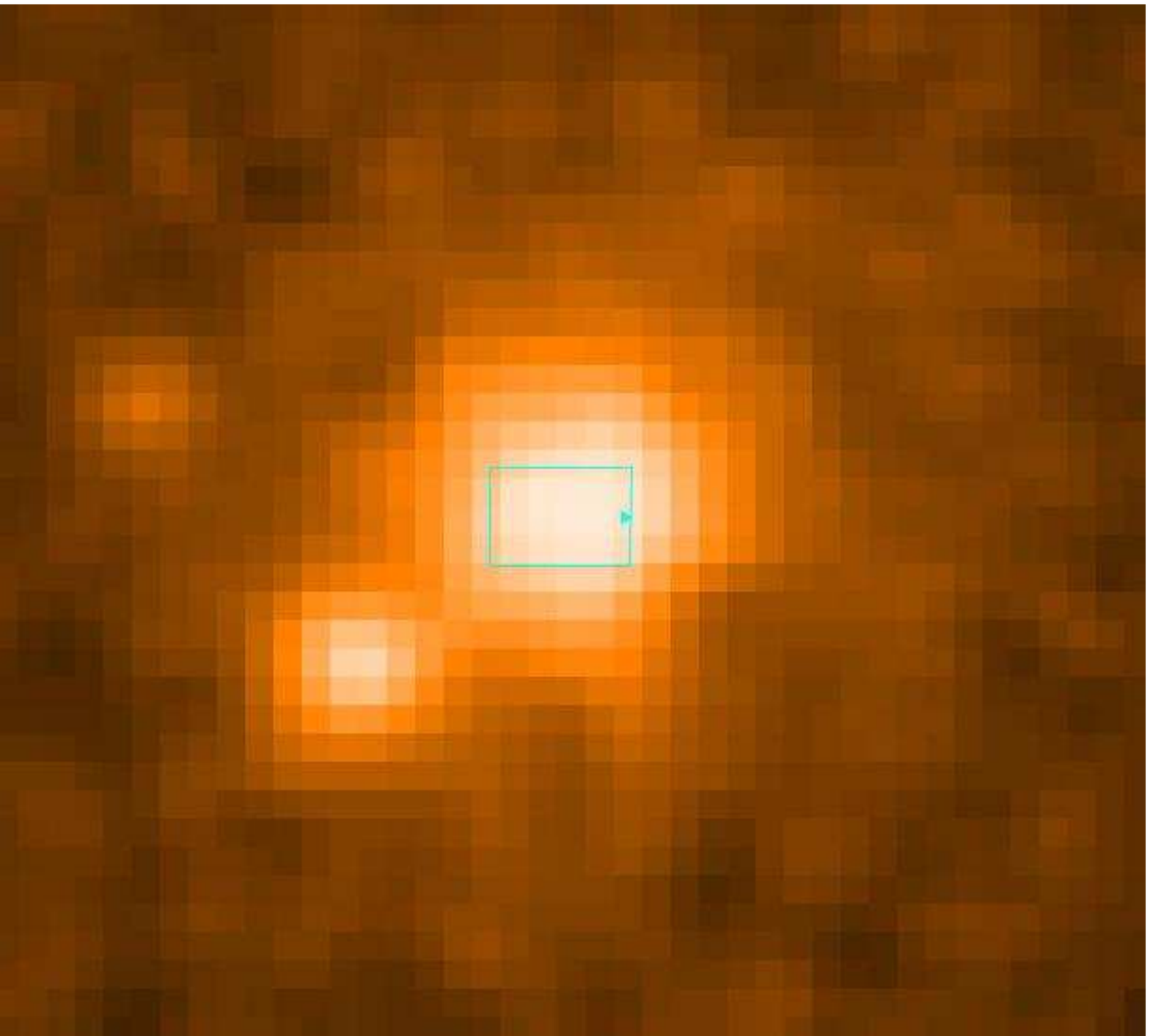}}\quad
         \subfigure[Abell 1644 -- PGC044257]{\includegraphics[width=4.2cm,height=4.2cm]{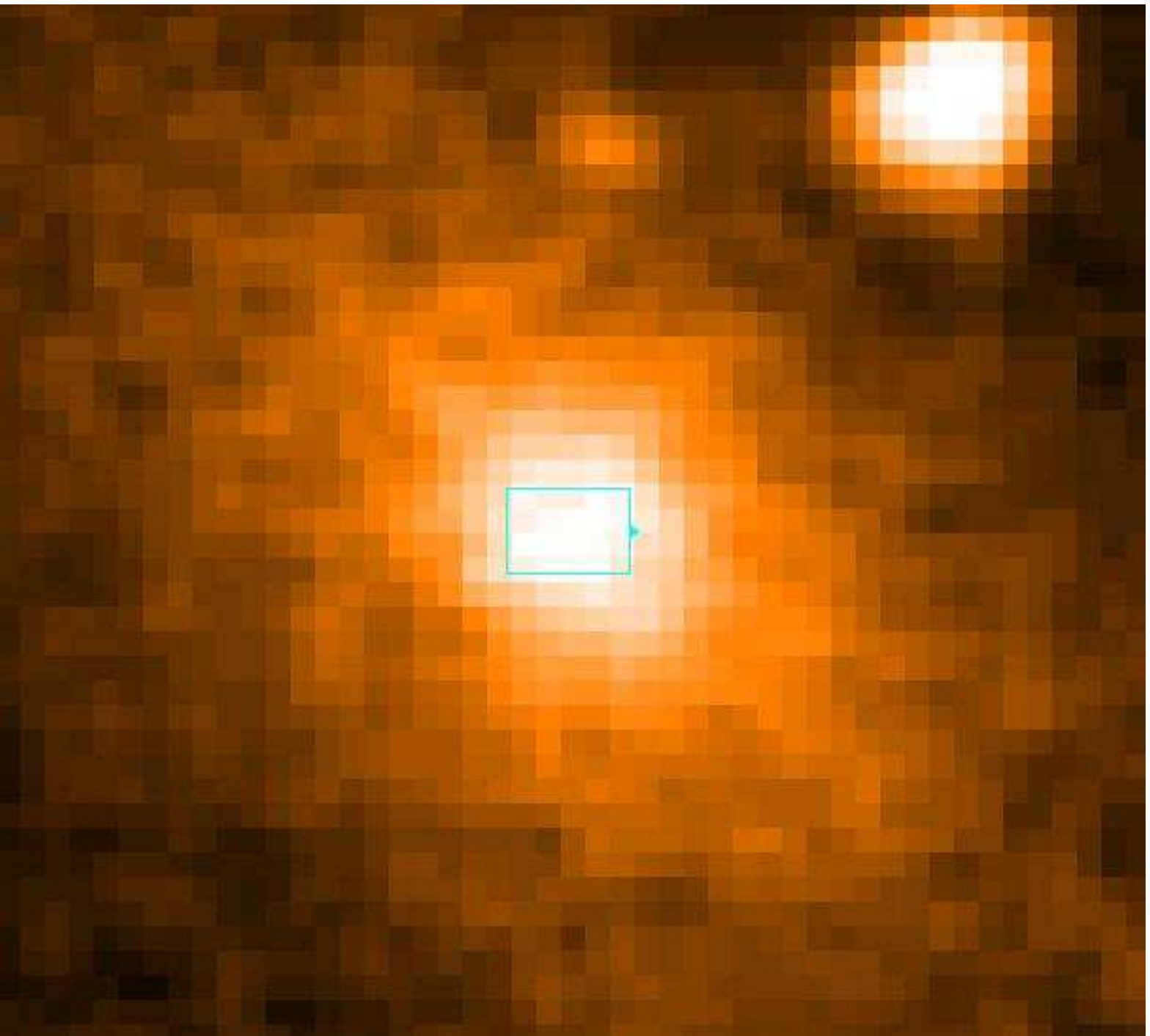}}\quad
\subfigure[Abell 2052 -- UGC09799]{\includegraphics[width=4.2cm,height=4.2cm]{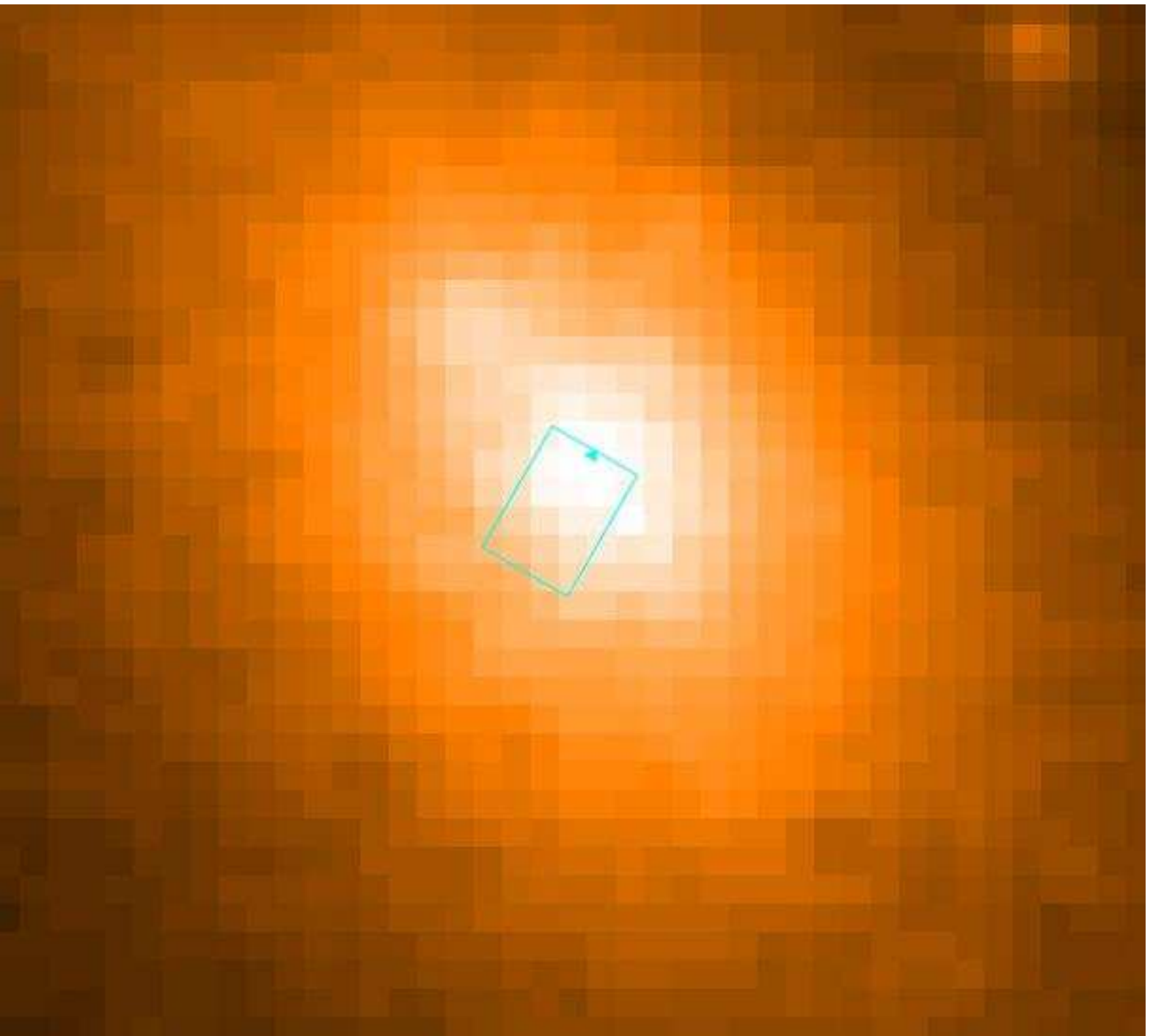}}}
   \mbox{\subfigure[Abell 0496 -- MCG-02-12-039]{\includegraphics[width=4.2cm,height=6.0cm]{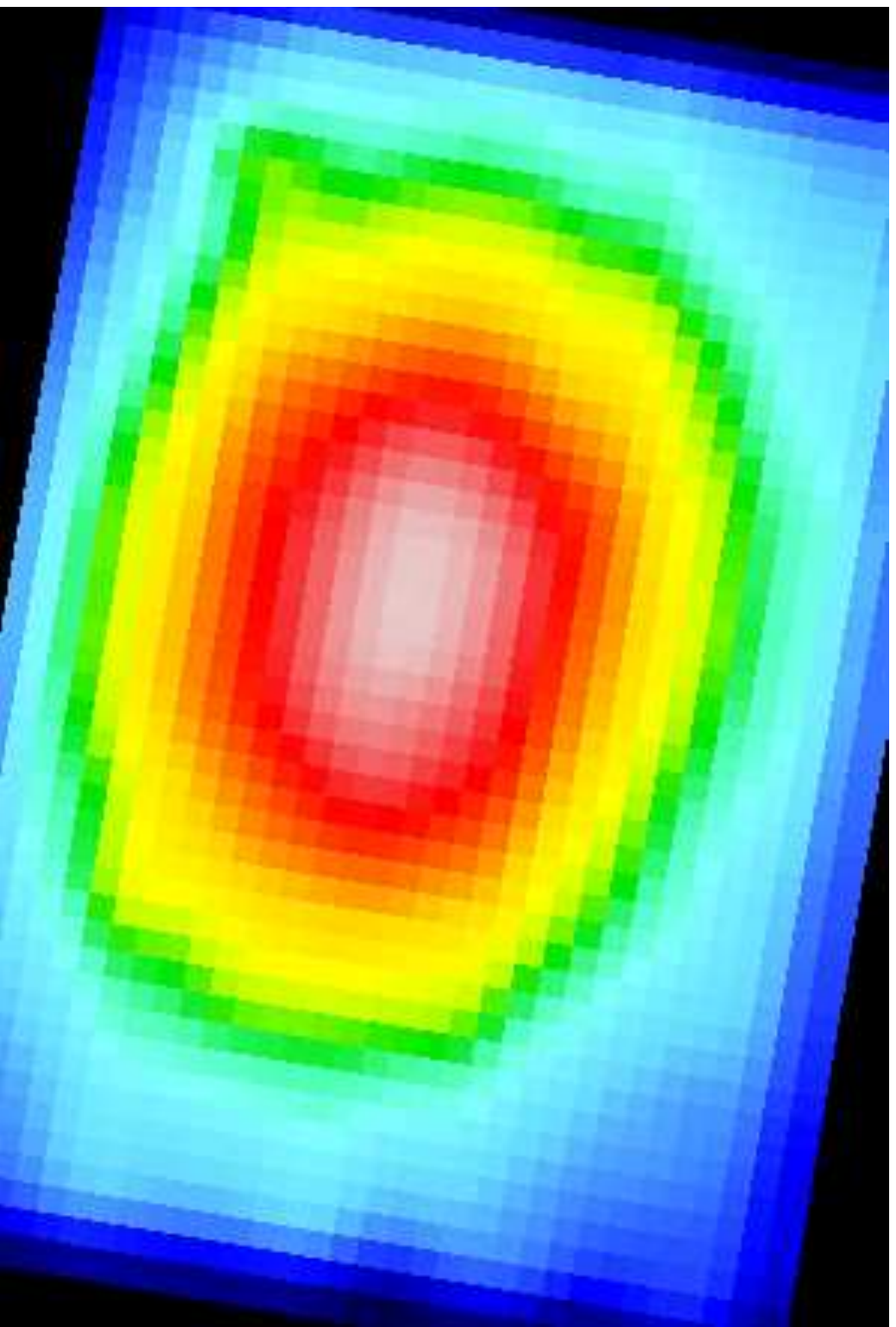}}\quad
         \subfigure[Abell 0780 -- PGC026269]{\includegraphics[width=4.2cm,height=6.0cm]{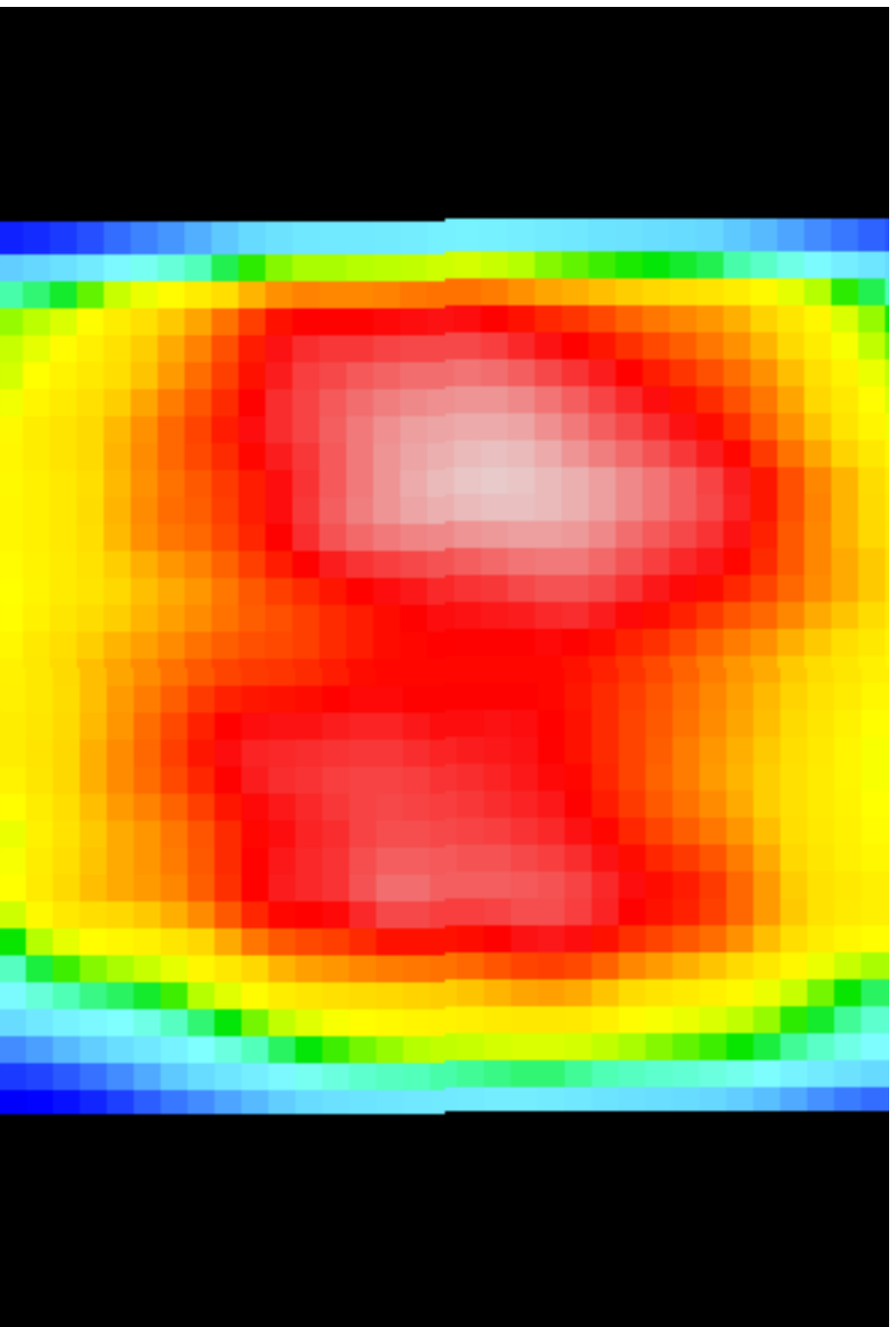}}\quad
         \subfigure[Abell 1644 -- PGC044257]{\includegraphics[width=4.2cm,height=6.0cm]{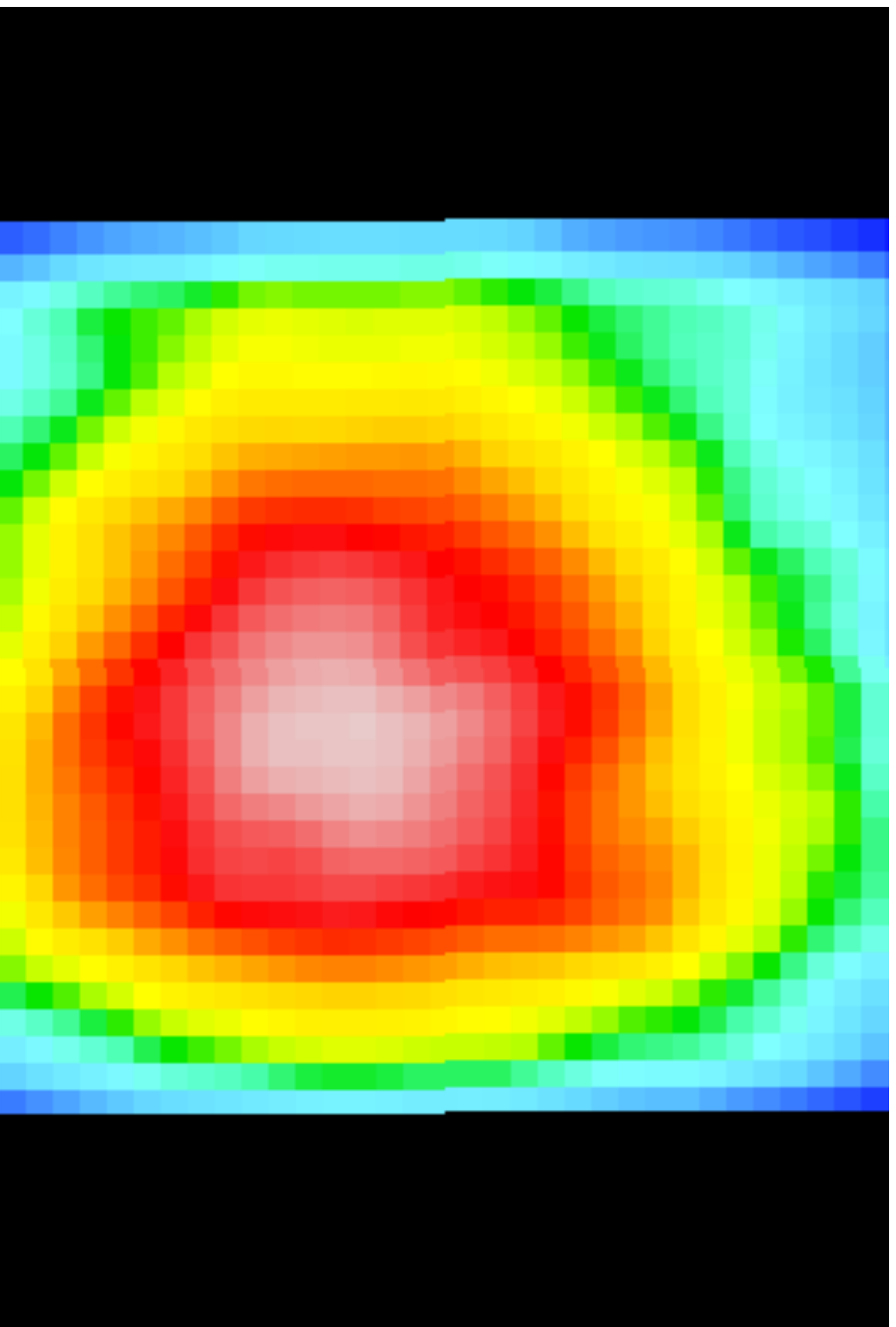}}\quad
\subfigure[Abell 2052 -- UGC09799]{\includegraphics[width=4.2cm,height=6.0cm]{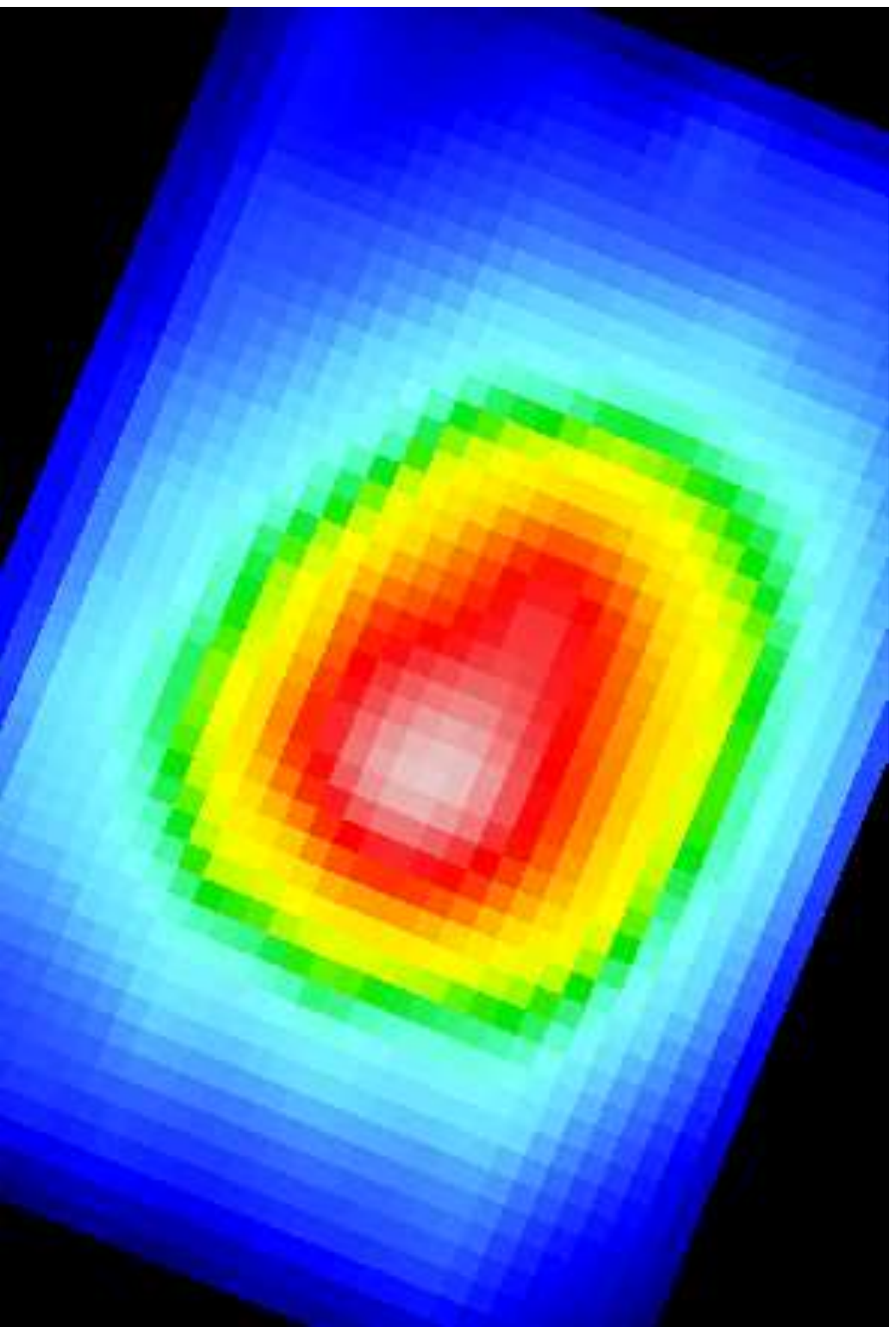}}}
\caption{DSS images of the four targets (north above, and east to the left for all images). The upper plots show 8 $\times$ 8 arcmin field of views, and the middle plots show the targets with the 5 $\times$ 3.5 arcsec IFU field of view overlayed. The top of the IFU FOV is indicated with a blue arrow. The lower plots show continuum images made from the IFU cubes (width of 50 \AA{} at 6350 \AA{}), smoothed spatially with a Gaussian with width 3 spaxels (which corresponds to 0.3 arcsec) and using the Sauron colourmap.}
\end{center}
\label{fig:Thumbnails} 
\end{figure*}

We have chosen our sample of active central cluster galaxies from the H$\alpha$ imaging presented in McDonald et al. (2010), who in turn, selected their sample from White, Jones $\&$ Forman (1997). McDonald et al.\ (2010) enforced the cuts: $\delta < +35 \deg$ and $0.025 < z < 0.092$, after which they selected 23 clusters to cover the full range of properties, from very rich clusters with high cooling rates to low-density clusters with small cooling flows. Their classical cooling rates range from 6.3 -- 431 M$_{\odot}$ yr$^{-1}$ which means that while covering a large range in properties, their sample consisted of only cooling flow clusters. From their 23 cooling flow clusters, we selected all the clusters with clearly detected H$\alpha$ in their centres (albeit filamentary, extended or nuclear emission). In addition, all of these central galaxies have optical imaging, near-IR (2MASS) and UV (Galex data) available. Thereafter, we selected all the central galaxies with detailed X-ray (Chandra) data, as well as VLA 1.4 GHz fluxes, available. This resulted in a sub-sample of 10 galaxies. We observed four of these galaxies with the GMOS IFU (as shown in Figure \ref{fig:Thumbnails}). We merely chose the objects with the most auxiliary information available. This additional information will be added in the future paper (where the underlying stellar populations will be analysed in detail) and might help to constrain the ionisation mechanisms.

The rest-wavelength range of the emission lines of interest is 4860--6731 \AA{} (H$\beta$ to [SII]$\lambda$6731). The ratio of the 
forbidden [NII]$\lambda$6584 to H$\alpha$ line will depend on the metallicity of the gas, the form of the ionising radiation, 
and the star formation rate. The relative strength of the [OIII]$\lambda$5007 and H$\beta$ lines reveals further excitation 
mechanism and gas metallicity information. The role of AGN photoionisation is confined to the central 2 -- 3" of 
active, massive nearby elliptical galaxies (Sarzi et al. 2006). Thus, IFU observations are ideal and will also allow us to 
study the 2D-distribution of the ionising radiation. In addition to the information from the emission lines, we are 
able to extract the underlying stellar absorption spectra using the improved GANDALF code (Sarzi et al. 2006). 
Thus, the kinematics and morphology of the hot ionised gas and stellar components can be correlated.

\begin{table*}
\begin{footnotesize}
\begin{tabular}{l c c c c c c c c} 
\hline Object & Cluster & Redshift & Linear scale & R$_{off}$ & T$_{X}$ & Classical cooling rates & Spectrally determined & Exposure Time \\
       &  &  \multicolumn{1}{c}{$z$} & \multicolumn{1}{c}{(kpc/arcsec)} & \multicolumn{1}{c}{(Mpc)} & \multicolumn{1}{c}{(keV)} & \multicolumn{1}{c}{(M$_{\odot}$/yr$^{-1}$)}& \multicolumn{1}{c}{(M$_{\odot}$/yr$^{-1}$)} &\multicolumn{1}{c}{(seconds)} \\
\hline					
MCG-02-12-039 & Abell 0496 & 0.0329 & 0.654 & 0.031 & 4.8 & 134 & 1.5 & 7 $\times$ 1800 \\
PGC026269 & Abell 0780 & 0.0539 & 1.059 & 0.015 & 4.7 & 222 & 7.5 & 6 $\times$ 1800 \\
PGC044257 & Abell 1644 & 0.0474 & 0.935 & 0.009 & 5.1 & 12 & 3.2 & 6 $\times$ 1800 \\
UGC09799 & Abell 2052 & 0.0345 & 0.685 & 0.038 & 3.4 & 94 & 2.6 & 6 $\times$ 1800 \\
\hline
\end{tabular}
\caption{Galaxies observed with the Gemini South telescope. All four galaxies show extended H$\alpha$ emission (McDonald et al.\ 2010). The cluster X-ray temperature (T$_{X}$) and classical cooling rates rates (\.{M}) are from White et al.\ (1997). The spectrally determined cooling rates are from McDonald et al.\ (2010). The values for R$_{off}$ are from Edwards et al.\ (2007), with the exception of PGC044257 which is from Peres et al.\ (1998).}
\label{table:objects}
\end{footnotesize}
\end{table*}

\begin{table*}
\centering
\begin{footnotesize}
\begin{tabular}{l c c c c c} 
\hline Object & Cluster & Rest wavelength range & Foreground extinction (mag) & Average extinction (mag) & Radio flux \\
       &  & \AA{} & E(B-V)$_{galactic}$ & measured E(B-V)$_{total}$ & mJy \\
\hline					
MCG-02-12-039 & Abell 0496 & 4648 -- 7540 & 0.140 & 0.425 & 121\\
PGC026269 & Abell 0780 & 4743 -- 7693 & 0.042 & 0.210 & 40800\\
PGC044257 & Abell 1644 & 4713 -- 7646 & 0.071 & 0.195 & 98\\
UGC09799 & Abell 2051 & 4655 -- 7552 & 0.037 & 0.460 & 5500\\
\hline
\end{tabular}
\caption{Further properties of the CCGs observed on Gemini South. Radio fluxes are from the NVSS.}
\label{table:objects2}
\end{footnotesize}
\end{table*}

\section{Observations and data reduction}
\label{reduction}
The data were obtained with the GMOS IFU on the Gemini South telescope in semester 2011A (February to July 2011).
The GMOS-IFU in 1-slit mode was used and allowed us to map at least a 3kpc wide region in the centre of the target galaxies with a simultaneus coverage of the 4600-6800 \AA{} range in the target rest frame (using the B600 grating) with a single pointing. This resulted in a spectral resolution of 1.5 \AA{}. This spectral resolution (81 km s$^{-1}$) is poorer than that of Edwards et al.\ (2007, who had a much shorter wavelength range), and much higher than that of Hatch et al.\ (2006) (223 -- 273 km s$^{-1}$).

The IFU field-of-view is 5 $\times$ 3.5 arcsec, and this area is divided into 500 lenslets (and another 250 lenslets offset for sky measurements). We obtained six exposures per galaxy, with the exception of MCG-02-12-039 where we obtained 7 exposures, resulting in a total of 12500 galaxy spectra. The targets and exposure times are shown in Table \ref{table:objects}. Our integration time is three times that of Edwards et al.\ (2009), and more than five times that of Hatch et al.\ (2007) with a much bigger instrument. In addition to the targets, the necessary bias, flat-fields, twilight flat-fields, and arcs frames at two different central wavelengths were also observed, as well as a spectrophotometric standard star for flux calibration. Two central wavelength settings were used to avoid losing crutial spectral information in the two CCD gaps. 

For more detail on the GMOS IFU data reduction process see Gerssen et al.\ (2006). The basic data reduction was done using the GMOS package in IRAF. The IFU sky-to-detector mapping was stored in the data array prior to data reduction. Several bias frames were averaged and subtracted directly from raw data, to correct the zero point for each pixel. Additional care was taken to avoid including raw bias frames in the mean bias frame that drifted measurably with time. Frames were mosaiced, and the overscan regions were trimmed. Flat-field and twilight flat-field frames were used to correct for differences in sensitivity both between detector pixels and across the IFU field. The majority of the cosmic rays were rejected in the individual frames before sky subtraction using the Gemini cosmic ray rejection routine. The remainder of the cosmic rays were eliminated using the LACosmic routine (van Dokkum 2001) with an IRAF script that retained the multi-extension fits format for further reductions. The sets of 2D spectra were calibrated in wavelength using the arc lamp spectra for the two different central wavelength settings. The IFU elements were, thereafter, extracted from the raw data format to a format more convenient for further processing. Sky emission lines and continuum were removed by averaging the sky spectrum over a number of spatial pixels (from the offset sky fibers on the edge of the science field) to reduce the noise level, before subtracting it from all the spatial pixels. Thus the process adds little extra noise to the result since the observations were obtained in dark time (the variability of the sky region was minimal), and a number of 250 spaxels were averaged in the sky-subtraction process. Thus, the error contribution of the sky-subtraction process is $\frac{1}{\sqrt{250}}\times $ the error on one sky spaxel. A spectrophotometric standard star (LTT4816) was used to correct the measured counts for the combined transmission of the instrument, telescope and atmosphere as a function of wavelength. We reduced the standard star observation with the same instrument configuration as the corresponding scientific data. A 1D spectrum was extracted by adding the central spatial pixels from the standard star observation, and it was used to convert the measured counts from the galaxy spectra into fluxes with erg cm$^{-2}$ s$^{-1}$ \AA{}$^{-1}$ units. 

The reduced 2D arrays were transformed back to a physical coordinate grid ($x, y, \lambda$ datacube) before scientific analysis, while also correcting for atmospheric dispersion. The latter (also called differential refraction) causes the position of a target within the IFU field to vary with wavelength. This correction was necessary as data were taken at different airmasses throughout the observing nights. The spatial offset as a function of wavelength was determined using an atmospheric refraction model (given the airmass, position angle and other parameters) from the SLALIB positional astronomy library. Each hexagonal spaxel (each spatial element) was 0.2 arcsec, and this was subsampled onto a rectangular grid of 0.1 arcsec per spaxel when the separate exposures were combined. The exposures were combined (median averaged) using a centroid algorithm to calculate the shifting in $x$ and $y$, and also shifting in $\lambda$ for the exposures at two different wavelength settings. No additional cosmic rays were visible after the LACosmic task was run on the individual cubes, so cosmic rays were not removed during this step. The cubes were also converted into RSS (row-stacked spectra) for further data reductions in IDL. Each spaxel was averaged with its eight neighbouring spaxels (improving the S/N by a factor of three), which is effectively smoothing over 0.3 arcsecs - this is still slightly undersampled compared to the average seeing ($\sim$1 arcsecs), but only larger regions will be analysed further.

\begin{figure}
   \centering
 \mbox{\subfigure[6548.2 \AA{}]{\includegraphics[height=3.3cm,width=2.3cm]{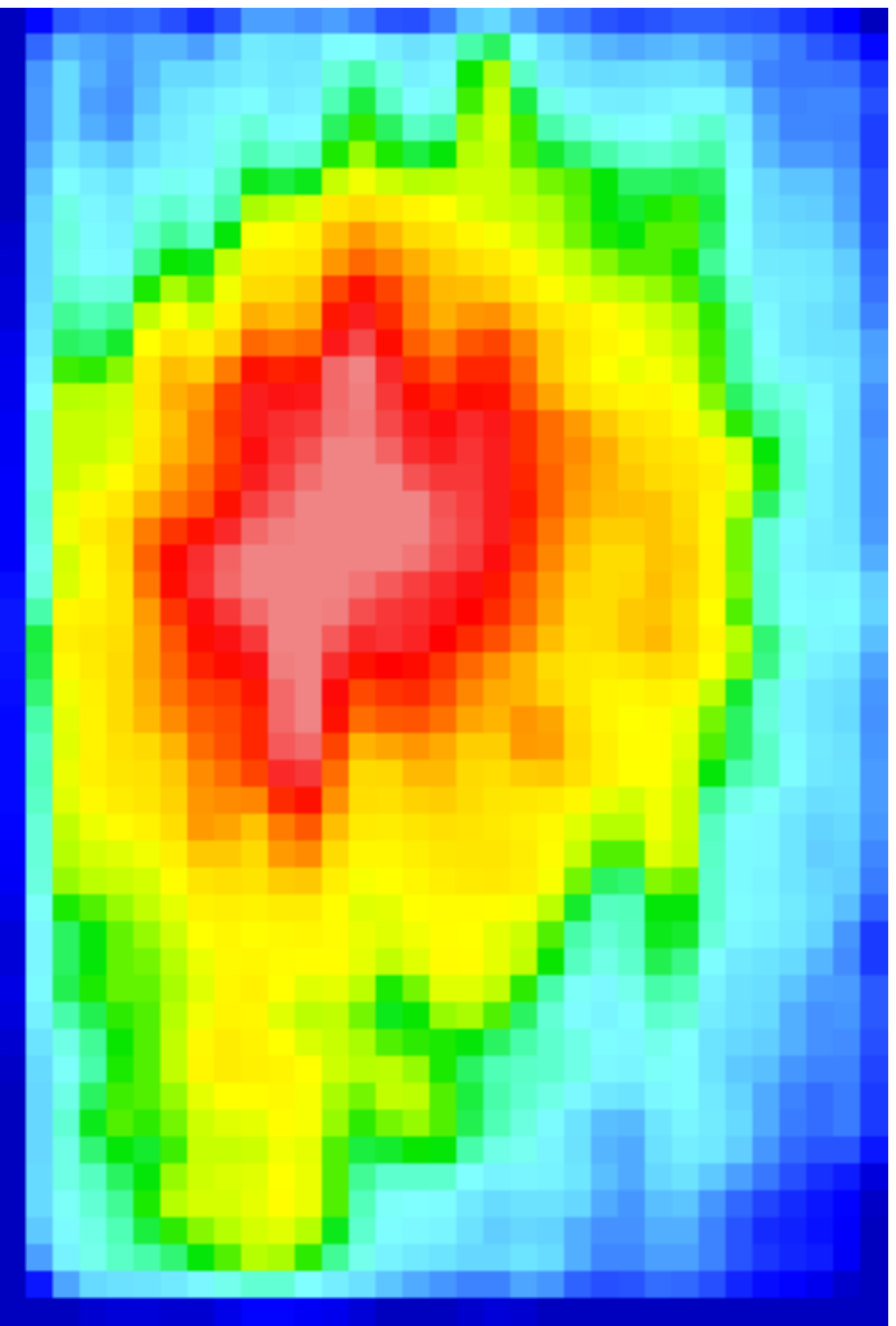}}\quad
         \subfigure[6550.9 \AA{}]{\includegraphics[height=3.3cm,width=2.3cm]{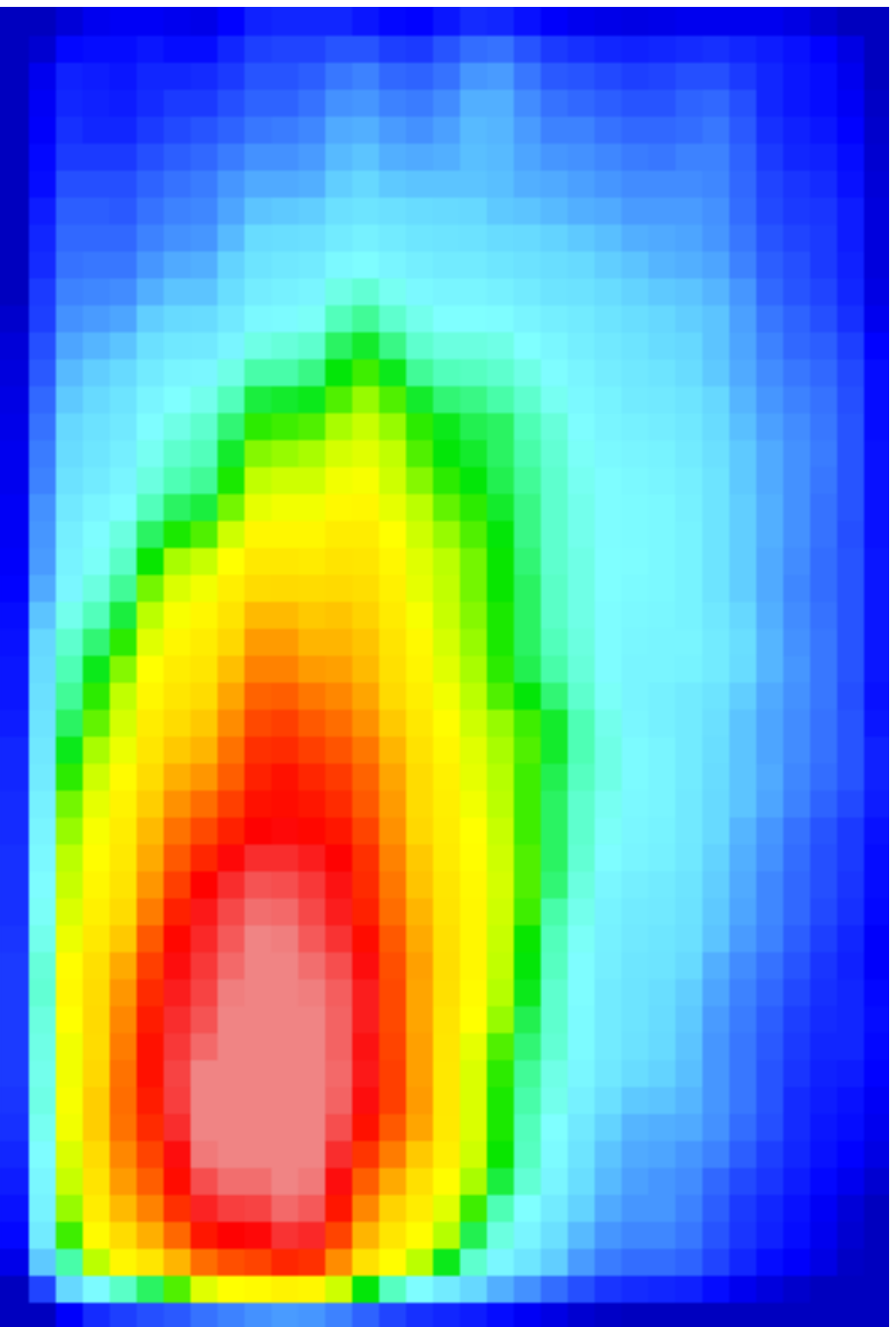}}\quad
         \subfigure[6553.5 \AA{}]{\includegraphics[height=3.3cm,width=2.3cm]{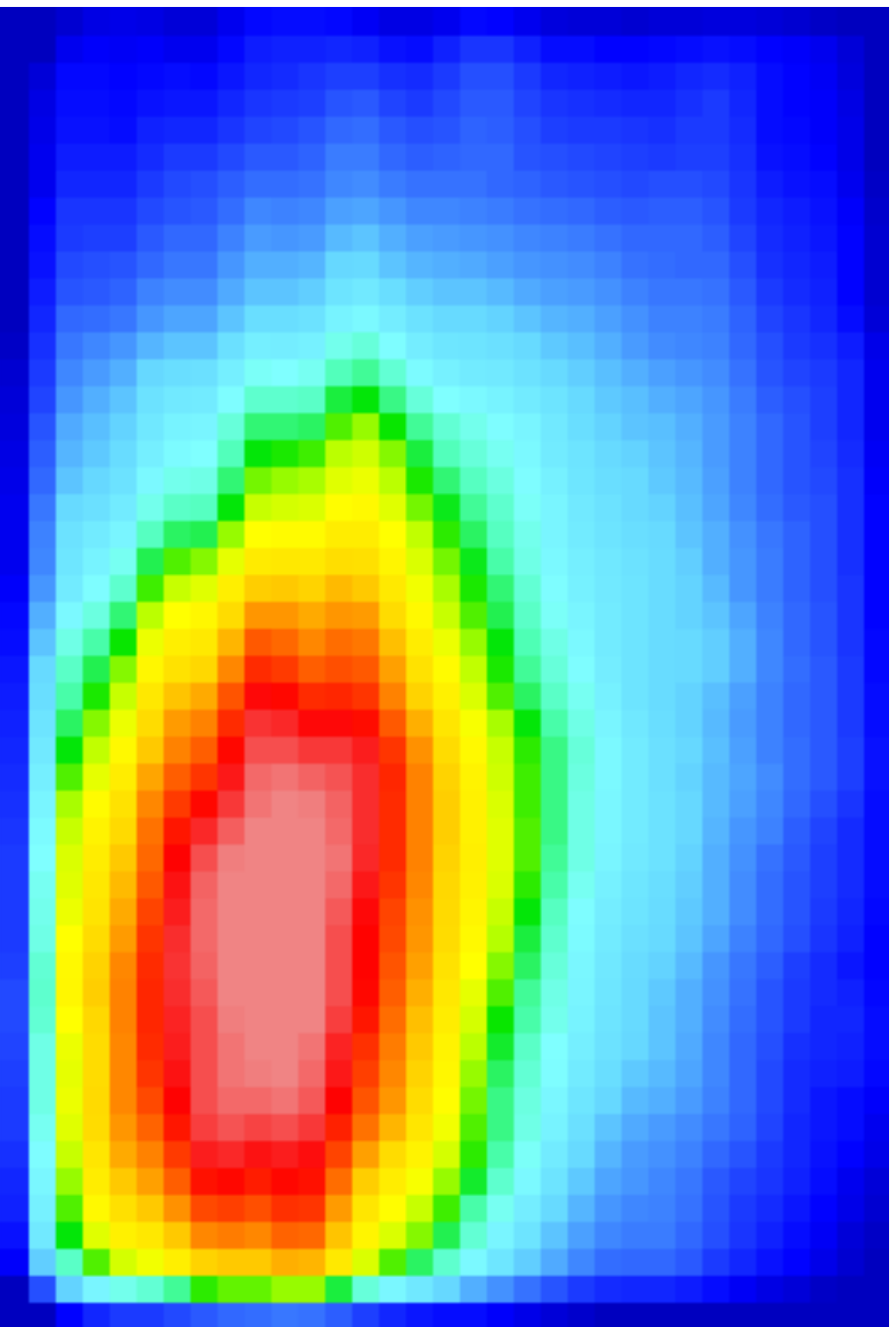}}}
\mbox{\subfigure[6556.2 \AA{}]{\includegraphics[height=3.3cm,width=2.3cm]{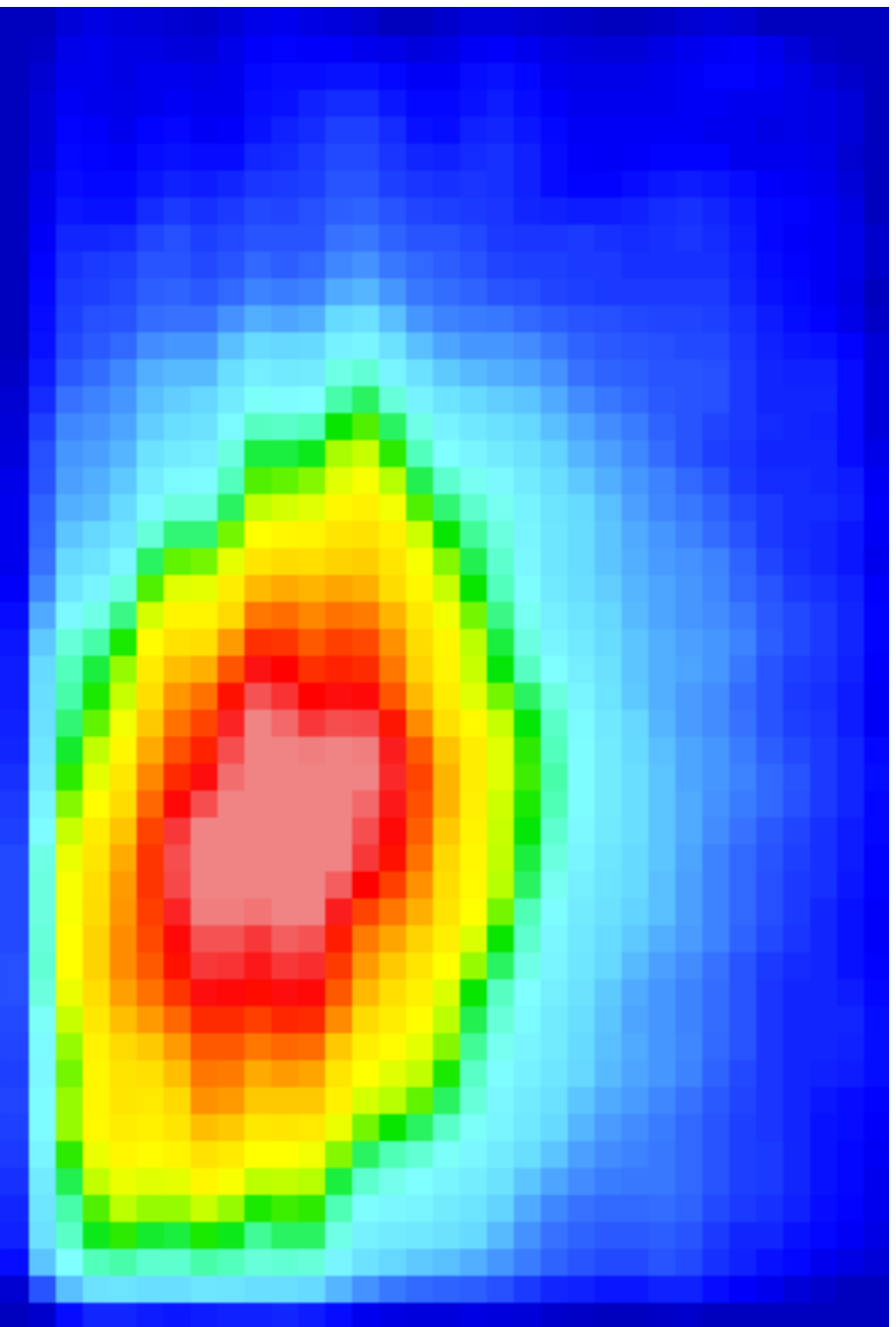}}\quad
\subfigure[6558.8 \AA{}]{\includegraphics[height=3.3cm,width=2.3cm]{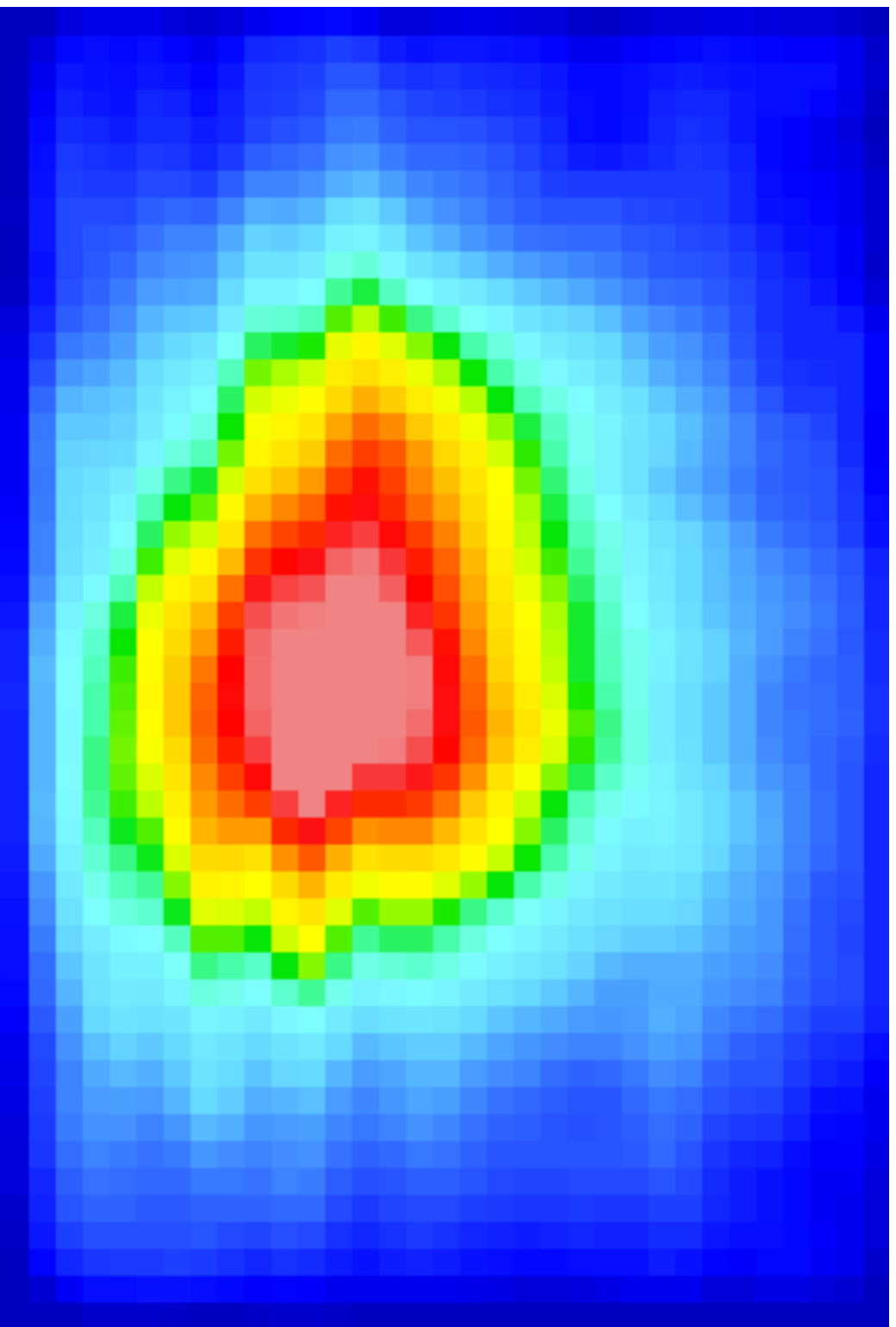}}\quad
         \subfigure[6561.4 \AA{}]{\includegraphics[height=3.3cm,width=2.3cm]{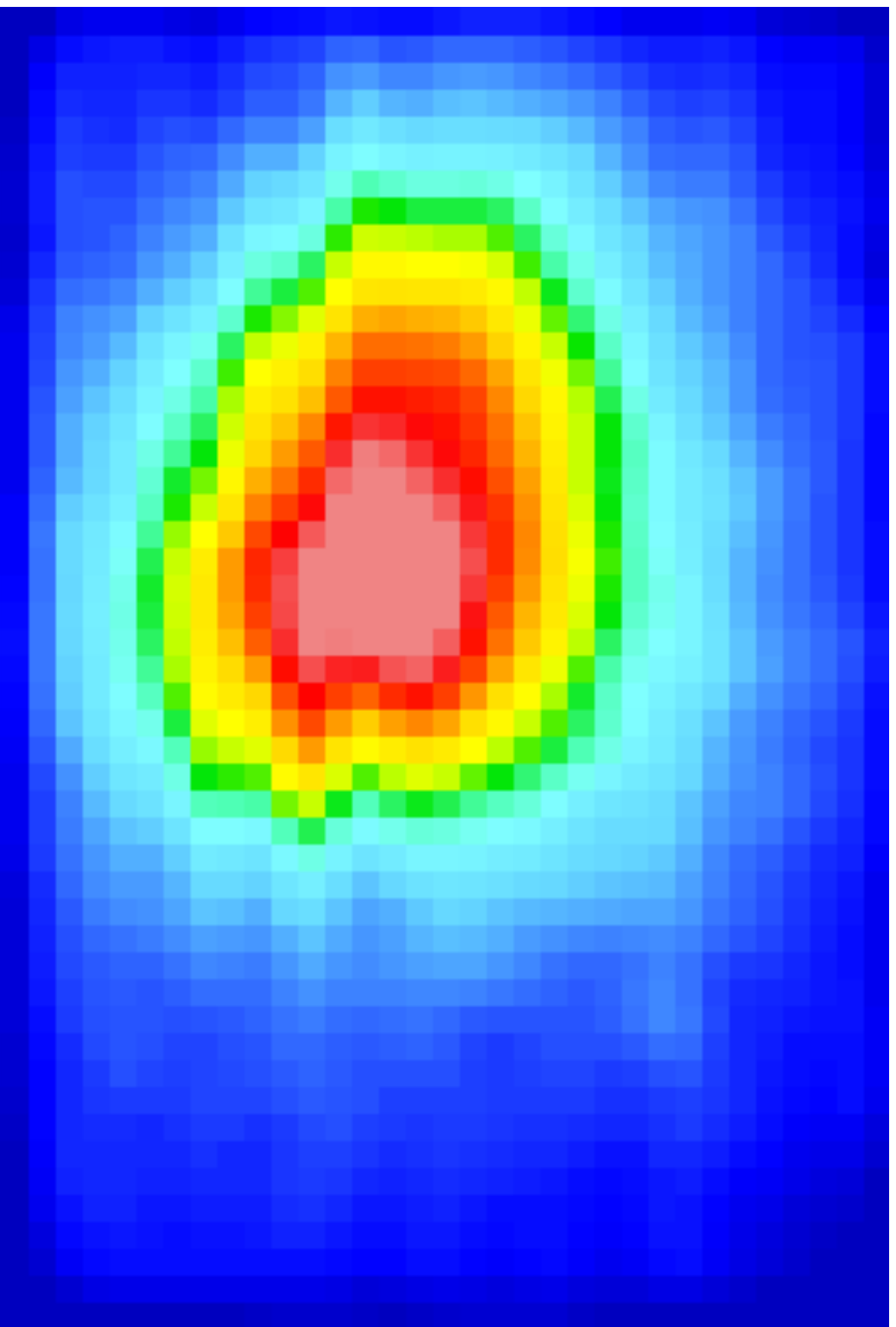}}}
\mbox{\subfigure[6564.1 \AA{}]{\includegraphics[height=3.3cm,width=2.3cm]{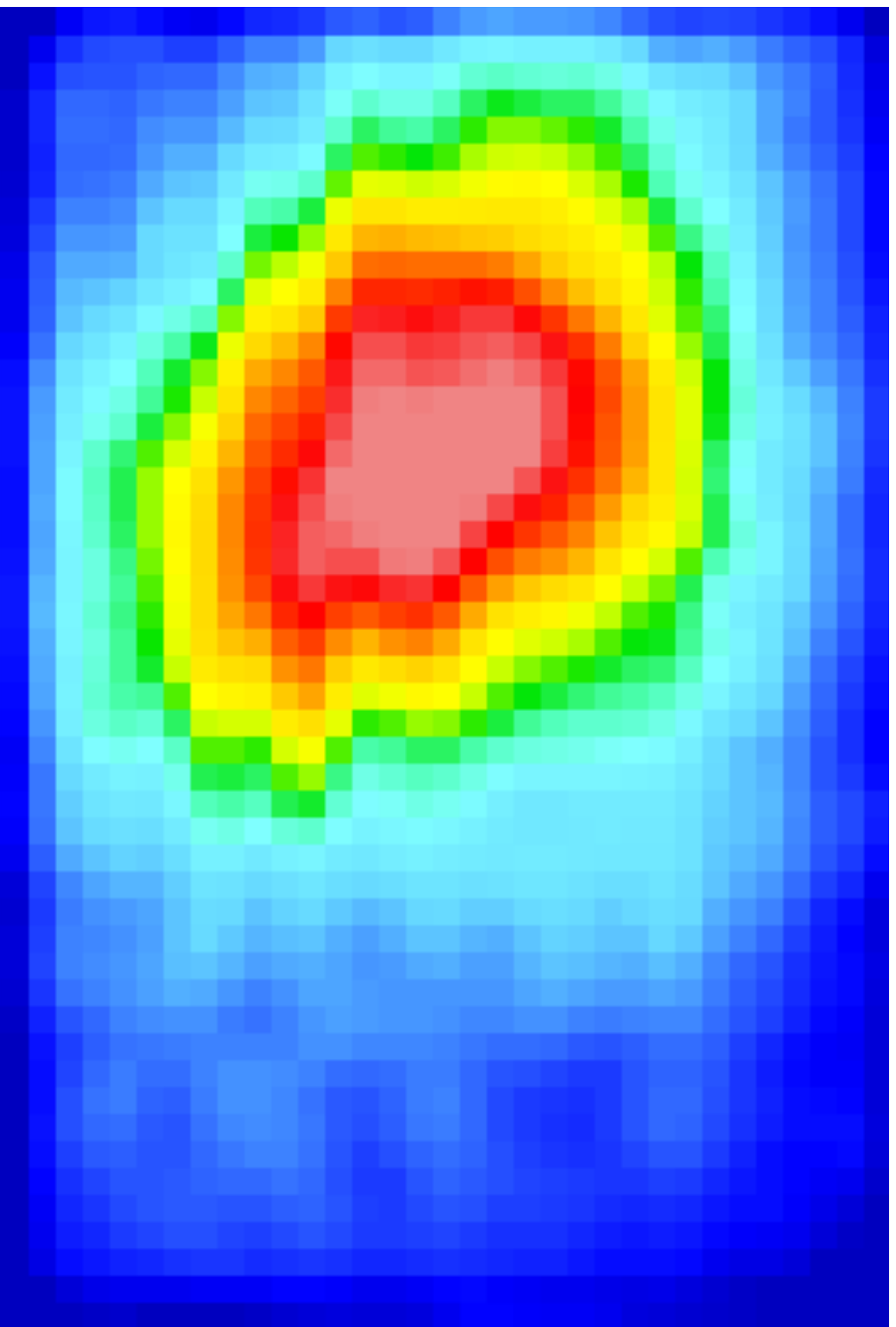}}\quad
\subfigure[6566.7 \AA{}]{\includegraphics[height=3.3cm,width=2.3cm]{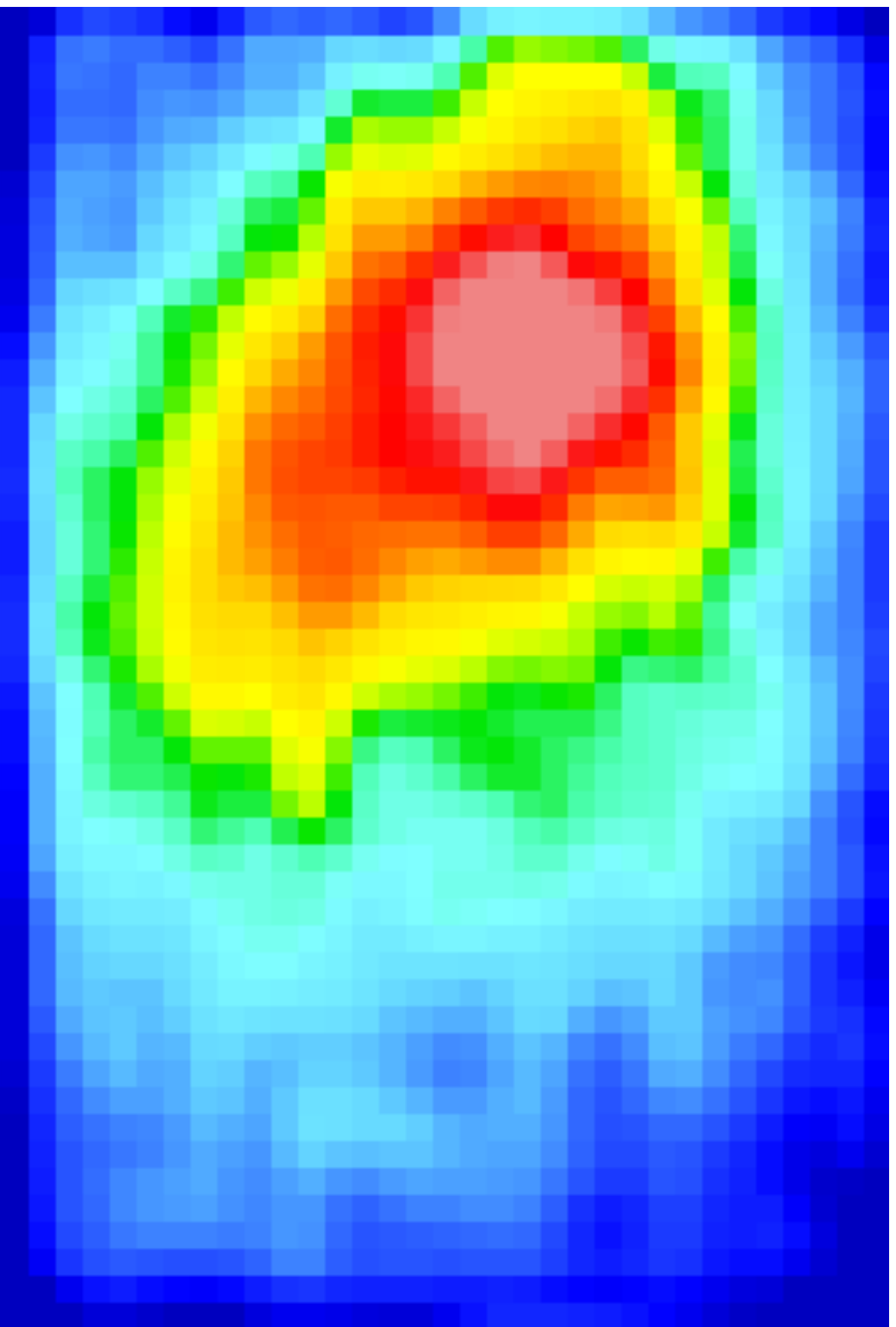}}\quad
\subfigure[6569.3 \AA{}]{\includegraphics[height=3.3cm,width=2.3cm]{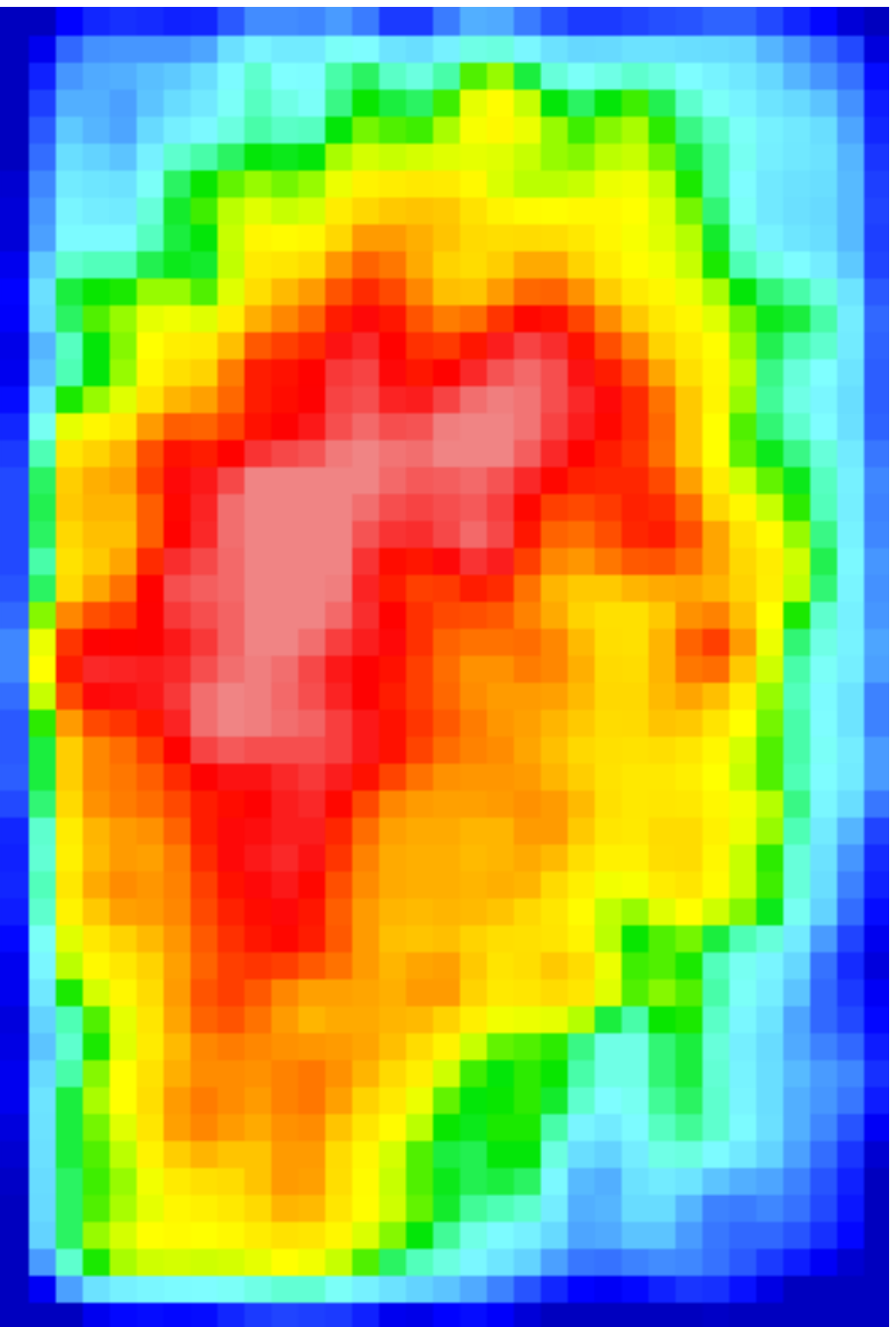}}}
\caption{Slices of width 1 \AA{} (smoothed spatially with Gaussian FWHM 1.5 spaxels and using a SAURON colourmap) of PGC026269 showing the wavelength and spatial scale over which the transition from [NII]$\lambda$6548 to H$\alpha$ emission occurs, and how the morphology of H$\alpha$ changes. Although the continuum emission is usually smooth, the morphologies of the line emission are not uniform.}
\label{fig:Thumbnails2}
\end{figure}

\section{Line measurements and internal extinction}
\label{extinction}
To accurately measure the emission-line fluxes of the CCG spectra, we use a combination of the \textsc{ppxf} (Cappellari $\&$ Emsellem 2004) and \textsc{gandalf} (Gas AND Absorption Line Fitting algorithm; Sarzi et al.\ 2006) routines\footnote{We make use of the corresponding \textsc{ppxf} and \textsc{gandalf IDL} (Interactive Data Language) codes which can be retrieved at http:/www.leidenuniv.nl/sauron/.}. Gandalf version 1.5 was used as it enables a reddening correction to be performed, and it encorporates errors. This code treats the emission lines as additional Gaussian templates, and solves linearly at each step for their amplitudes and the optimal combination of stellar templates, which are convolved by the best stellar line-of-sight velocity distribution. The stellar continuum and emission lines are fitted simultaneously. All 985 stars of the MILES stellar library (S\'{a}nchez-Bl\'{a}zquez et al.\ 2006) were used as stellar templates to automatically include $\alpha$-enhancement in the derived optimal template. The emission lines were masked while the optimal template was derived. The H$\alpha$ and [NII]$\lambda$6583 lines were fitted first, and the kinematics of all the other lines were tied to these lines, following the procedure described in Sarzi et al.\ (2006). However, in cases where the emission of the other lines were strong enough to measure velocity and velocity dispersion (as was mainly the case except for the extreme edges of the data cube), this was calculated independently as there is no a priori reason to expect the kinematics measured from all the lines to be the same (as they can originate in different regions). After the kinematics are fixed, a Gaussian template is constructed for each emission line at each iteration, and the best linear combination of both stellar and emission-line templates (with positive weights) is determined. This is done without assuming line ratios, except in the case of doublets where their relative strength is fixed by the ratio of the corresponding transition probabilities. We have adapted the gandalf code to apply it to the GMOS IFU cubes for a longer wavelength range. All 1617 spaxels were collapsed together to obtain a 1D spectrum per cube, thereafter all 985 stars for the MILES library were used to create a global optimal template for each galaxy. This global optimal template (and the stars it consisted of -- to account for varying weights over the spatial region) was then applied to all 1617 spectra per cube. The spectral types of the stars that the global optimal templates consisted of (from the MILES library) are shown in Table \ref{absorption}.

\begin{table*}
\centering
\begin{footnotesize}
\begin{tabular}{l c c c} 
\hline Object & Spectral types in the global optimal template\\
\hline					
MCG-02-12-039 & A0 Ia, G0, G1 Ib, G5, G8 III, K0, K0 III, K IIvw \\
PGC026269 & A0 Ia, K0 Ibpvar, M4 III \\
PGC044257 & B3 III, A5, F3 V, F6, F7 V, G0 Vw, G1 Ib, G5, G5 V, G8 III/IVw, K1 III \\
UGC09799 & B8 Ib, A5, F3 III, G0 Vw, G1 Ib, K0 III\\
\hline
\end{tabular}
\caption{Properties of the underlying stellar populations.}
\label{absorption}
\end{footnotesize}
\end{table*}

Some ellipticals contain dust in the centre that can be patchy, uniform or filamentary (Laine et al.\ 2003). The long wavelength range of the spectra allows us to constrain the amount of reddening using the observed decrement of the Balmer lines, which can be set to have an intrinsic decrement consistent with the recombination theory by treating the lines as a multiplet. The physical constraints on the emission from the higher-order Balmer lines also ensures the strength of the corresponding absorption features is correctly estimated. 

\begin{figure*}
\mbox{\subfigure{\includegraphics[scale=0.7, trim=1mm 1mm 50mm 1mm, clip]{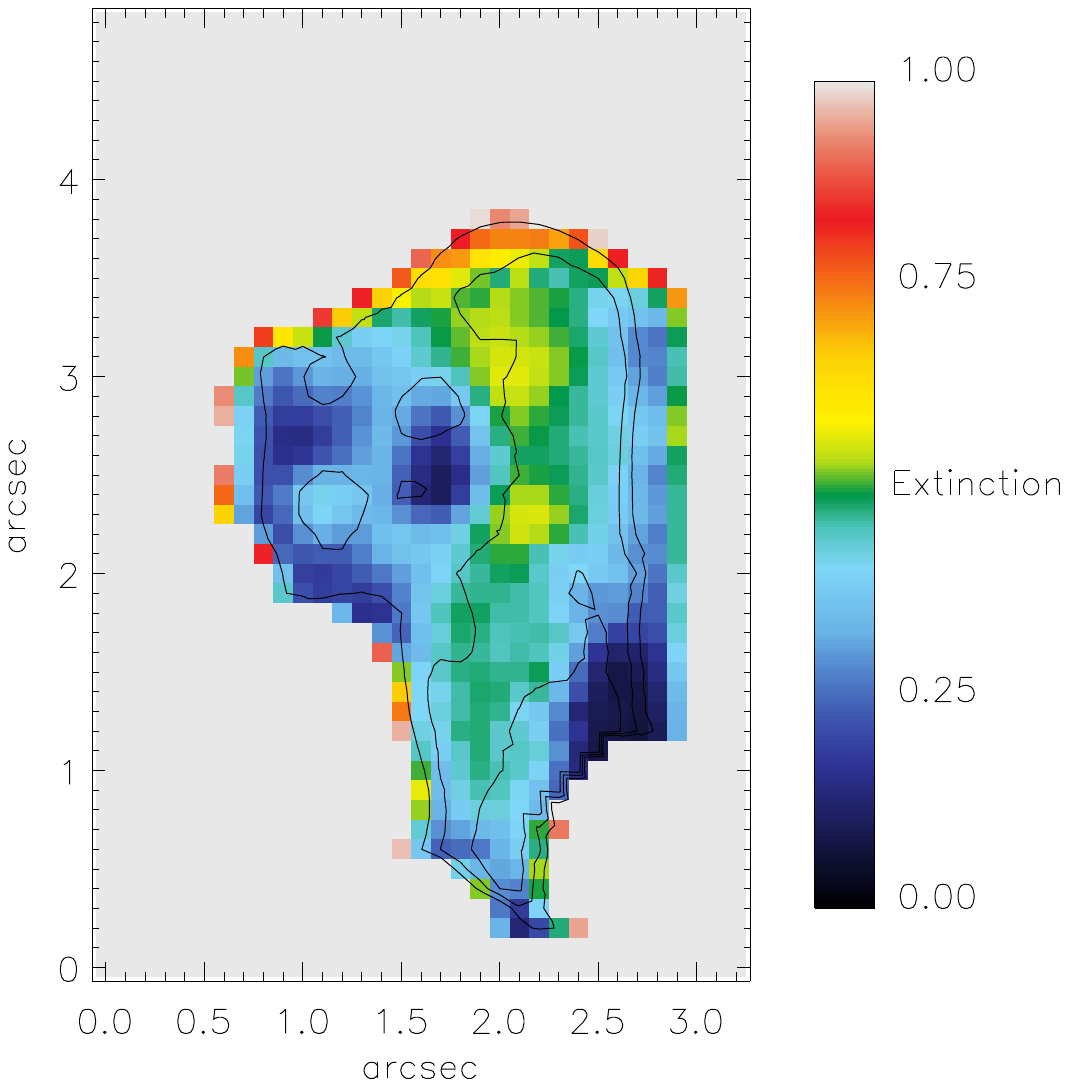}}\quad
\subfigure{\includegraphics[scale=0.7, trim=1mm 1mm 50mm 1mm, clip]{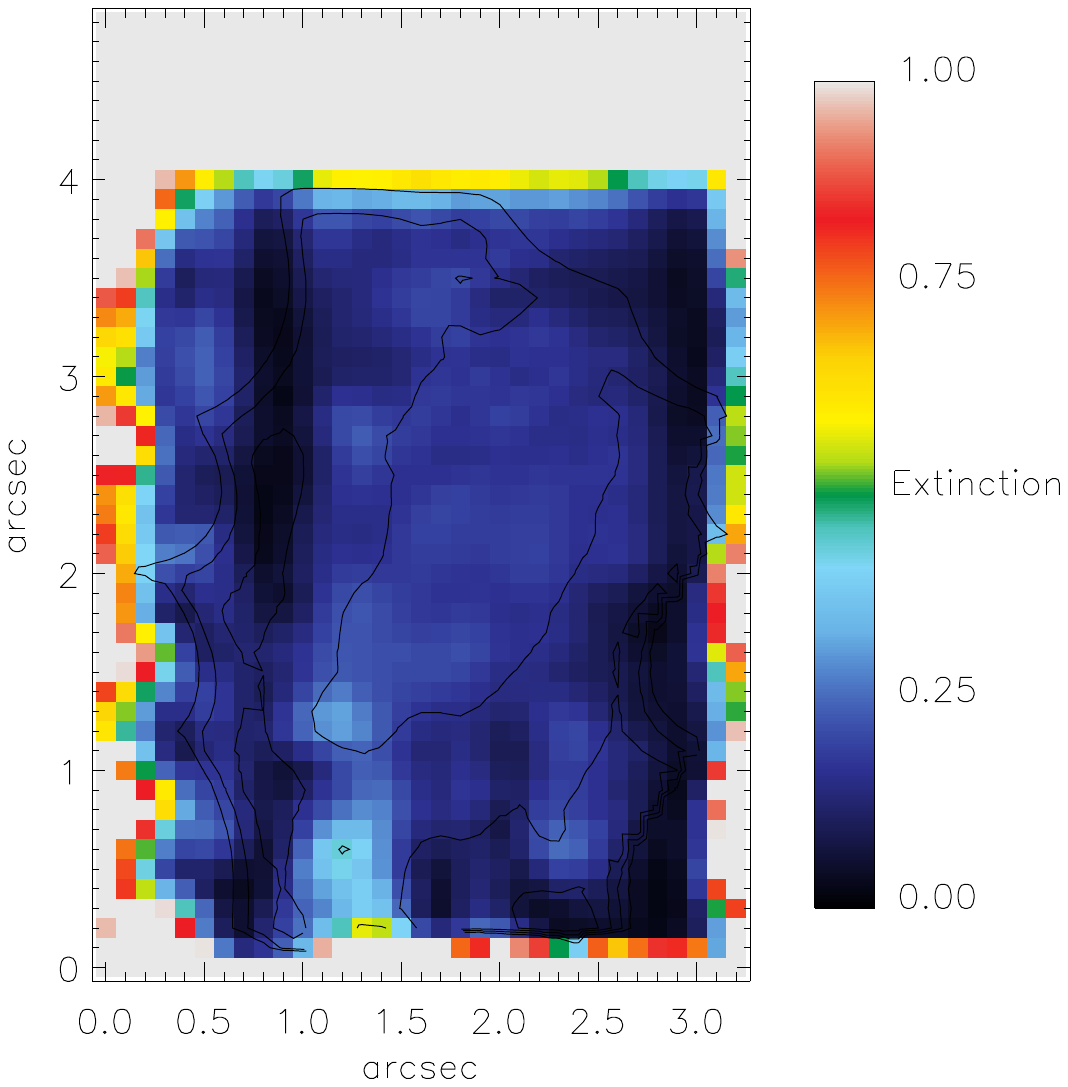}}}
\mbox{\subfigure[MCG-02-12-039]{\includegraphics[scale=0.3]{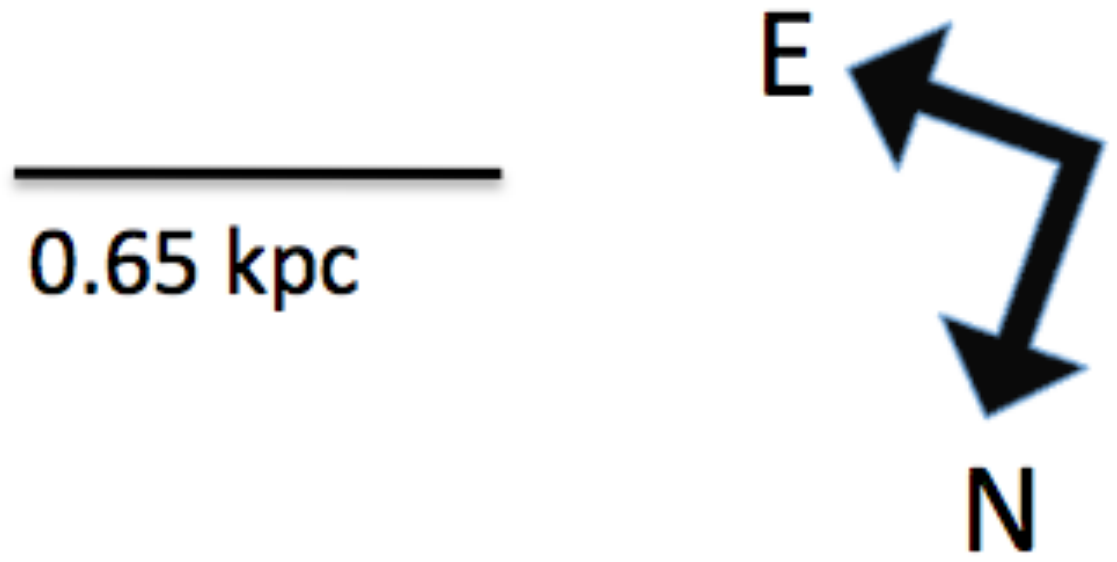}}\quad
\subfigure[PGC026269]{\includegraphics[scale=0.3]{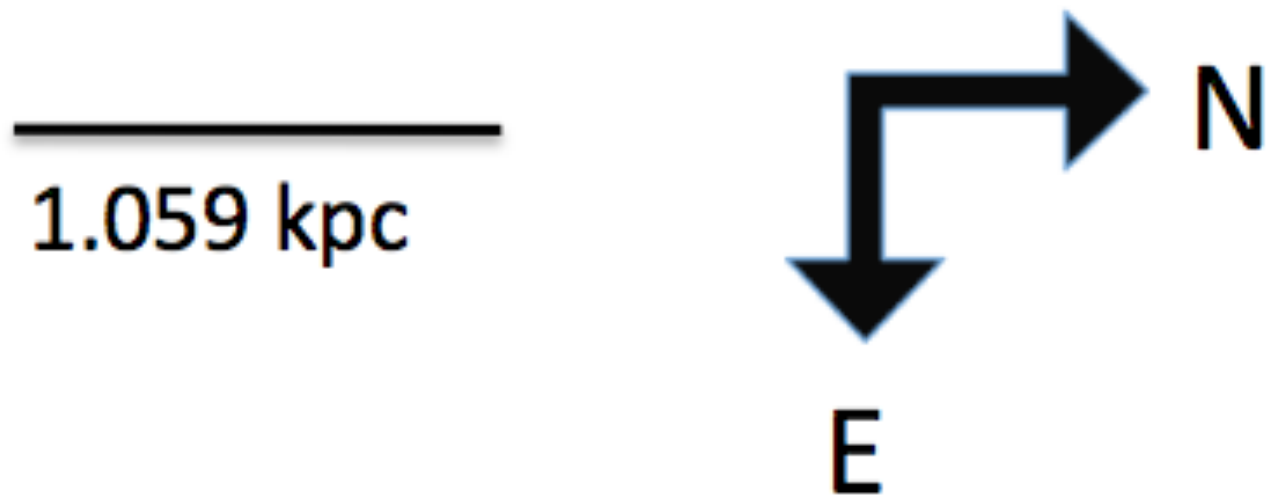}}}
\mbox{\subfigure{\includegraphics[scale=0.7, trim=1mm 1mm 50mm 1mm, clip]{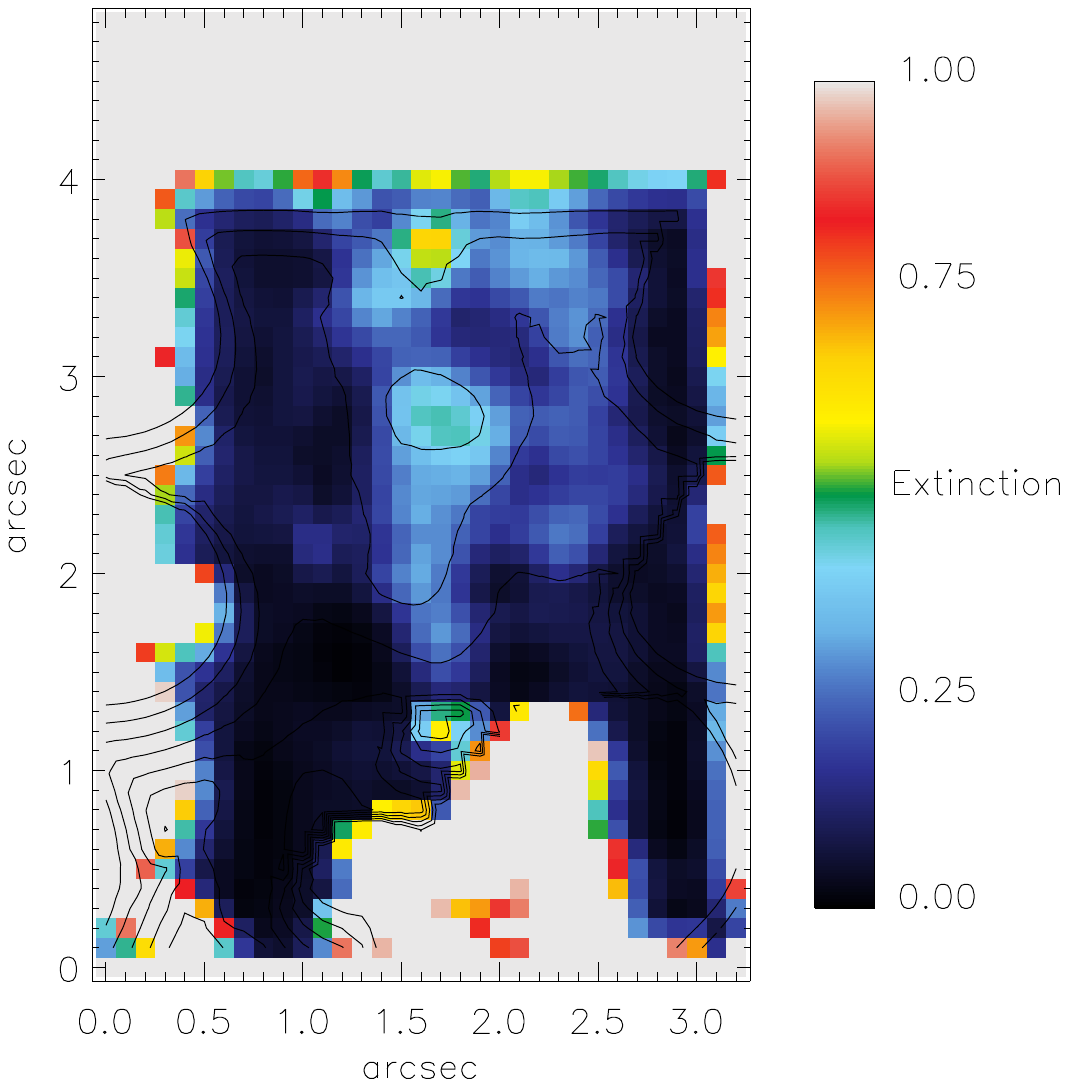}}\quad
\subfigure{\includegraphics[scale=0.7, trim=1mm 1mm 50mm 1mm, clip]{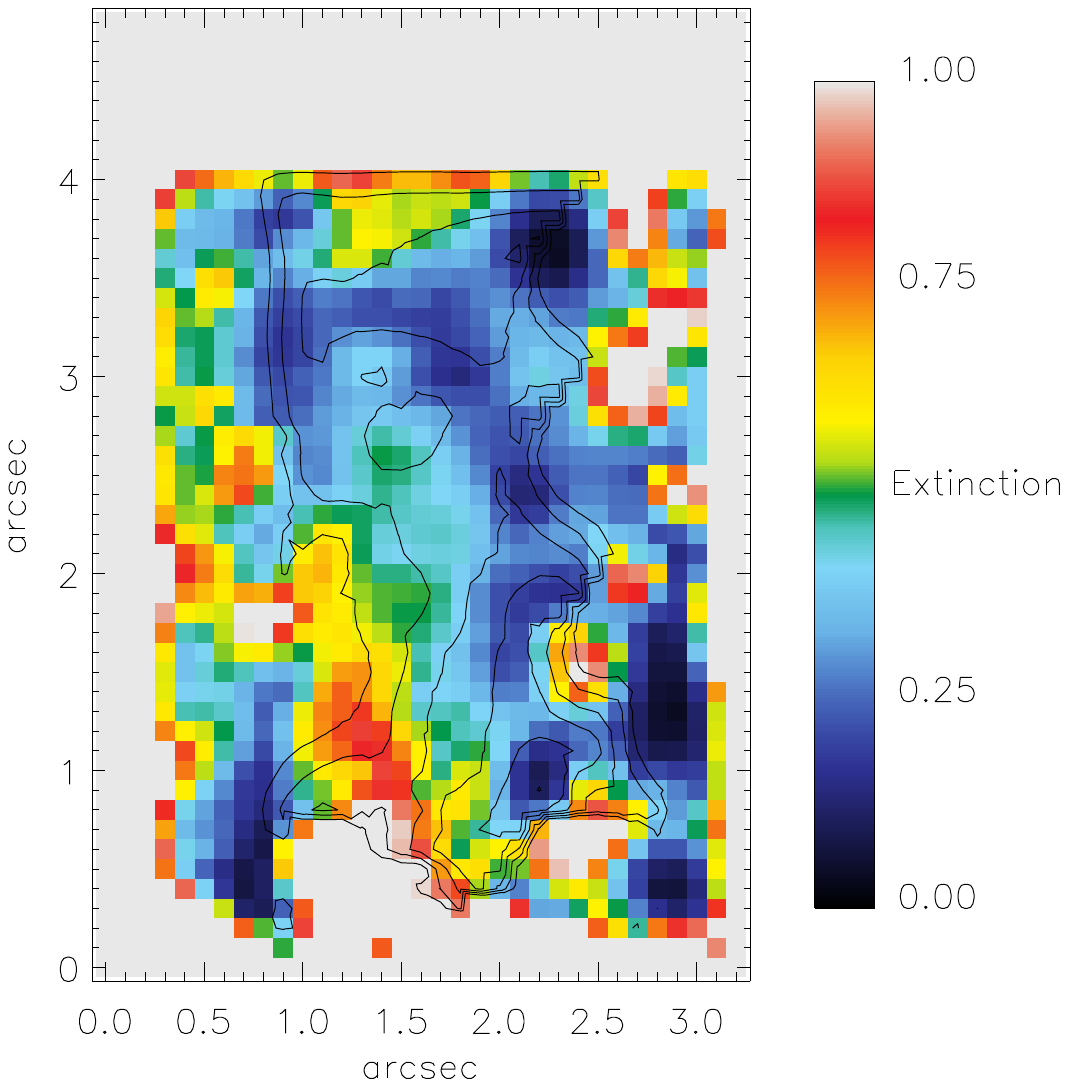}}}
\mbox{\subfigure[PGC044257]{\includegraphics[scale=0.3]{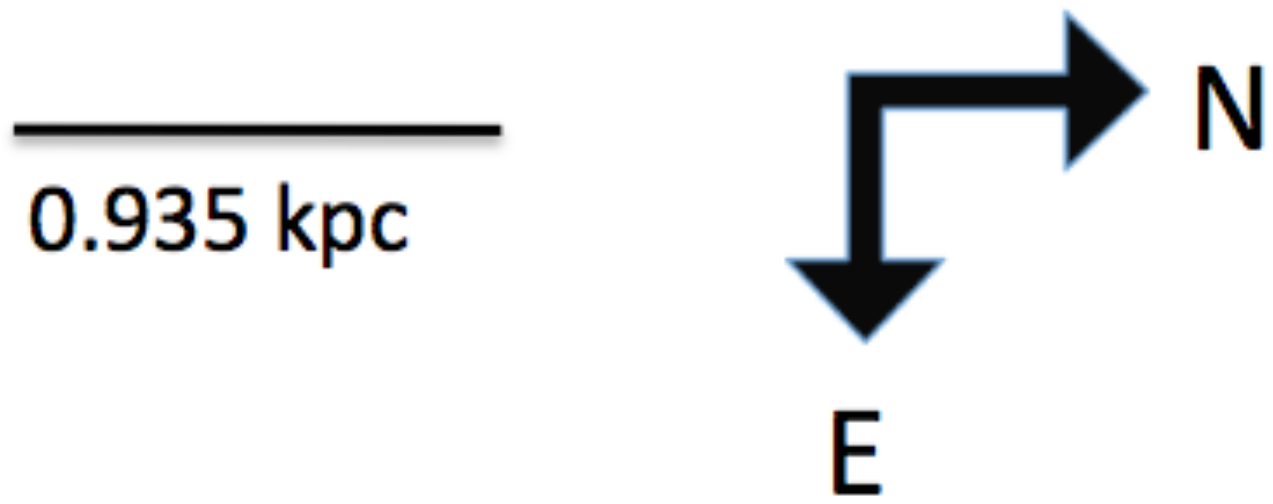}}\quad
\subfigure[UGC09799]{\includegraphics[scale=0.3]{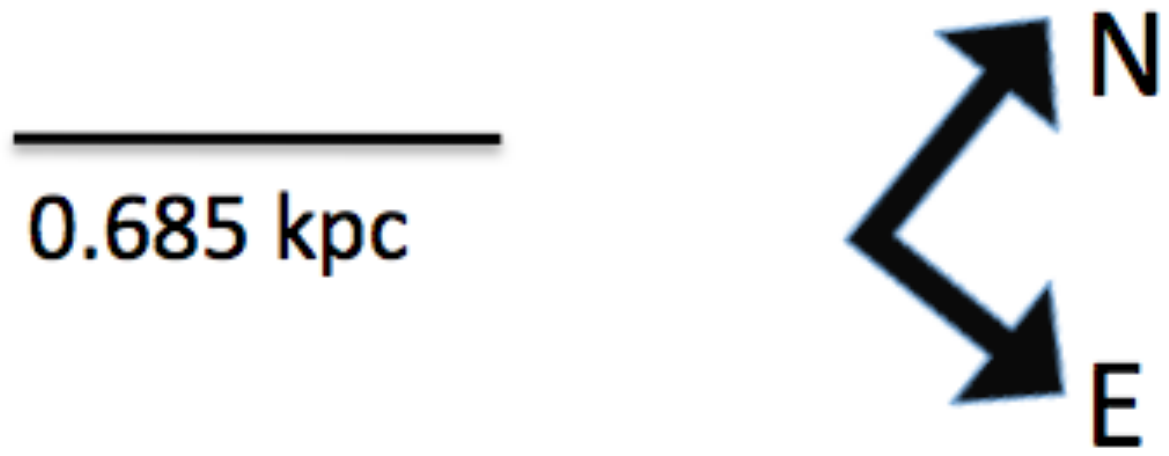}}}
\caption{Total extinction of the four CCGs. The extinction was smoothed over 0.3 arcseconds, and is only plotted where the velocity dispersion of the H$\alpha$ line is less than 500 km s$^{-1}$ (to avoid spaxels where the H$\alpha$ line could not be separated from the [NII] lines), and it is only plotted where the A/N ratio of the H$\alpha$ line is higher than 3. Overplotted is the H$\alpha$ flux contours one magnitude apart. The grey regions in these, and all other, IFU maps are where the A/N is too low to plot.}
   \label{fig:MCG_extinct}
\end{figure*}

We used the dust models by Calzetti et al.\ (2000) to calculate the flux attenuation values at the desired wavelength for any given $E(B-V)$ value (optional see below). The Balmer decrement assumes a case B recombination for a density of 100 cm$^{-3}$ and a temperature of 10$^{4}$ K, resulting in the predicted H$\alpha$/H$\beta$ ratio of 2.86 (Osterbrock 1989). The code can adopt either a single dust component, affecting both the stellar continuum and the emission-line fluxes, or in addition include a second dust component that affects only the emission-line templates. We did not specify Galactic extinction (from the NED database, Schlegel, Finkbeiner $\&$ Davis 1998), but we measured a total diffuse component which gave the total extinction (hence including the foreground galactic extinction). This total extinction measured from the Balmer decrement are shown in Figure \ref{fig:MCG_extinct} and averages are noted in Table \ref{table:objects2} (the galactic extinction noted in Table \ref{table:objects2} was not subtracted). The extinction was smoothed over 0.3 arcseconds, and is only plotted where the velocity dispersion of the H$\alpha$ line is less than 500 km s$^{-1}$ (to avoid spaxels where the H$\alpha$ line could not be separated from the [NII] lines), and it is only plotted where the amplitude-to-noise (A/N) ratio of the H$\alpha$ line is higher than 3 (as defined in Sarzi et al.\ 2006)\footnote{The A/N is related to the S/N: $EW=\frac{F}{S}=\frac{A/N \times \sqrt{2 \pi \sigma}}{S/N}$ where $EW$ is the equivalent width of the line, and $\sigma$ is the line width.}. The A/N maps of the H$\alpha$ measurements are shown in Section \ref{figures_NVSS} which gives a direct indication of the errors on the extinction.

The parameter $E(B-V)$, i.e.\ the colour excess between 4350 \AA{} and 5550 \AA{}, for the galactic extinction for each of the four galaxies was taken from the NED database (Schlegel et a.\ 1998), and ranged between 0.037 and 0.140 mag (Table \ref{table:objects2}). The parameter R$_{\rm V}$, i.e.\ the ratio of the absolute extinction at 5550 \AA{} (A$_{\rm V}$) to the colour excess $E(B-V)$, was taken as 3.1 for the interstellar medium (Cardelli, Clayton $\&$ Mathis 1989). The total extinction is given by:

\[E(B-V)_{total}=\frac{2.177}{-0.37 R_{V}} \times (\log {\frac{H\alpha_{0}}{H\beta_{0}}}-\log {\frac{H\alpha_{Obs}}{H\beta_{Obs}}})\]

The theoretical H$\alpha$/H$\beta$ flux ratio of 2.86 may not be the ideal value to use for Seyfert-type galaxies, but the actual value is debated. It is often assumed that the H$\alpha$ emission in these systems is enhanced due to collisional processes, and several authors use a value $R_{V}$ of 3.1 (Gaskell $\&$ Ferland 1984; Osterbrock $\&$ Ferland 2006), although other values have also been determined (Binette et al.\ 1990 calculate a value of 3.4). The total extinction maps are presented in Figure \ref{fig:MCG_extinct}, and shows mostly very low extinction, but some morphological features can be seen in MCG-02-12-039 and UGC09799. Particularly high values of $E(B-V)_{internal}$ can be seen in UGC09799. The galactic extinction of PGC044257 is $E(B-V)_{gal}$ = 0.071 mag, and from long-slit spectra, Crawford et al.\ (1999) derived the total extinction as 0.46 to 0.63 mag. This agrees with the extinction we derived for the very centre of the galaxy in Figure \ref{fig:MCG_extinct}, but on average our spatially resolved extinction is slightly lower. The galactic extinction $E(B-V)_{gal}$ of UGC09799 is 0.037 mag, and Crawford et al.\ (1999) derived an integrated internal extinction of $E(B-V)_{internal}$ of 0.22 mag for the centre of this galaxy. This corresponds very well to what we derived and plotted in Figure \ref{fig:MCG_extinct}, although we find that some regions show much higher internal extinction. The values of extinction determined here may be slightly overestimated due to the choice of intrinsic H$\alpha$/H$\beta$ flux used.

Figure \ref{fig:Thumbnails2} show slices with width 1 \AA{} (smoothed spatially with Gaussian FWHM 1.5 spaxels and using the Sauron colourmap) of PGC026269 showing the wavelength and spatial scale over which the morphology of H$\alpha$ changes. The morphology of the nuclear region changes rapidly across the emission lines. This is illustrated in the figure showing a series of monochromatic slices. Although the continuum emission is usually smooth, the morphologies of the line emission are not uniform.

We are able to measure H$\alpha$ line fluxes with 3 A/N accuracy at around 5 $\times$ 10$^{-18}$ erg cm$^{-2}$ s$^{-1}$. After correcting for extinction, we are able to measure the following lines within our wavelength range: [ArIV]$\lambda$4740 \AA{}; H$\beta \lambda$4861 \AA{}; [OIII]$\lambda \lambda$4958,5007 \AA{}; [NI]$\lambda \lambda$5198,5200 \AA{}; HeI$\lambda$5876 \AA{}; [OI]$\lambda \lambda$6300,6364 \AA{}; [NII]$\lambda \lambda$6548,6583 \AA{}; H$\alpha$ at 6563 \AA{}; [SII]$\lambda \lambda$6716,6731 \AA{} (see Figures \ref{fig:MCGkinematics2} to \ref{fig:UGCkinematics3} and \ref{Hbeta} for the A/N ratios of the individual emission lines). H$\alpha$ and [NII]$\lambda \lambda$6548,6583 \AA{} has already been measured for MCG-02-12-039 and UGC09799, and H$\beta$ and [OIII]$\lambda \lambda$4958,5007 \AA{} has also been measured for UGC09799. All the other line measurements are new.

\renewcommand*{\thesubfigure}{}.pdf

\renewcommand*{\thesubfigure}{(\alph{subfigure})}

\section{Individual galaxies}
\label{figures_NVSS}

\begin{figure*}
   \centering
   \includegraphics[scale=0.8]{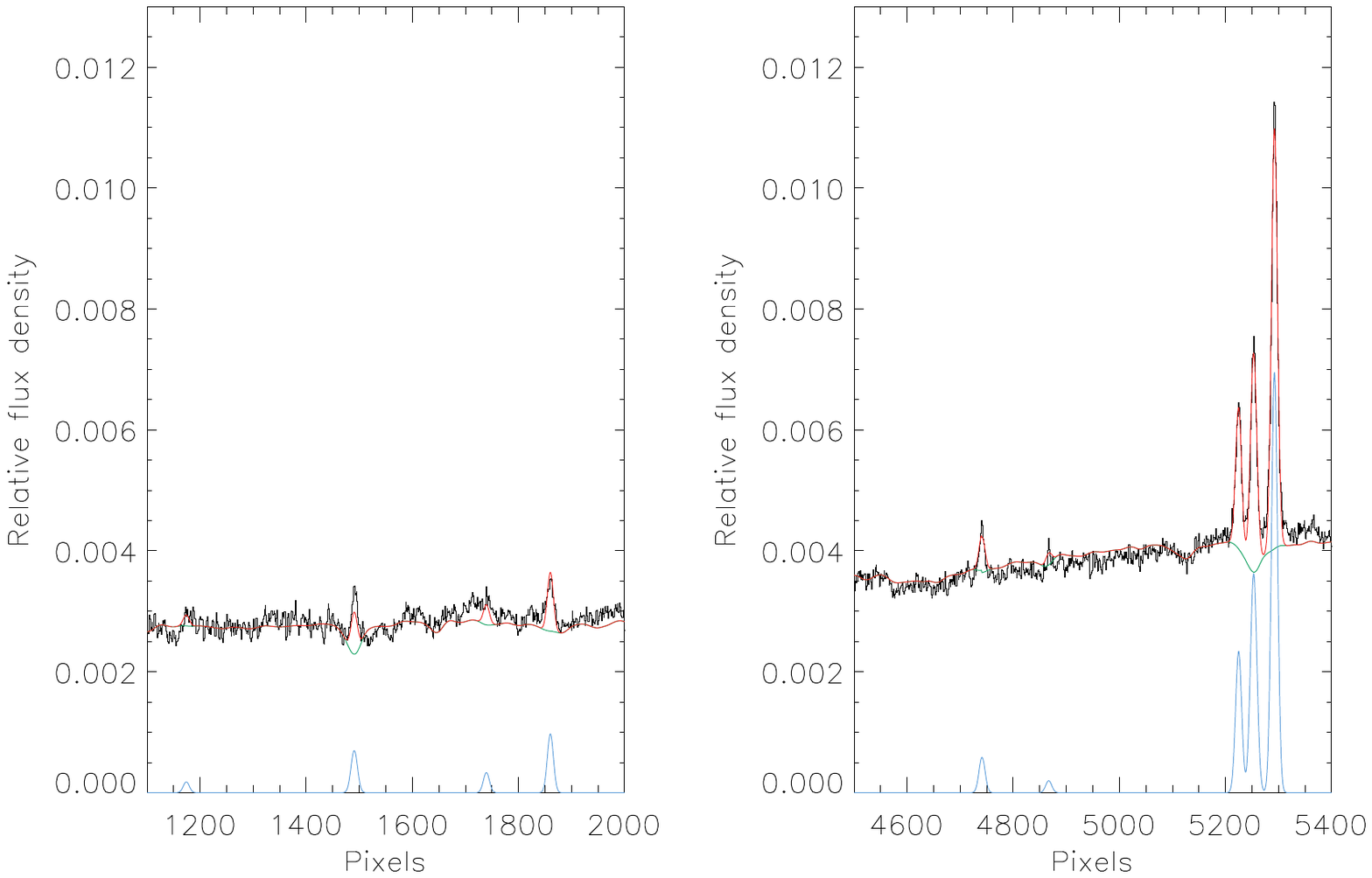}
   \caption{A random spectrum in the central region of MCG-02-12-039. The red line indicates the best-fitting stellar template and Gaussians at the emission lines, and the green line indicates the best-fitting stellar templates with the emission lines subtracted. The blue line indicates the relative flux for the measured emission lines. The lines are (from left to right) in the left plot: [ArIV], H$\beta$, and the [OIII] doublet, and in the right plot: [OI] doublet, [NII] doublet, and H$\alpha$. The H$\alpha$ A/N of this pixel was 15.}
   \label{fig:MCG_gandalf}
\end{figure*}

The continuum emission of the galaxies are shown in Figure \ref{fig:Thumbnails}. Figure \ref{fig:MCG_gandalf}, \ref{fig:PGC026_gandalf}, \ref{fig:PGC044_gandalf} and \ref{fig:UGC_gandalf} shows a random spectrum in the central region of MCG-02-12-039, PGC026269, PGC044257 and UGC09799 respectively. The best-fitting stellar template, Gaussians at the emission lines, best-fitting stellar templates with the emission lines subtracted and the relative flux for the measured emission lines are also shown. The absorption-extracted, dereddened maps of the H$\alpha$, [NII]$\lambda$6583/H$\alpha$, and [OIII]$\lambda$5007/H$\beta$ emission are shown in Figure \ref{Haflux}.

For comparison purposes, the kinematics were extracted from the IFU image along the same slit position as the long-slit data in Loubser et al.\ (2008). This comparison for MCG-02-12-039, PGC026269 and PGC044257 is showed in Section \ref{stellar_kinematics_comparison}, and the data points compare satisfactory.

The absorption-extracted, dereddened maps of the H$\alpha$ emission are shown in Figure \ref{Haflux} in units of 10$^{-15}$ erg cm$^{-2}$ s$^{-1}$. We also show the useful ratios [NII]$\lambda$6583/H$\alpha$, and [OIII]$\lambda$5007/H$\beta$. Figures \ref{fig:MCGkinematics2} to \ref{fig:UGCkinematics3} shows the A/N ratios of the H$\alpha$, [NII]$\lambda$6583 \AA{}; [SII]$\lambda \lambda$6716,6731 \AA{}; [OIII]$\lambda$5007 \AA{}; [OI]$\lambda$6300 \AA{} lines. As mentioned in the Introduction, MCG-02-12-039 and UGC09799 already has previous IFU observations which is compared (in terms of S/N and lines measurable) in the beginning of Section 3 and the last paragraph of Section 4. The stellar kinematics in Figures \ref{fig:MCGkinematics}, \ref{fig:PGC026kinematics}, \ref{fig:PGC044kinematics} and \ref{fig:UGCkinematics} were measured using all the absorption lines within our wavelength range.

For a sample of $\sim$50 elliptical galaxies, Sarzi et al.\ (2006) find ionised gas velocities (estimated using the [OIII]$\lambda5007$ line) between --250 and 250 km s$^{-1}$, and gas velocity dispersion as high as 250 km s$^{-1}$. Using the same line, we find gas velocities from $\pm 100$ to $\pm 350$ km s$^{-1}$, and line widths from 200 to 420 km s$^{-1}$. Using H$\alpha$, we find gas velocities from $\pm 125$ to $\pm 350$ km s$^{-1}$, and line widths from 200 to 400 km s$^{-1}$. 


\subsection{MCG-02-12-039}

The central galaxy (MCG-02-12-039) of Abell 0496 is a fairly weak line emitter (Fabian et al. 1981; Hu, Cowie $\&$ Wang 1985) and is host to the compact radio source MSH 04-112 (Markovic, Owen $\&$ Eilek 2004). Abell 0496 is a relaxed cluster with a cool core with a central metal abundance enhancement (Tamura et al. 2001). At a redshift of 0.0329, the galaxy has a linear scale of 0.656 kpc arcsec$^{-1}$ (from Hatch et al.\ 2007). This cluster has an interesting H$\alpha$ morphology (McDonald et al.\ 2010). There are at least 5 distinct filaments, with various shapes and directions. The two longest filaments run parallel to each other for $\sim$ 12 kpc (18.3 arcsec, whereas our observations only cover 3.5 by 5 arcsecs in the centre).

Figure \ref{Haflux} shows that the peak H$\alpha$ emission is also where the dust (extinction) is the highest in Figure \ref{fig:MCG_extinct}. This is similar to the finding of Hatch et al.\ (2007), where they found that the peak H$\alpha$ + [NII] emission corresponds to the area where the three dust lanes meet in the galaxy centre. We confirm the observation by Hatch et al.\ (2007), that the line emission follows the general path of the dust (extinction) features, but is not as filamentary.

\begin{figure*}
   \centering
 \mbox{\subfigure[Stellar velocity]{\includegraphics[scale=0.6, trim=1mm 1mm 50mm 1mm, clip]{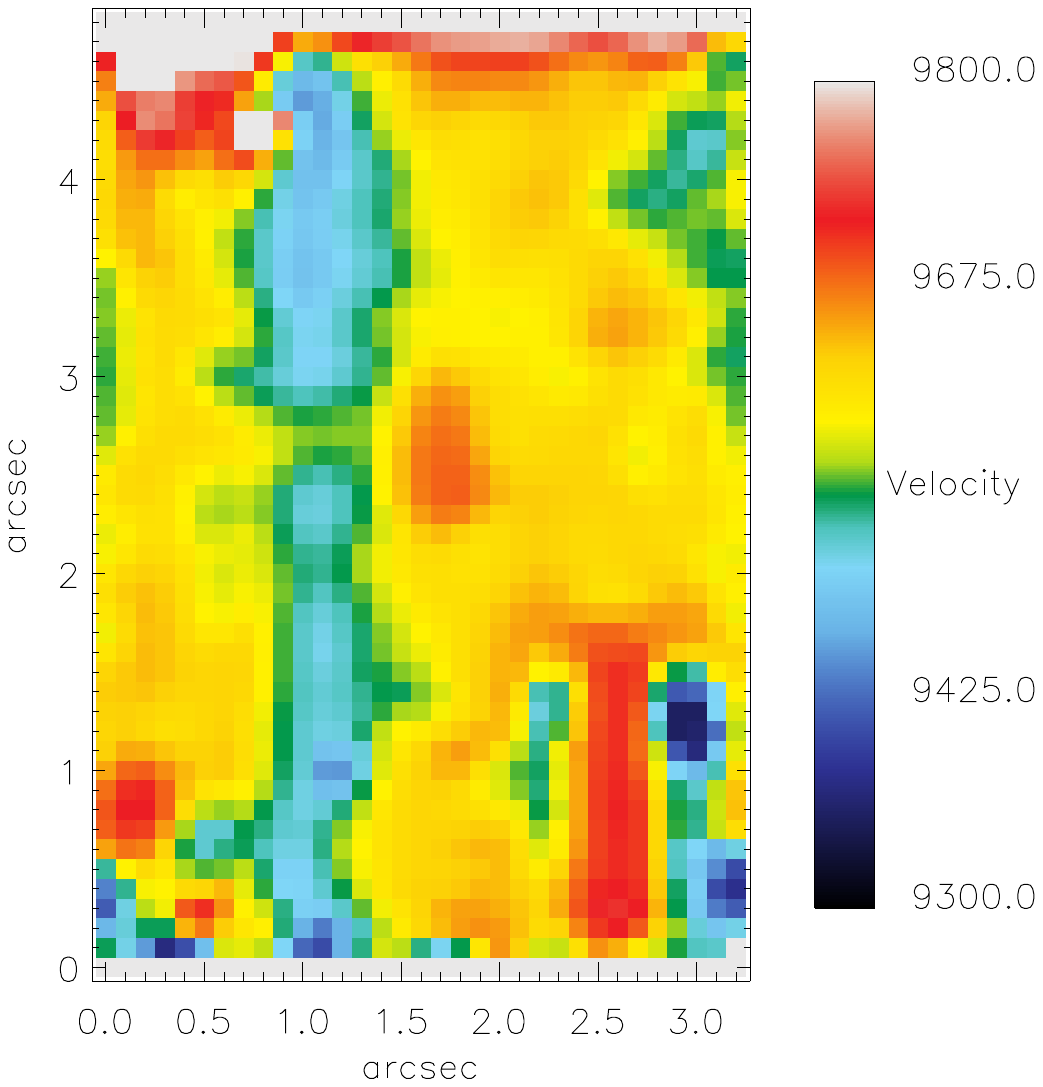}}\quad
\subfigure[Stellar velocity dispersion]{\includegraphics[scale=0.6, trim=1mm 1mm 50mm 1mm, clip]{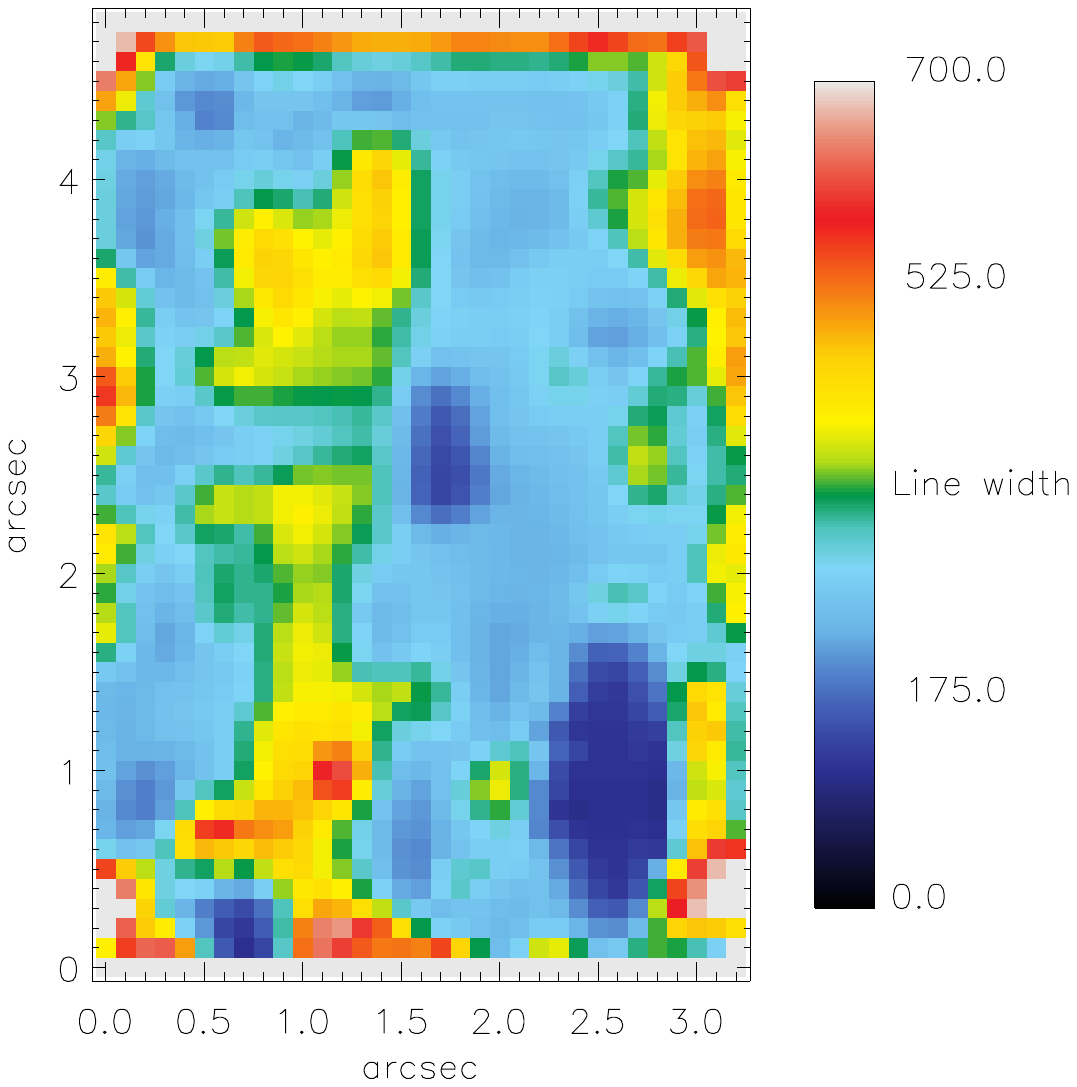}}}
\mbox{\subfigure{\includegraphics[scale=0.25]{MCG_arrow.pdf}}}
\caption{MCG-02-12-039: Velocity and velocity dispersion of the absorption lines in km s$^{-1}$.}
\label{fig:MCGkinematics} 
\end{figure*}

\begin{figure*}
 \mbox{\subfigure[H$\alpha$ velocity]{\includegraphics[scale=0.5, trim=1mm 1mm 50mm 1mm, clip]{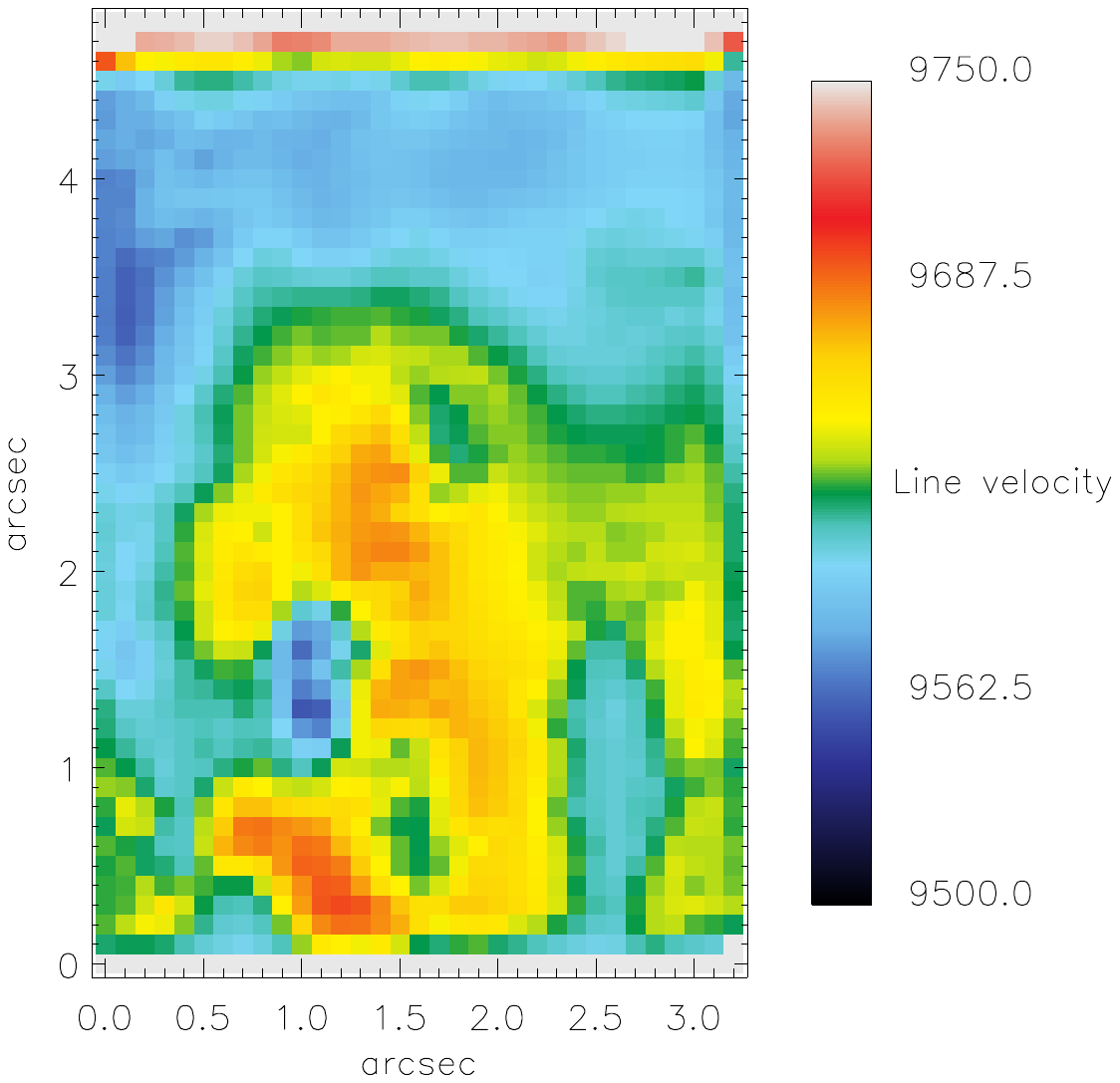}}\quad
\subfigure[{[NII]}$\lambda$6583 velocity]{\includegraphics[scale=0.5, trim=1mm 1mm 50mm 1mm, clip]{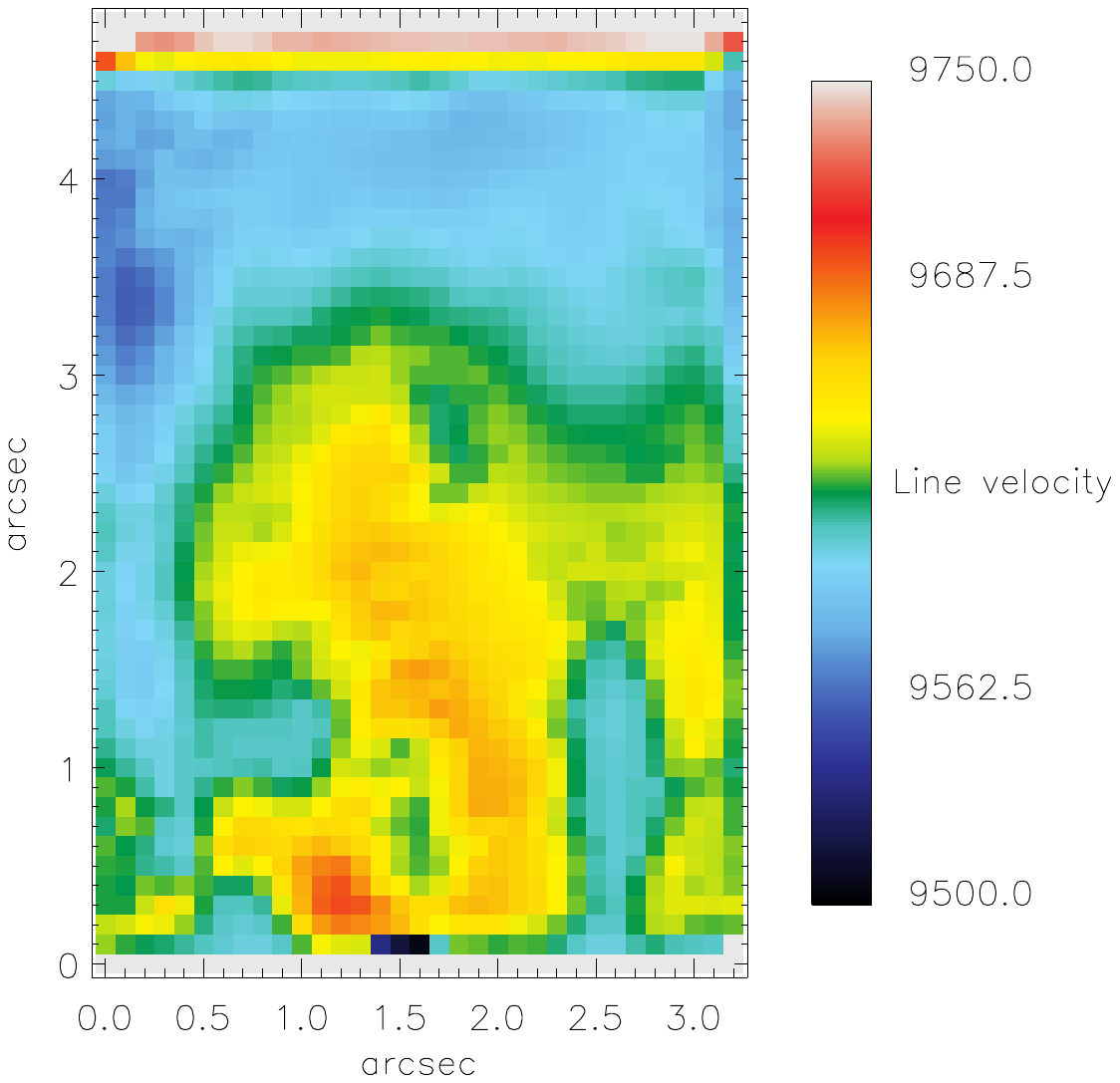}}\quad
\subfigure[{[SII]}$\lambda\lambda$6731,6717 velocity]{\includegraphics[scale=0.5, trim=1mm 1mm 50mm 1mm, clip]{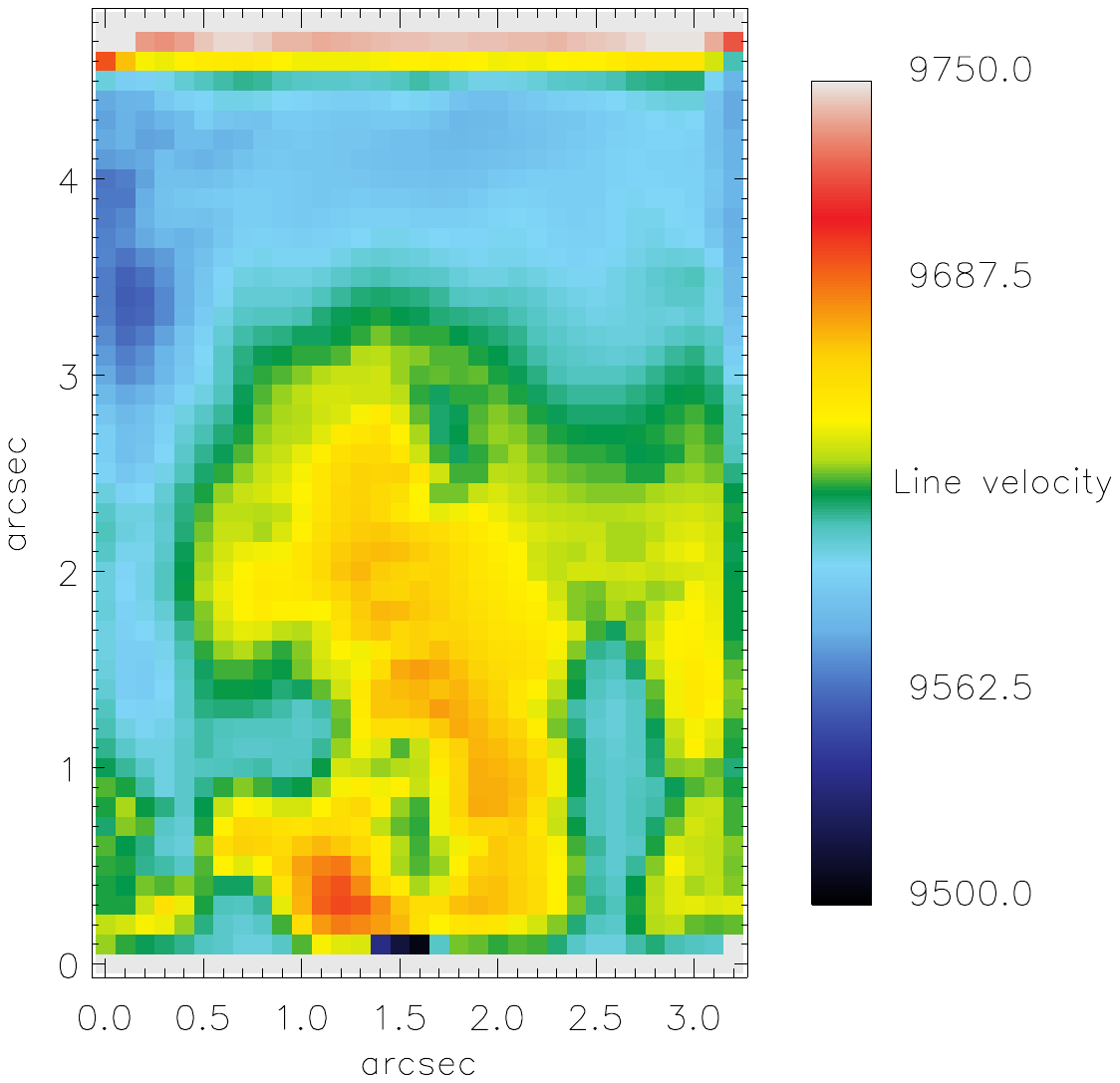}}}
   \mbox{\subfigure[H$\alpha$ line width]{\includegraphics[scale=0.5, trim=1mm 1mm 50mm 1mm, clip]{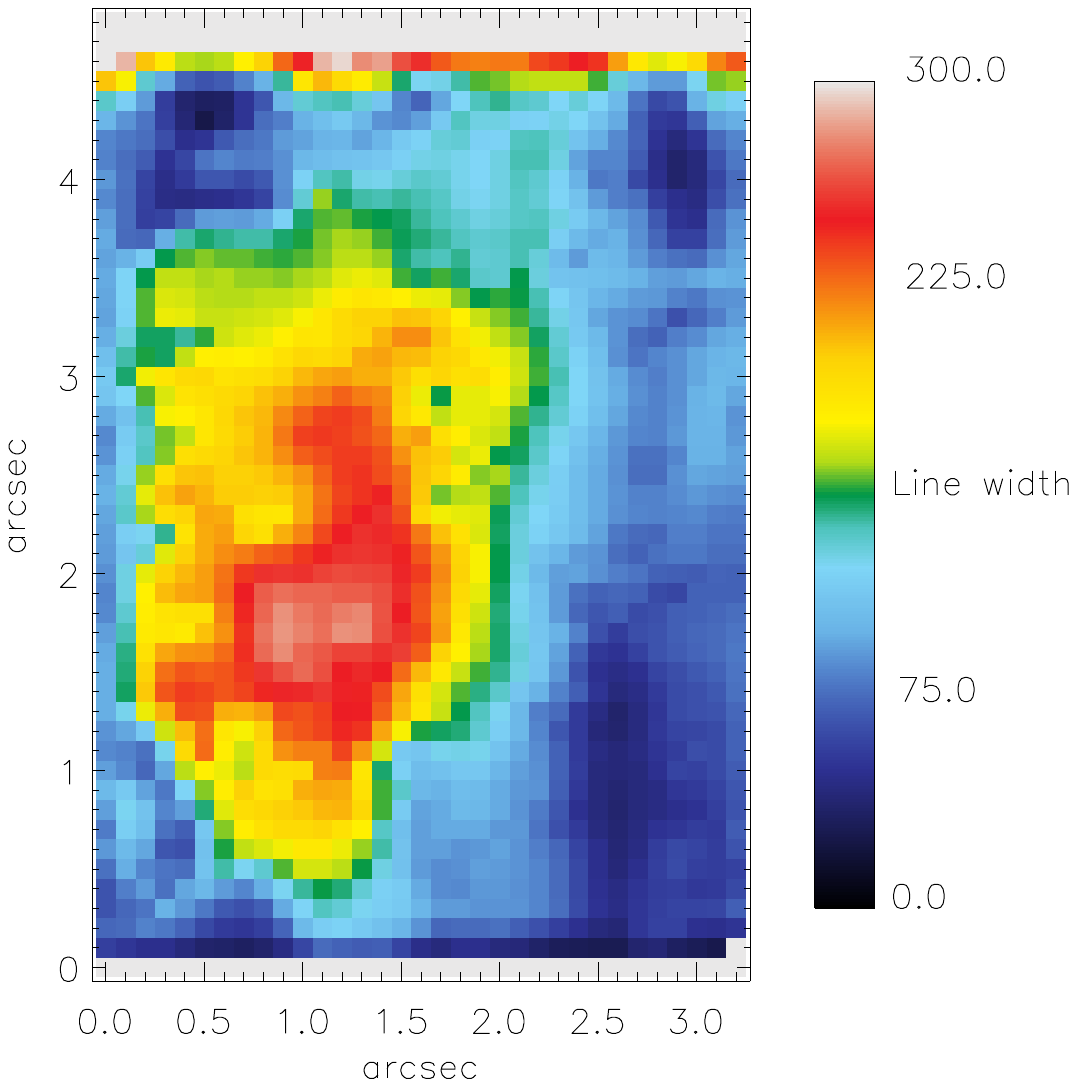}}\quad
\subfigure[{[NII]}$\lambda$6583 line width]{\includegraphics[scale=0.5, trim=1mm 1mm 50mm 1mm, clip]{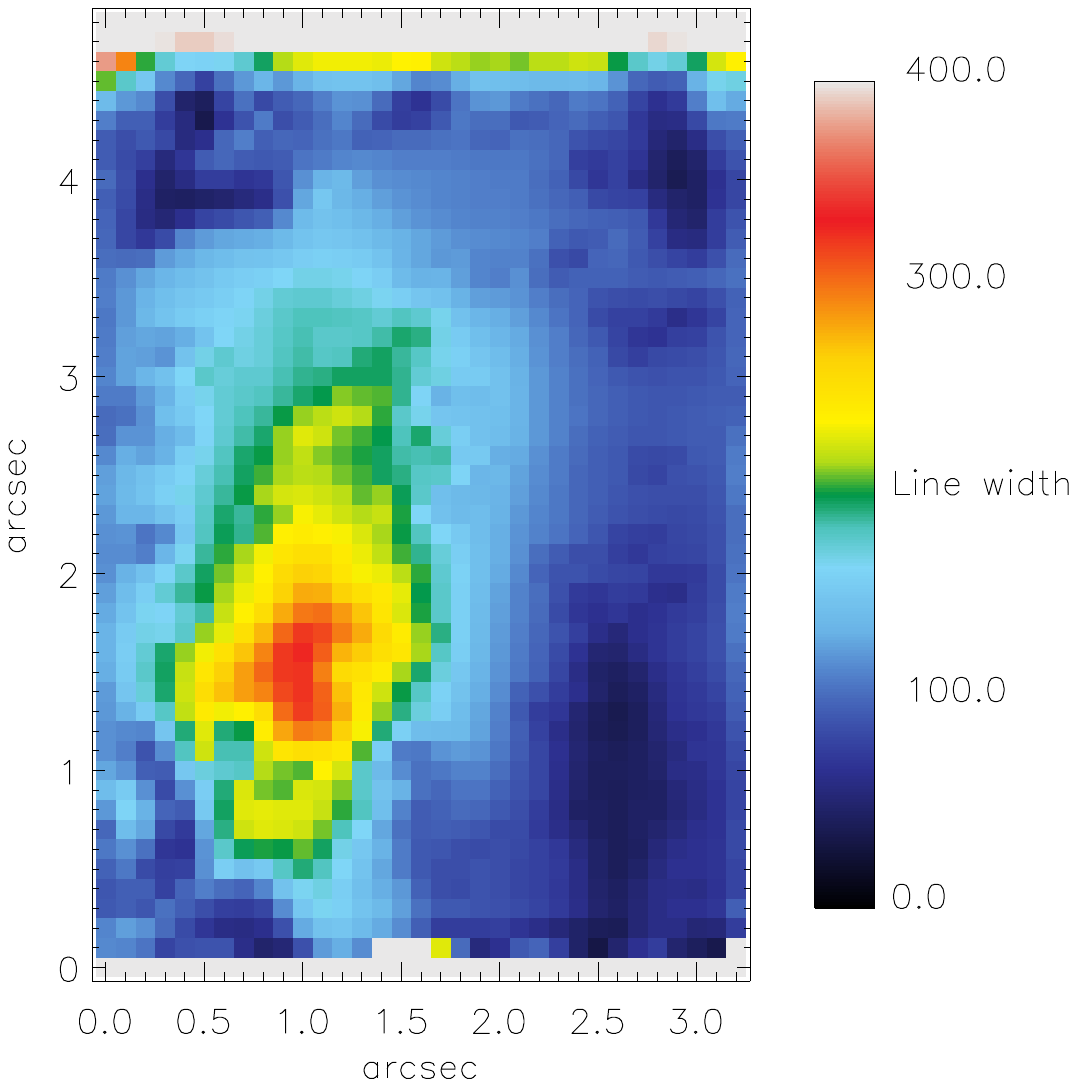}}\quad
\subfigure[{[SII]}$\lambda\lambda$6731,6717 line width]{\includegraphics[scale=0.5, trim=1mm 1mm 50mm 1mm, clip]{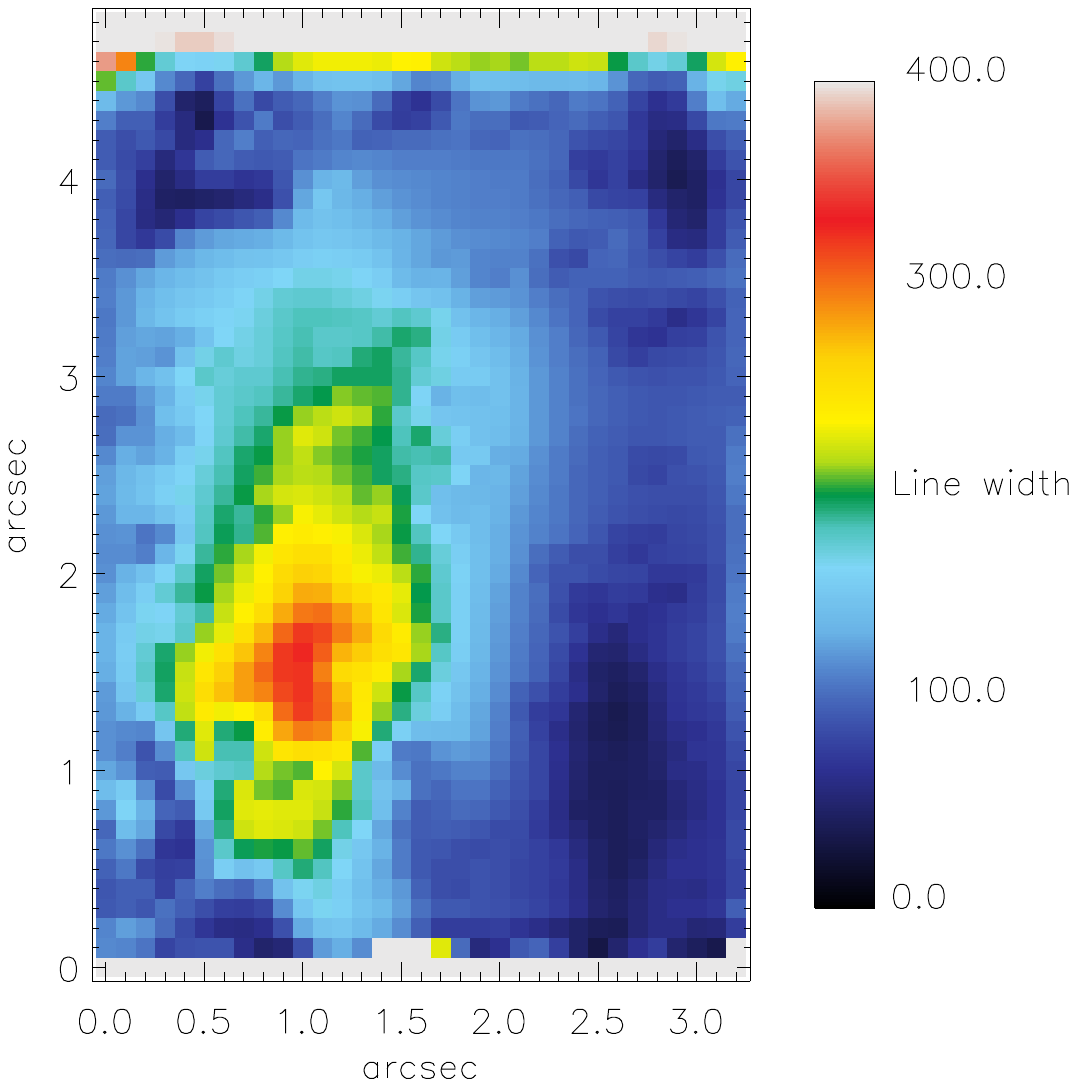}}}
\mbox{\subfigure[H$\alpha$ A/N]{\includegraphics[scale=0.5, trim=1mm 1mm 50mm 1mm, clip]{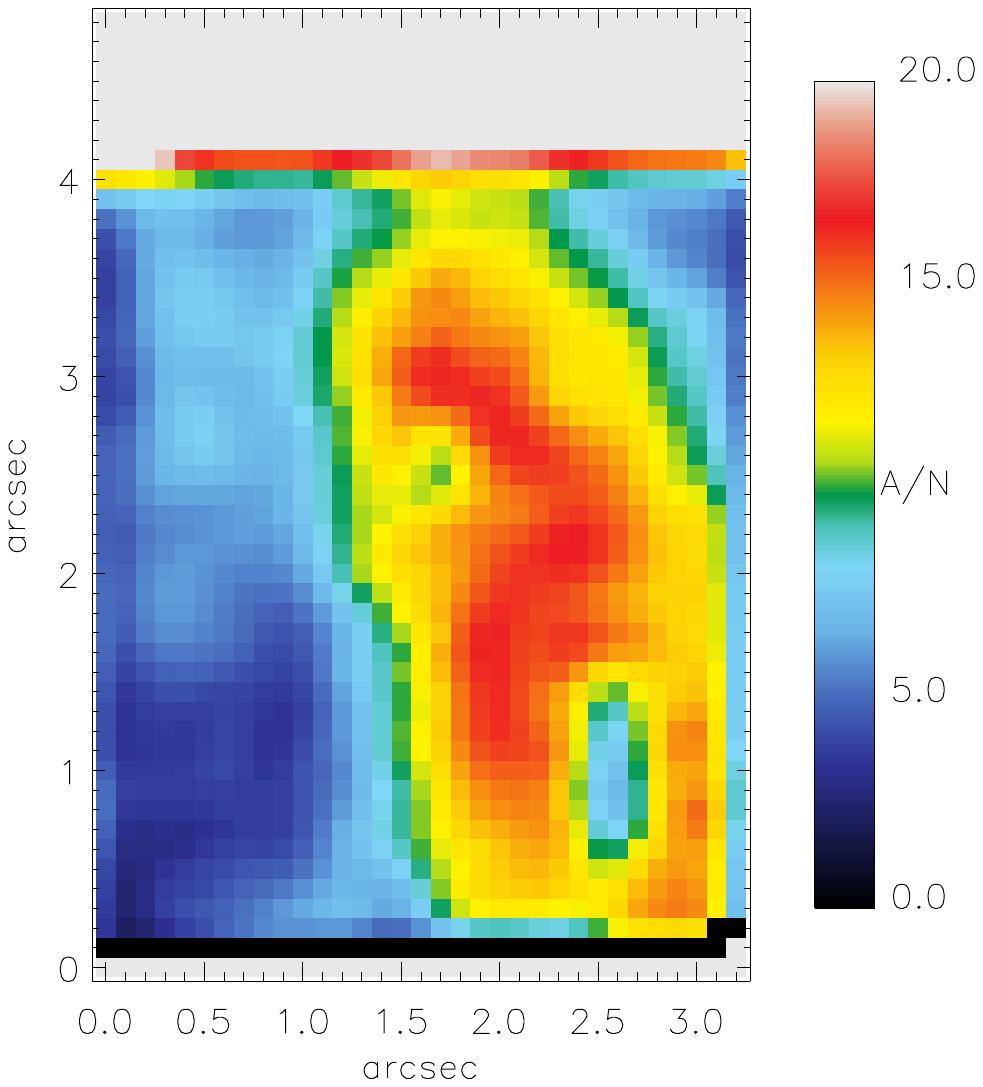}}\quad
\subfigure[{[NII]}$\lambda$6583 line A/N]{\includegraphics[scale=0.5, trim=1mm 1mm 50mm 1mm, clip]{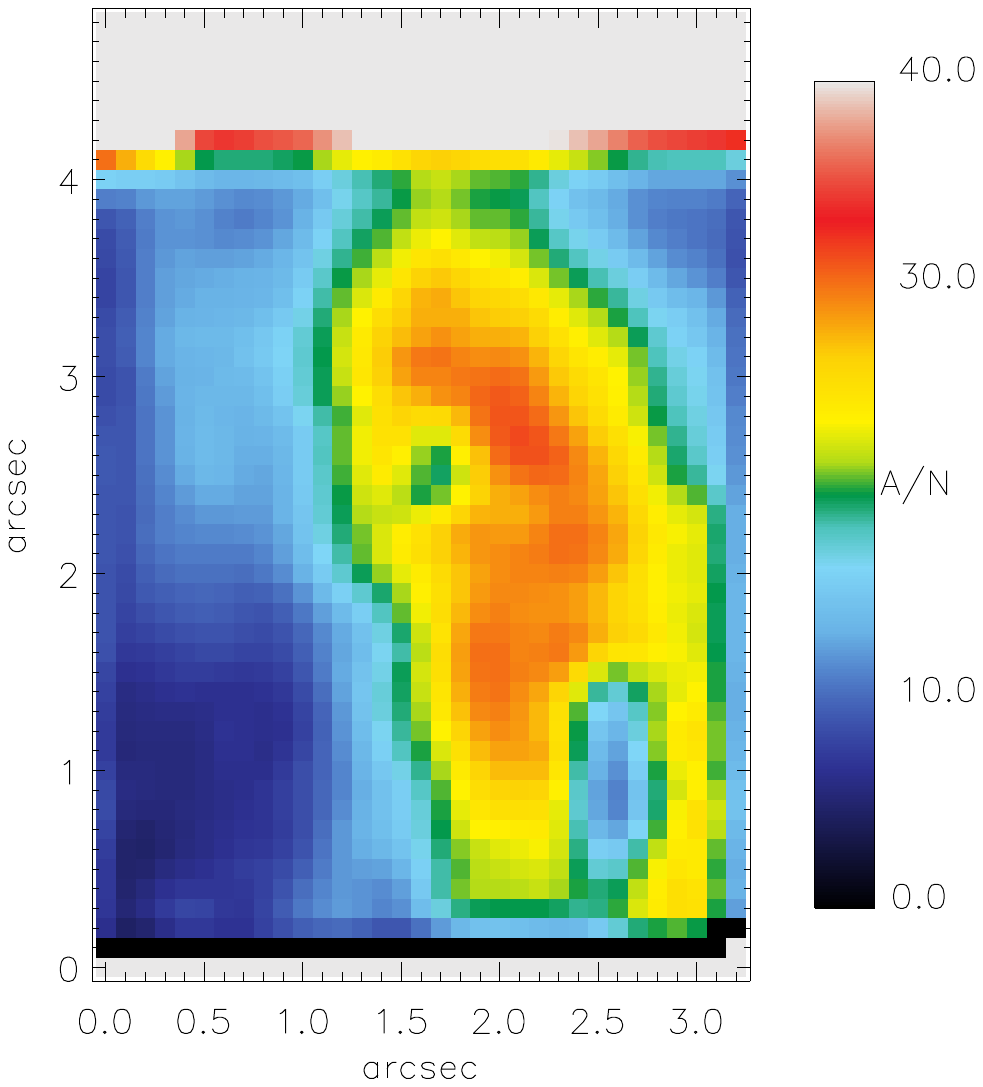}}\quad
\subfigure[{[SII]}$\lambda\lambda$6731,6717 line A/N]{\includegraphics[scale=0.5, trim=1mm 1mm 50mm 1mm, clip]{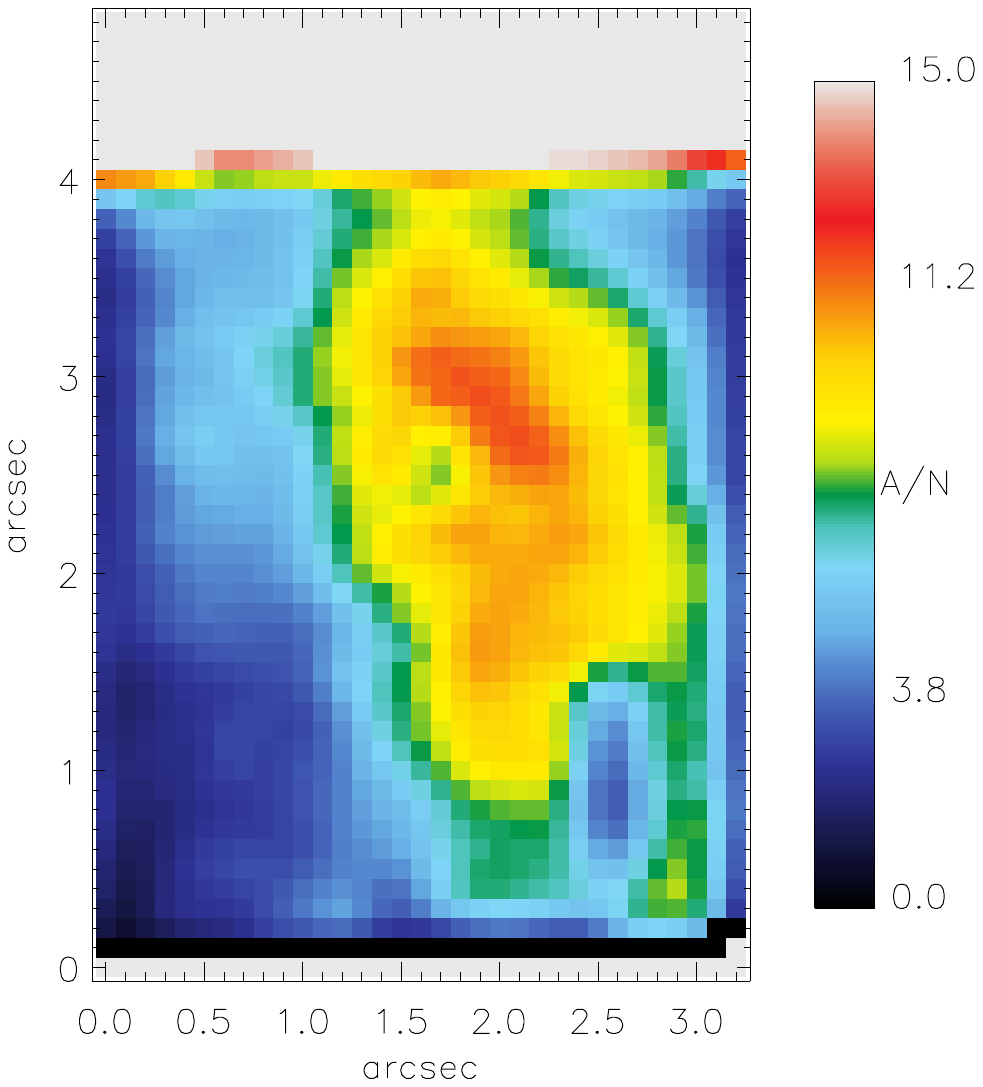}}}
\mbox{\subfigure{\includegraphics[scale=0.28]{MCG_arrow.pdf}}}
\caption{MCG-02-12-039: Velocity (in km s$^{-1}$), line width (in km s$^{-1}$) and A/N of the H$\alpha$, [NII]$\lambda$6583 and [SII]$\lambda\lambda$6731,6717 lines.}
\label{fig:MCGkinematics2} 
\end{figure*}

\begin{figure*}
 \mbox{\subfigure[{[OIII]}$\lambda$5007 velocity]{\includegraphics[scale=0.5, trim=1mm 1mm 50mm 1mm, clip]{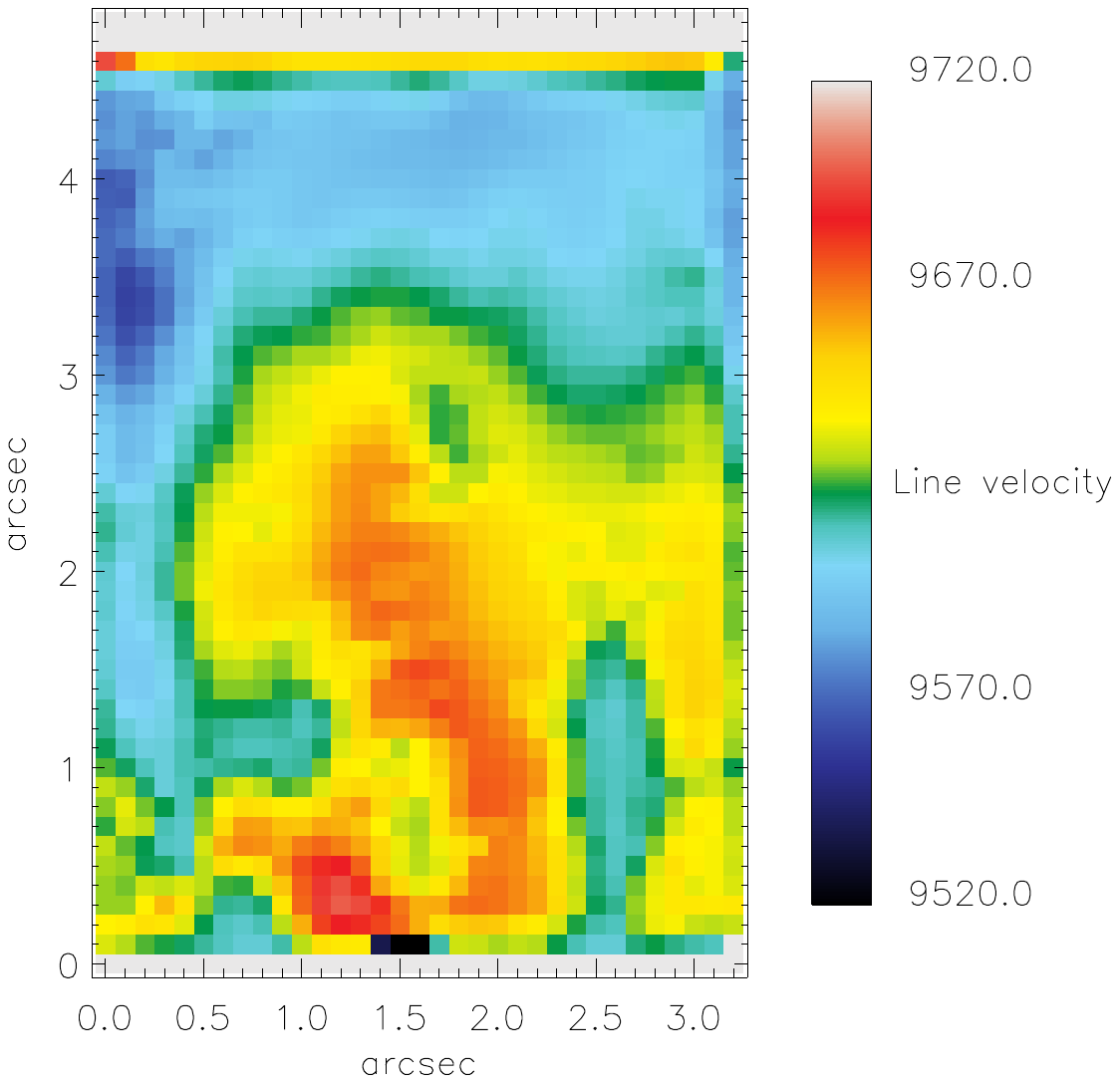}}\quad
\subfigure[{[OI]}$\lambda$6300 velocity]{\includegraphics[scale=0.5, trim=1mm 1mm 50mm 1mm, clip]{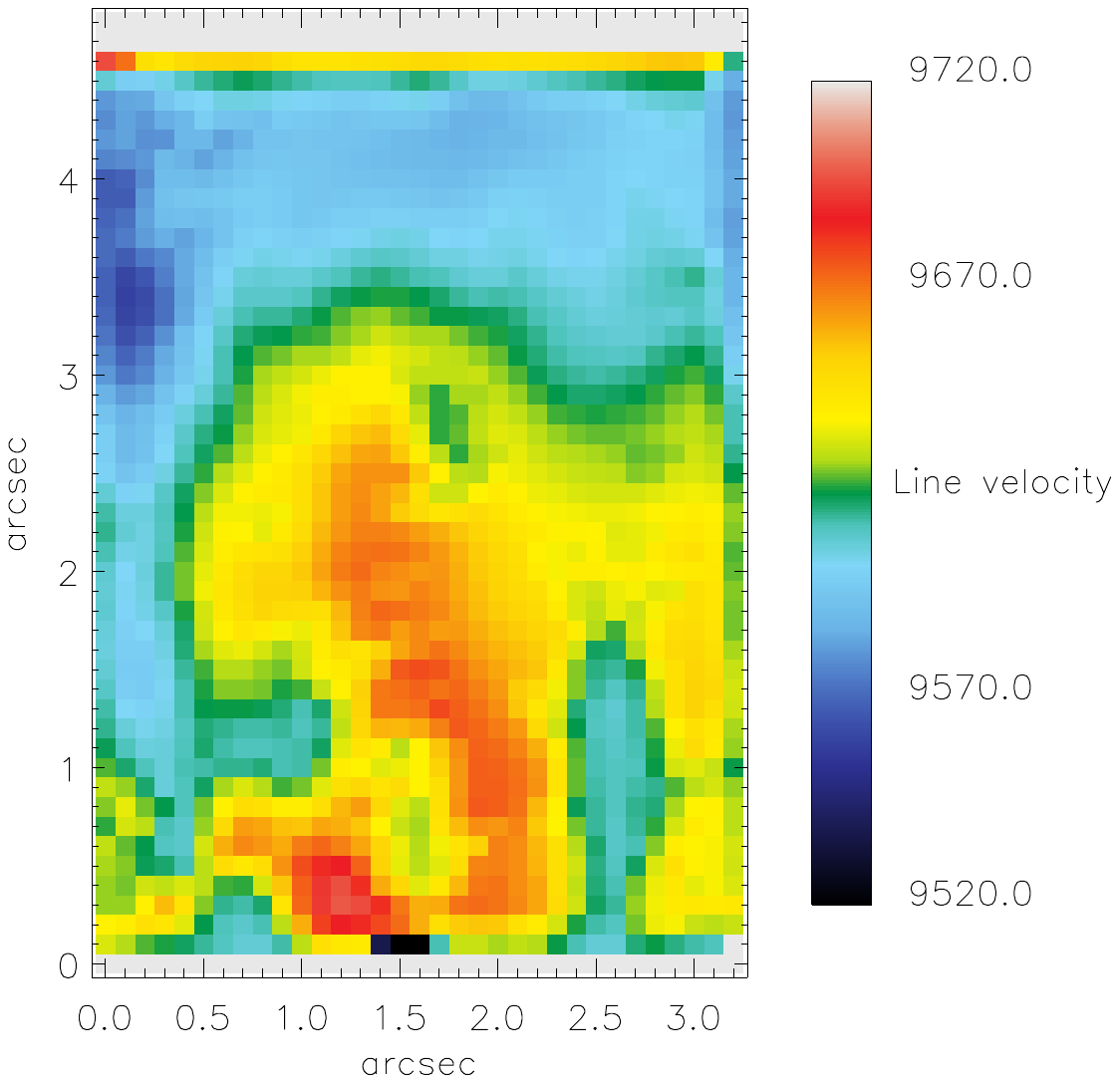}}}
   \mbox{\subfigure[{[OIII]}$\lambda$5007 line width]{\includegraphics[scale=0.5, trim=1mm 1mm 50mm 1mm, clip]{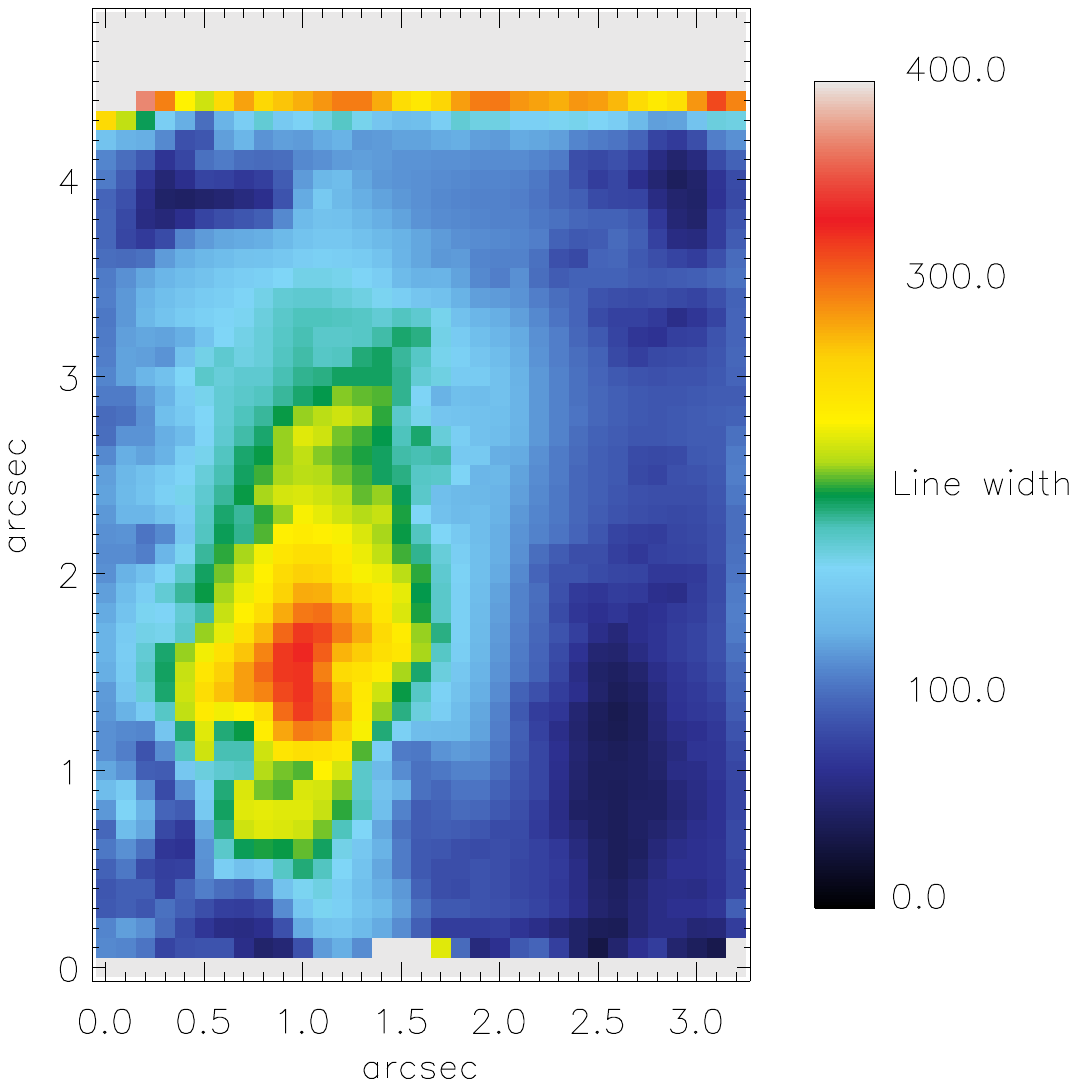}}\quad
\subfigure[{[OI]}$\lambda$6300 line width]{\includegraphics[scale=0.5, trim=1mm 1mm 50mm 1mm, clip]{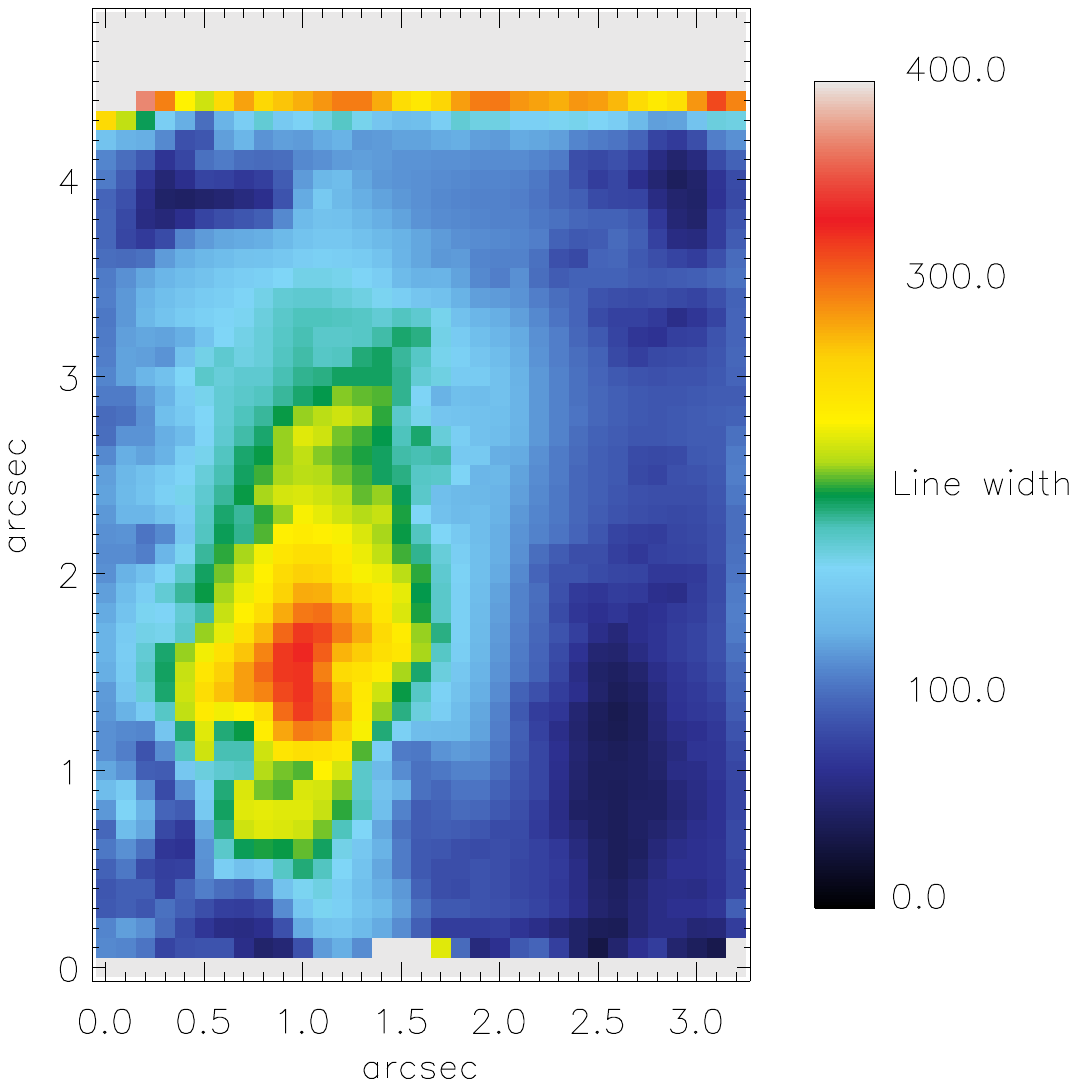}}}
\mbox{\subfigure[{[OIII]}$\lambda$5007 A/N]{\includegraphics[scale=0.5, trim=1mm 1mm 50mm 1mm, clip]{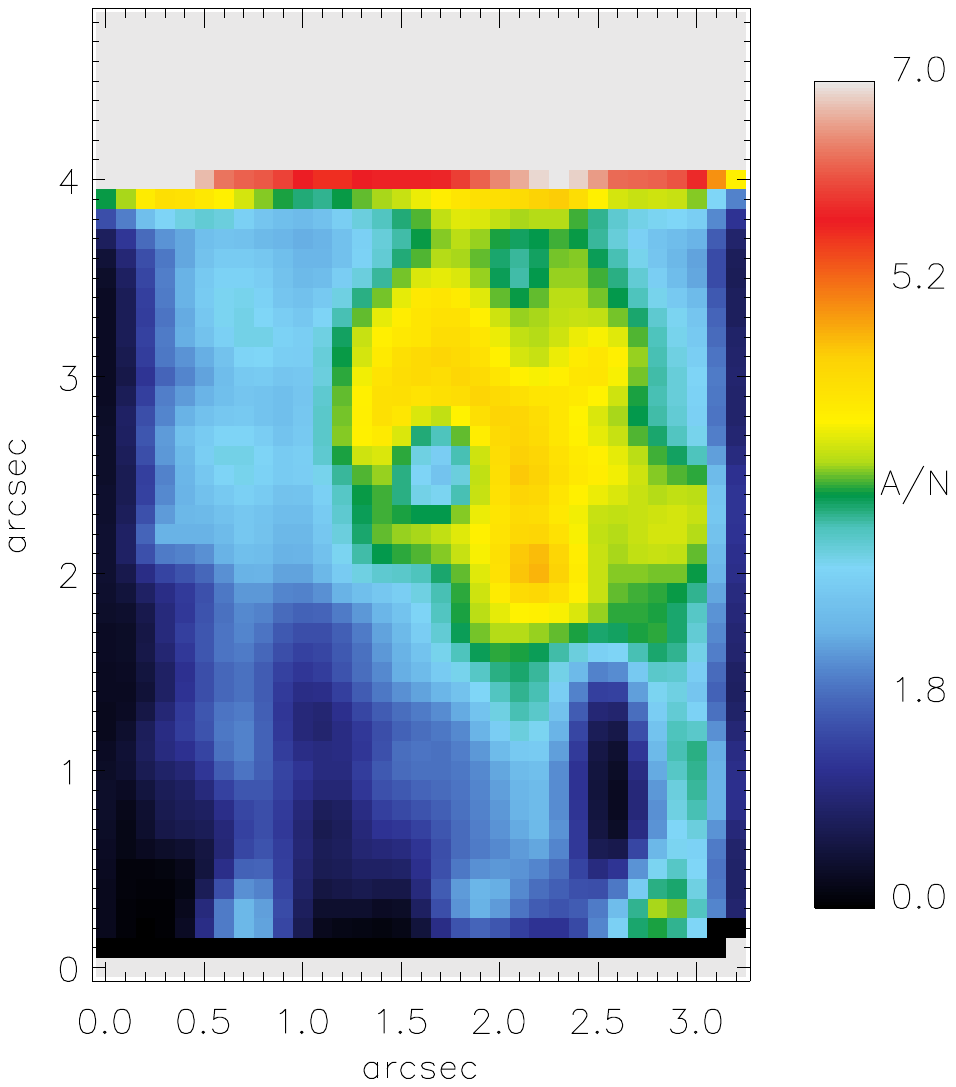}}\quad
\subfigure[{[OI]}$\lambda$6300 line A/N]{\includegraphics[scale=0.5, trim=1mm 1mm 50mm 1mm, clip]{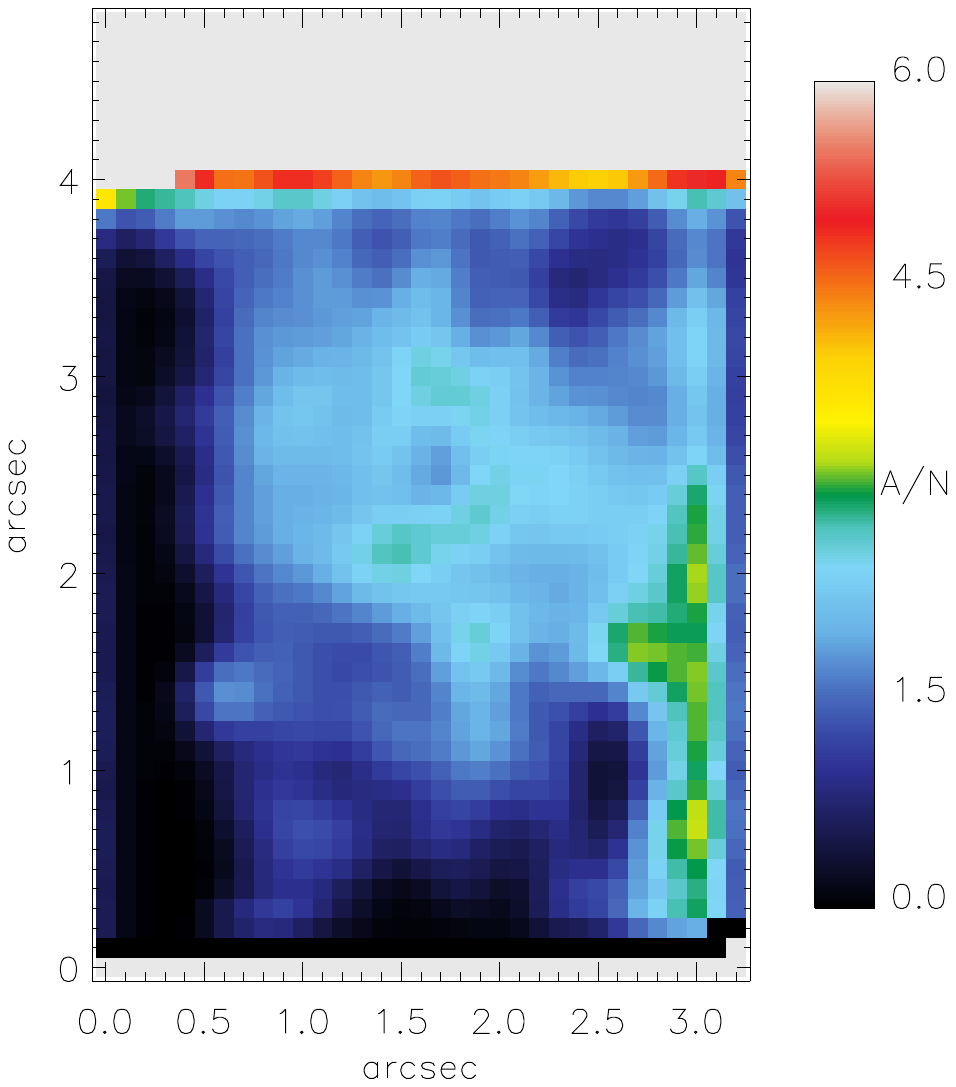}}}
\mbox{\subfigure{\includegraphics[scale=0.28]{MCG_arrow.pdf}}}
\caption{MCG-02-12-039: Velocity (in km s$^{-1}$), line width (in km s$^{-1}$) and A/N of the [OIII]$\lambda$5007 and [OI]$\lambda$6300 lines.}
\label{fig:MCGkinematics3} 
\end{figure*}

\begin{figure*}
   \centering
   \includegraphics[scale=0.8]{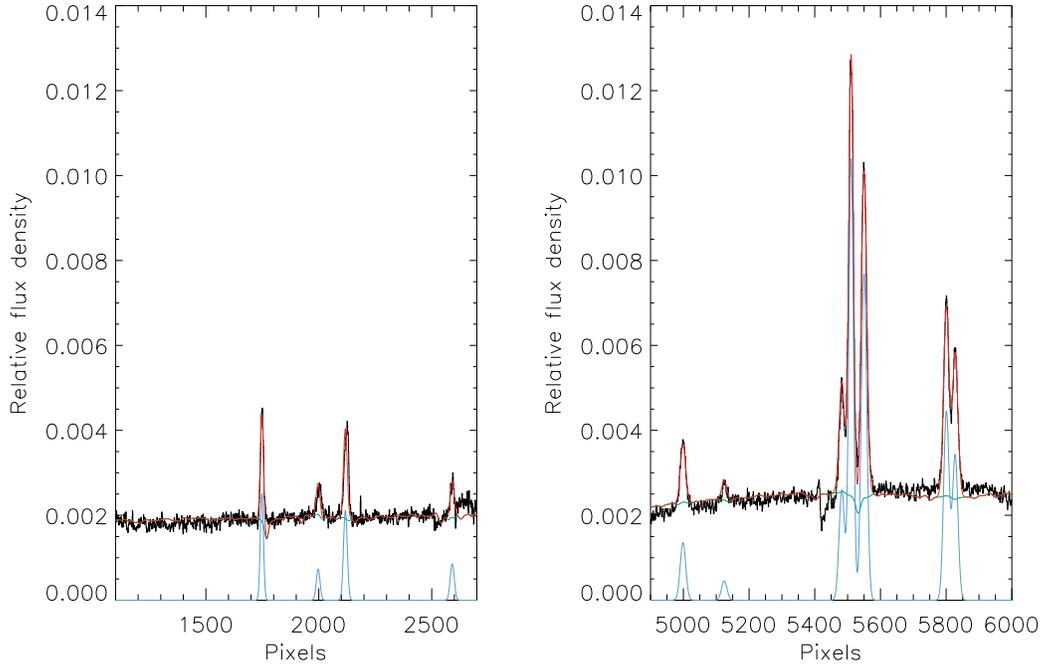}
   \caption{A random spectrum in the central region of PGC026269. The red line indicates the best-fitting stellar template and Gaussians at the emission lines, and the green line indicates the best-fitting stellar templates with the emission lines subtracted. The blue line indicates the relative flux for the measured emission lines. The lines are (from left to right) in the left plot: H$\beta$, the [OIII] doublet, and [NI], and in the right plot: [OI] doublet, [NII], H$\alpha$, [NII], and the [SII] doublet.The sharp feature to the left of the first [NII] line is one of the two CCD chip gaps and was masked during the pPXF and GANDALF fitting processes.}
   \label{fig:PGC026_gandalf}
\end{figure*}

IFU observations of this galaxy was also taken by Hatch et al.\ (2007), but they only measured the H$\alpha$ and [NII]$\lambda \lambda$6548, 6583 \AA{} lines. Their spectral resolution is much poorer at 223 - 273 km s$^{-1}$ than ours at 81 km s$^{-1}$. Our S/N is also more than five times that of Hatch et al.\ (2007), observed with a much bigger instrument. We plot H$\alpha$, [NII], [SII], [OIII] and [OI] velocities, line width and A/N in Figure \ref{fig:MCGkinematics2} and \ref{fig:MCGkinematics3}.
The stellar component displays a elongated morphology in the kinematics plotted in Figure \ref{fig:MCGkinematics}. The gas components also appear elongated (kinematics plotted in Figure \ref{fig:MCGkinematics2}), but to a lesser extend. It suggests that the stars and gas are kinematically decoupled. Comparison of the gas kinematics (Figures \ref{fig:MCGkinematics2} and \ref{fig:MCGkinematics3}) with the stellar kinematics presented in Figure \ref{fig:MCGkinematics}, and Figure \ref{fig:MCG_kinematics}, as well as by Fisher, Illingworth $\&$ Franx (1995) confirms the Hatch et al.\ (2007) observations that the two components are disconnected. Comparison of the plots in Figure \ref{fig:MCGkinematics2} showed that, to the degree that our spatial resolution reveals, it appears that all the optical forbidden and hydrogen recombination lines originate in the same gas.

We find a maximum emission line width of 400 km s$^{-1}$ as shown in Figures \ref{fig:MCGkinematics2} and \ref{fig:MCGkinematics3}. This is again similar to the finding of Hatch et al.\ (2007). They found a maximum linewidth of 600 km s$^{-1}$ in the dust-free central region to the north-east of the galaxy centre, and the rest of the nebula to have a linewidth of 100 -- 250 km s$^{-1}$. 

The stellar line-of-sight velocity reveals that some parts of the nucleas are blueshifted by $\sim$125 km s$^{-1}$ and some parts are redshifted by the same amount. Hatch et al.\ (2007) found the southern part of the galaxy to be blueshifted by --200 km s$^{-1}$ whilst the northern section is marginally redshifted up to +150 km s$^{-1}$. No clear kinematic pattern is associated with the dust structures.

The peak- to-peak gas velocity of 250 km s$^{-1}$ is fairly low compared to the other CCGs, and the stellar component of Abell 0496 has a mean rotation of 59 km s$^{-1}$ (Paper 1). 

\subsection{PGC026269}

This galaxy coincides with the well-known luminous radio counterpart Hydra A. Abell 0780 is a poor cluster with an associated cooling flow nebulae. IFU observations of this galaxy has not been presented in the literature before this study.

The stellar line-of-sight velocity (from the IFU data) is blue-and redshifted by $\sim$175 km s$^{-1}$ which is slightly higher than that derived from long slit spectroscopy (51 $\pm$ 20 km s$^{-1}$ Paper 1).

\begin{figure*}
   \centering
 \mbox{\subfigure[Stellar velocity]{\includegraphics[scale=0.6, trim=1mm 1mm 50mm 1mm, clip]{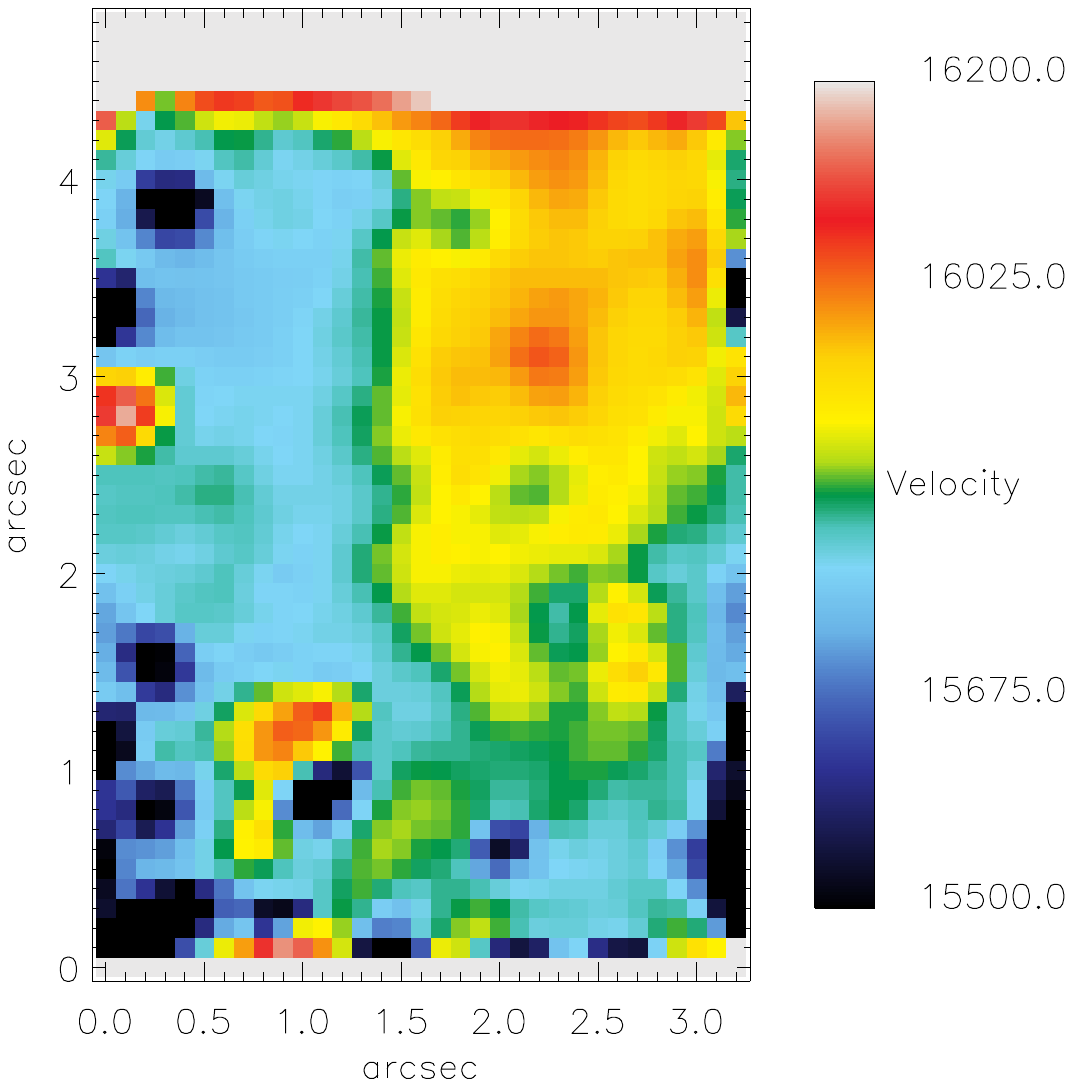}}\quad
\subfigure[Stellar velocity dispersion]{\includegraphics[scale=0.6, trim=1mm 1mm 50mm 1mm, clip]{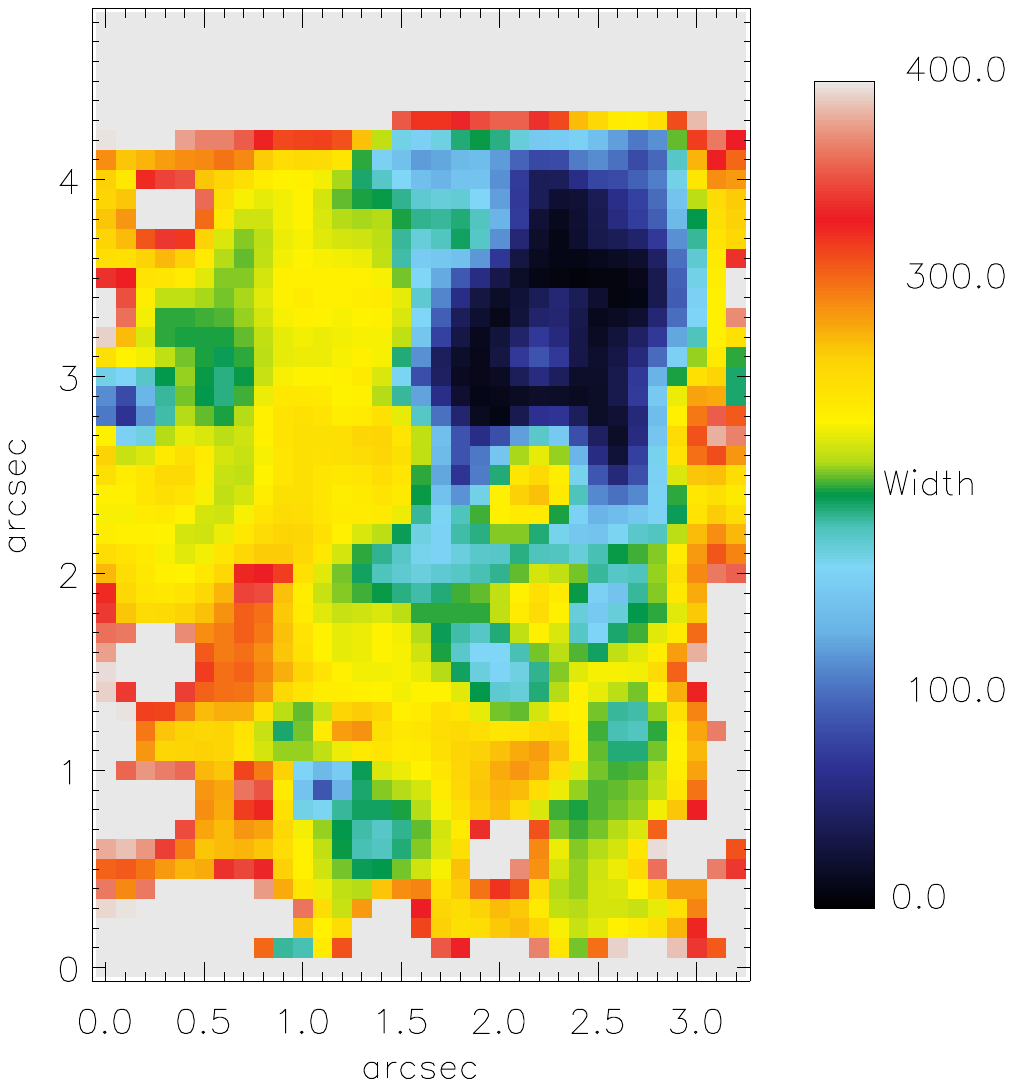}}}
\mbox{\subfigure{\includegraphics[scale=0.25]{PGC026_arrow.pdf}}}
\caption{PGC026269: Velocity and velocity dispersion of the absorption lines in km s$^{-1}$.}
\label{fig:PGC026kinematics} 
\end{figure*}

\begin{figure*}
 \mbox{\subfigure[H$\alpha$ velocity]{\includegraphics[scale=0.5, trim=1mm 1mm 50mm 1mm, clip]{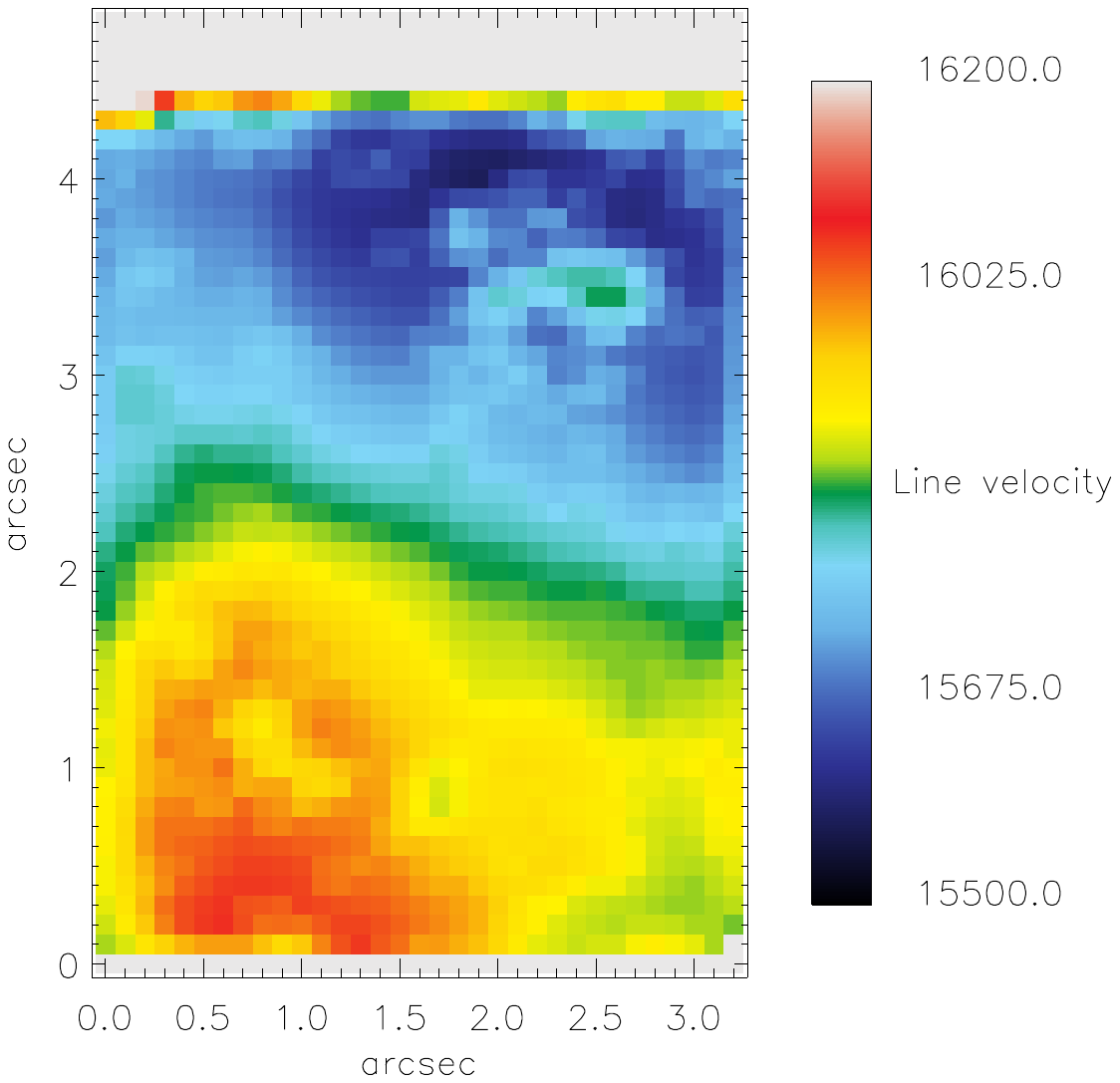}}\quad
\subfigure[{[NII]}$\lambda$6583 velocity]{\includegraphics[scale=0.5, trim=1mm 1mm 50mm 1mm, clip]{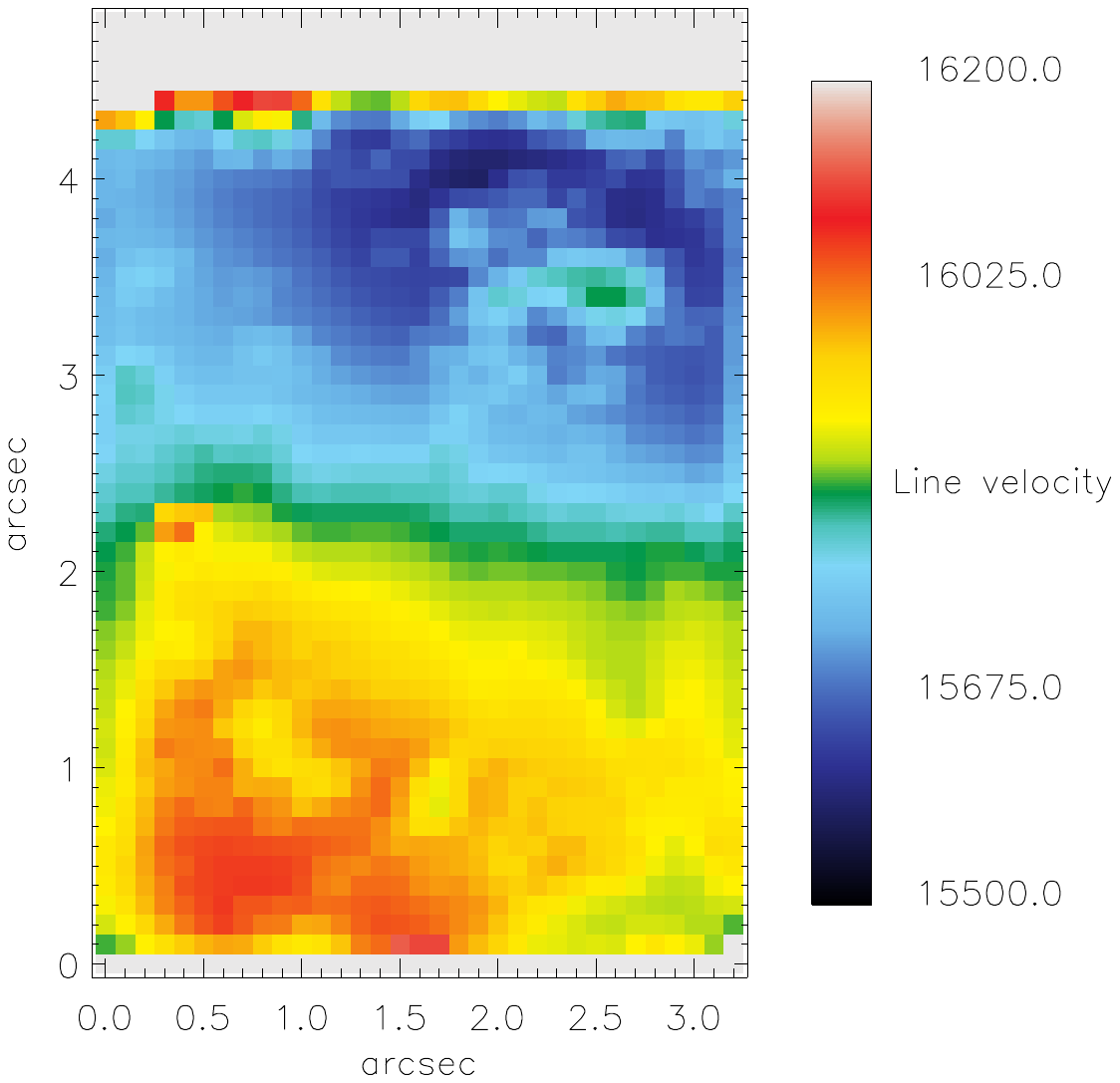}}\quad
\subfigure[{[SII]}$\lambda\lambda$6731,6717 velocity]{\includegraphics[scale=0.5, trim=1mm 1mm 50mm 1mm, clip]{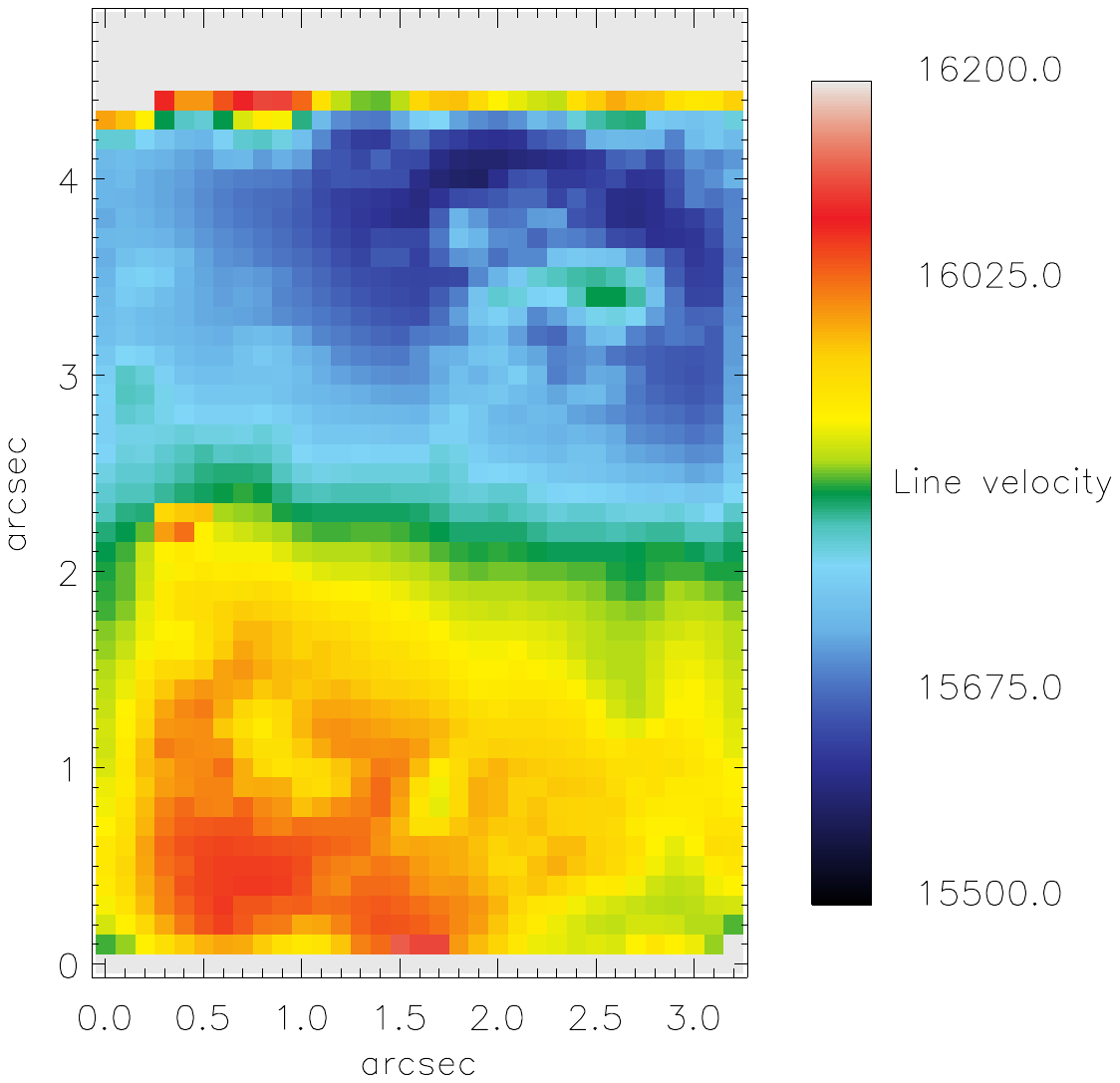}}}
   \mbox{\subfigure[H$\alpha$ line width]{\includegraphics[scale=0.5, trim=1mm 1mm 50mm 1mm, clip]{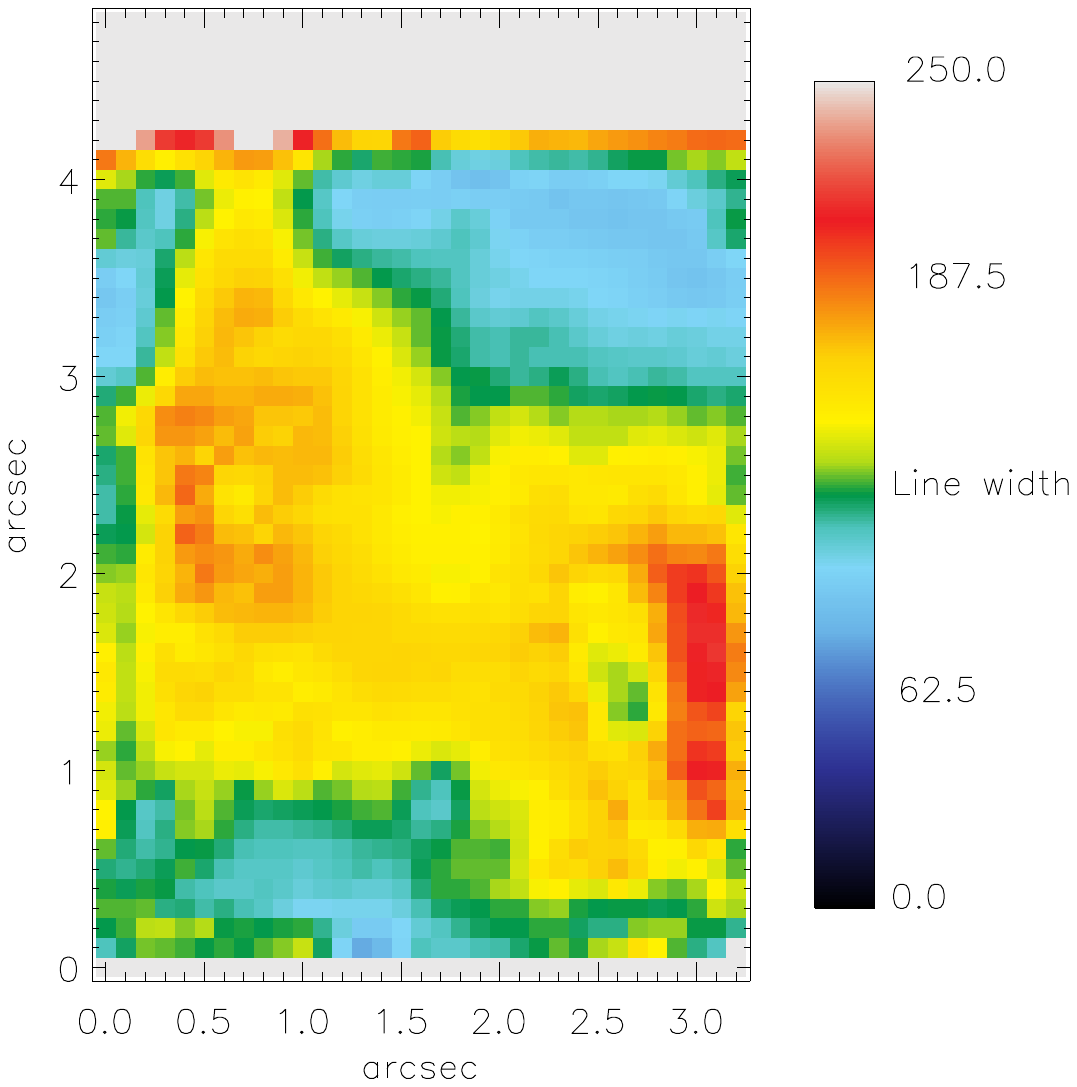}}\quad
\subfigure[{[NII]}$\lambda$6583 line width]{\includegraphics[scale=0.5, trim=1mm 1mm 50mm 1mm, clip]{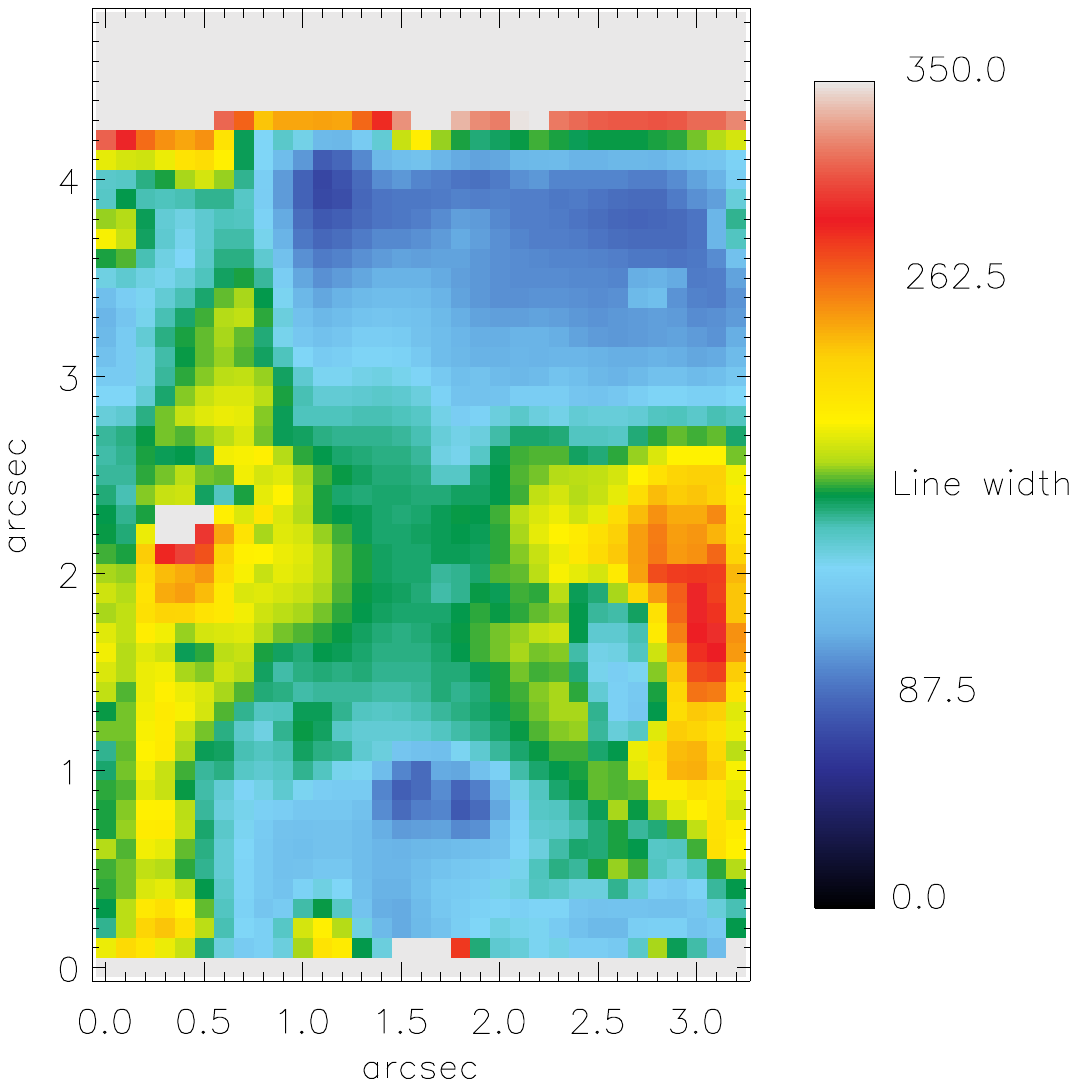}}\quad
\subfigure[{[SII]}$\lambda\lambda$6731,6717 line width]{\includegraphics[scale=0.5, trim=1mm 1mm 50mm 1mm, clip]{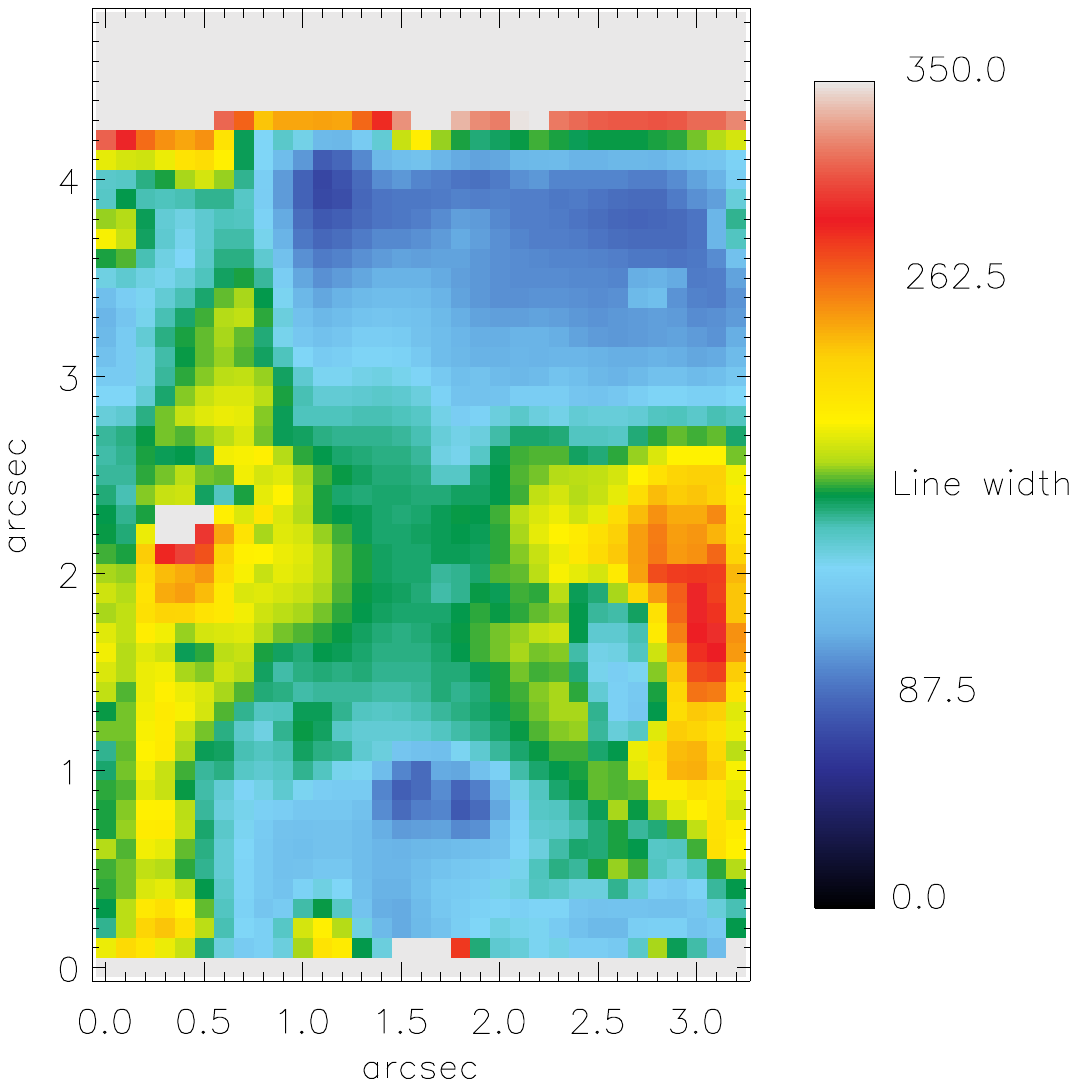}}}
\mbox{\subfigure[H$\alpha$ A/N]{\includegraphics[scale=0.5, trim=1mm 1mm 50mm 1mm, clip]{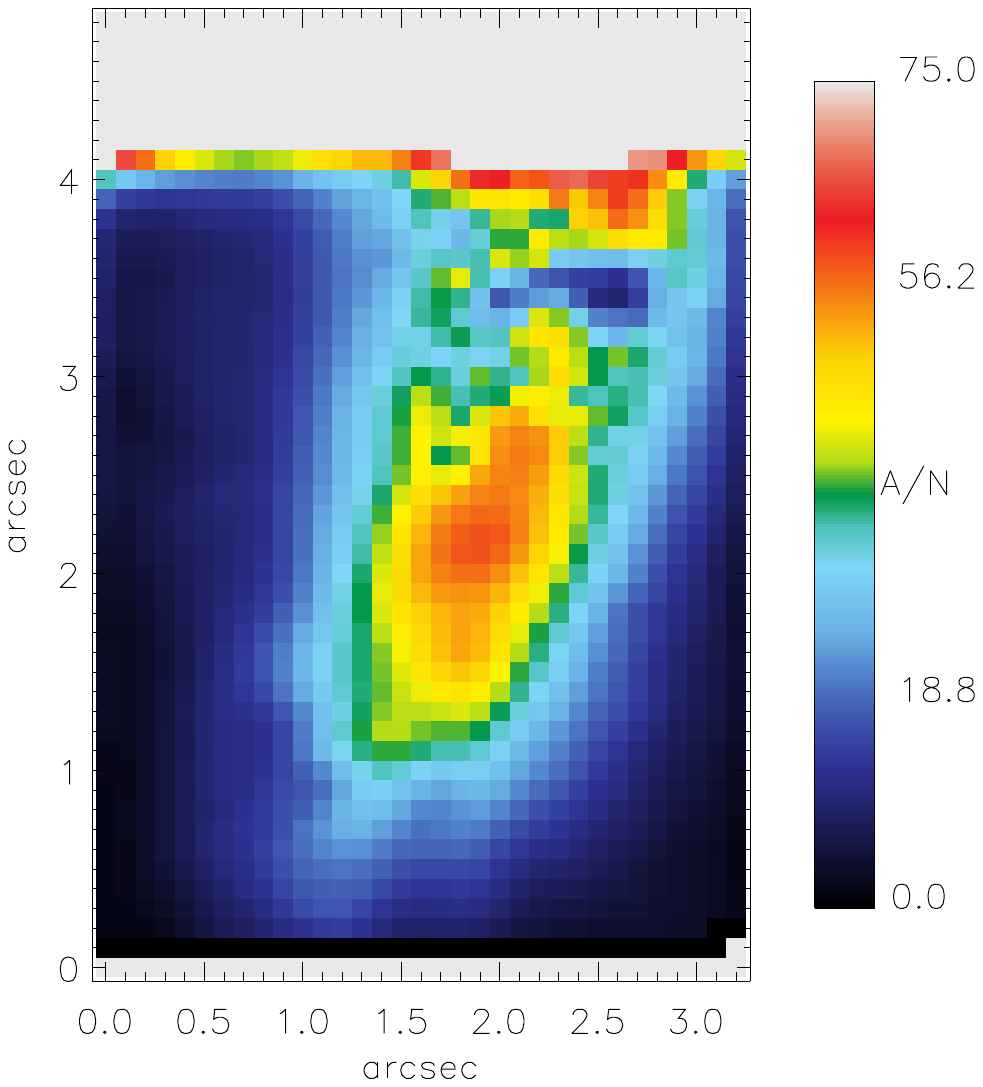}}\quad
\subfigure[{[NII]}$\lambda$6583 line A/N]{\includegraphics[scale=0.5, trim=1mm 1mm 50mm 1mm, clip]{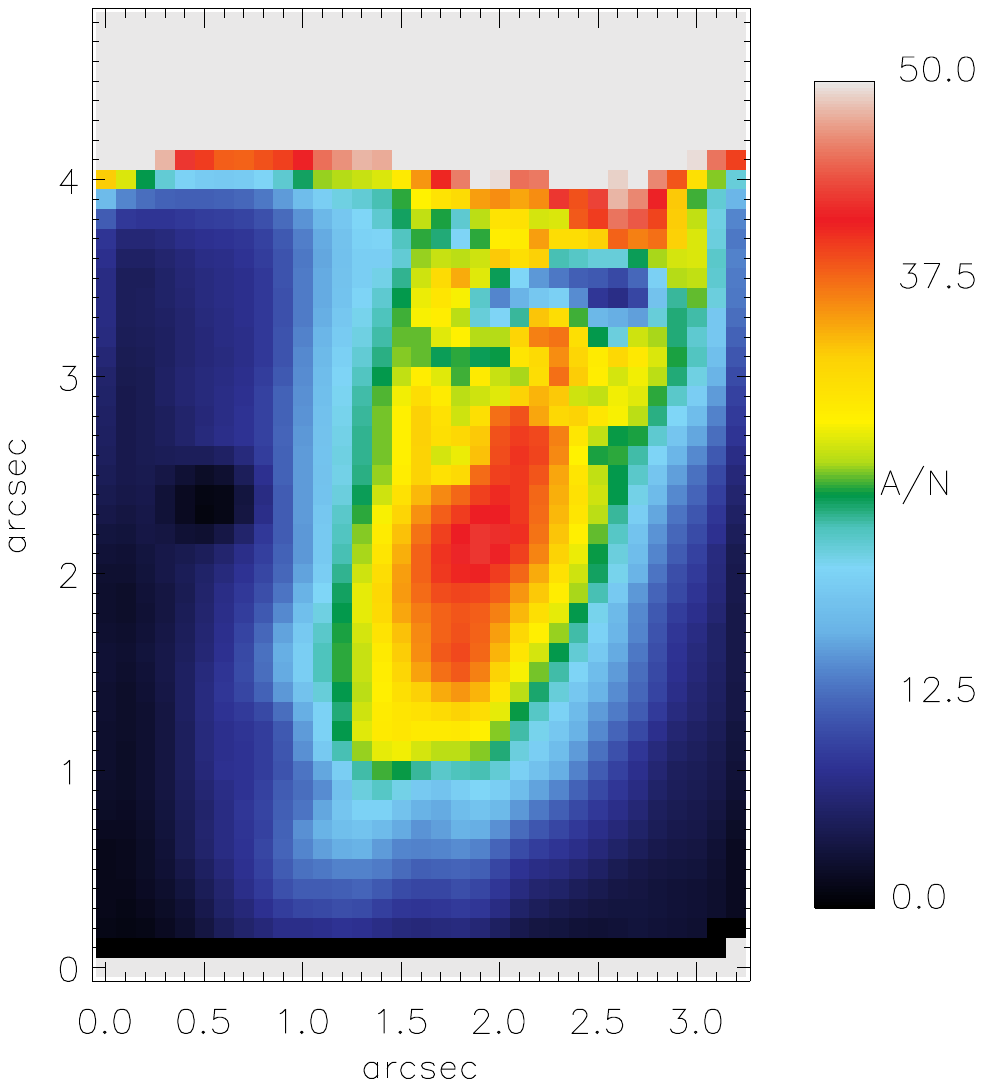}}\quad
\subfigure[{[SII]}$\lambda\lambda$6731,6717 line A/N]{\includegraphics[scale=0.5, trim=1mm 1mm 50mm 1mm, clip]{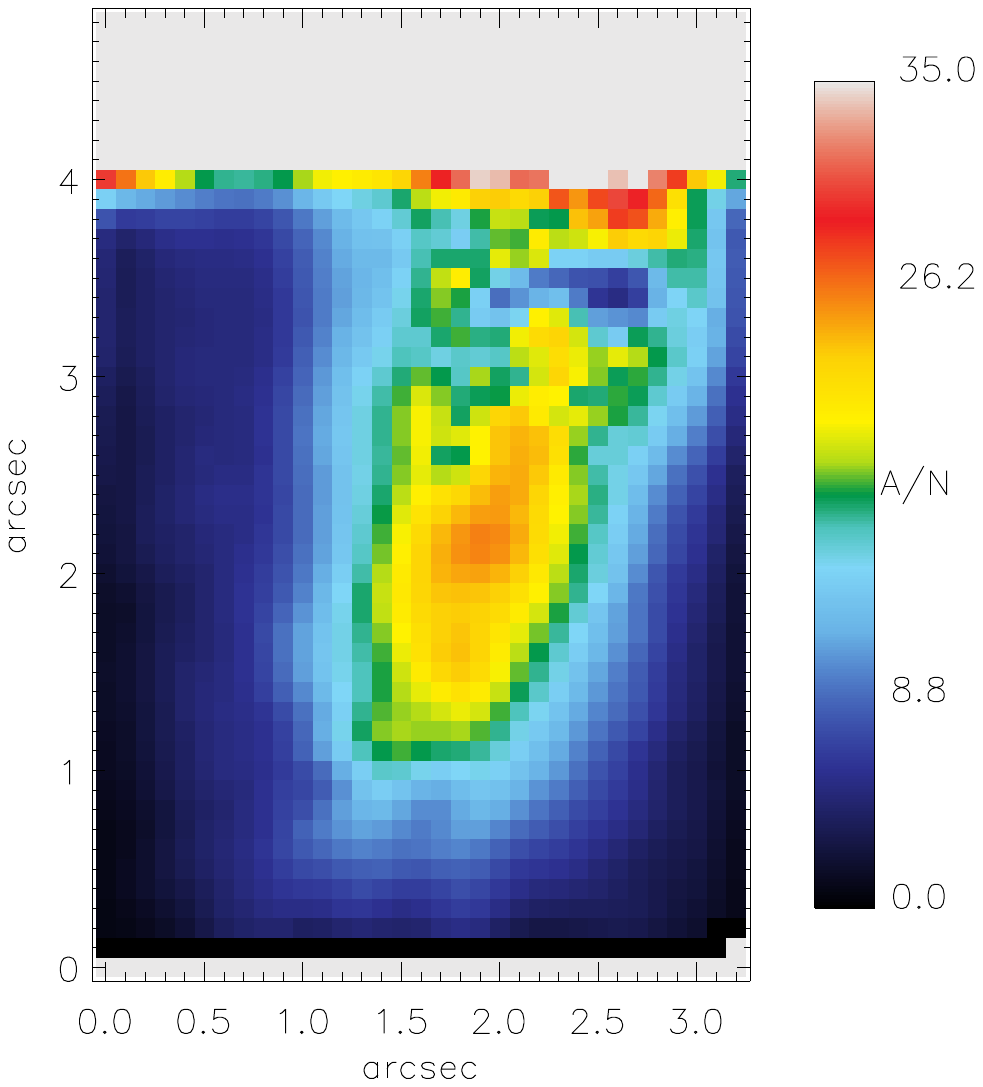}}}
\mbox{\subfigure{\includegraphics[scale=0.28]{PGC026_arrow.pdf}}}
\caption{PGC026269: Velocity (in km s$^{-1}$), line width (in km s$^{-1}$) and A/N of the H$\alpha$, [NII]$\lambda$6583 and [SII]$\lambda\lambda$6731,6717 lines.}
\label{fig:PGC026kinematics2} 
\end{figure*}

We plot the kinematics of the stellar and gaseous components in Figures \ref{fig:PGC026kinematics} and \ref{fig:PGC026kinematics2}. Both the stellar and gaseous components show clear rotation in Figures \ref{fig:PGC026kinematics} and \ref{fig:PGC026kinematics2}, although it also seems to be decoupled. This system bears resemblance to the BCG NCG 3311 in Abell 1060, studied by Edwards et al.\ (2009) which also showed striking rotation. The rotation of the warm gas agrees with the long-slit observations by McDonald et al.\ (2012).

\begin{figure*}
 \mbox{\subfigure[{[OIII]}$\lambda$5007 velocity]{\includegraphics[scale=0.5, trim=1mm 1mm 50mm 1mm, clip]{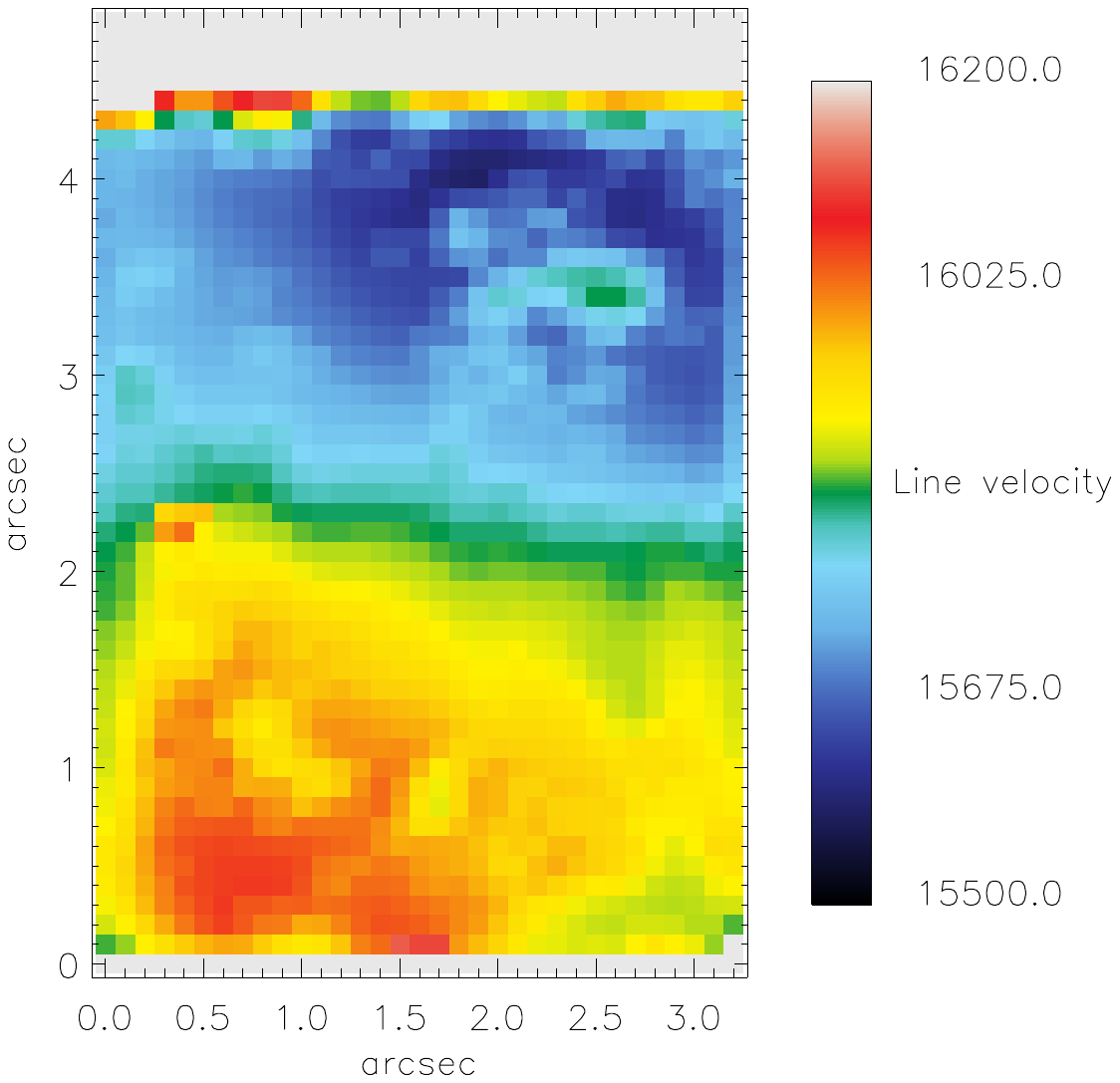}}\quad
\subfigure[{[OI]}$\lambda$6300 velocity]{\includegraphics[scale=0.5, trim=1mm 1mm 50mm 1mm, clip]{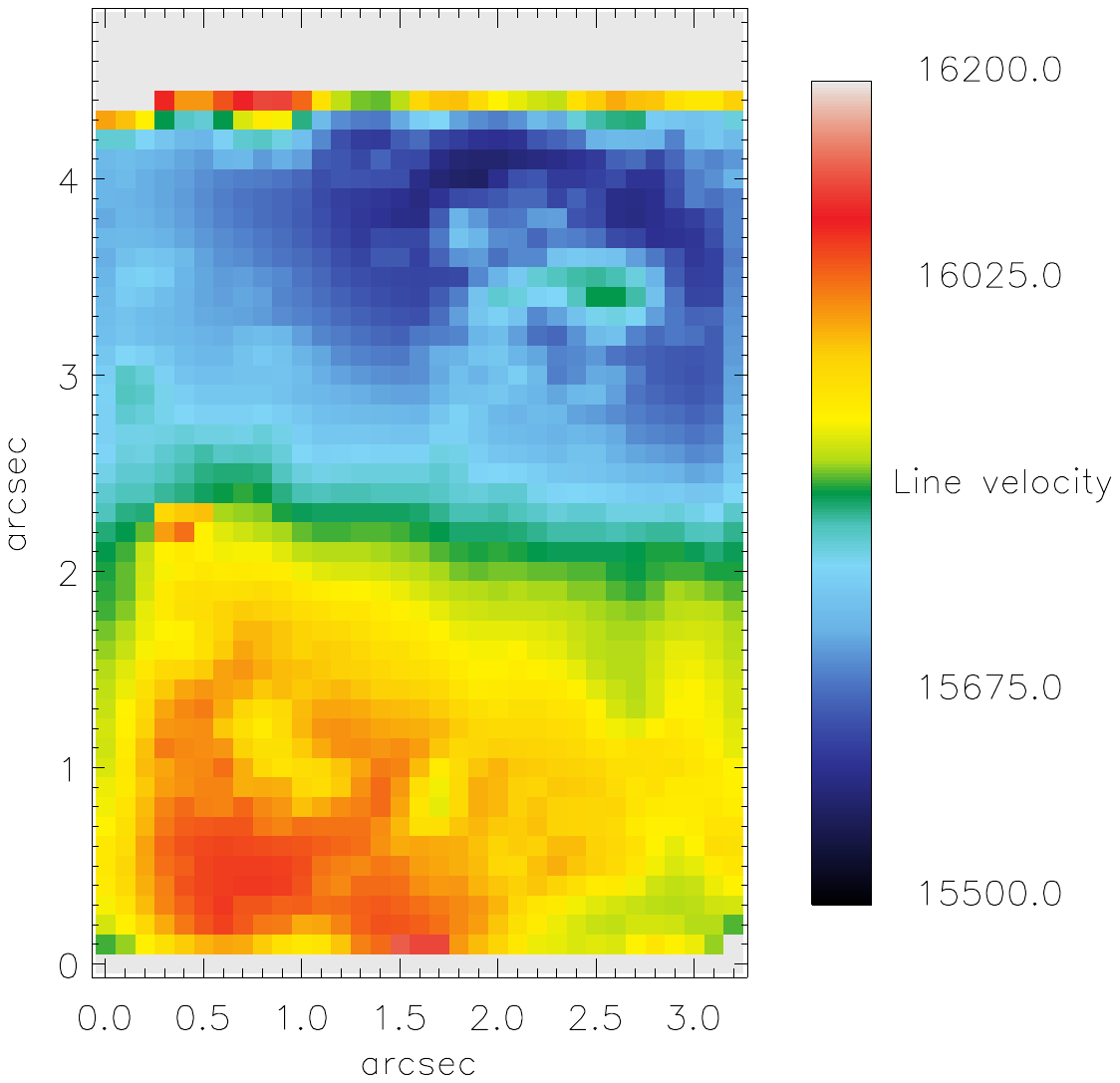}}}
   \mbox{\subfigure[{[OIII]}$\lambda$5007 line width]{\includegraphics[scale=0.5, trim=1mm 1mm 50mm 1mm, clip]{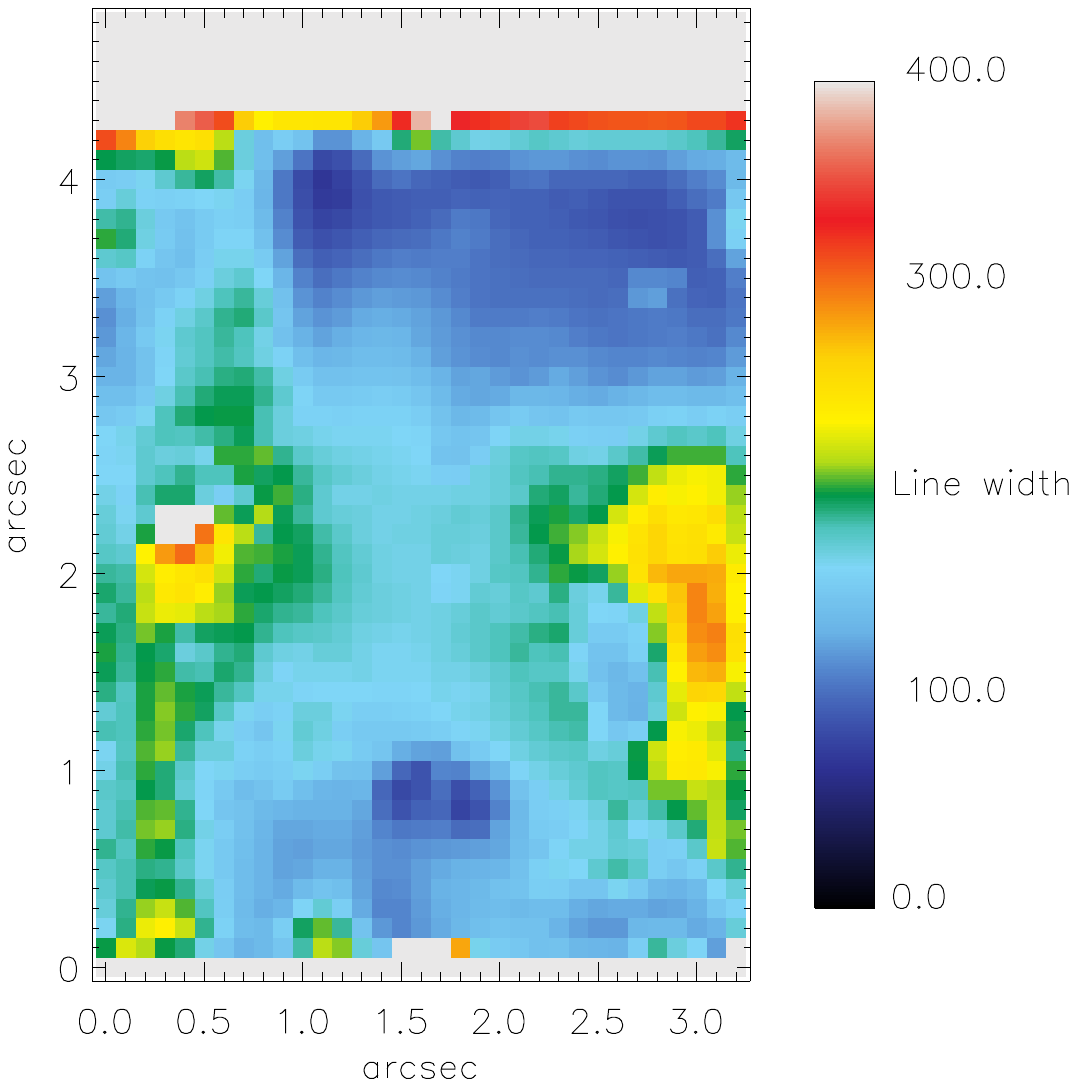}}\quad
\subfigure[{[OI]}$\lambda$6300 line width]{\includegraphics[scale=0.5, trim=1mm 1mm 50mm 1mm, clip]{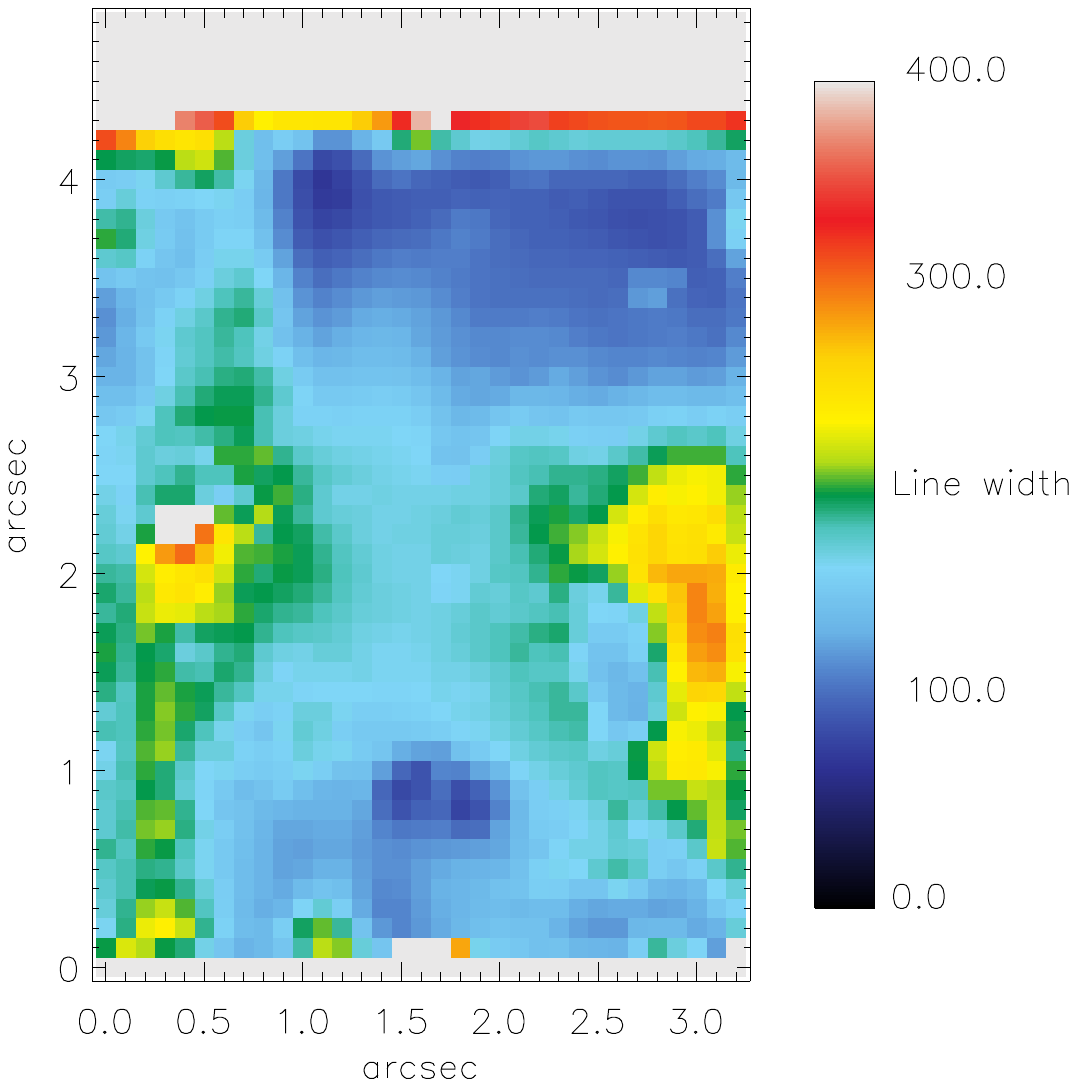}}}
\mbox{\subfigure[{[OIII]}$\lambda$5007 A/N]{\includegraphics[scale=0.5, trim=1mm 1mm 50mm 1mm, clip]{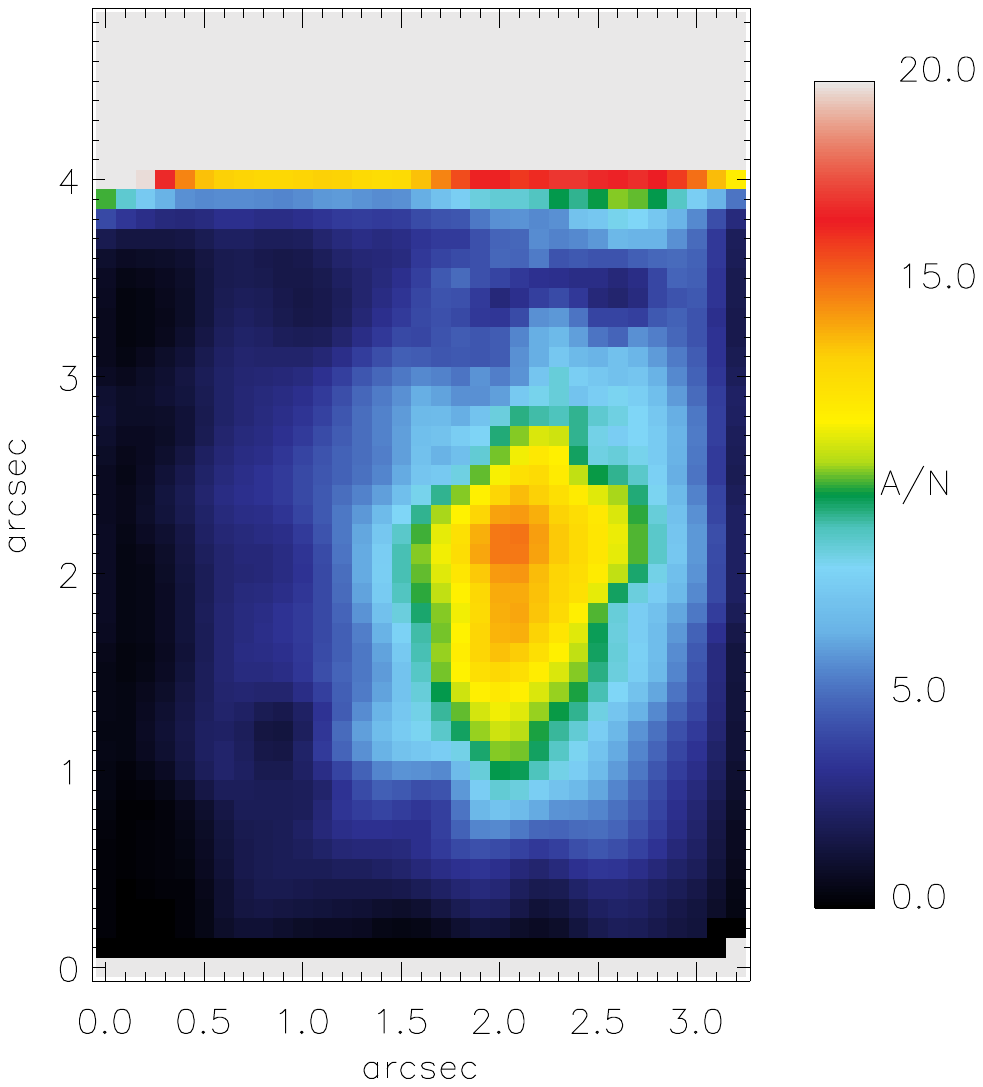}}\quad
\subfigure[{[OI]}$\lambda$6300 line A/N]{\includegraphics[scale=0.5, trim=1mm 1mm 50mm 1mm, clip]{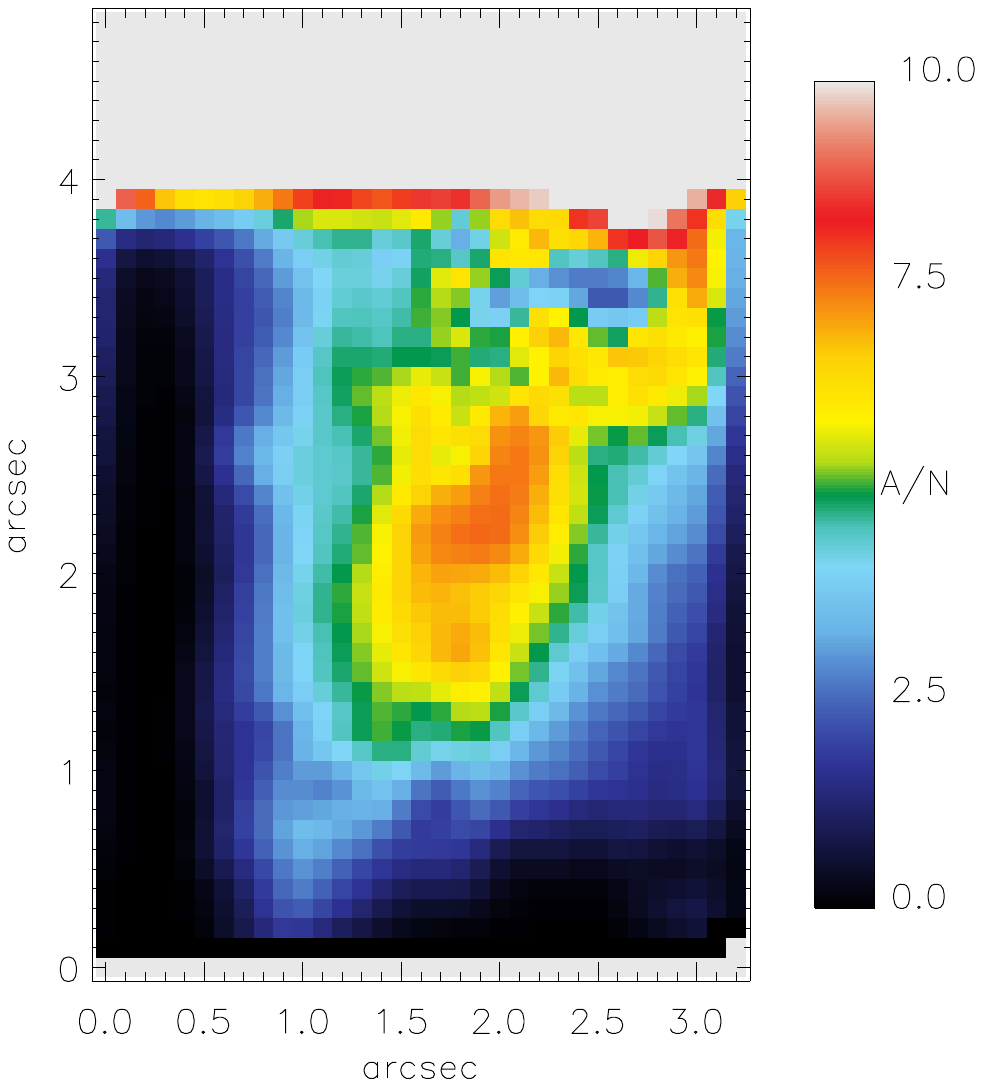}}}
\mbox{\subfigure{\includegraphics[scale=0.28]{PGC026_arrow.pdf}}}
\caption{PGC026269: Velocity (in km s$^{-1}$), line width (in km s$^{-1}$) and A/N of the [OIII]$\lambda$5007 and [OI]$\lambda$6300 lines.}
\label{fig:PGC026kinematics3} 
\end{figure*}

\begin{figure*}
   \centering
   \includegraphics[scale=0.8]{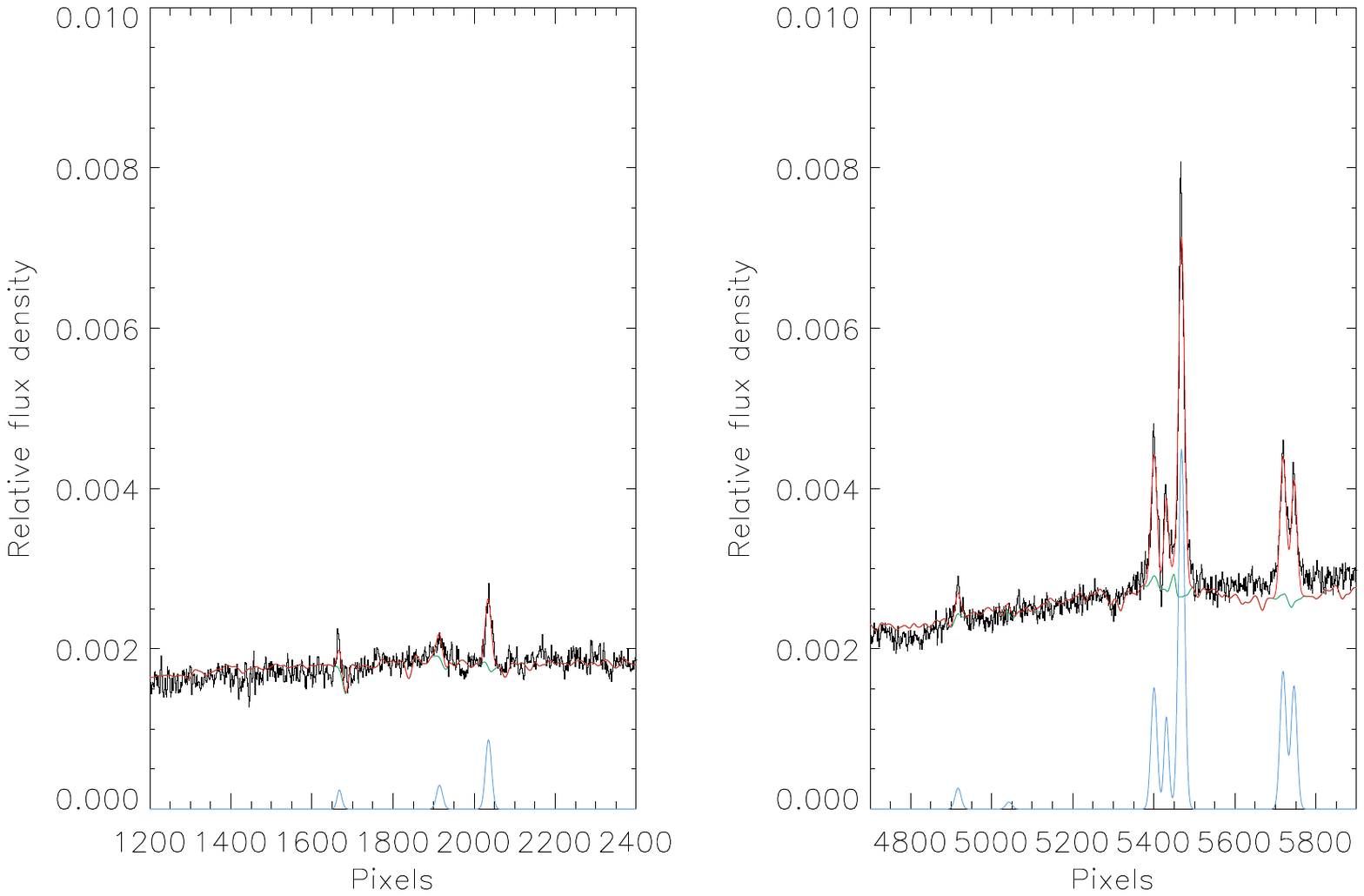}
   \caption{A random spectrum in the central region of PGC044257. The red line indicates the best-fitting stellar template and Gaussians at the emission lines, and the green line indicates the best-fitting stellar templates with the emission lines subtracted. The blue line indicates the relative flux for the measured emission lines. The lines are (from left to right) in the left plot: H$\beta$, and the [OIII] doublet, and in the right plot: [OI] doublet, [NII], H$\alpha$, [NII], and the [SII] doublet.}
   \label{fig:PGC044_gandalf}
\end{figure*}

We plot H$\alpha$, [NII], [SII], [OIII] and [OI] velocities, line width and A/N in Figure \ref{fig:PGC026kinematics2} and \ref{fig:PGC026kinematics3}. Comparison of the plots in Figure \ref{fig:MCGkinematics2} showed that, to the degree that our spatial resolution reveals, it appears that all the optical forbidden and hydrogen recombination lines originate in the same gas.

Wise et al. (2007) present a summary of the X-ray and radio properties of this cluster, showing the excellent correlation between the radio jets and the X-ray cavities. The arcing H$\alpha$ filament that McDonald et al.\ (2010) detect north of the CCG (on a larger scale as this study) appears to be spatially correlated with the radio jet. Our Figure \ref{fig:BPTs} (d --  f) show that our optical observations confirms that the photoionisation is caused by an AGN (the green and red curves), and are therefore consistent with the X-ray and radio studies of this system on larger scales.

\subsection{PGC044257}

This CCG is one of the rare cases (occurs in $\sim$ 3 per cent of CCGs; Hamer et al.\ 2012) where the CCG is slightly offset from the optical line emission (Johnson et al.\ 2010). The rarity of such offsets points to a large event in cluster evolution, a major cluster merger or possibly a powerful AGN outburst. Whatever the reason for the separation the gas cooling at the X-ray peak will continue and cooled gas will be deposited away from the CCG (Hamer et al.\ 2012). 

IFU observations of this galaxy has not been presented in the literature before this study. The stellar line-of-sight velocity is blue-and redshifted by $\sim$225 km s$^{-1}$ which is higher than derived from long slit spectroscopy (20 $\pm$ 16 km s$^{-1}$ Paper 1).

\begin{figure*}
   \centering
 \mbox{\subfigure[Stellar velocity]{\includegraphics[scale=0.6, trim=1mm 1mm 50mm 1mm, clip]{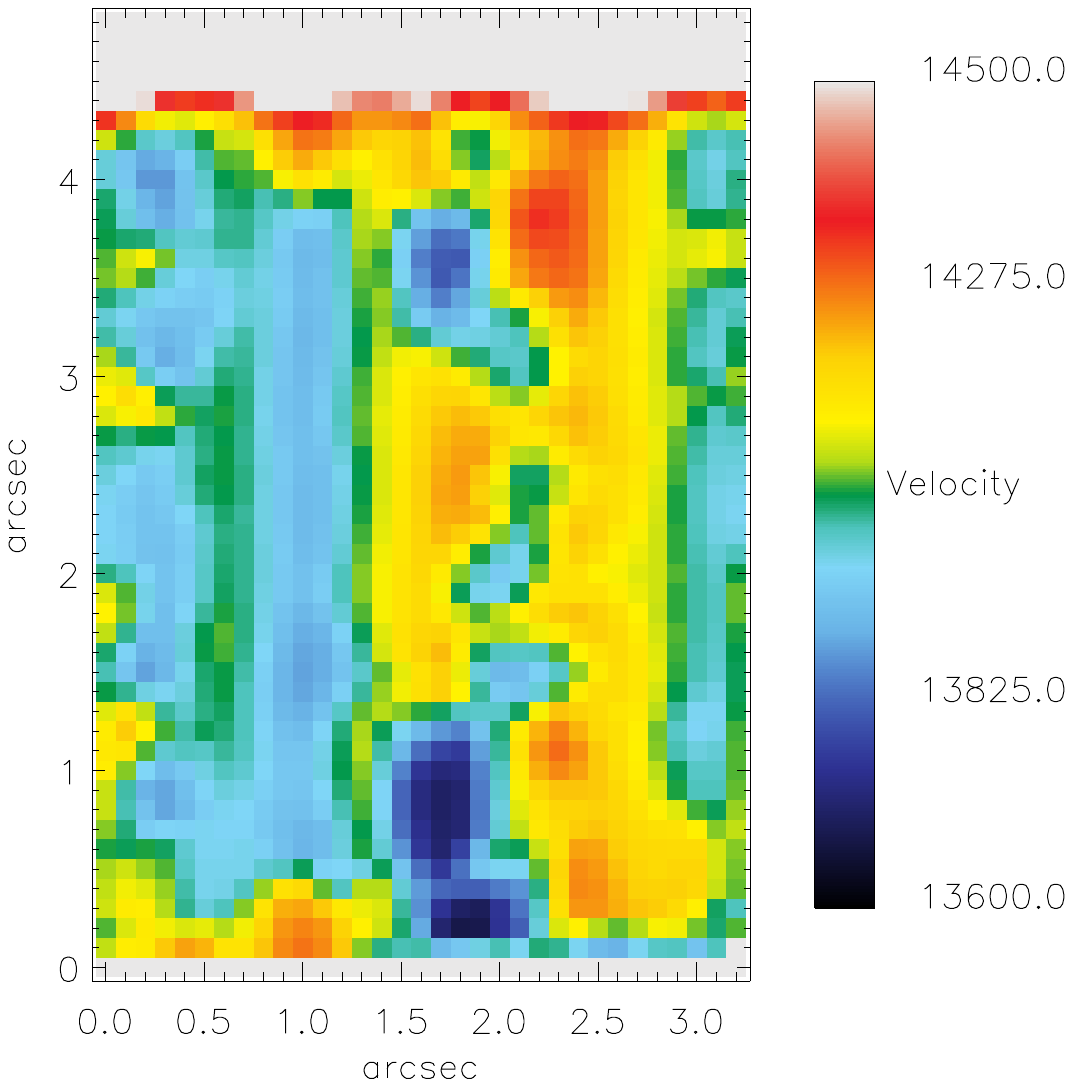}}\quad
\subfigure[Stellar velocity dispersion]{\includegraphics[scale=0.6, trim=1mm 1mm 50mm 1mm, clip]{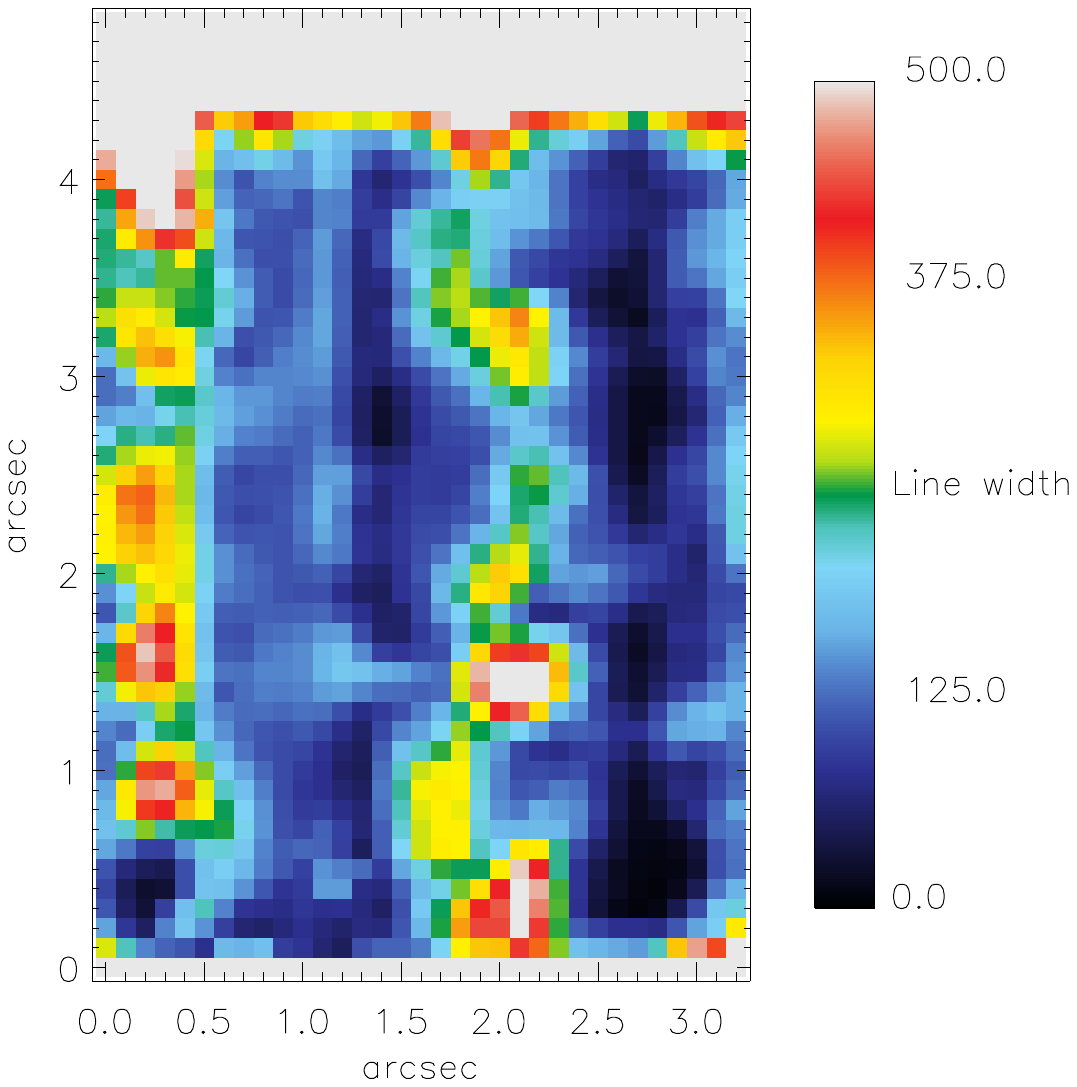}}}
\mbox{\subfigure{\includegraphics[scale=0.25]{PGC044_arrow.pdf}}}
\caption{PGC044257: Velocity and velocity dispersion of the absorption lines in km s$^{-1}$.}
\label{fig:PGC044kinematics} 
\end{figure*}

\begin{figure*}
 \mbox{\subfigure[H$\alpha$ velocity]{\includegraphics[scale=0.5, trim=1mm 1mm 50mm 1mm, clip]{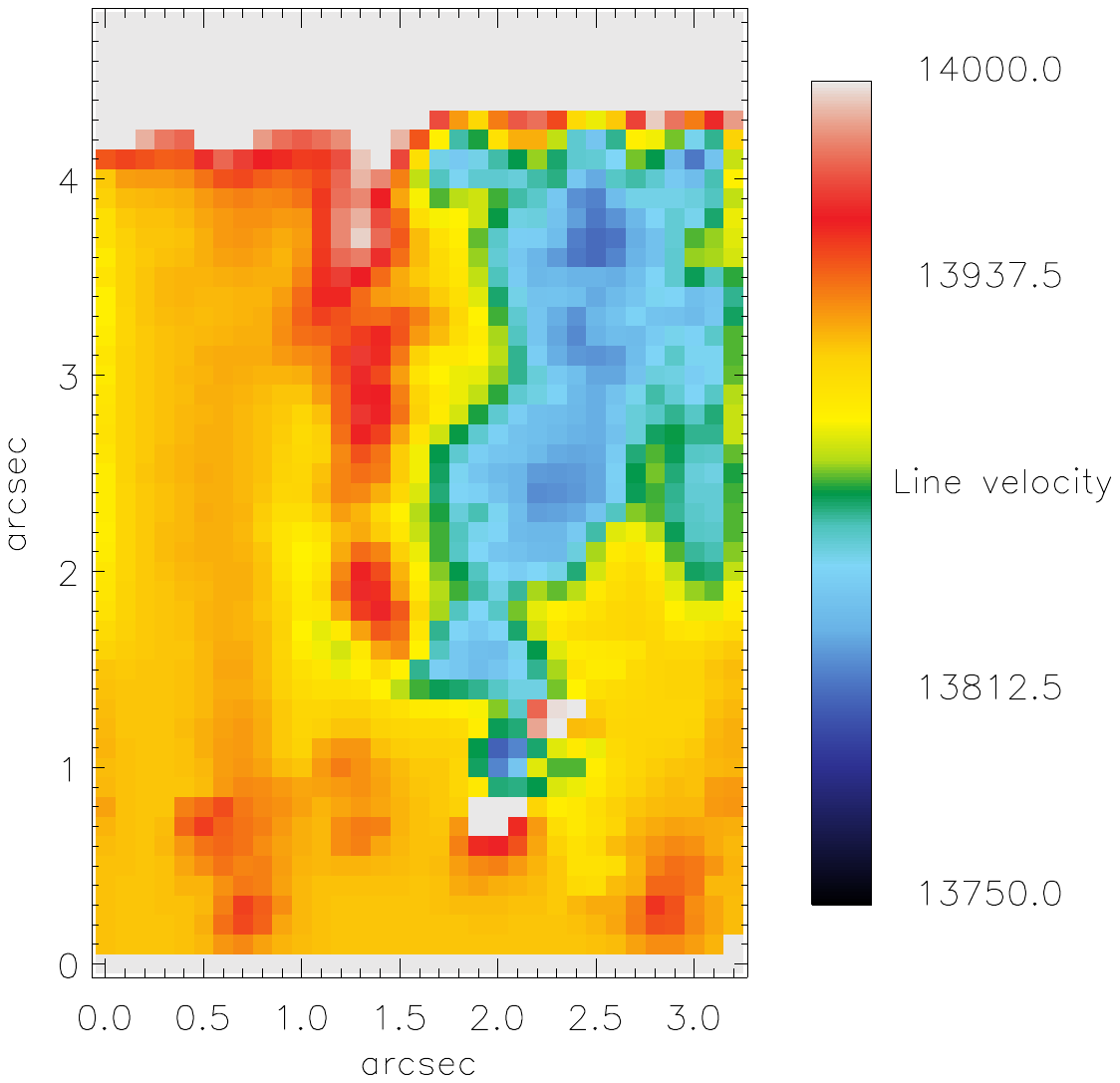}}\quad
\subfigure[{[NII]}$\lambda$6583 velocity]{\includegraphics[scale=0.5, trim=1mm 1mm 50mm 1mm, clip]{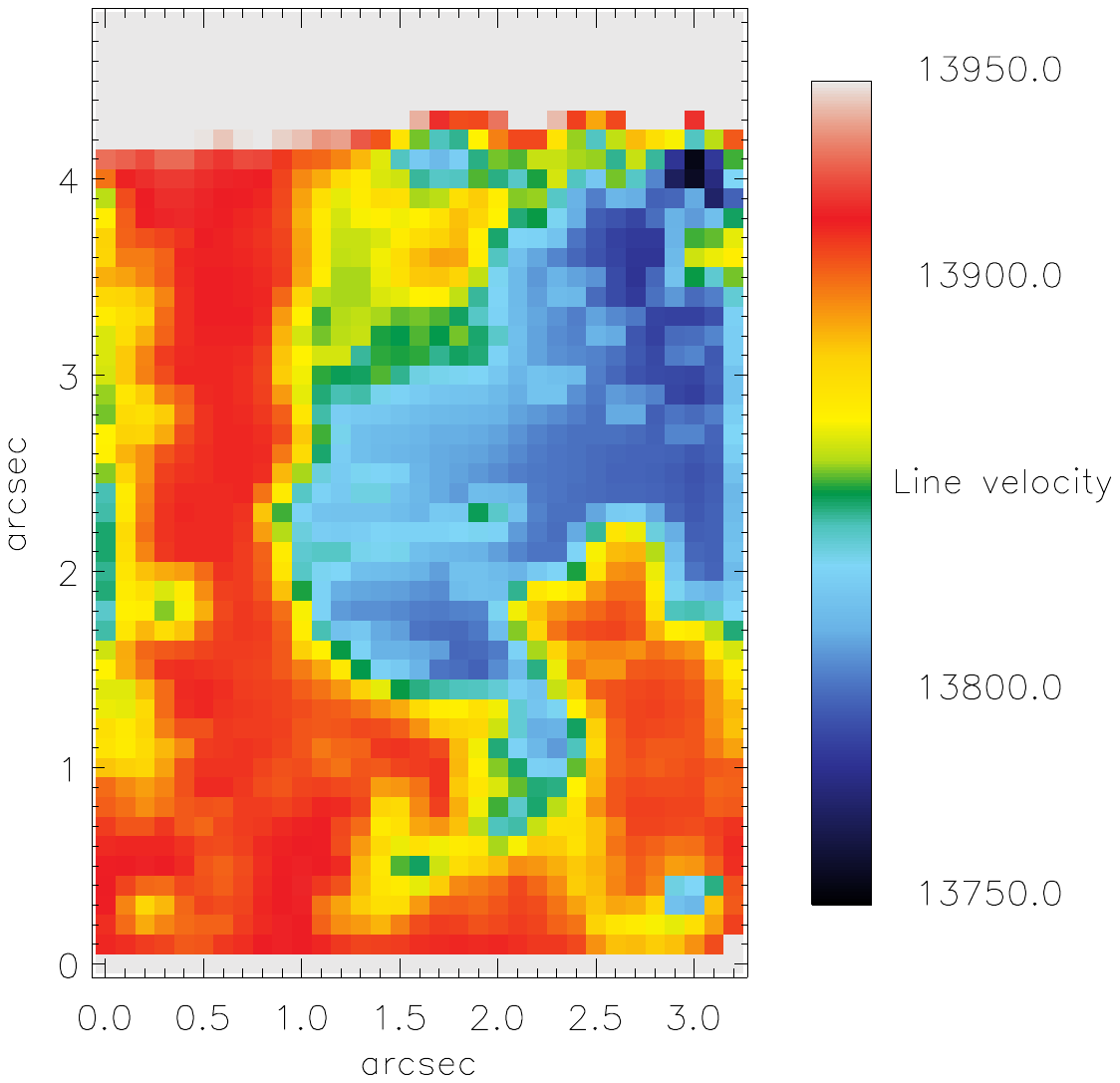}}\quad
\subfigure[{[SII]}$\lambda\lambda$6731,6717 velocity]{\includegraphics[scale=0.5, trim=1mm 1mm 50mm 1mm, clip]{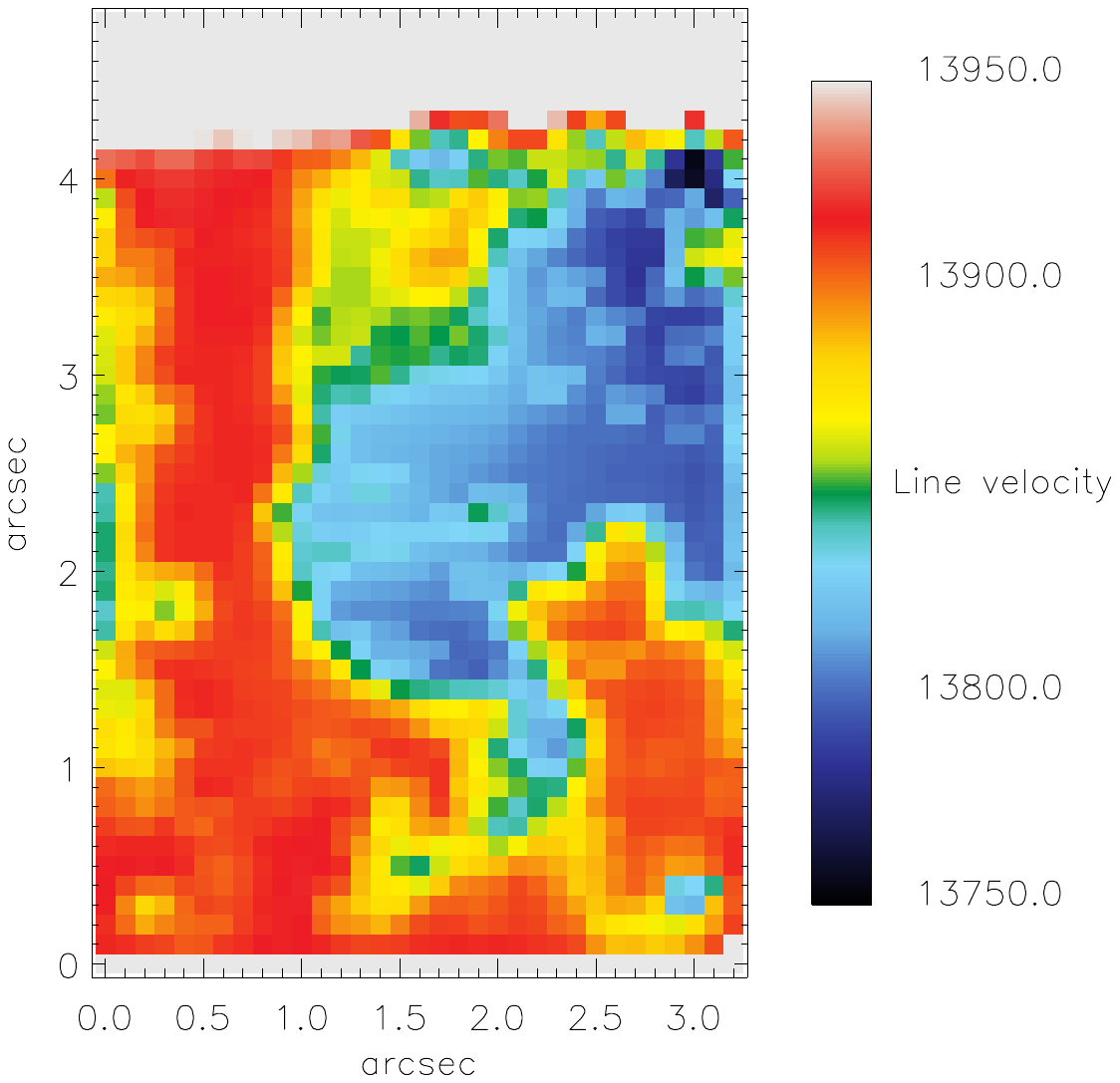}}}
   \mbox{\subfigure[H$\alpha$ line width]{\includegraphics[scale=0.5, trim=1mm 1mm 50mm 1mm, clip]{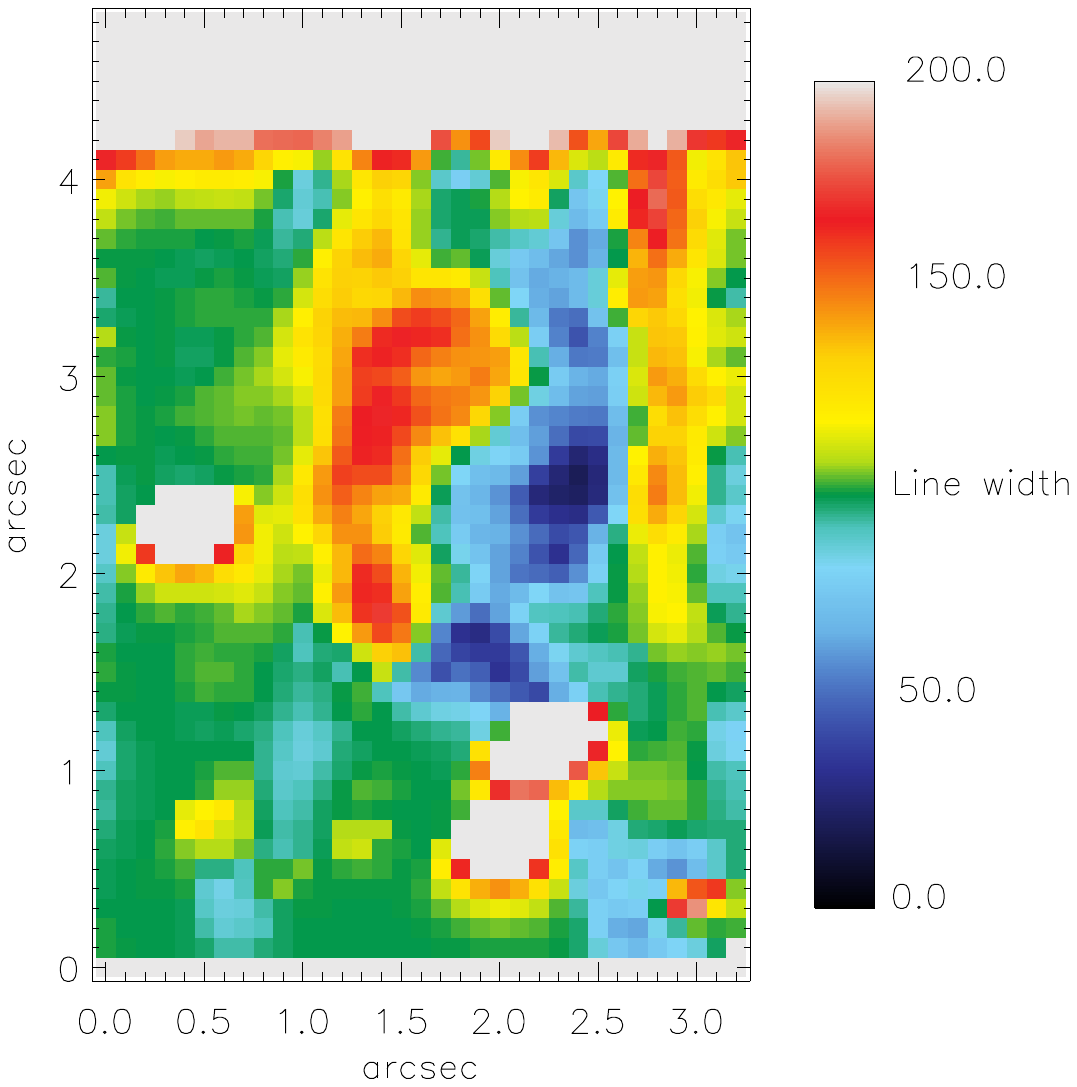}}\quad
\subfigure[{[NII]}$\lambda$6583 line width]{\includegraphics[scale=0.5, trim=1mm 1mm 50mm 1mm, clip]{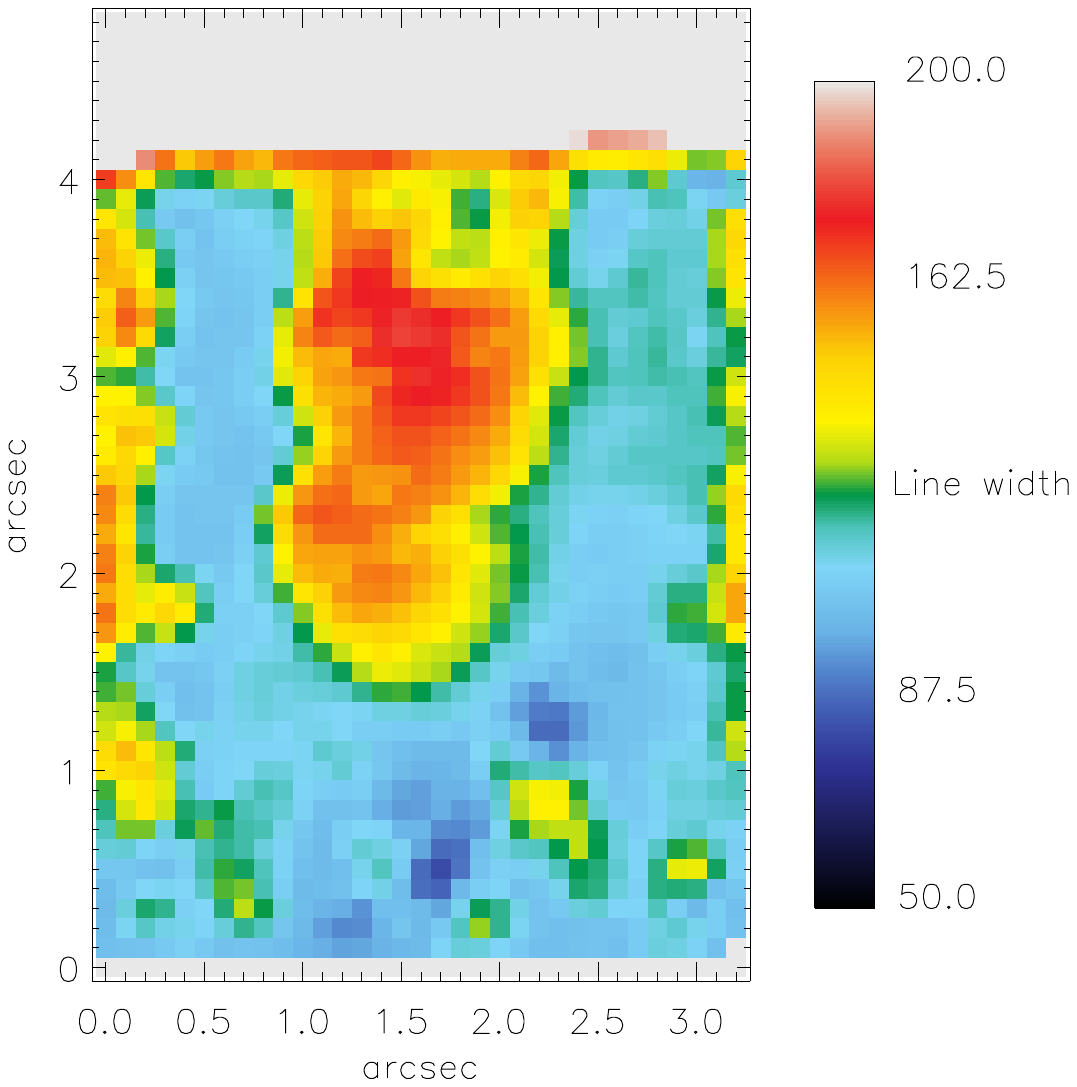}}\quad
\subfigure[{[SII]}$\lambda\lambda$6731,6717 line width]{\includegraphics[scale=0.5, trim=1mm 1mm 50mm 1mm, clip]{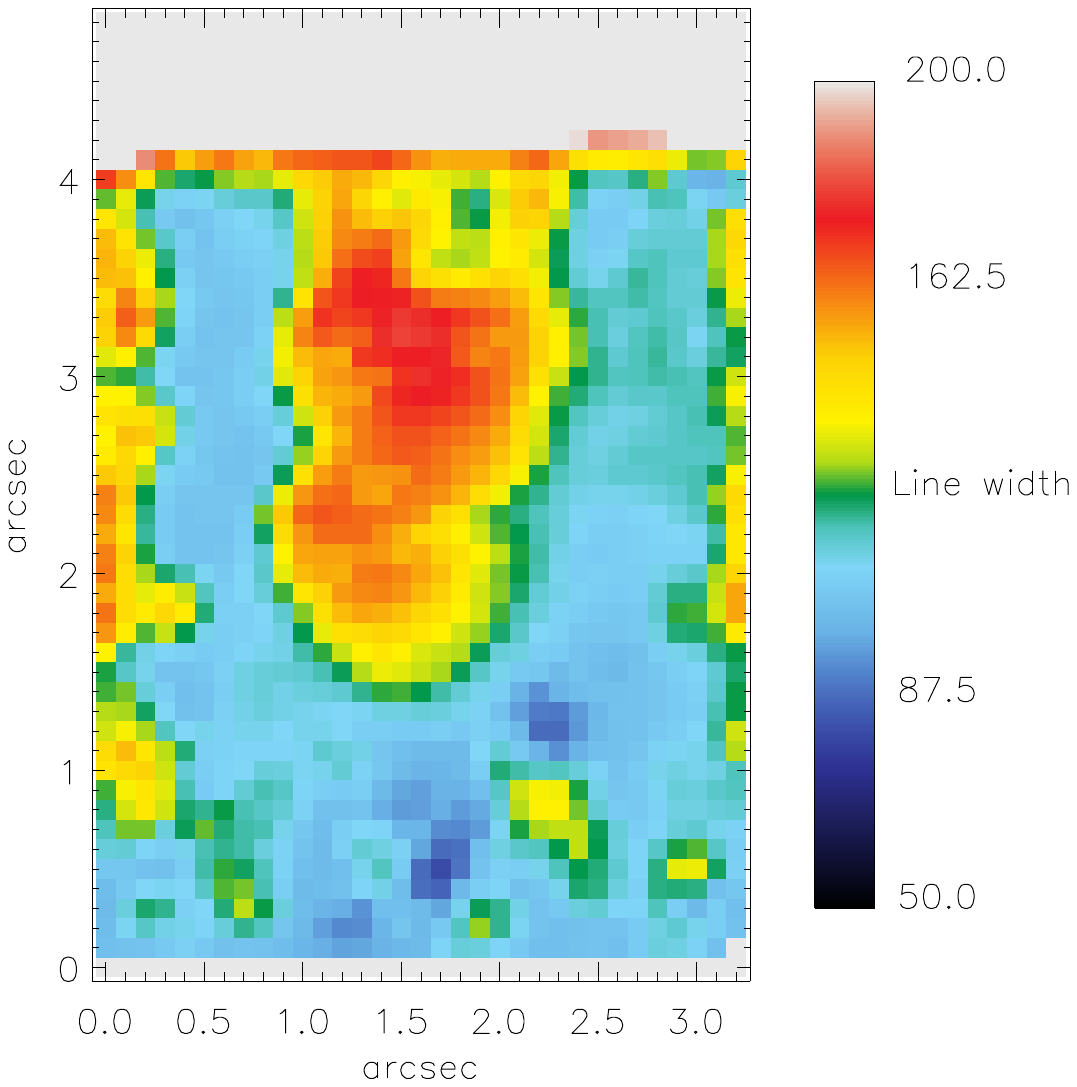}}}
\mbox{\subfigure[H$\alpha$ A/N]{\includegraphics[scale=0.5, trim=1mm 1mm 50mm 1mm, clip]{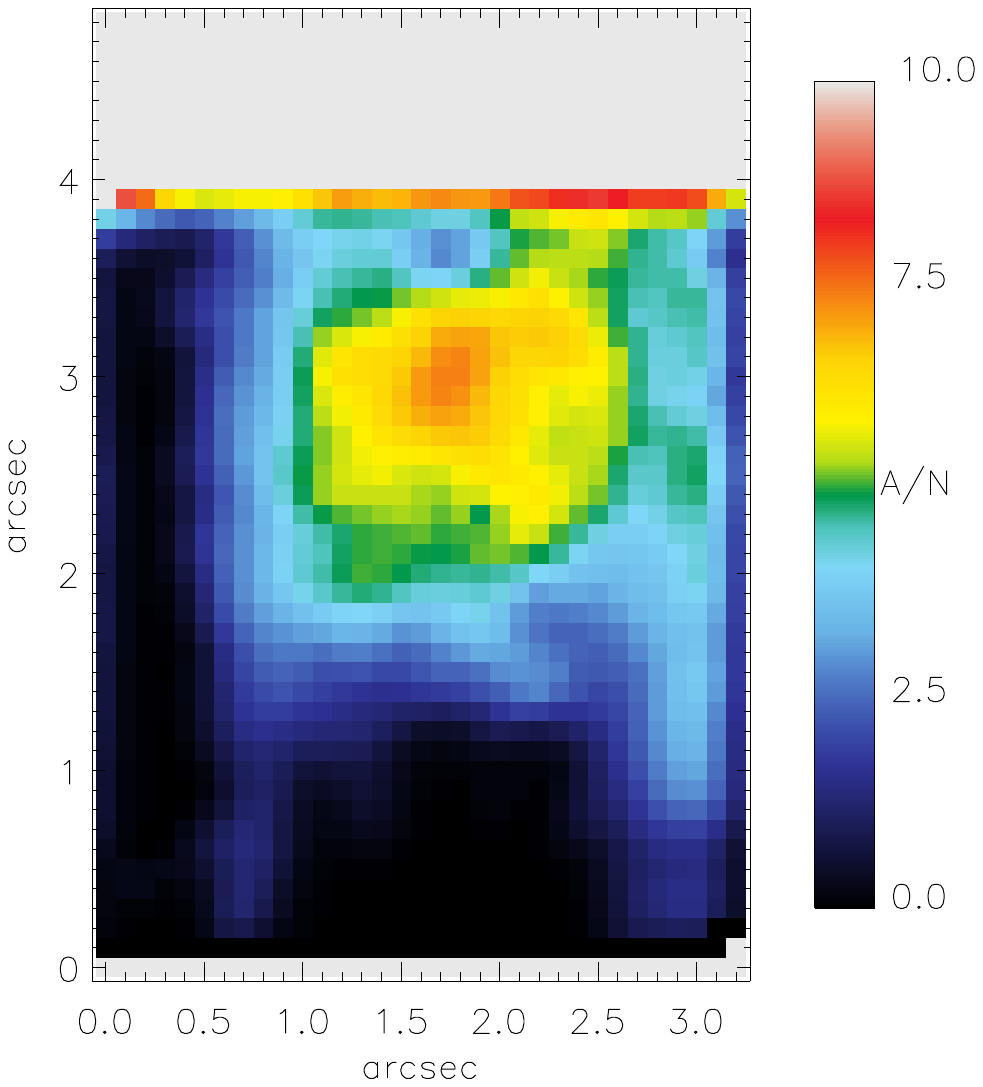}}\quad
\subfigure[{[NII]}$\lambda$6583 line A/N]{\includegraphics[scale=0.5, trim=1mm 1mm 50mm 1mm, clip]{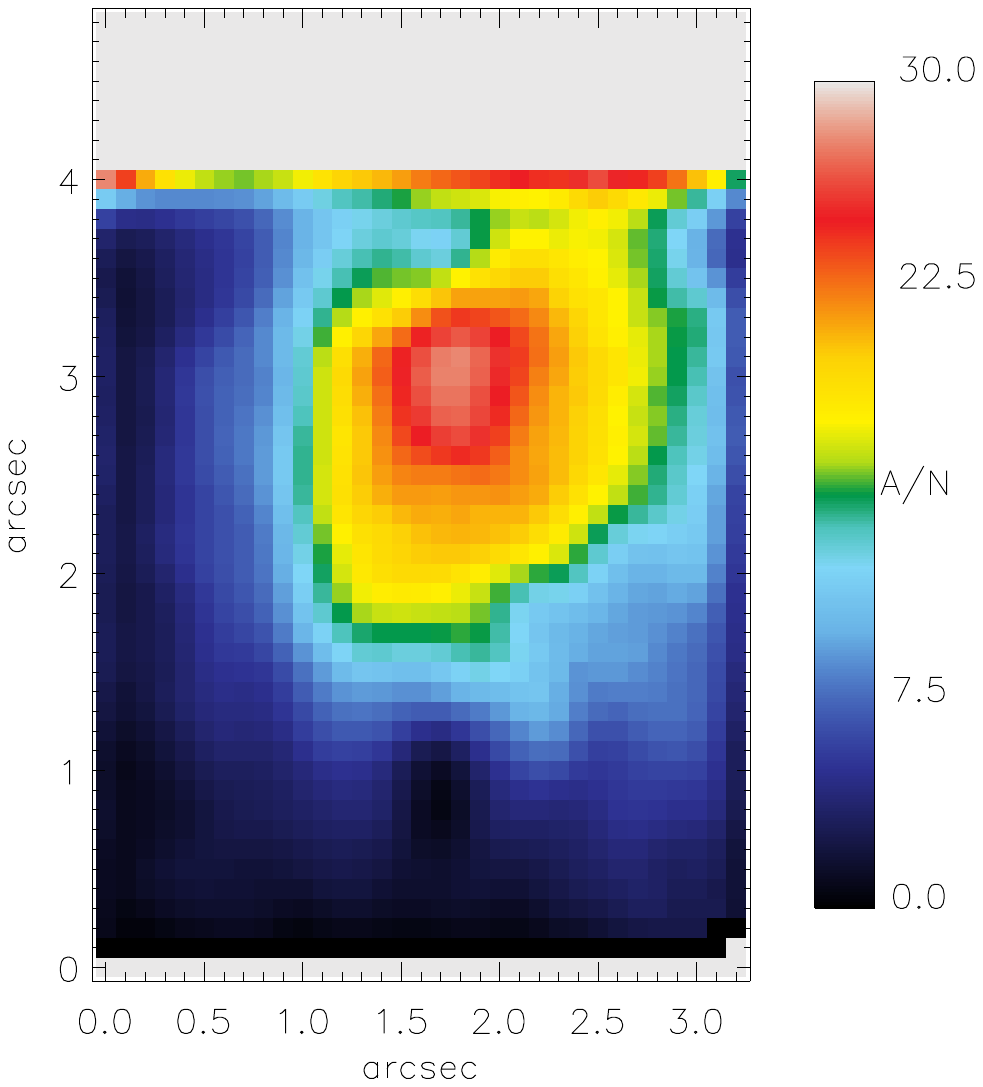}}\quad
\subfigure[{[SII]}$\lambda\lambda$6731,6717 line A/N]{\includegraphics[scale=0.5, trim=1mm 1mm 50mm 1mm, clip]{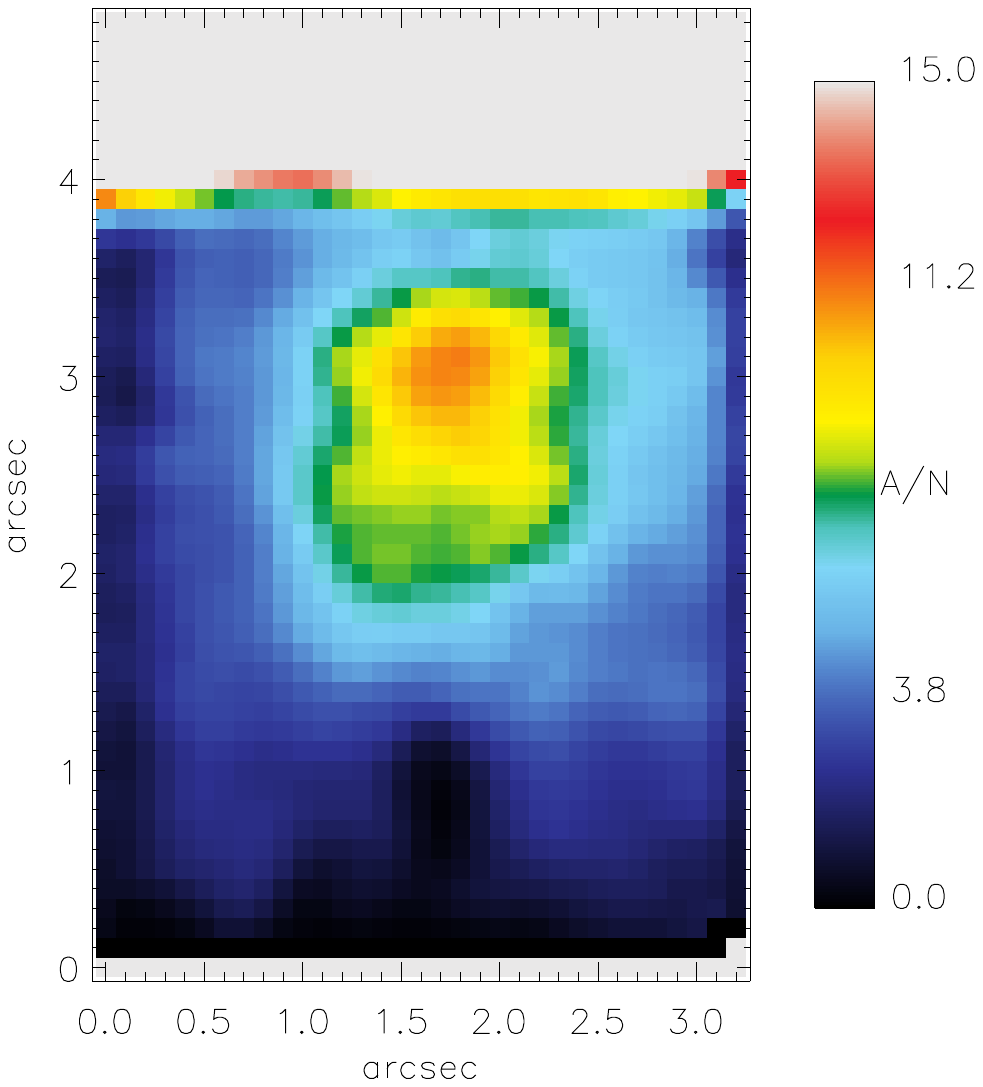}}}
\mbox{\subfigure{\includegraphics[scale=0.28]{PGC044_arrow.pdf}}}
\caption{PGC044257: Velocity (in km s$^{-1}$), line width (in km s$^{-1}$) and A/N of the H$\alpha$, [NII]$\lambda$6583 and [SII]$\lambda\lambda$6731,6717 lines.}
\label{fig:PGC044kinematics2} 
\end{figure*}

We plot the kinematics of the stellar and gaseous components in Figures \ref{fig:PGC044kinematics} and \ref{fig:PGC044kinematics2}. Both the stellar and gaseous components show morphologies (Figures \ref{fig:PGC044kinematics} and \ref{fig:PGC044kinematics2}) that are elongated and aligned. However the recession velocity of the gas is $\sim$ 200 km s$^{-1}$, which is higher than that of the stars, suggesting that the gaseous and stellar components are decoupled. The rotation of the warm gas agrees with the long-slit observations by McDonald et al.\ (2012).

\begin{figure*}
 \mbox{\subfigure[{[OIII]}$\lambda$5007 velocity]{\includegraphics[scale=0.5, trim=1mm 1mm 50mm 1mm, clip]{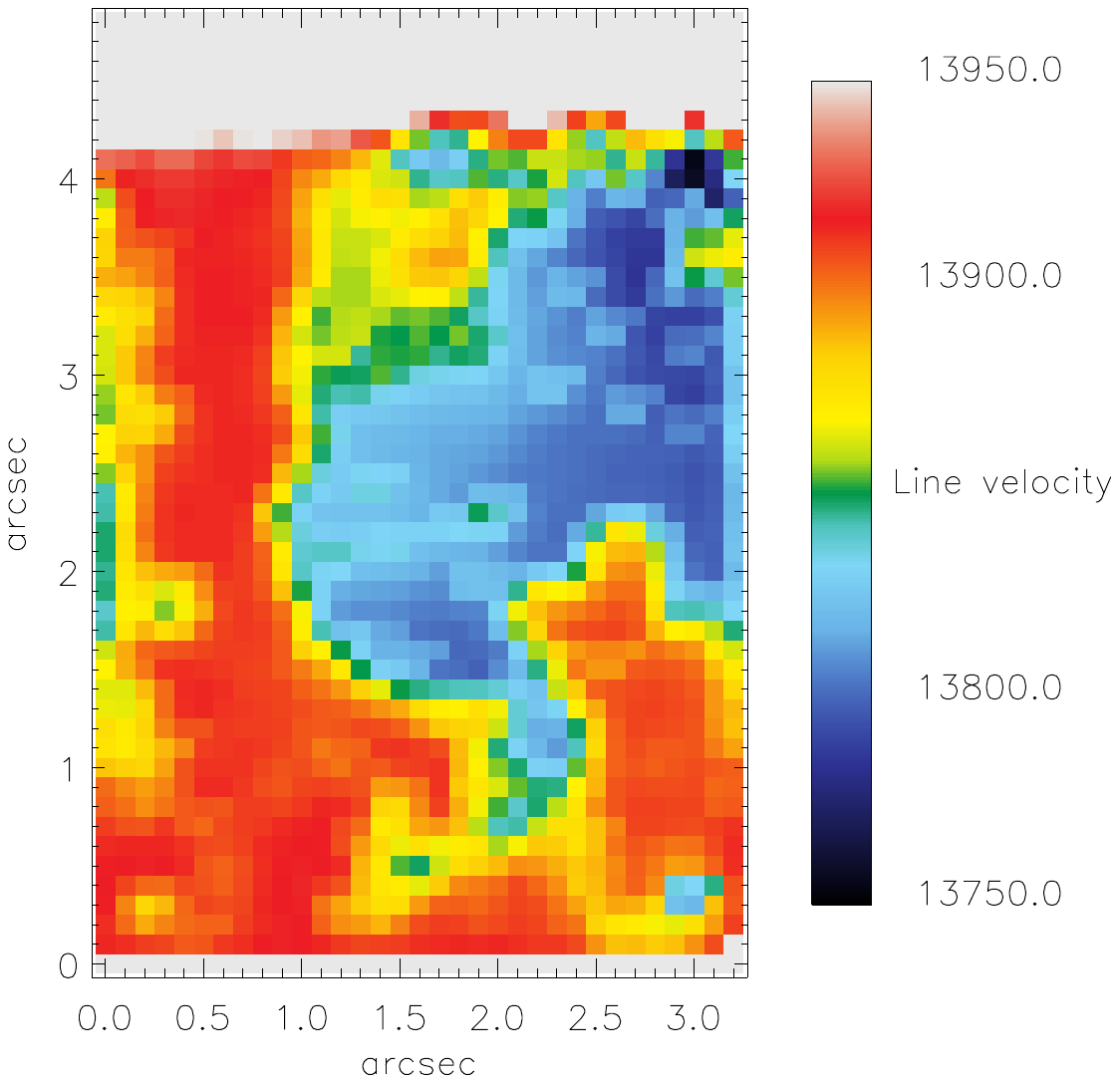}}\quad
\subfigure[{[OI]}$\lambda$6300 velocity]{\includegraphics[scale=0.5, trim=1mm 1mm 50mm 1mm, clip]{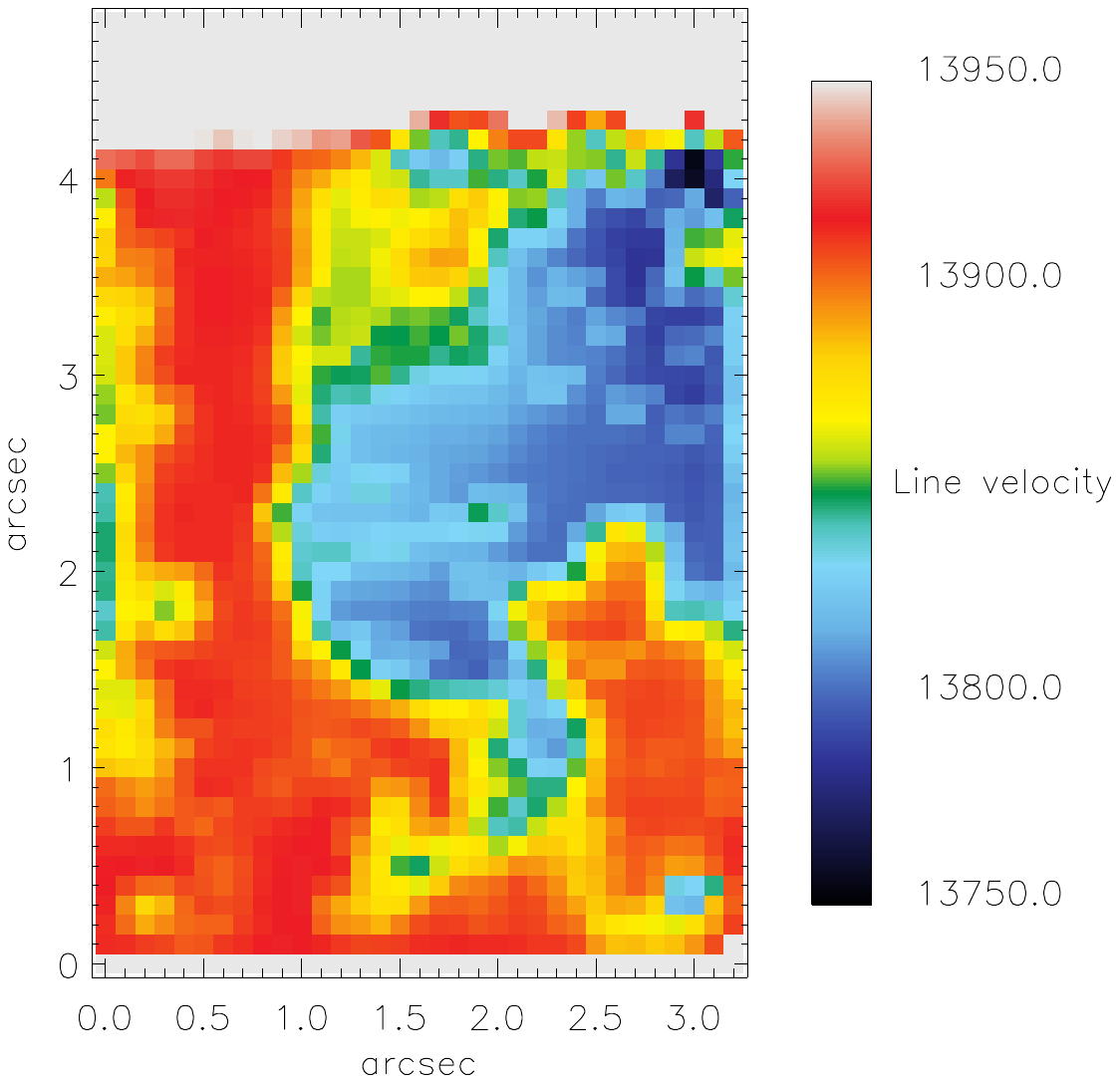}}}
   \mbox{\subfigure[{[OIII]}$\lambda$5007 line width]{\includegraphics[scale=0.5, trim=1mm 1mm 50mm 1mm, clip]{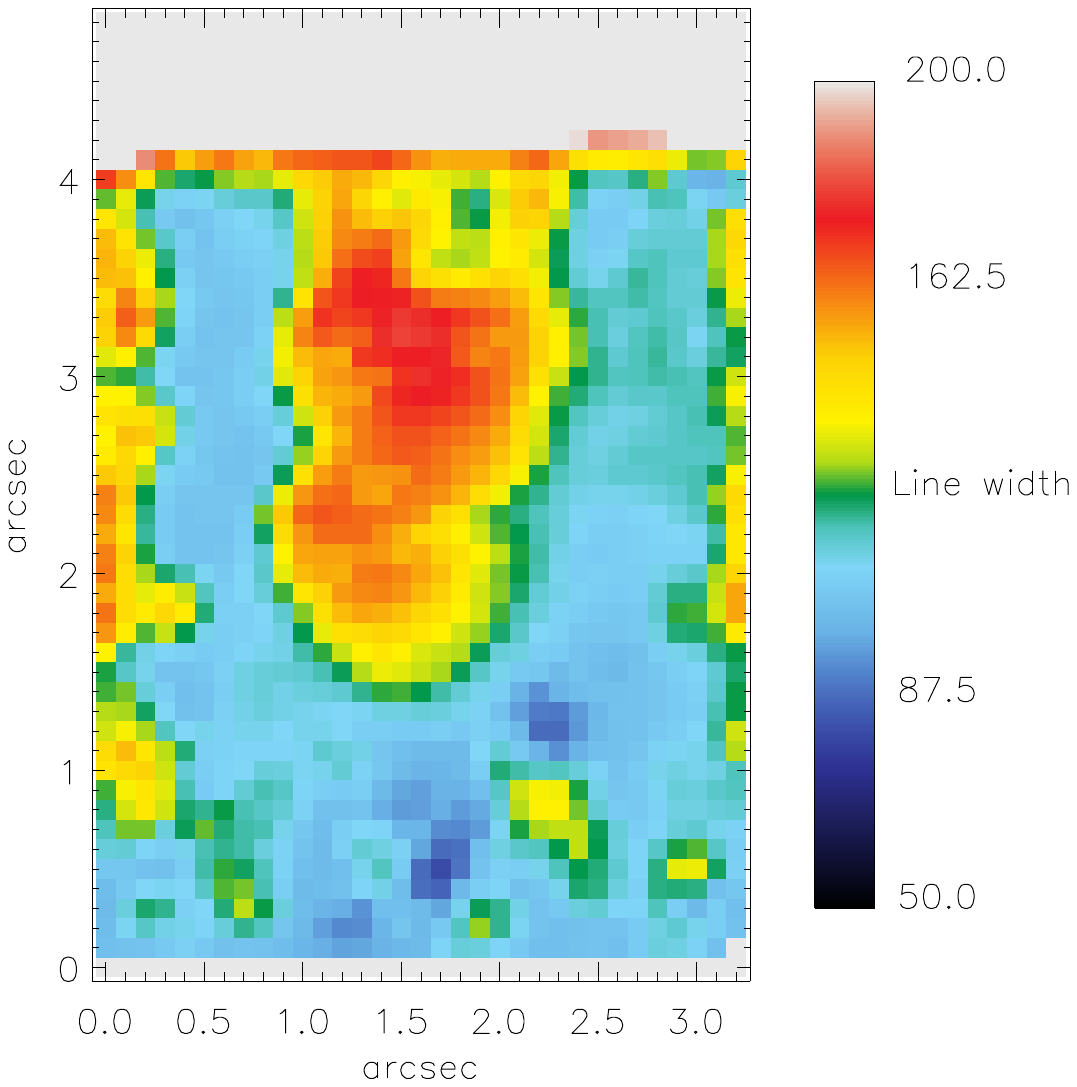}}\quad
\subfigure[{[OI]}$\lambda$6300 line width]{\includegraphics[scale=0.5, trim=1mm 1mm 50mm 1mm, clip]{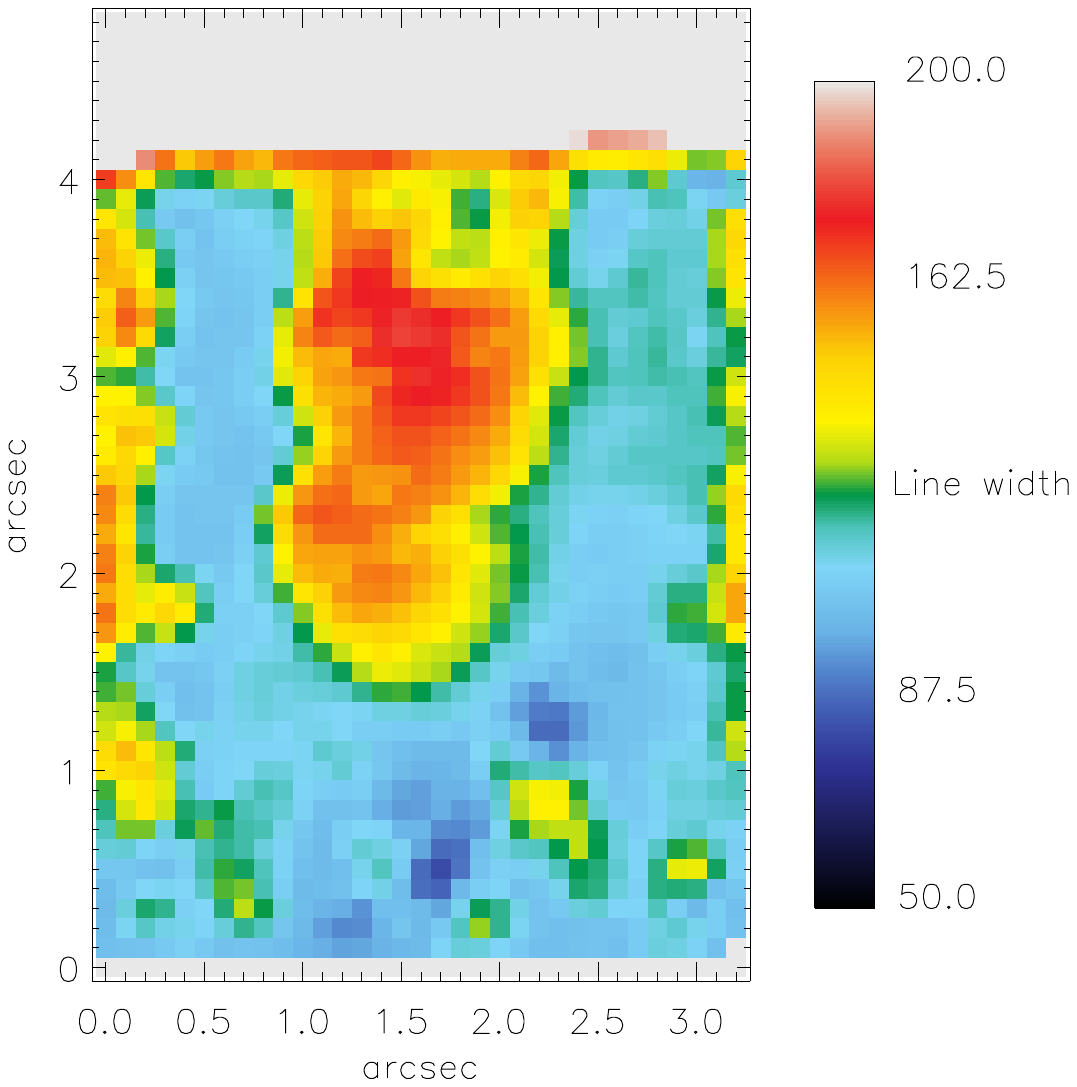}}}
\mbox{\subfigure[{[OIII]}$\lambda$5007 A/N]{\includegraphics[scale=0.5, trim=1mm 1mm 50mm 1mm, clip]{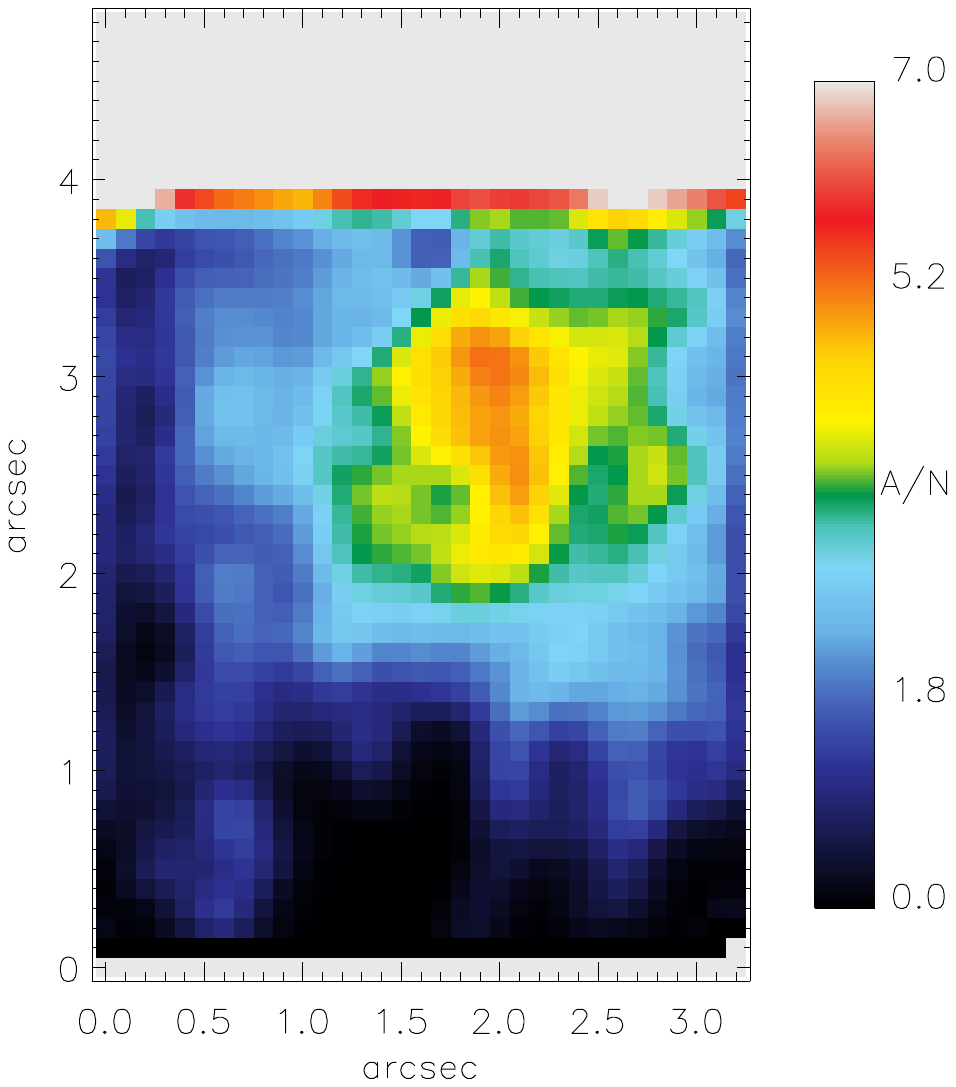}}\quad
\subfigure[{[OI]}$\lambda$6300 line A/N]{\includegraphics[scale=0.5, trim=1mm 1mm 50mm 1mm, clip]{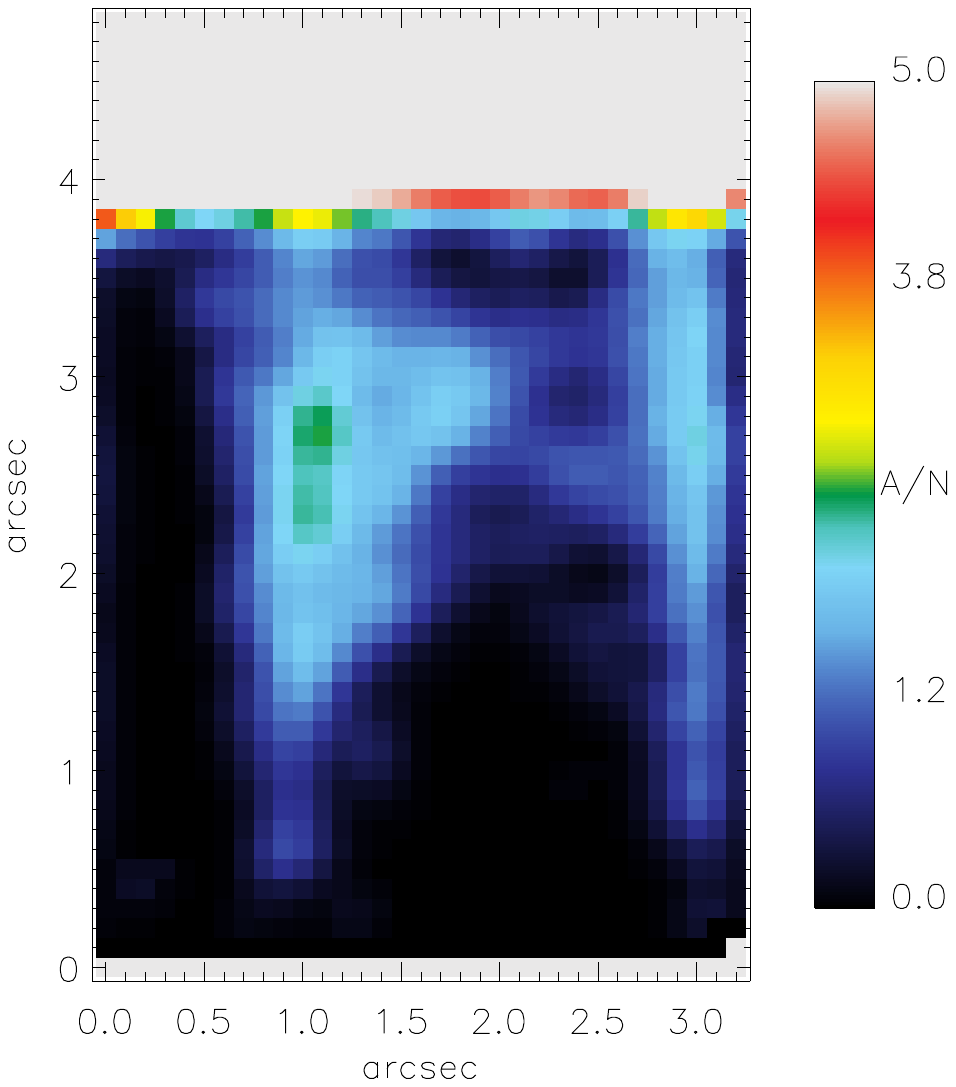}}}
\mbox{\subfigure{\includegraphics[scale=0.28]{PGC044_arrow.pdf}}}
\caption{PGC044257: Velocity (in km s$^{-1}$), line width (in km s$^{-1}$) and A/N of the [OIII]$\lambda$5007 and [OI]$\lambda$6300 lines.}
\label{fig:PGC044kinematics3} 
\end{figure*}

\begin{figure*}
   \centering
   \includegraphics[scale=0.8]{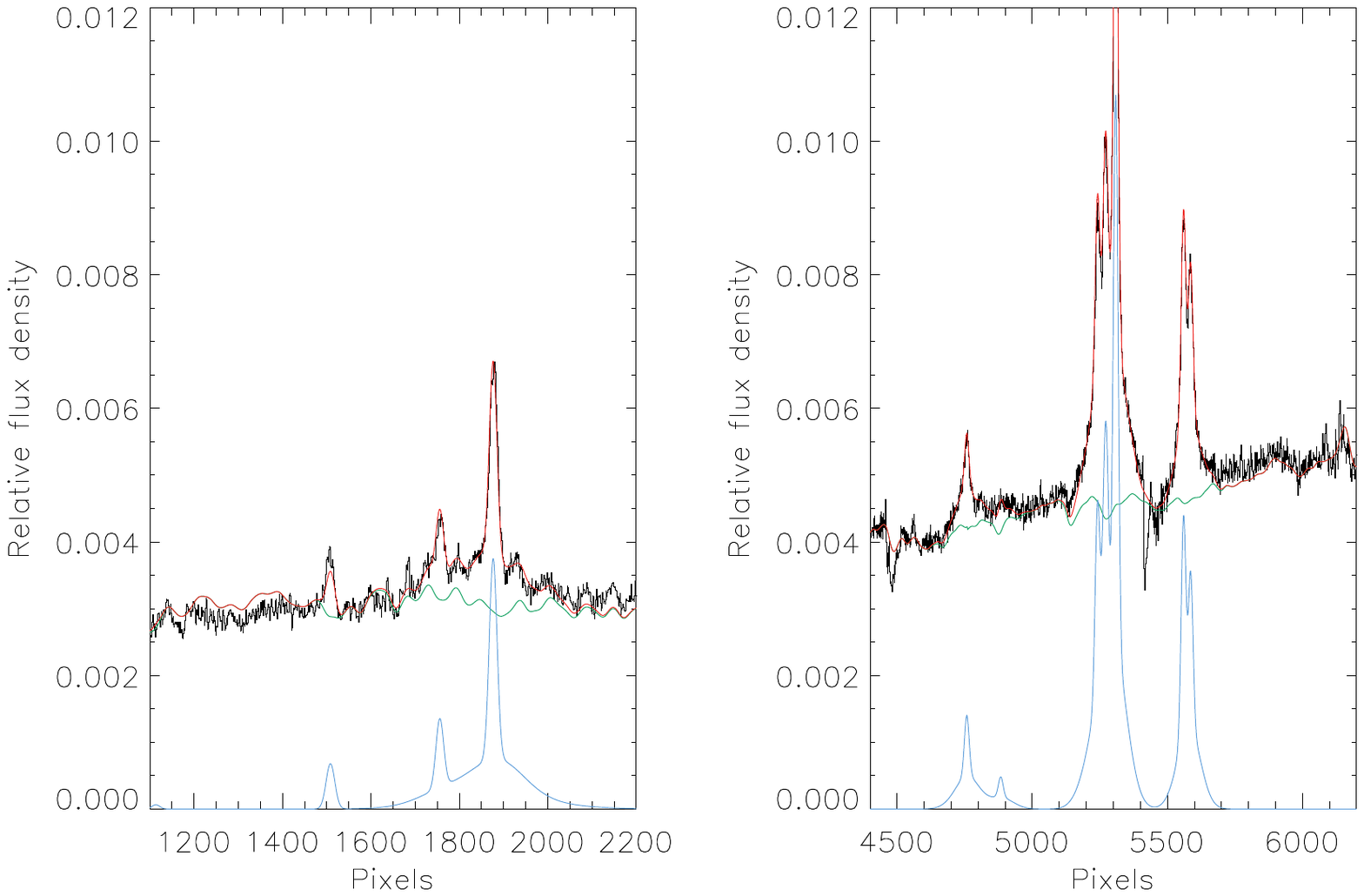}
   \caption{A random spectrum in the central region of UGC09799. The red line indicates the best-fitting stellar template and Gaussians at the emission lines, and the green line indicates the best-fitting stellar templates with the emission lines subtracted. The blue line indicates the relative flux for the measured emission lines. The lines are (from left to right) in the left plot: H$\beta$, and the [OIII] doublet, and in the right plot: [OI] doublet, [NII], H$\alpha$, [NII], and the [SII] doublet. The sharp feature between the [NII] and [SII] lines is one of the two CCD chip gaps and was masked during the pPXF and GANDALF fitting processes.}
   \label{fig:UGC_gandalf}
\end{figure*}

We plot H$\alpha$, [NII], [SII], [OIII] and [OI] velocities, line width and A/N in Figure \ref{fig:PGC044kinematics2} and \ref{fig:PGC044kinematics3}. To the degree that our spatial resolution reveals, it appears that all the optical forbidden and hydrogen recombination lines originate in the same gas.
 
We find the H$\alpha$ flux to be uniform in Figure \ref{Haflux} suggesting that all of the gas had the same origin, but the H$\alpha$ flux is also quite low. If Figure \ref{Haflux} is compared to Figure \ref{fig:PGC044kinematics2}, then it can be seen that some structure is visible where the A/N of H$\alpha$ is the highest. The H$\alpha$ appear to be quite core-dominated. McDonald et al.\ (2010) found that the direction of the H$\alpha$ filaments (from narrow band imaging) correlate with the position of nearby galaxies. The filaments in Abell 1644 also appear to be star-forming (McDonald et al.\ 2011). The star forming filaments that McDonald et al.\ (2011) detected are on a bigger scale than the region we probe in this study. Our results show that the very centre of the system display properties of a LINER (see Section \ref{ionisation}), and therefore both mechanisms may be part of this system at different scales.

\subsection{UGC09799}

IFU observations of this galaxy (around H$\alpha$ and H$\beta$) was previously presented by Edwards et al.\ (2009). The observations presented in this study are improved in that it was observed with triple the integration time of the previous observations. The morphology of the continuum emission of this galaxy (see Figure \ref{fig:Thumbnails}) is centrally concentrated and condensed, as was also found by Edwards et al.\ (2009).

\begin{figure*}
  \centering
 \mbox{\subfigure[Single Gaussians]{\includegraphics[scale=0.5]{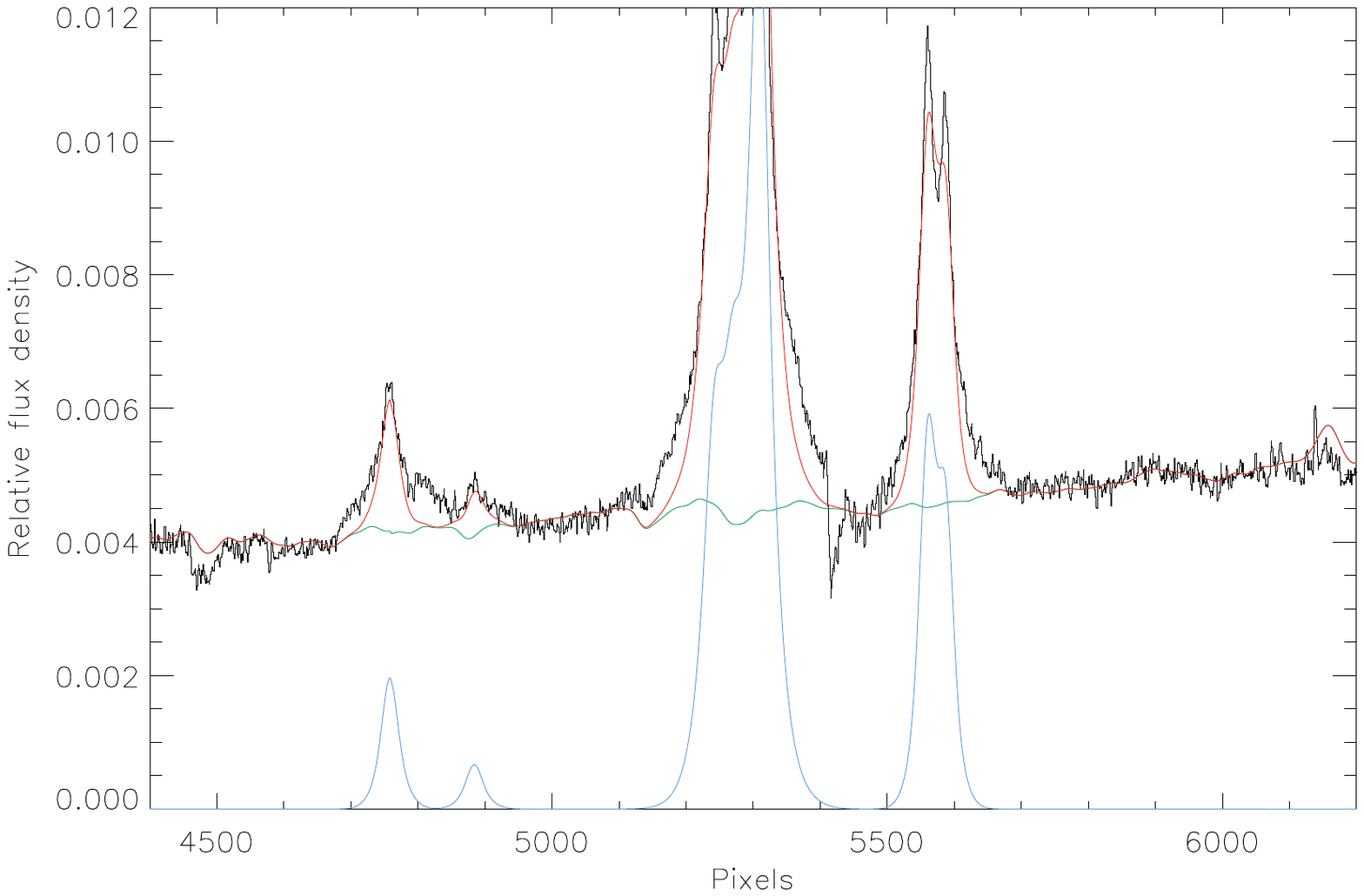}}\quad
         \subfigure[Two Gaussians]{\includegraphics[scale=0.5]{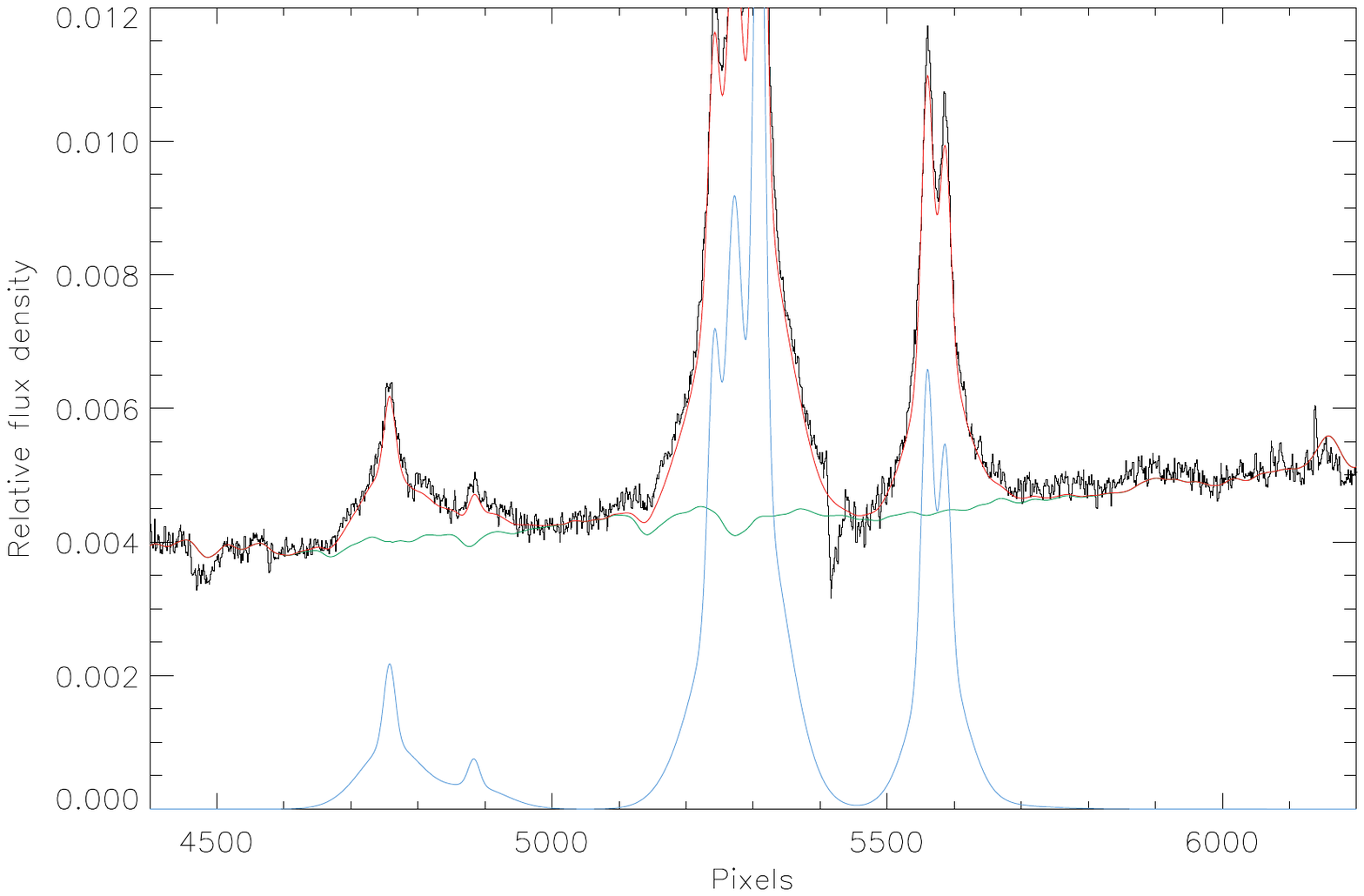}}}
 \caption{UGC09799 showing single Gaussian fits and double Gaussian fits (one narrow and one broad) to the [OI] doublet, [NII], H$\alpha$, [NII], and the [SII] doublet.}
   \label{fig:UGC_before_after}
\end{figure*}

Just fitting one single Gaussian per emission line resulted in very poor fits (as shown in Figure \ref{fig:UGC_before_after}). We also tried fitting Voigt profiles to the [OIII] lines, for example, as well as slightly offset blue and red velocity wings. The best fits were consistently achieved using 2 Gaussians for each individual line in the [OIII], [OI], [NII] and [SII] doublets, of which one of the Gaussians were broader than the other. The H$\beta$ lines (and therefore also H$\alpha$) however required no additional Gaussian to achieve a good fit. Edwards et al.\ (2009) could not detect H$\beta$ emission above the 1$\sigma$ level for this galaxy. We do detect H$\beta$ above this level and plot the flux ratios in Figure \ref{Haflux}. Figure 16 in Edwards et al.\ (2009) can be compared with Figures \ref{fig:UGCkinematics2} and \ref{Haflux} here -- both show smooth centrally condensed emission.

\begin{figure*}
   \centering
 \mbox{\subfigure[Stellar velocity]{\includegraphics[scale=0.6, trim=1mm 1mm 50mm 1mm, clip]{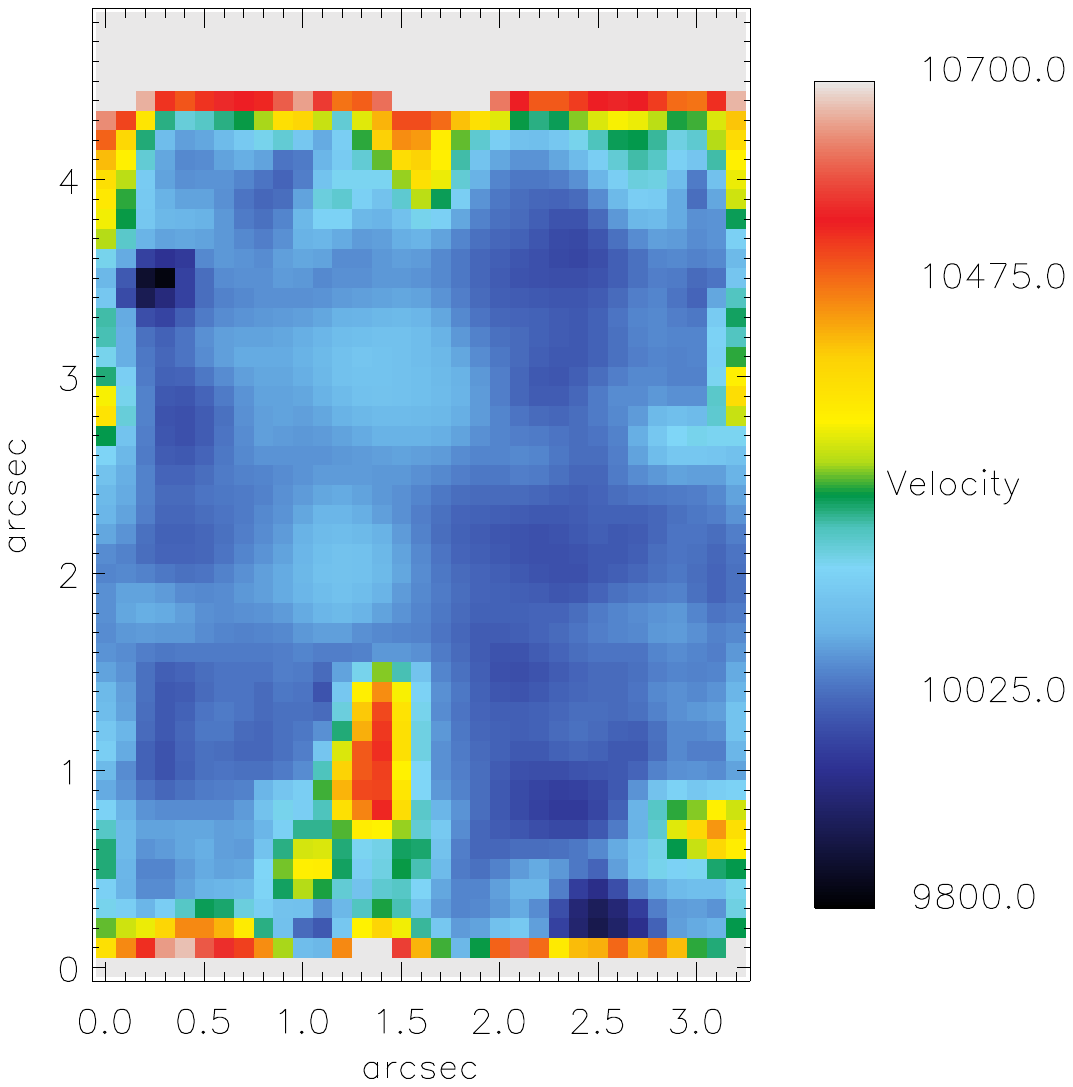}}\quad
\subfigure[Stellar velocity dispersion]{\includegraphics[scale=0.6, trim=1mm 1mm 50mm 1mm, clip]{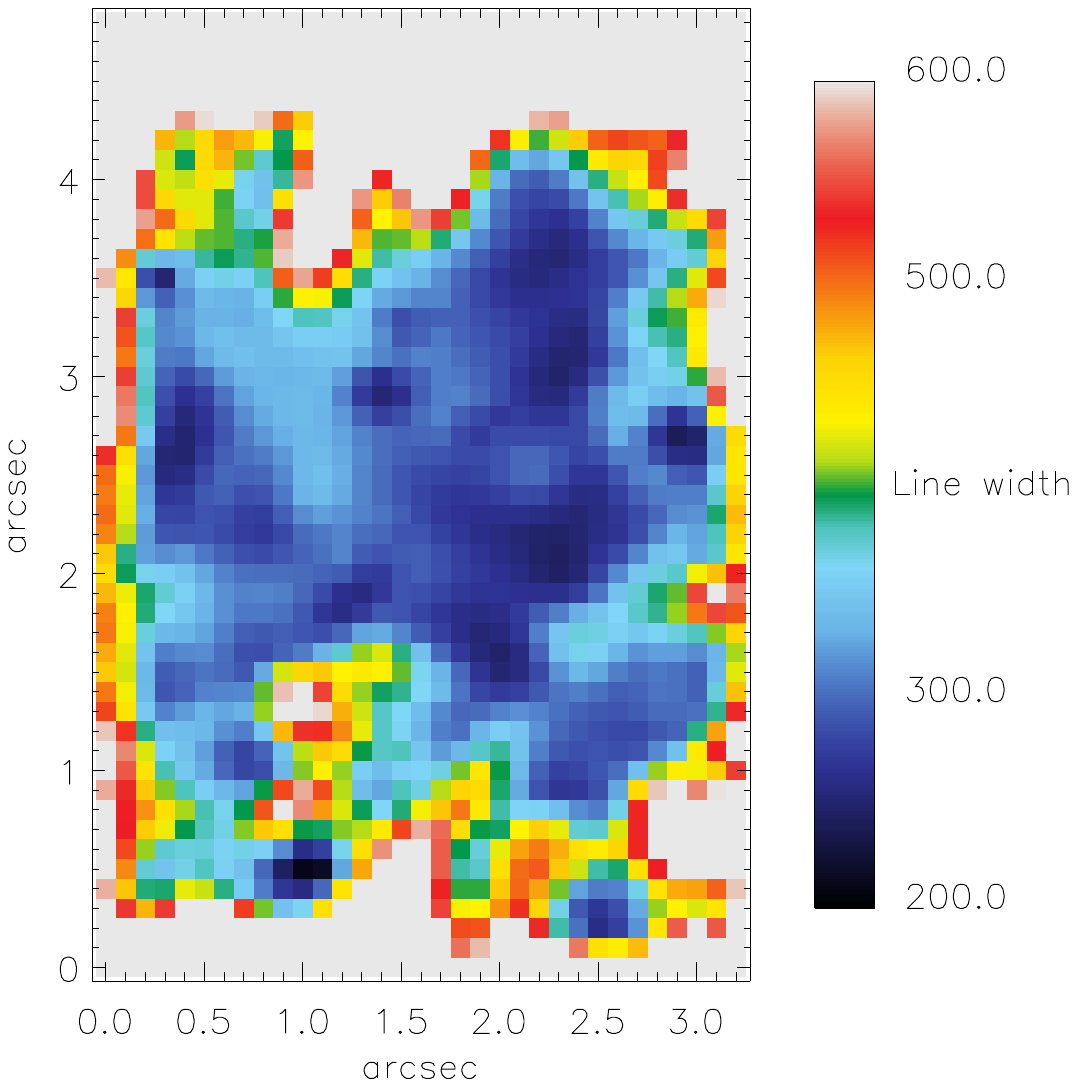}}}
\mbox{\subfigure{\includegraphics[scale=0.25]{UGC_arrow.pdf}}}
\caption{UGC09799: Velocity and velocity dispersion of the absorption lines in km s$^{-1}$.}
\label{fig:UGCkinematics} 
\end{figure*}

\begin{figure*}
 \mbox{\subfigure[H$\alpha$ velocity]{\includegraphics[scale=0.5, trim=1mm 1mm 50mm 1mm, clip]{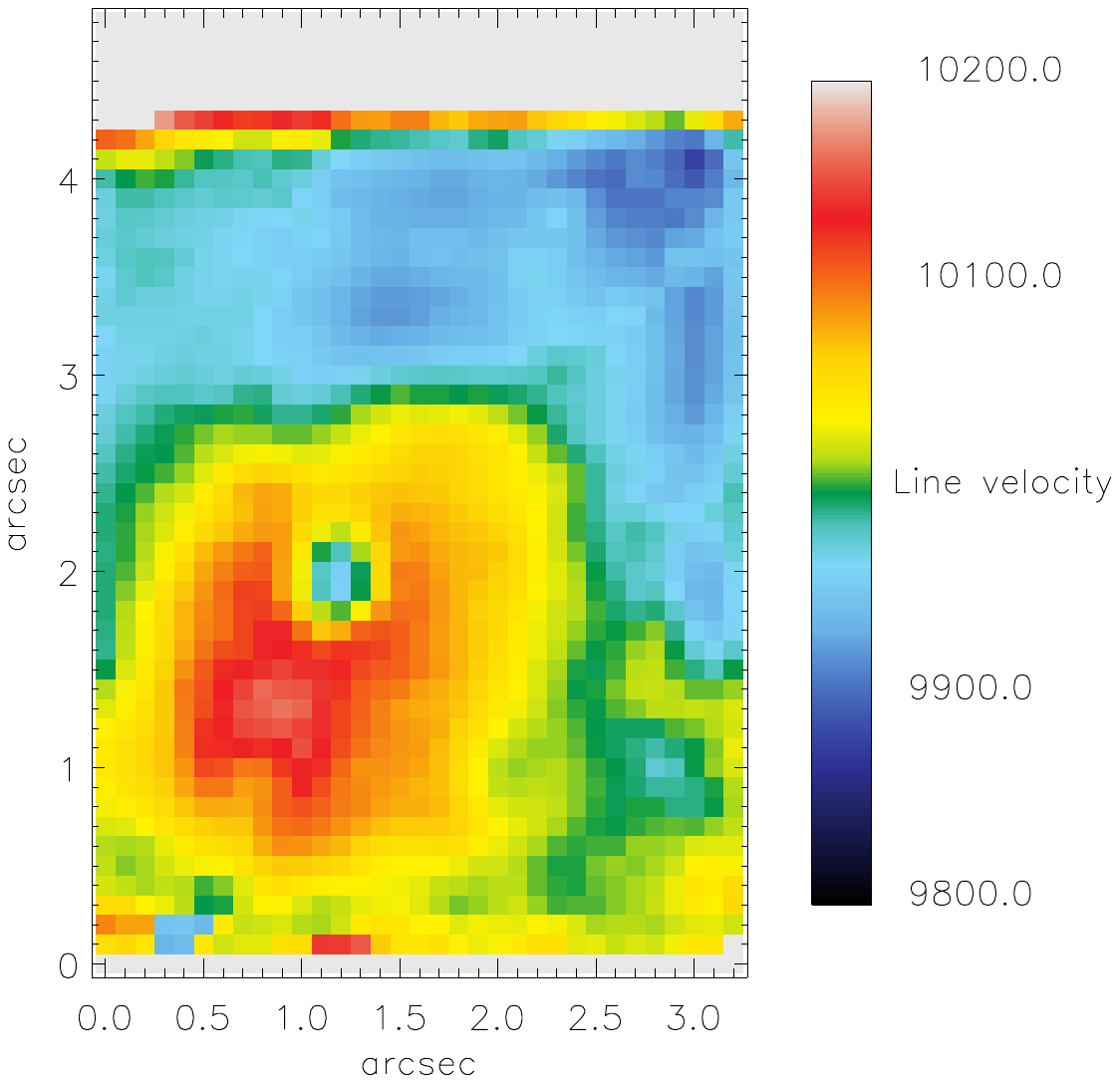}}\quad
\subfigure[{[NII]}$\lambda$6583 velocity]{\includegraphics[scale=0.5, trim=1mm 1mm 50mm 1mm, clip]{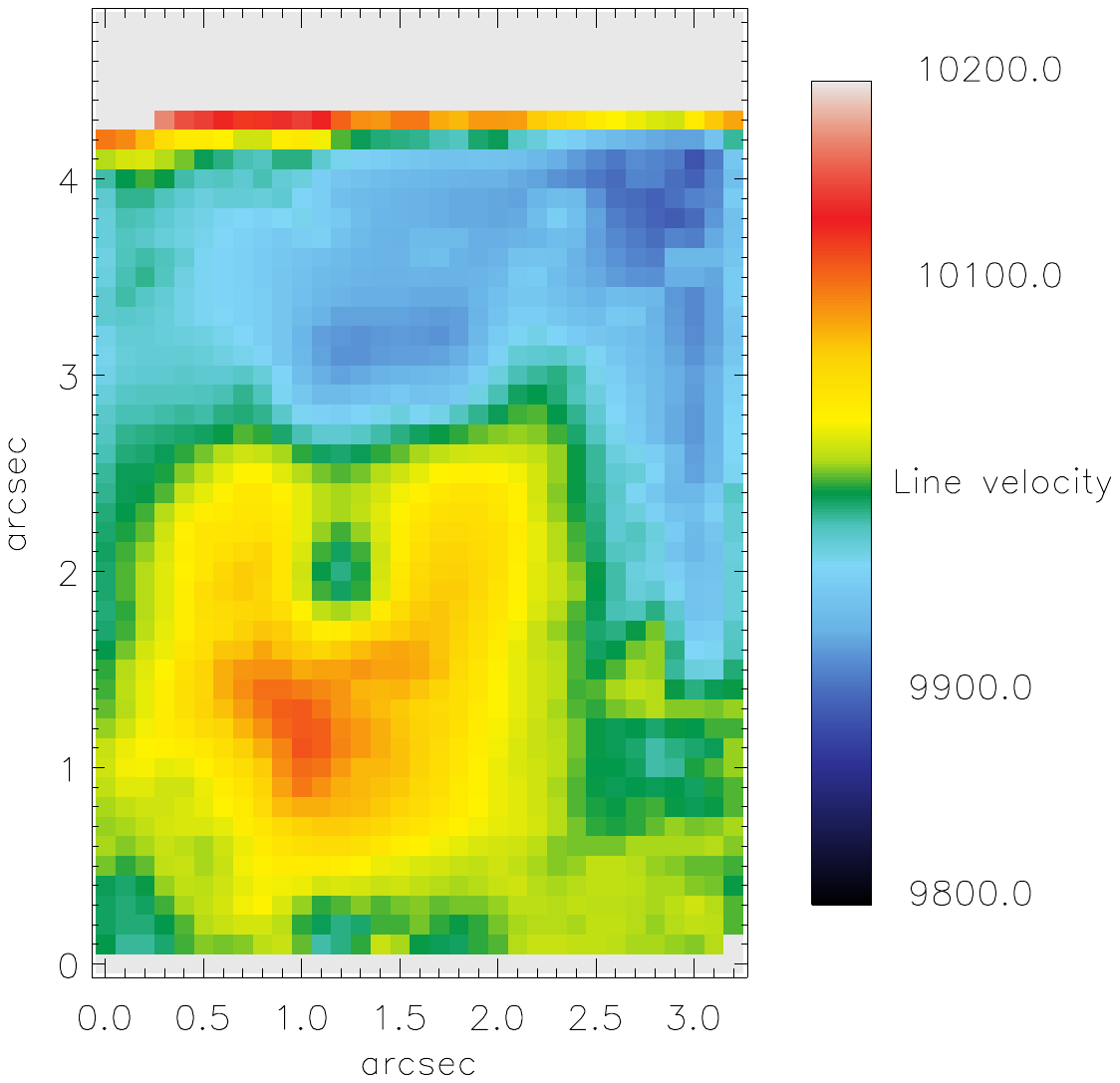}}\quad
\subfigure[{[SII]}$\lambda\lambda$6731,6717 velocity]{\includegraphics[scale=0.5, trim=1mm 1mm 50mm 1mm, clip]{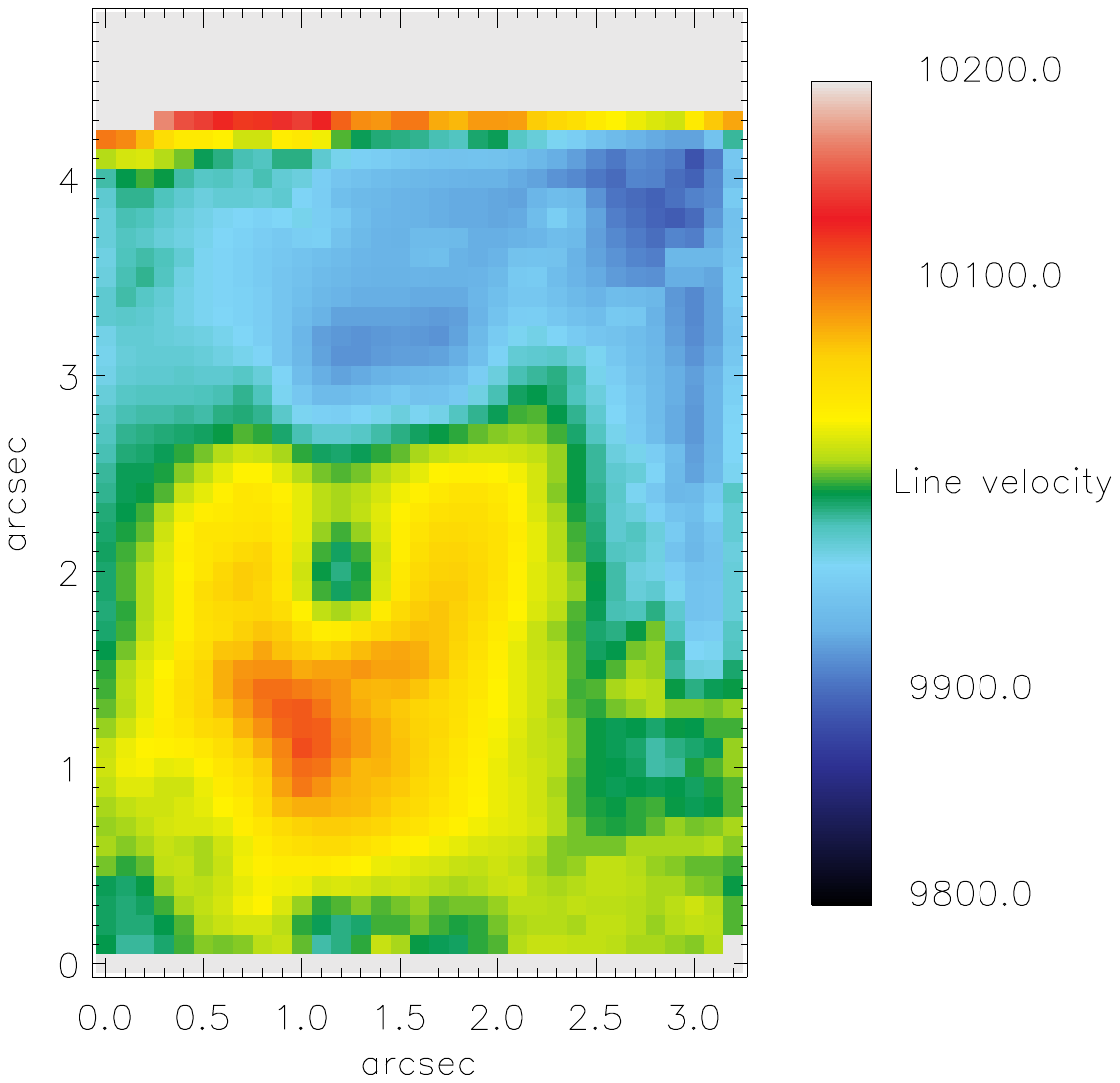}}}
   \mbox{\subfigure[H$\alpha$ line width]{\includegraphics[scale=0.5, trim=1mm 1mm 50mm 1mm, clip]{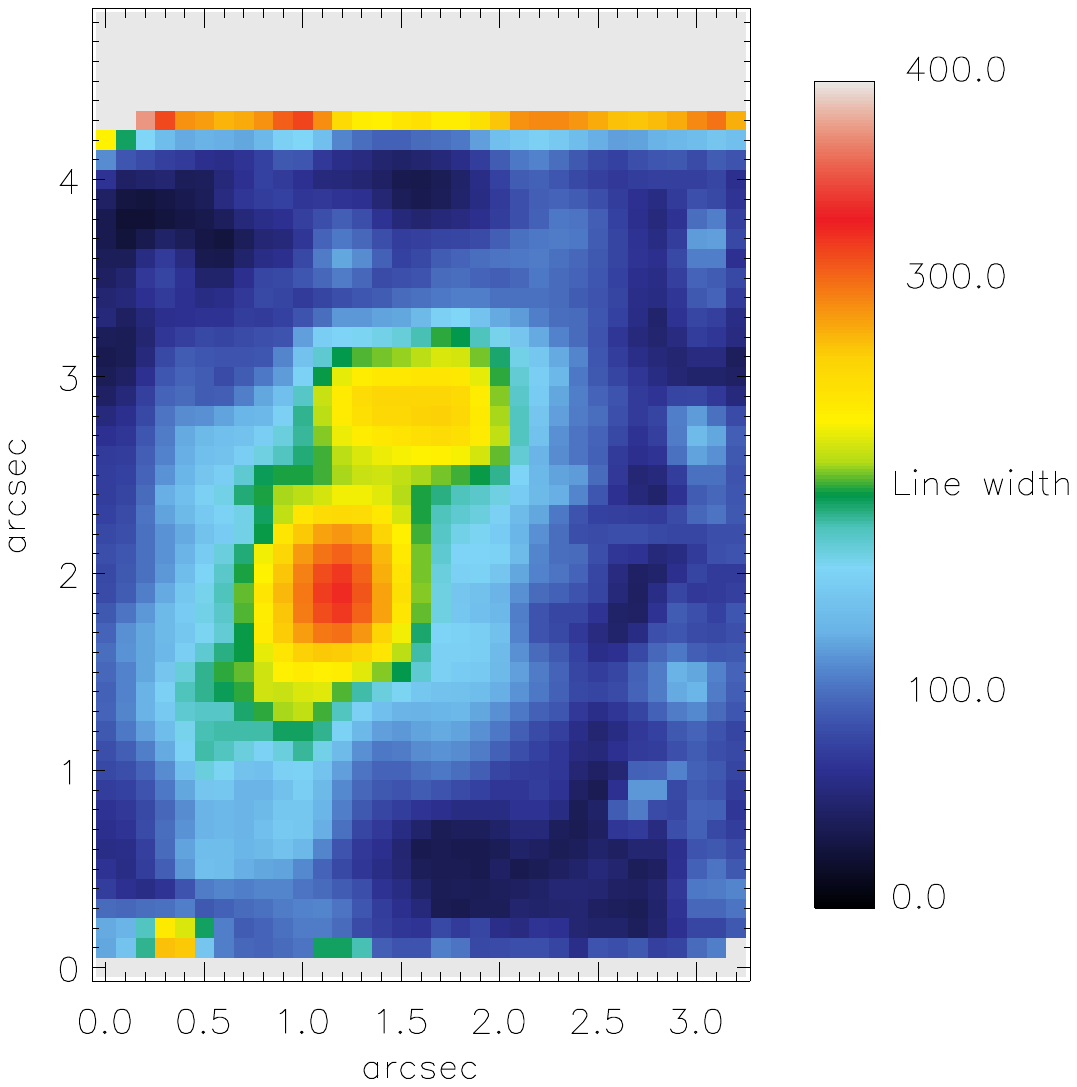}}\quad
\subfigure[{[NII]}$\lambda$6583 line width]{\includegraphics[scale=0.5, trim=1mm 1mm 50mm 1mm, clip]{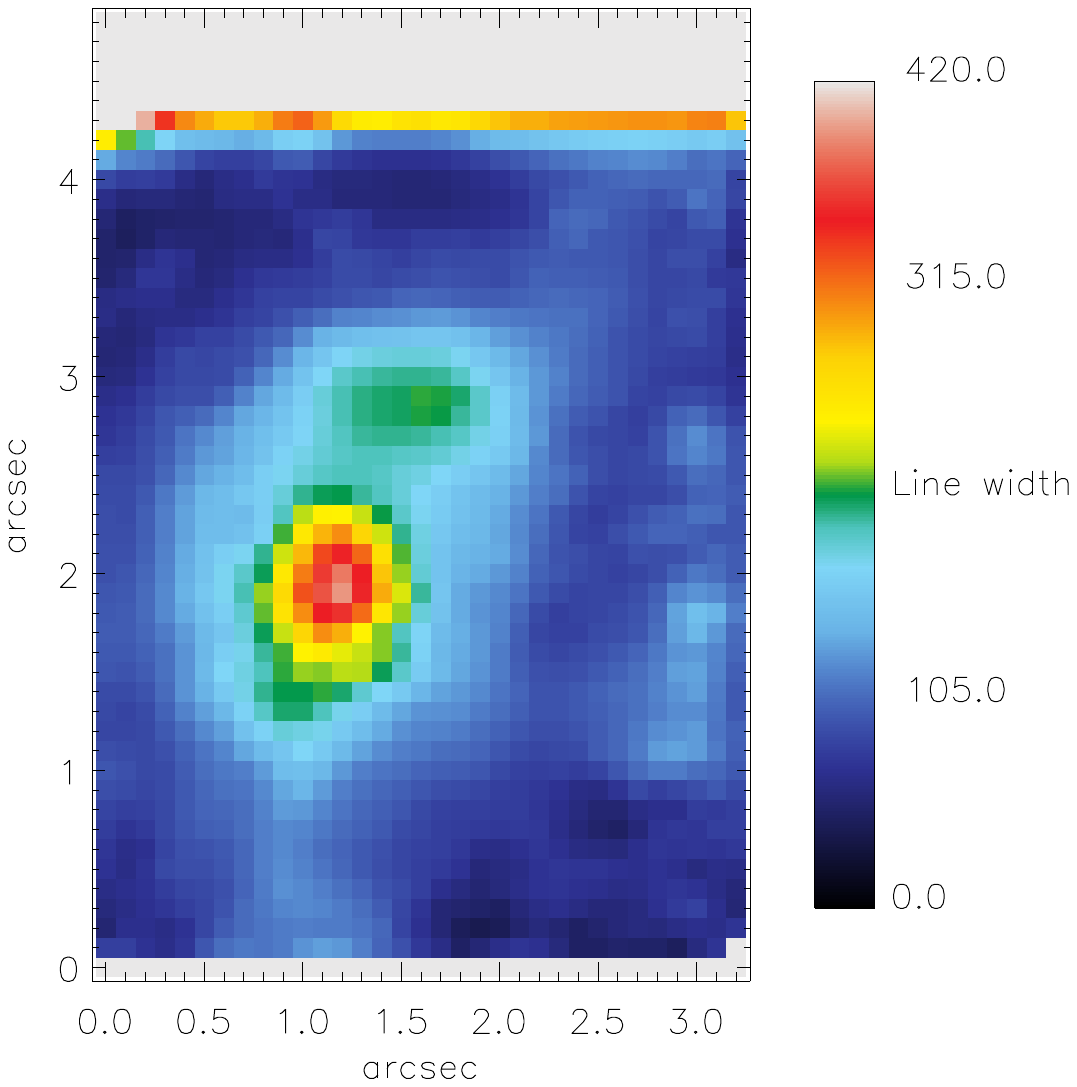}}\quad
\subfigure[{[SII]}$\lambda\lambda$6731,6717 line width]{\includegraphics[scale=0.5, trim=1mm 1mm 50mm 1mm, clip]{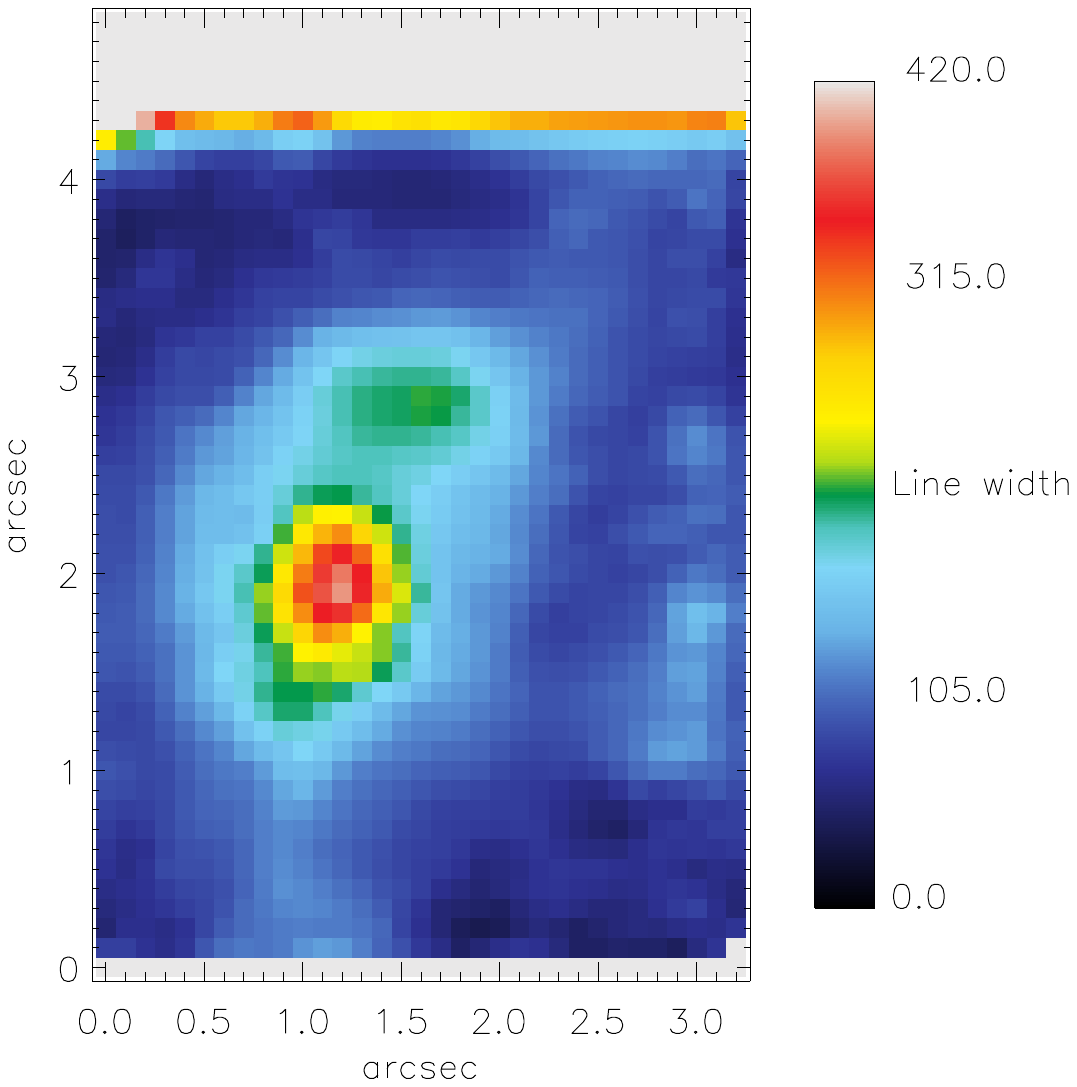}}}
\mbox{\subfigure[H$\alpha$ A/N]{\includegraphics[scale=0.5, trim=1mm 1mm 50mm 1mm, clip]{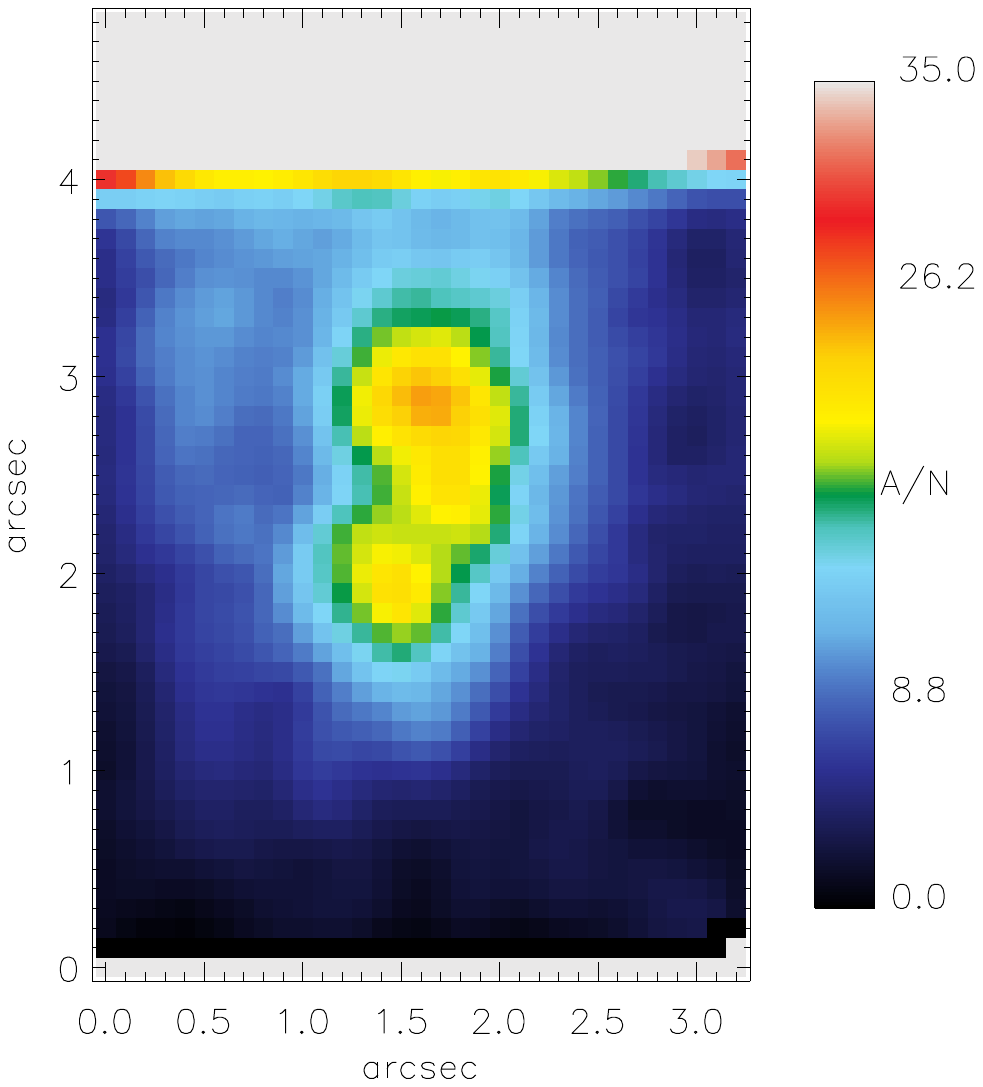}}\quad
\subfigure[{[NII]}$\lambda$6583 line A/N]{\includegraphics[scale=0.5, trim=1mm 1mm 50mm 1mm, clip]{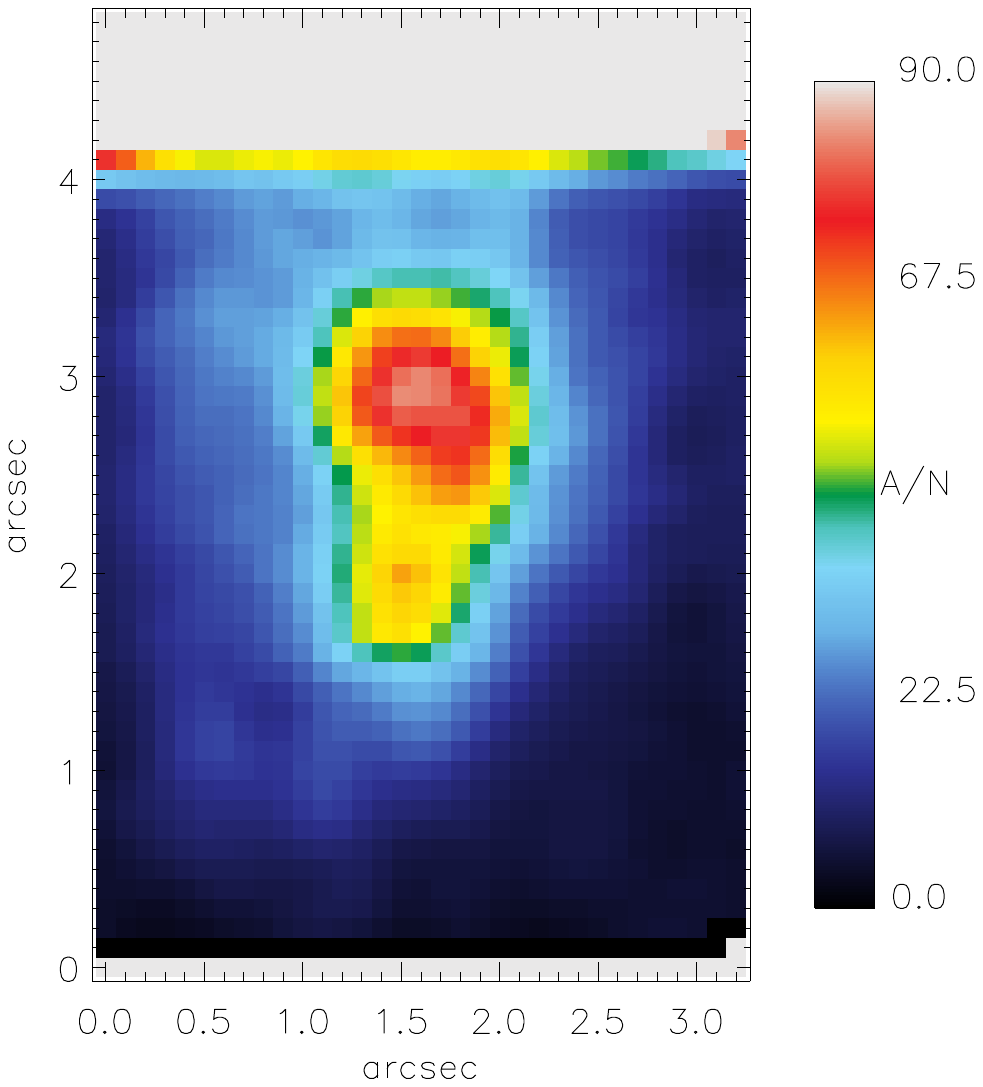}}\quad
\subfigure[{[SII]}$\lambda\lambda$6731,6717 line A/N]{\includegraphics[scale=0.5, trim=1mm 1mm 50mm 1mm, clip]{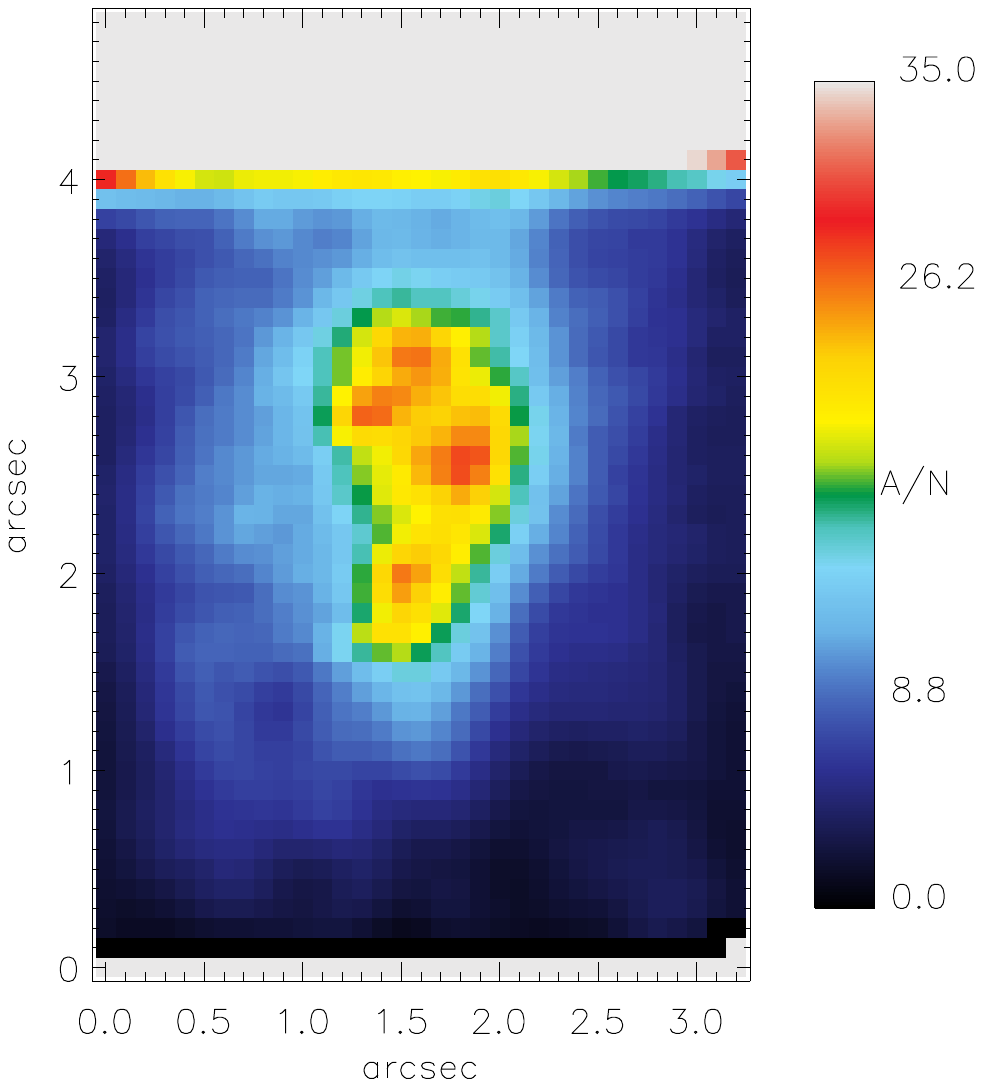}}}
\mbox{\subfigure{\includegraphics[scale=0.28]{UGC_arrow.pdf}}}
\caption{UGC09799: Velocity (in km s$^{-1}$), line width (in km s$^{-1}$) and A/N of the H$\alpha$, [NII]$\lambda$6583 and [SII]$\lambda\lambda$6731,6717 lines.}
\label{fig:UGCkinematics2} 
\end{figure*}

\begin{figure*}
 \mbox{\subfigure[{[OIII]}$\lambda$5007 velocity]{\includegraphics[scale=0.5, trim=1mm 1mm 50mm 1mm, clip]{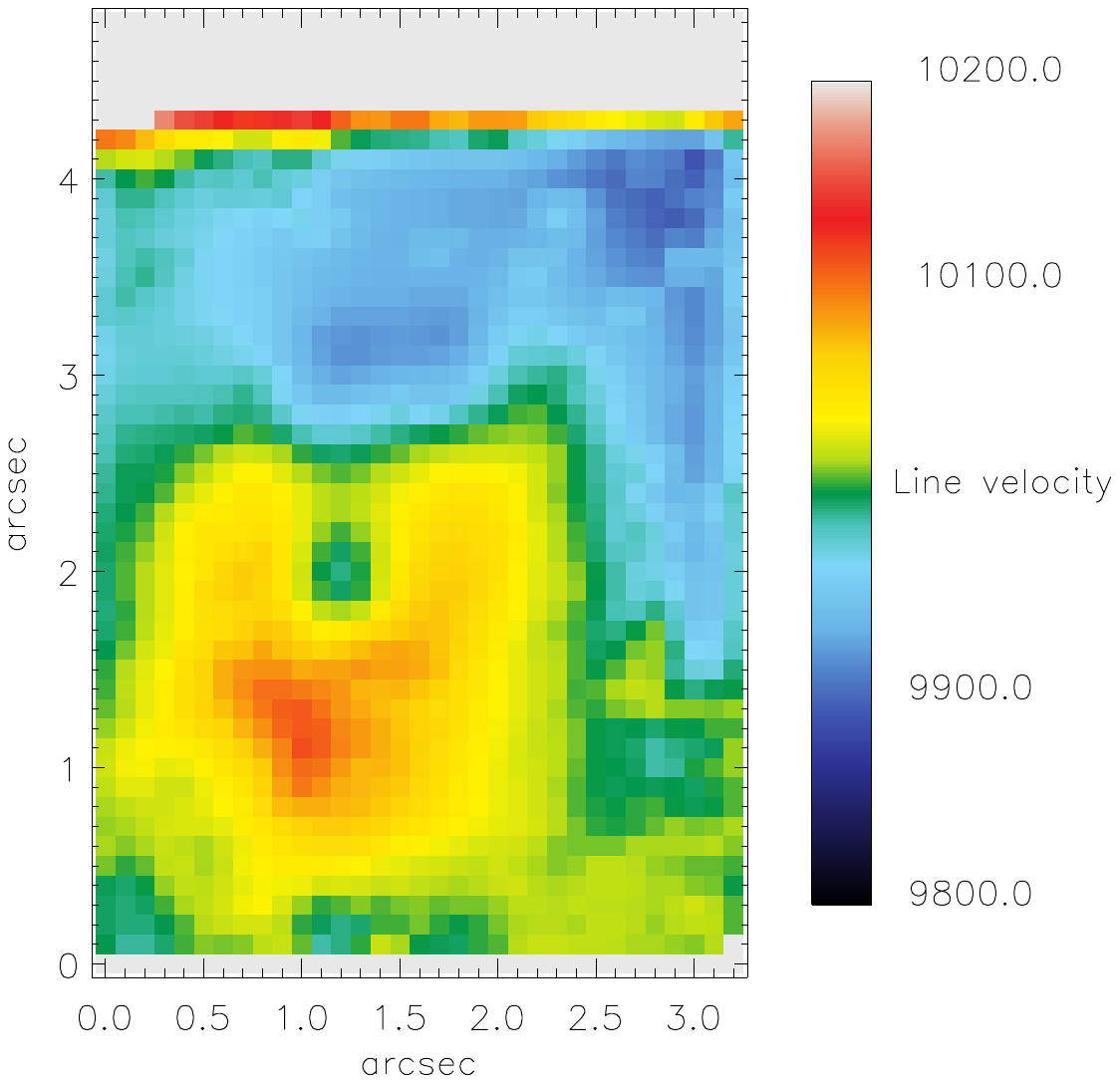}}\quad
\subfigure[{[OI]}$\lambda$6300 velocity]{\includegraphics[scale=0.5, trim=1mm 1mm 50mm 1mm, clip]{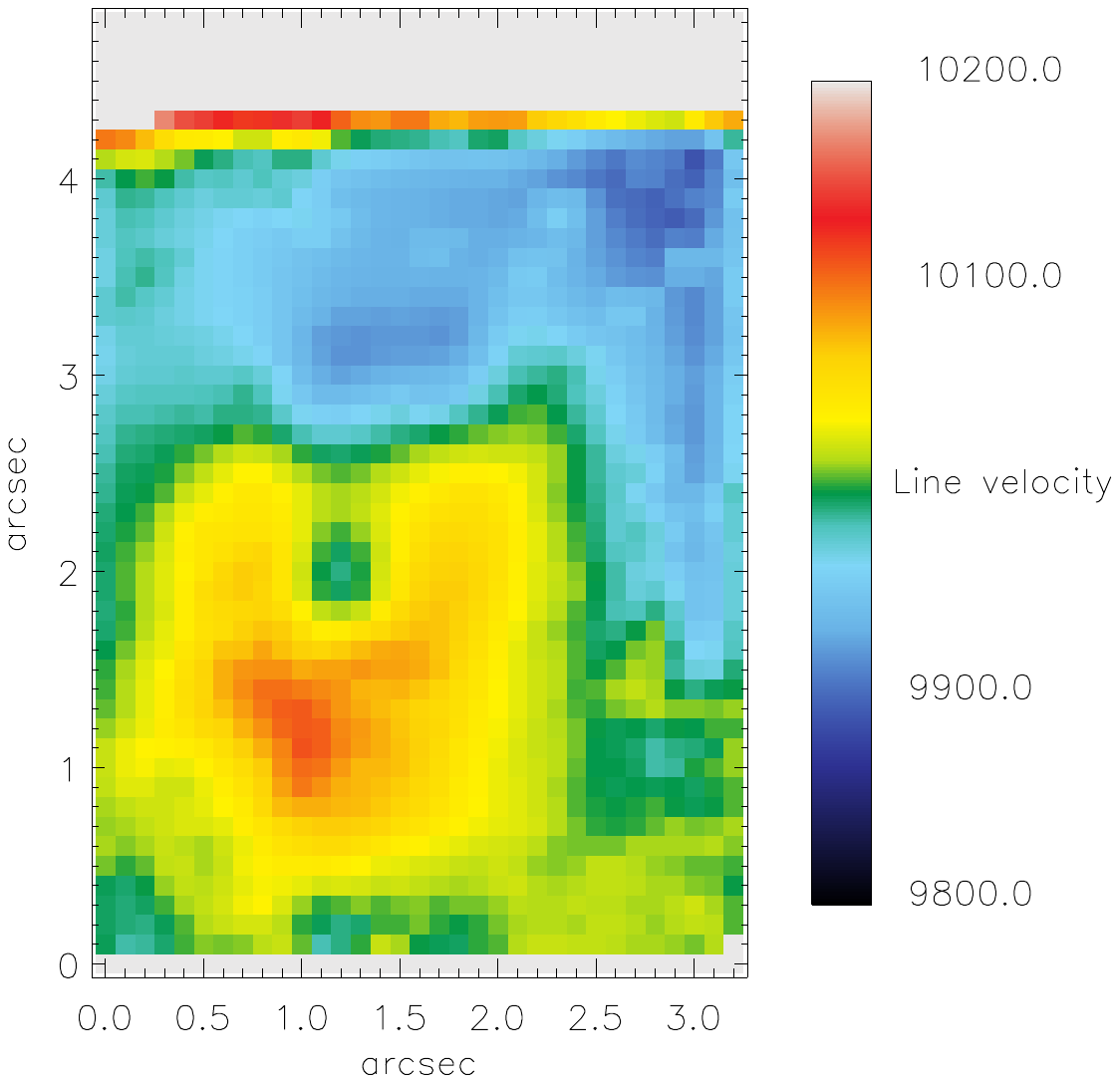}}}
   \mbox{\subfigure[{[OIII]}$\lambda$5007 line width]{\includegraphics[scale=0.5, trim=1mm 1mm 50mm 1mm, clip]{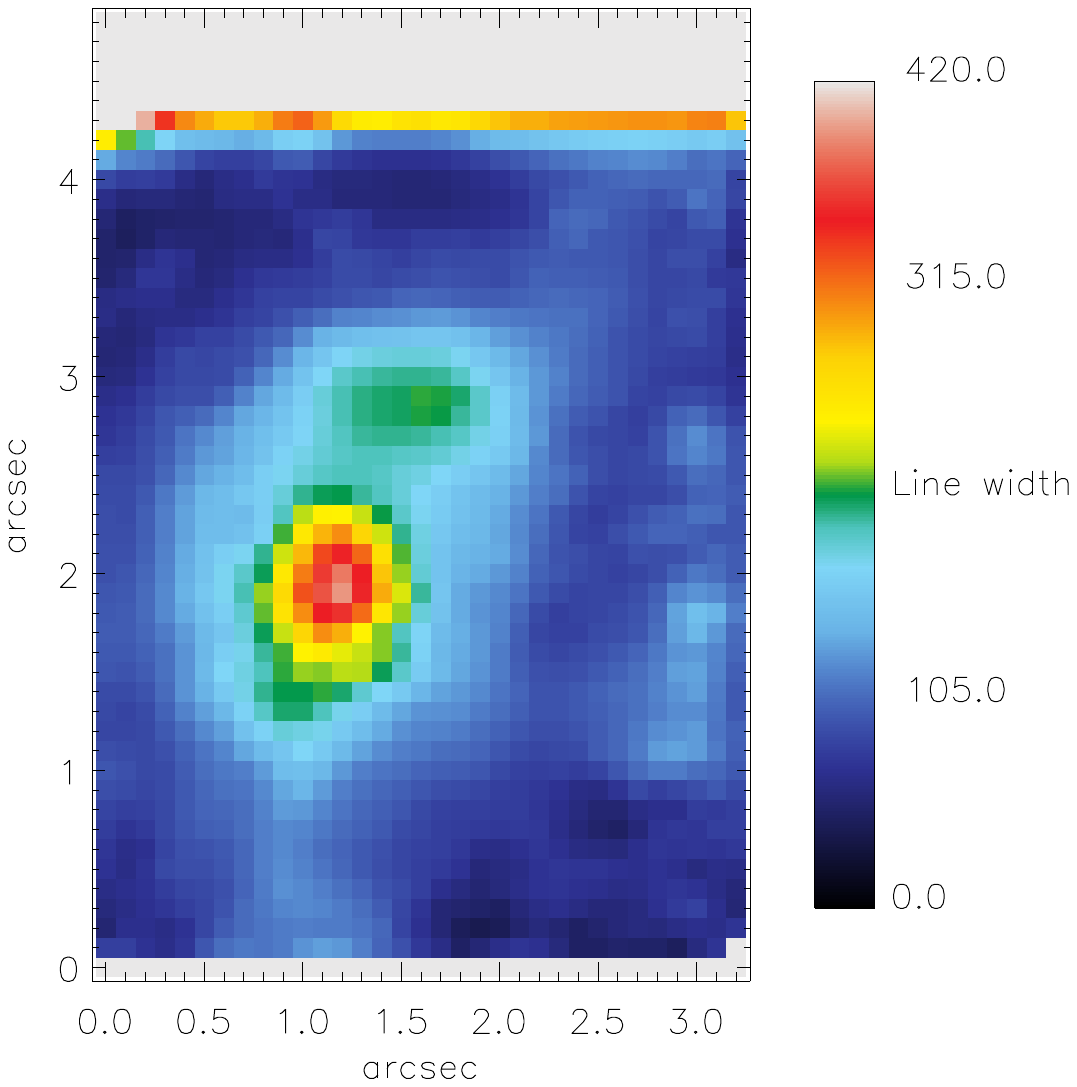}}\quad
\subfigure[{[OI]}$\lambda$6300 line width]{\includegraphics[scale=0.5, trim=1mm 1mm 50mm 1mm, clip]{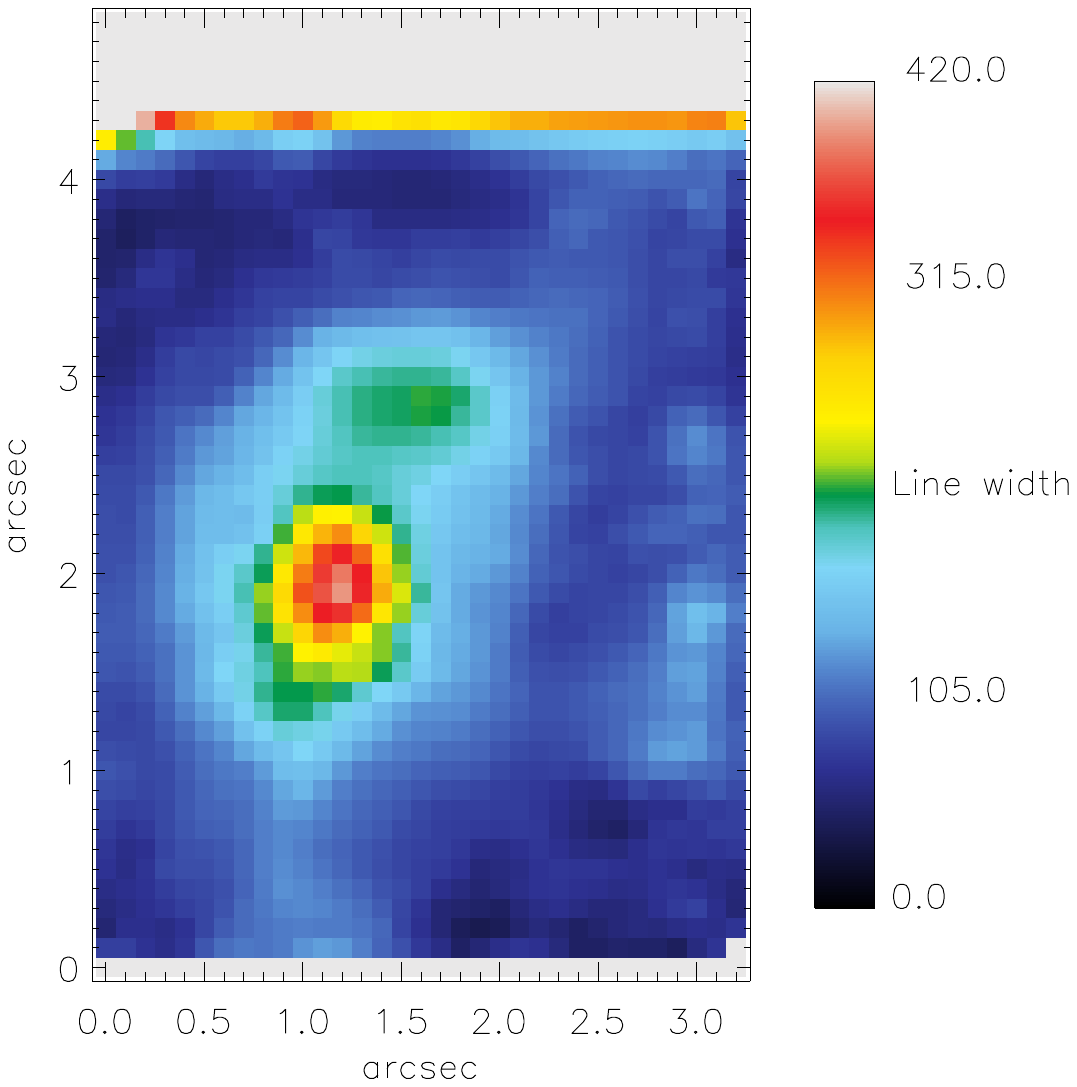}}}
\mbox{\subfigure[{[OIII]}$\lambda$5007 A/N]{\includegraphics[scale=0.5, trim=1mm 1mm 50mm 1mm, clip]{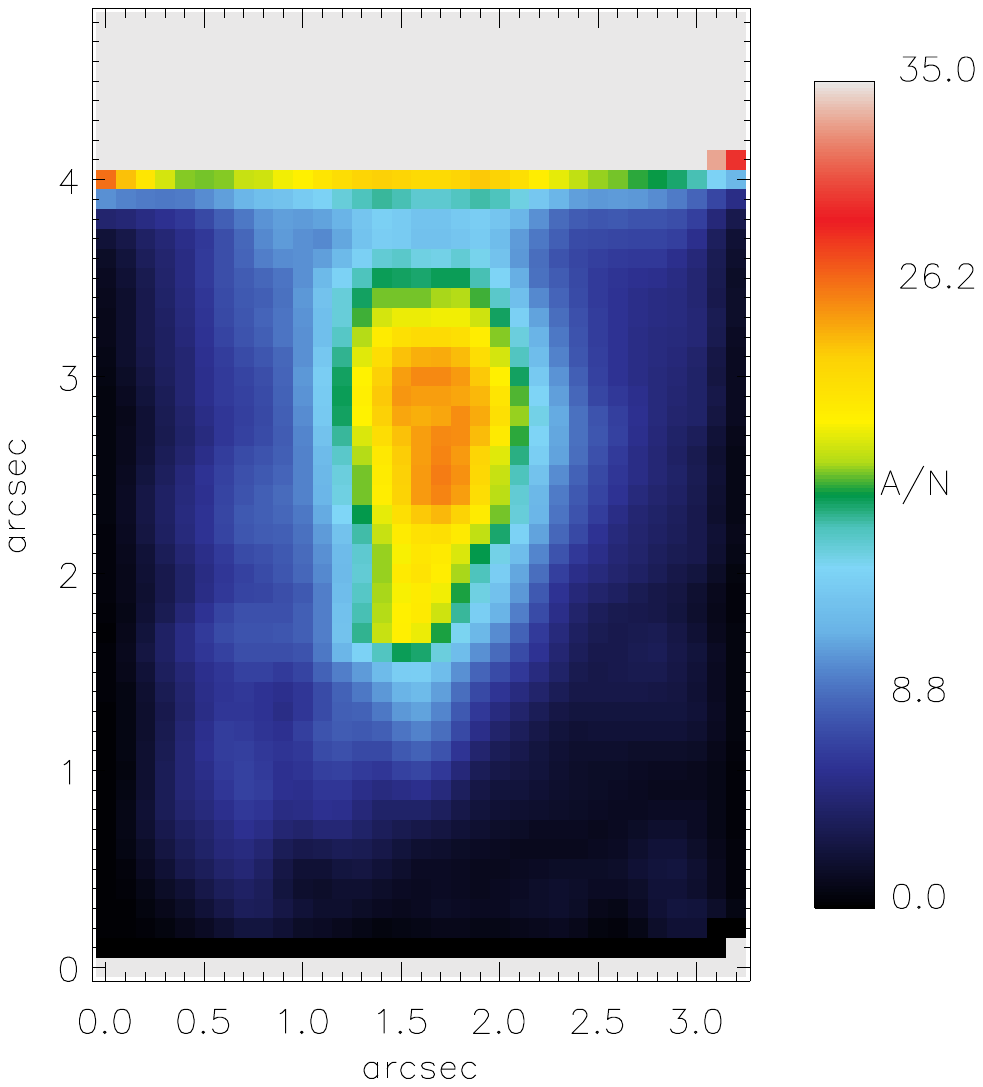}}\quad
\subfigure[{[OI]}$\lambda$6300 line A/N]{\includegraphics[scale=0.5, trim=1mm 1mm 50mm 1mm, clip]{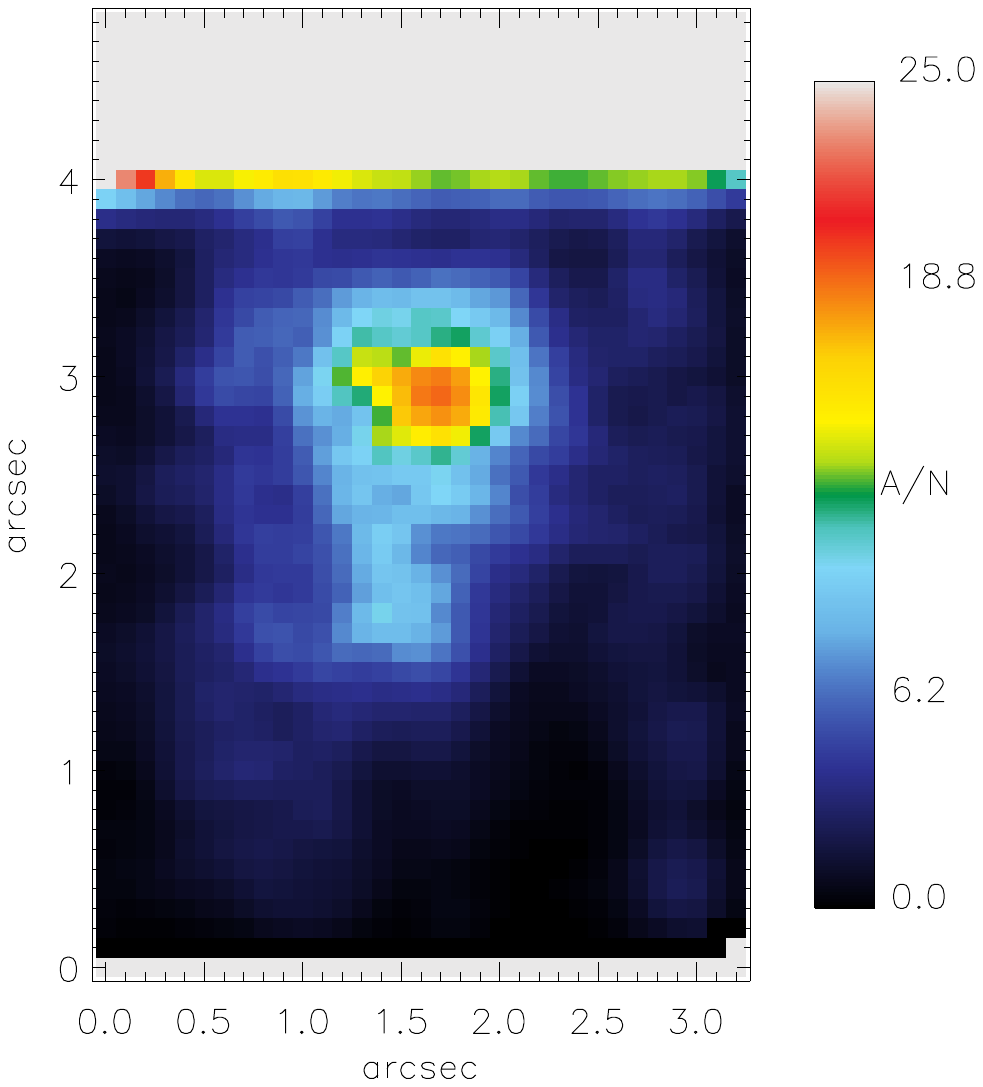}}}
\mbox{\subfigure{\includegraphics[scale=0.28]{UGC_arrow.pdf}}}
\caption{UGC09799: Velocity (in km s$^{-1}$), line width (in km s$^{-1}$) and A/N of the [OIII]$\lambda$5007 and [OI]$\lambda$6300 lines.}
\label{fig:UGCkinematics3} 
\end{figure*}

\begin{figure*}
 \mbox{\subfigure[MCG-02-12-039 H$\beta$ A/N]{\includegraphics[scale=0.5, trim=1mm 1mm 50mm 1mm, clip]{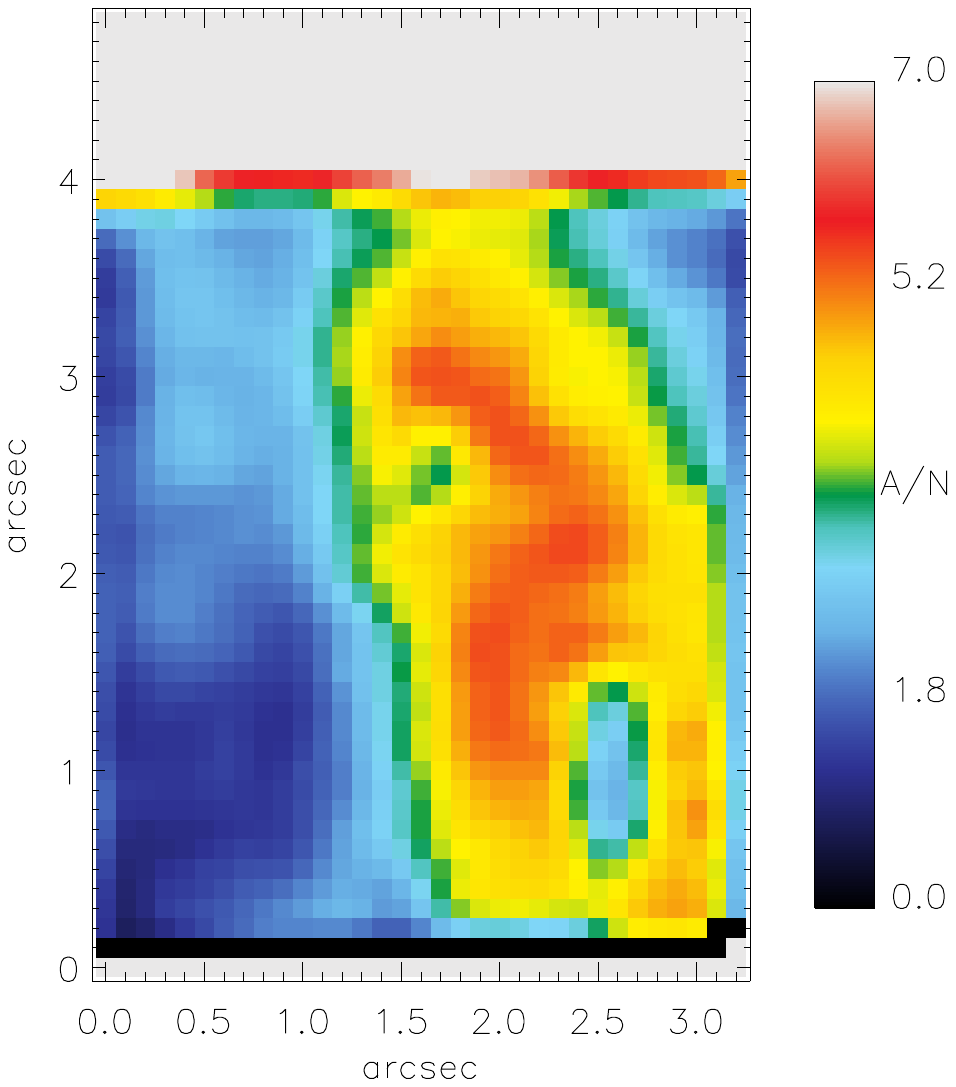}}\quad
\subfigure[PGC026269 H$\beta$ A/N]{\includegraphics[scale=0.5, trim=1mm 1mm 50mm 1mm, clip]{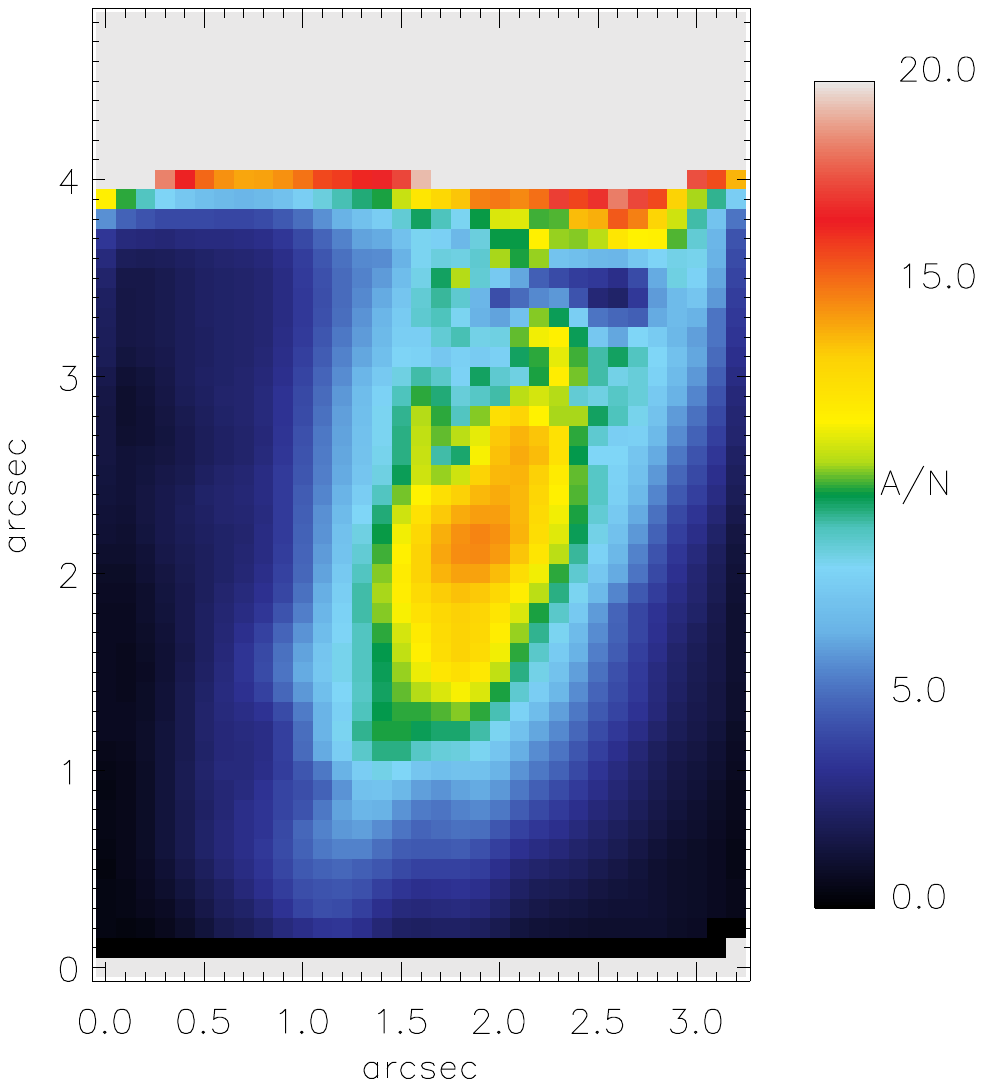}}}
\mbox{\subfigure{\includegraphics[scale=0.28]{MCG_arrow.pdf}}\quad
\subfigure{\includegraphics[scale=0.28]{PGC026_arrow.pdf}}}
\mbox{\subfigure[PGC044257 H$\beta$ A/N]{\includegraphics[scale=0.5, trim=1mm 1mm 50mm 1mm, clip]{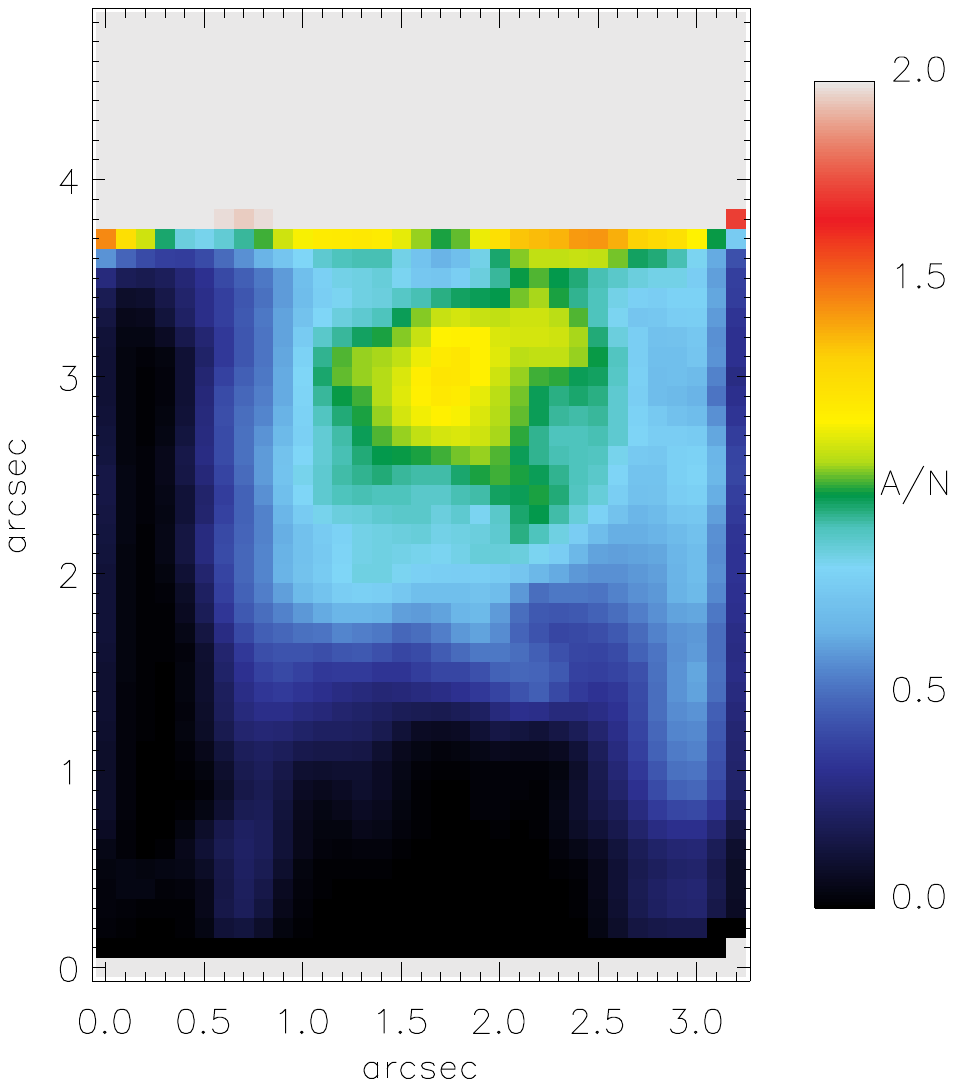}}\quad
\subfigure[UGC09799 H$\beta$ A/N]{\includegraphics[scale=0.5, trim=1mm 1mm 50mm 1mm, clip]{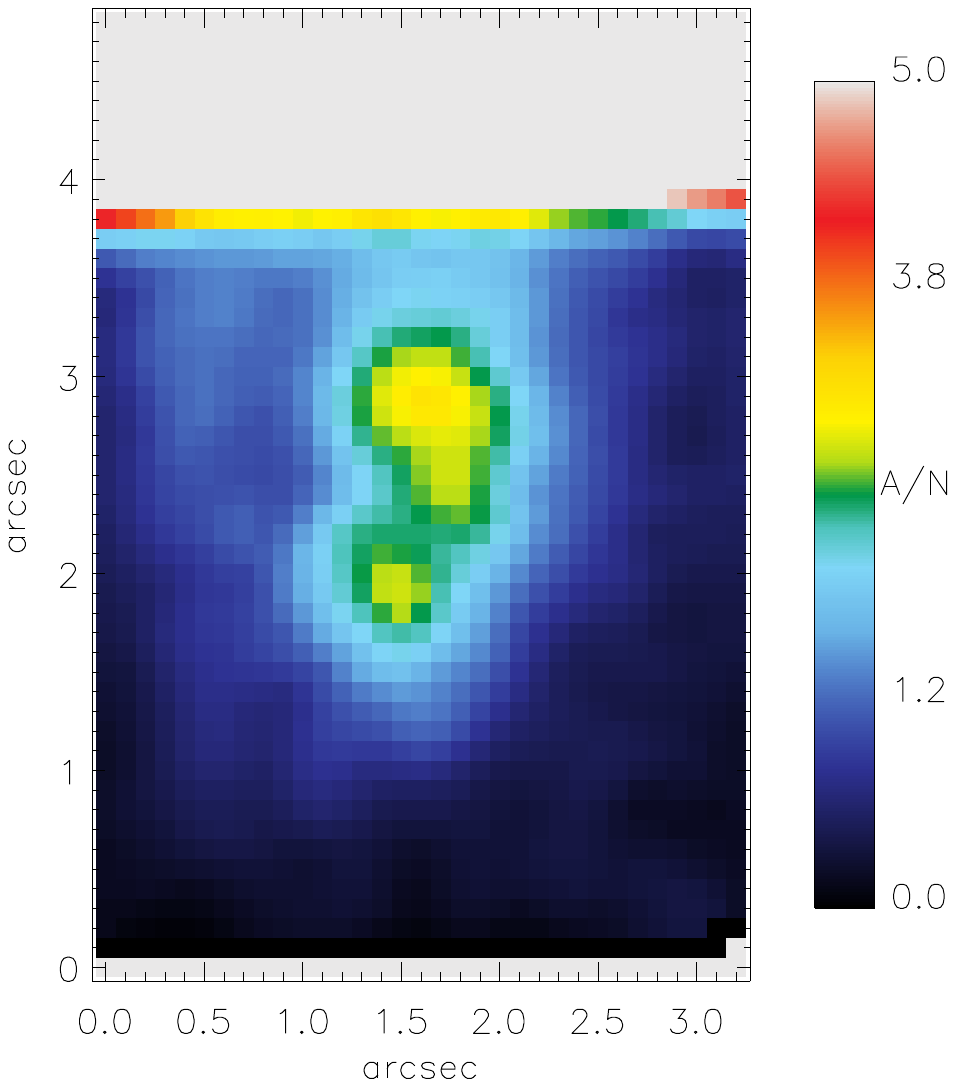}}}
\mbox{\subfigure{\includegraphics[scale=0.28]{PGC044_arrow.pdf}}
\subfigure{\includegraphics[scale=0.28]{UGC_arrow.pdf}}}
\caption{A/N of the H$\beta$ lines.}
\label{Hbeta} 
\end{figure*}

We plot the kinematics of the stellar and gaseous components in Figures \ref{fig:UGCkinematics} and \ref{fig:UGCkinematics2}. We plot H$\alpha$, [NII], [SII], [OIII] and [OI] velocities, line width and A/N in Figure \ref{fig:UGCkinematics2} and \ref{fig:UGCkinematics3}. To the degree that our spatial resolution reveals, it appears that all the optical forbidden and hydrogen recombination lines originate in the same gas.
Our kinematic analysis can be compared to that of Edwards et al.\ (2009), who detected a gradient in the velocity but could not differentiate between a rotation or outflow. Our stellar kinematics is shown in Figure \ref{fig:UGCkinematics}, however we do not detect rotational kinematics. Edwards et al.\ (2009) also found H$\alpha$ to have a velocity range of --250 km s$^{-1}$ to +150 km s$^{-1}$. Our H$\alpha$ kinematics are shown in \ref{fig:UGCkinematics2} and show also show a range ($\sim$ 400 km s$^{-1}$) of velocities for H$\alpha$. The gaseous components show rotation in Figure \ref{fig:UGCkinematics2} even though no rotation is apparent in the stellar components (Figure \ref{fig:UGCkinematics}). 

This galaxy has patchy dust in the centre (Laine et al.\ 2003, also shown in Figure \ref{fig:MCG_extinct}), and Hicks $\&$ Mushotzky (2005) have deduced star formation from the excess UV--IR emission. The FUV/H$\alpha$ ratio suggests heating by fast shocks or some other source of hard ionisation (e.g., cosmic rays, AGN). The filaments of Abell 2052 have ratios which are consistent with this picture (McDonald et al.\ 2011). The highest extinction (in Figure \ref{fig:MCG_extinct}) does not coincide with the H$\alpha$ peak (Appendix B1), and is slightly off centre. This does not necessarily contrast with the results quoted by Laine et al.\ (2003) that the dust is patchy in the centre, as the centre of the observations might not necessarily be exactly the same as our IFU placement (see Figure 1). Figure 1 also shows that our continuum emission is smooth, similar to the conclusion reached by Edwards et al.\ (2009).

Venturi, Dallacasa $\&$ Stefanachi (2004) found a parsec-scale bipolar radio source, and Chandra X-ray emission show two bubbles in the ICM on a larger scale (Blanton, Sarazin $\&$ McNamara 2003). Our results are consistent with the central emission being that of a LINER (see Section \ref{ionisation}). 

\section{Discussion -- Ionisation mechanisms}
\label{ionisation}

Various mechanisms have been proposed as sources of excitation in the nebulae, including photoionisation by radiation from the AGN, cluster X-rays, or hot stars, collisional heating by high-energy particles, shocks or cloud-cloud collisions, and conduction of heat from the X-ray corona (Peterson et al.\ 2003). However, none so far satisfactorily describes all the characteristic spectra, energetics, and kinematics of the extended emission-line regions (Wilman et al.\ 2006; Hatch et al.\ 2007).

We reiterate that the FOV of the observations are only 3.5 $\times$ 5 arcsec (corresponding to the central few parsec -- see Table \ref{table:objects}). These results only look at the very heart of these sources. Therefore, it is really only the emission mechanisms for the gas at the very center of these systems that is being studied here.

The forbidden lines such as [NII]$\lambda$6583 result from the excitation of N$^{+}$ through collisions with electrons liberated through photoionisation. The H$\alpha$ emission results from the recombination of the hydrogen ion. [NII]$\lambda$6583 flux depends on the N$^{+}$ abundance, the strength of the radiation field, and the form of the radiation field: a harder ionising source will produce a greater flux. The H$\alpha$ flux also depends on the strength of the radiation field. Therefore, the ratio will depend on the metallicity of the gas and the form of the ionising radiation. The form of the ionising radiation and/or the gas metallicity are not uniform but must vary within each galaxy and between the whole sample. Different excitation mechanisms may act in different regions (Hatch et al.\ 2007, Edwards et al.\ 2009), or the H$\alpha$ emission might be disturbed by the presence of companion galaxies (Wilman et al.\ 2006). 

\begin{figure*}
  \centering
 \mbox{\subfigure[MCG-02-12-039]{\includegraphics[scale=0.5]{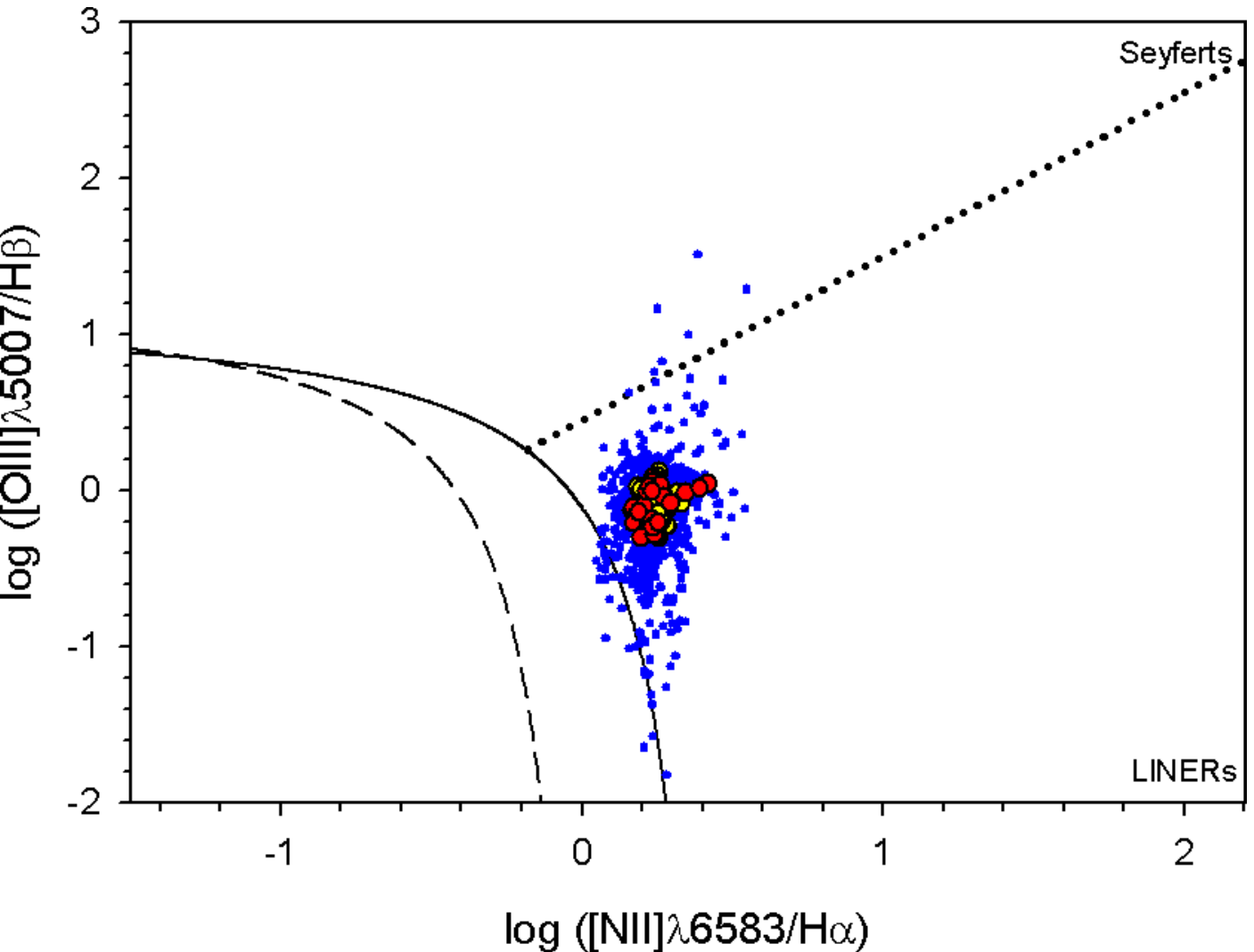}}\quad
         \subfigure[PGC026269]{\includegraphics[scale=0.5]{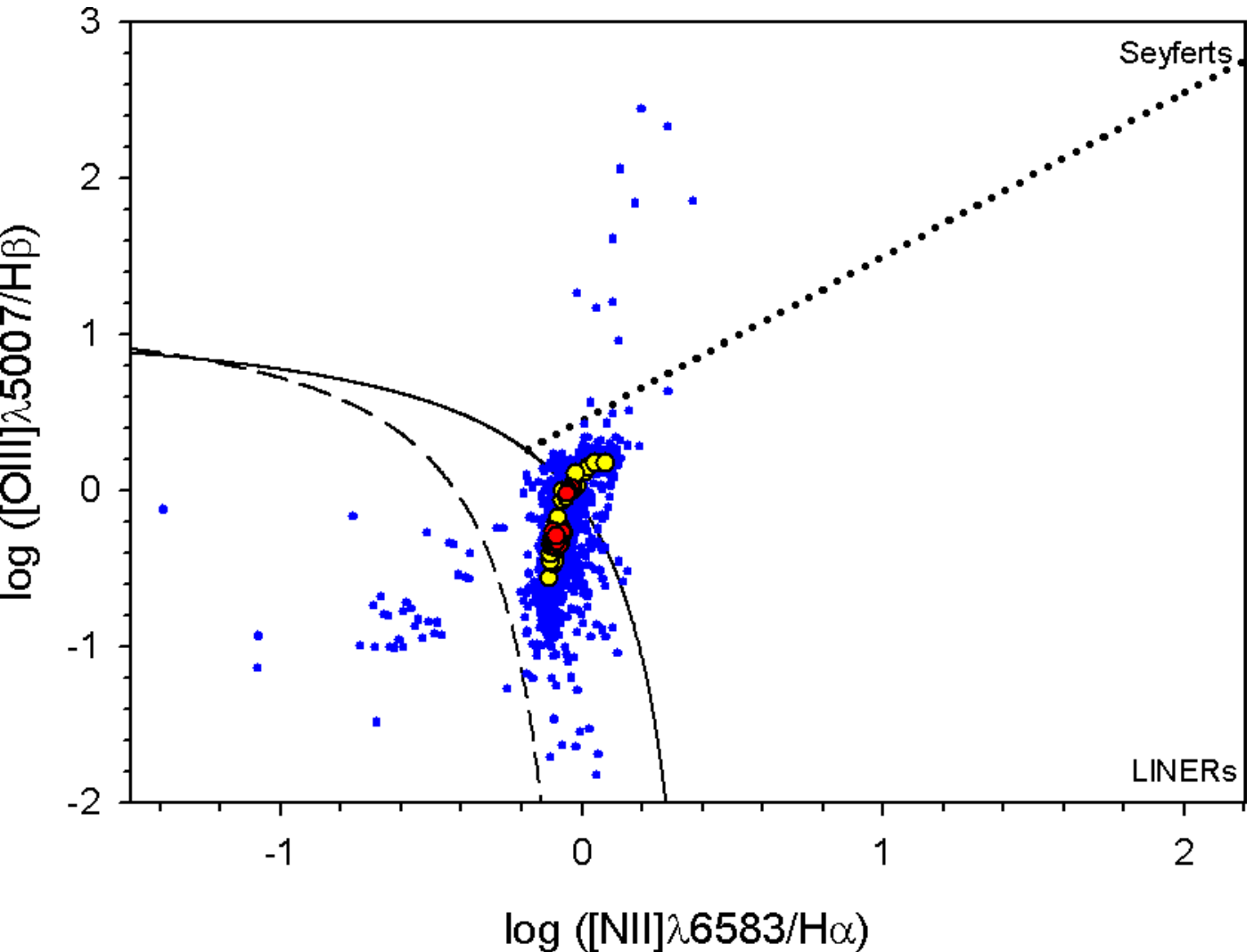}}}
 \caption{BPT diagram for MCG-02-12-039 and PGC026269. The red circles indicate the central 0.5 $\times$ 0.5 arcsec of the galaxy, the yellow circles 1.0 $\times$ 1.0 arcsec, and the blue circles the full 3.5 $\times$ 5.0 arcsec. The centre of the galaxy was determined as the luminosity peak in the continuum images in Figure \ref{fig:Thumbnails}. The black solid curve is the theoretical maximum starburst model from Kewley et al.\ (2001), devised to isolate objects whose emission line ratios can be accounted for by the photoionisation by massive stars (below and to the left of the curve) from those where some other source of ionisation is required. The black-dotted curve in the diagram represent the Seyfert-LINER dividing line from Kewley et al.\ (2006) and transposed to the [NII]$\lambda$6584/H$\alpha$ diagram by Schawinski et al.\ (2007).}
   \label{fig:MCG_BPT}
\end{figure*}

\begin{figure*}
   \centering
 \mbox{\subfigure[PGC044257]{\includegraphics[scale=0.5]{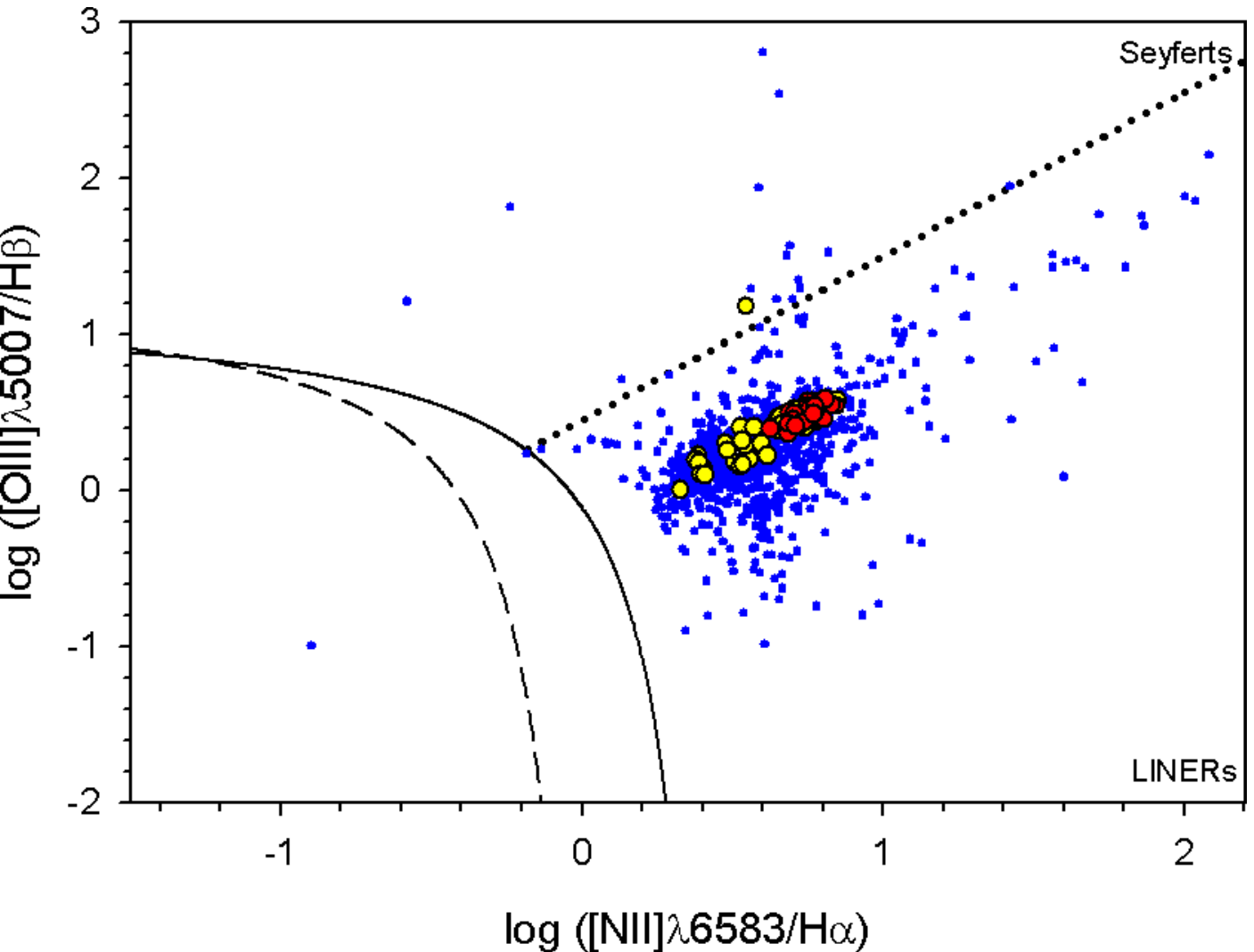}}\quad
\subfigure[UGC09799]{\includegraphics[scale=0.5]{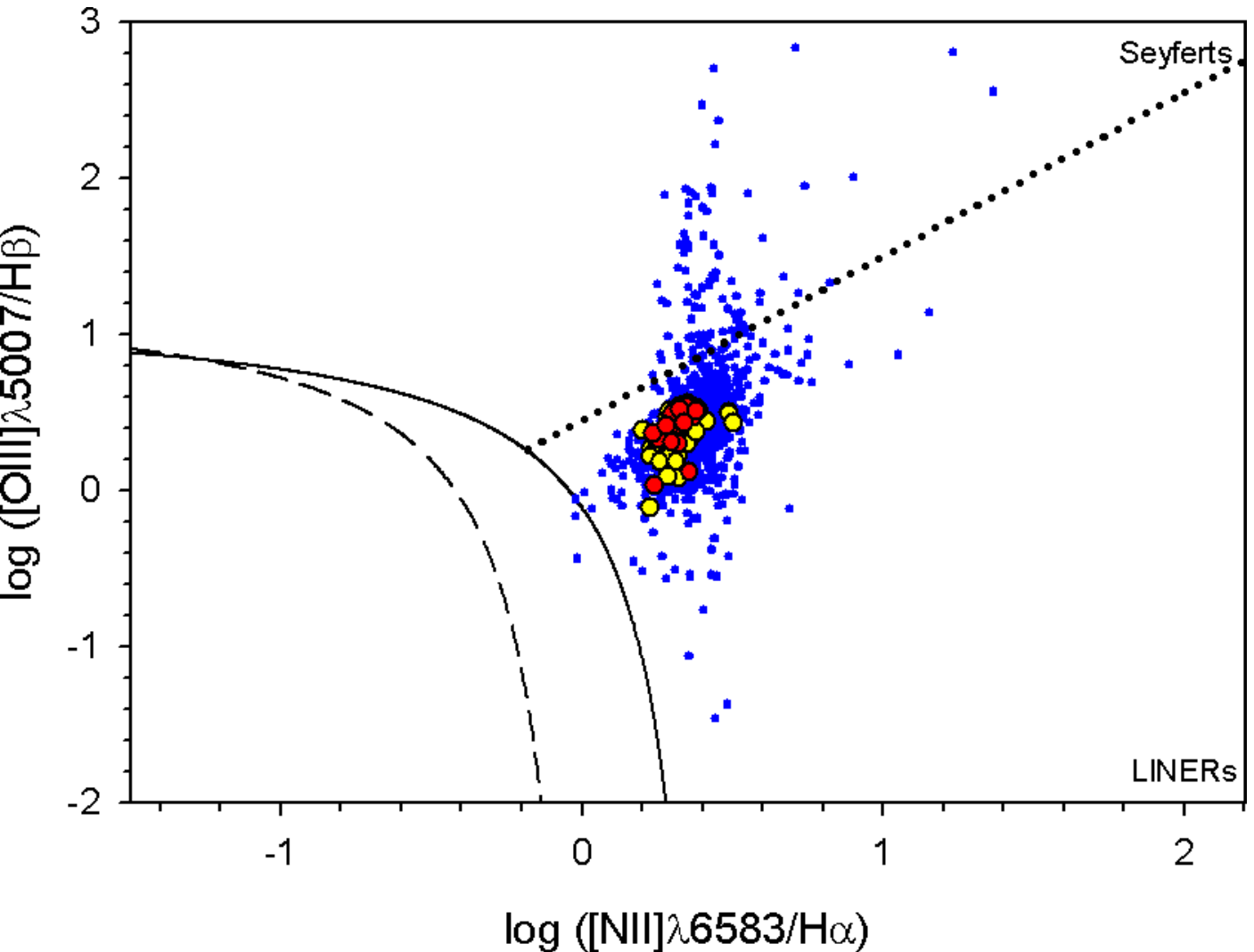}}}
\caption{BPT diagram for PGC044257 and UGC09799. The red circles indicate the central 0.5 $\times$ 0.5 arcsec of the galaxy, the yellow circles 1.0 $\times$ 1.0 arcsec, and the blue circles the full 3.5 $\times$ 5.0 arcsec. See caption of Figure \ref{fig:MCG_BPT}.}
\label{fig:UGC_BPT} 
\end{figure*}

A commonly used method to distinguish between the sources of ionisation uses the emission-line diagrams pioneered by Baldwin, Phillips $\&$ Terlevich (1981, hereafter BPT diagram) which separate the two major origins of emission: star formation and AGN. The diagrams use pairs of emission line ratios, of which the most commonly used is [OIII]$\lambda$5007/H$\beta$ against [NII]$\lambda$6584/H$\alpha$. Extinction-corrected emission-line measurements were used (even though the BPT diagram is almost insensitive to reddening). We plot the BPT diagrams in Figures \ref{fig:MCG_BPT} and \ref{fig:UGC_BPT} using the following criteria:

Kewley criteria:
Galaxies above this line are AGN. Kewley et al.\ (2001) used a combination of photoionisation and stellar population synthesis models to place a theoretical upper limit on the location of star forming galaxies on the BPT diagram, $\log (\frac{[OIII]}{H\beta} = 0.61/{\log([NII]/H\alpha)-0.47}+1.19)$.

Kauffmann et al.\ (2003) criteria:
Galaxies below this line are purely star forming. Between Kewley and Kauffmann criteria is composite objects. Kauffmann et al.\ (2003) revised the criteria as follows: A galaxy is defined to be an AGN if $\log (\frac{[OIII]}{H\beta} > 0.61/{\log([NII]/H\alpha)-0.05}+1.3)$.

LINERS--Seyffert line (Schawinsky et al.\ 2007):
This line distinguishes LINERS from Seyffert galaxies $\log (\frac{[OIII]}{H\beta}) = 1.05 \log(\frac{[NII]}{H\alpha})+0.45$.

SDSS emission-line galaxies occupy a well-defined region shaped like the wings of a seagull (Stasinska et al.\ 2008). Although the exact location of the line dividing the star forming and active galactic nuclei galaxies is still controversial (Kewley et al 2001; Kauffmann et al 2003; Stasinska et al 2006). 

All galaxies show important LINER emission, but that at least one has significant Seyfert emission areas, and at least one other has significant HII like emission line ratios for many pixels as shown in Figures \ref{fig:MCG_BPT} and \ref{fig:UGC_BPT}. This is in agreement with the long-slit data of these sources plotted on BPT diagrams in figure 3 of McDonald et al.\ (2012).

However, there is a debate about the ionisation mechanism in LINERS (Low-Ionisation Nuclear Emission-line Region). The most viable excitation mechanisms are: a low accretion-rate AGN (Kewley et al.\ 2006), photoionisation by old post-asymptotic giant branch (pAGB) stars (Stasinska et al.\ 2008), and fast shocks (Dopita $\&$ Sutherland 1995). Sarzi et al.\ (2010) investigate the ionising sources for the gas in elliptical galaxies based on SAURON intergral-field spectroscopy whose spectra are limited to a relatively narrow wavelength range. They conclude that pAGB stars are the main source of ionisation. In contrast, Annibali et al.\ (2010) analyse long-slit spectra of 65 ellipticals and claim that their nuclear line emission can be explained by excitation from the hard ionising continuum from an AGN and/or fast shocks. However, they can not completely rule out a contribution from pAGB stars at large radii. Voit $\&$ Donahue (1997) suggested that sources of supplementary heating produce the LINER-like properties of the spectra, though not necessarily through the same mechanism in all systems. 

Annibali et al.\ (2010) found that from the centre outward, galaxies move left and down in the BPT diagram for their study of 65 early-type galaxies. Thus, the hardness of the ionising continuum decreases with galactocentric distance (up to half the half-light radius in the Annibali et al.\ (2010) sample). Figures \ref{fig:MCG_BPT} and \ref{fig:UGC_BPT} also show the flux ratios as function of distance from the galaxy centre. The red circles indicate the central 0.5 $\times$ 0.5 arcsec of the galaxy, the yellow circles 1.0 $\times$ 1.0 arcsec, and the blue circles the full 3.5 $\times$ 5.0 arcsec. The centre of the galaxy was determined as the luminosity peak in the continuum images in Figure \ref{fig:Thumbnails}. In our case, the hardness of the ionising continuum stay mostly uniform with galactocentric distance over out limited spatial extend. PGC044257 show an interesting core separation of the emission in the very centre of the galaxy in Figure \ref{fig:UGC_BPT}a.

There is also the debate about which galaxies are LINERS (AGN) and which have just LINER-like emission (non-AGN). The division is usually made by looking at the extend of the LINER signature: core dominated mean true LINER and diffuse means LINER-like. The arising problem is that the scale of the SDSS fibers is already too large to make the distinction for most galaxies. For the nearby galaxies, the centroiding of the fibres is not accurate enough to be sure if the measured spectra cover the core of the galaxy. We have core dominated LINER  emission for at least three out of the four galaxies. Additional confirmation comes from the fact that at least one of our galaxies (PGC026269) is a strong radio source (AGN, see \ref{table:objects2}). Similarly, our findings for UGC09799 agree with Edwards et al.\ (2009) who found line ratios consistent with Seyffert or LINER activity in most of their central spaxels for UGC09799. Edwards (2009) also found that the Seyffert signature dominated the central spaxels of the CCG (UGC09799). The H$\alpha$ emission surrounding the BCG in this cluster is coincident with radio-blown bubbles in the central region of the cluster. These bubbles to the north and south of the cluster core are filled with radio emission, which likely originated from the AGN within the CCG (Blanton et al.\ 2003). Since the H$\alpha$ emission seen by McDonald et al.\ (2010) is primarily along the edges of the northern bubble, they suspect that shocks may be responsible for the heating in this case. However, for PGC044257, McDonald et al.\ (2010) find very little evidence for an AGN in the cluster hosting the galaxy, in terms of the radio power, X-ray morphology and hard X-ray flux. Two of the galaxies in the current study have detected X-ray point sources in the CCG (PGC026269 and UGC09799) with large cavities in their X-ray haloes, suggesting that the AGN is influencing the surrounding medium.

\begin{figure*}
   \centering
 \mbox{\subfigure[MCG-02-12-039]{\includegraphics[scale=0.38]{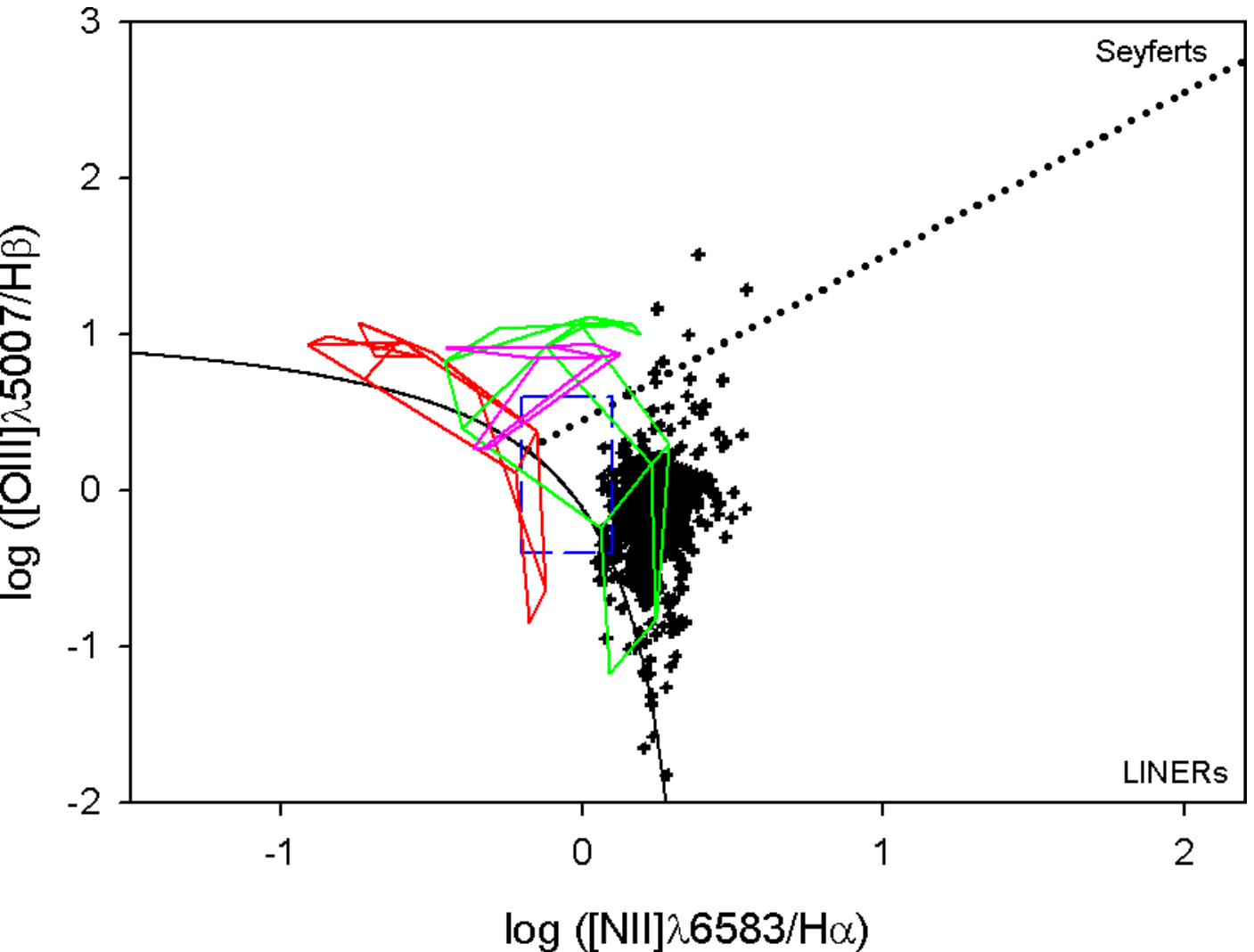}}\quad
         \subfigure[MCG-02-12-039]{\includegraphics[scale=0.38]{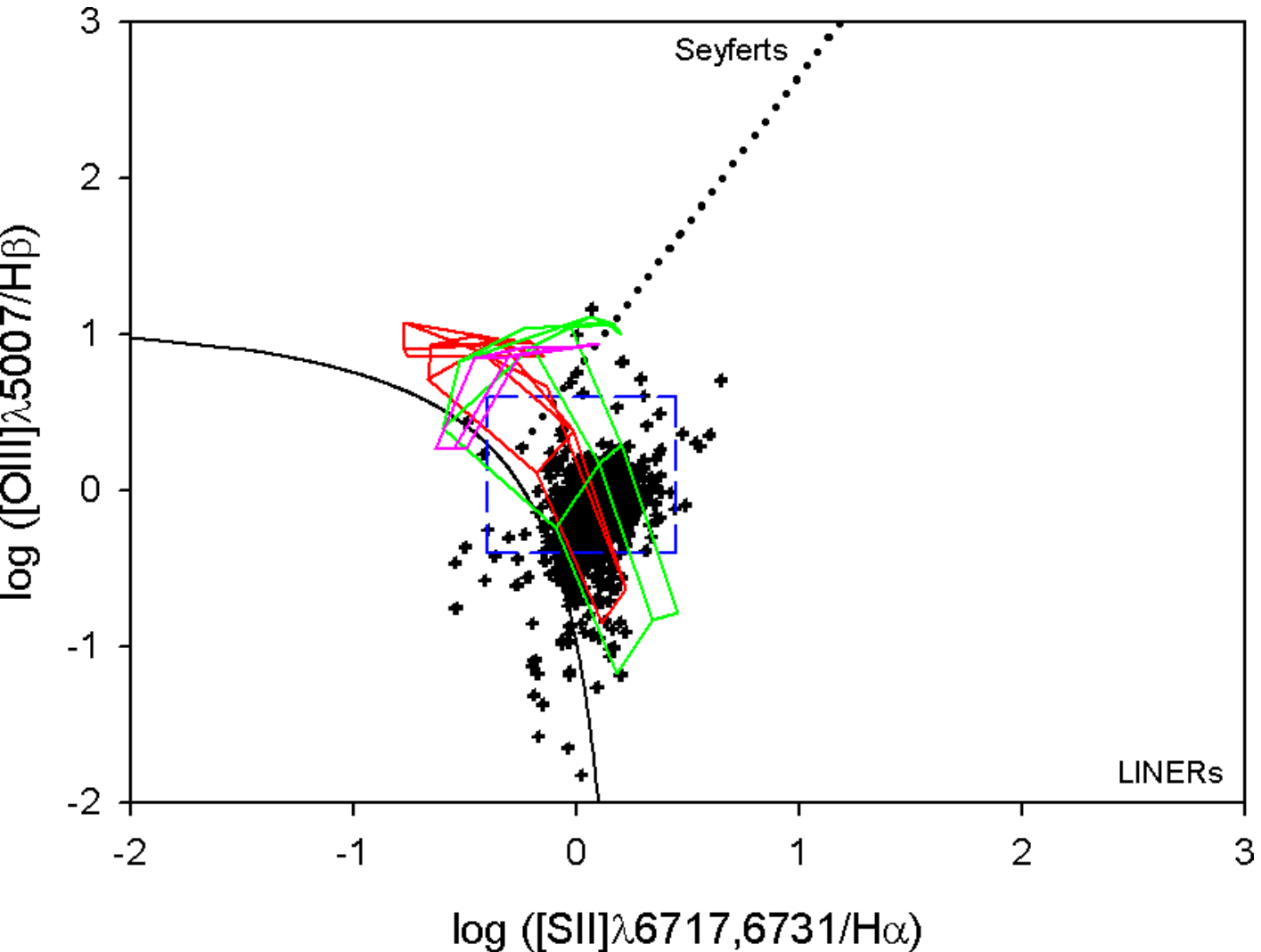}}\quad
         \subfigure[MCG-02-12-039]{\includegraphics[scale=0.38]{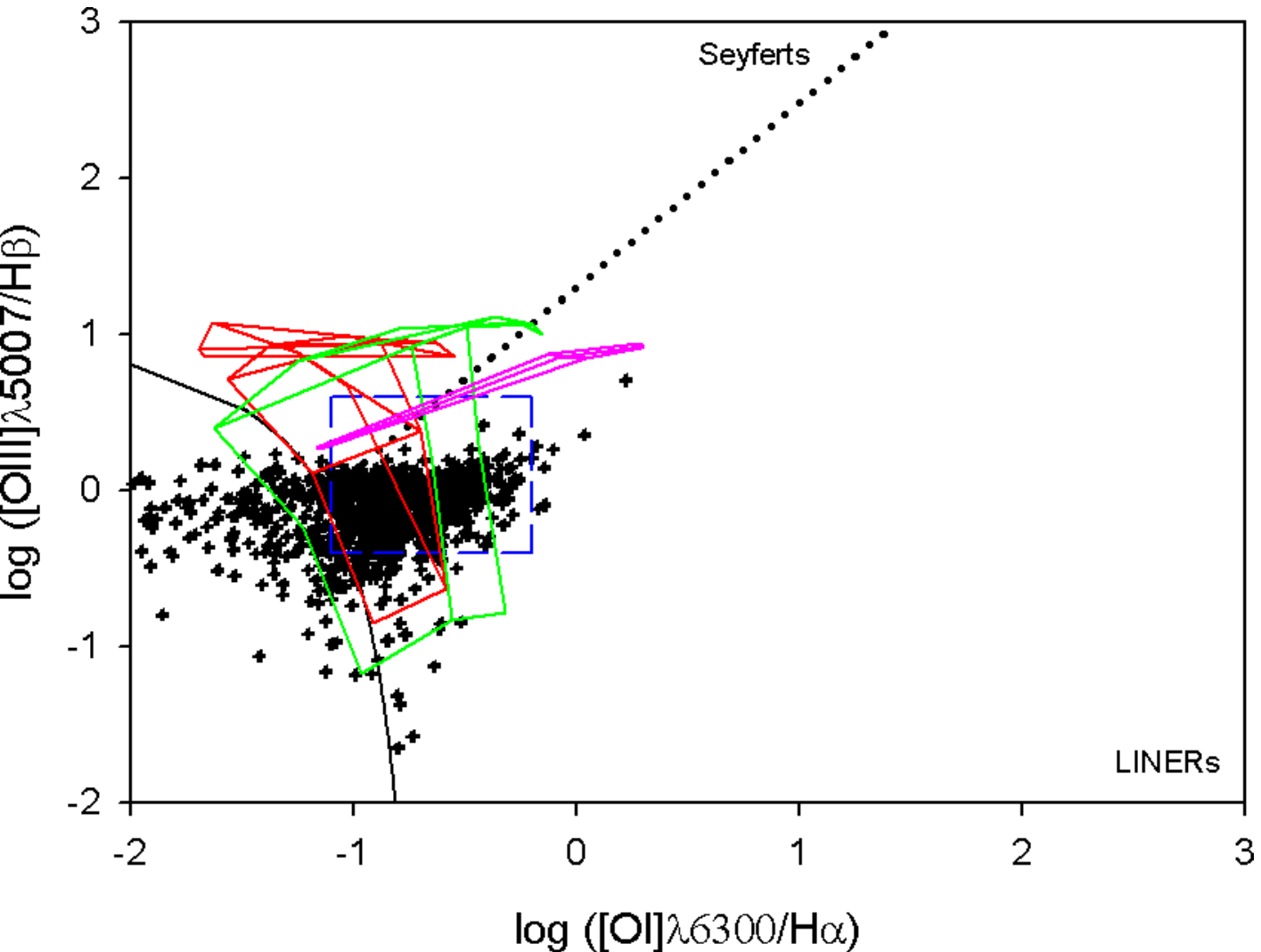}}}
   \mbox{\subfigure[PGC026269]{\includegraphics[scale=0.38]{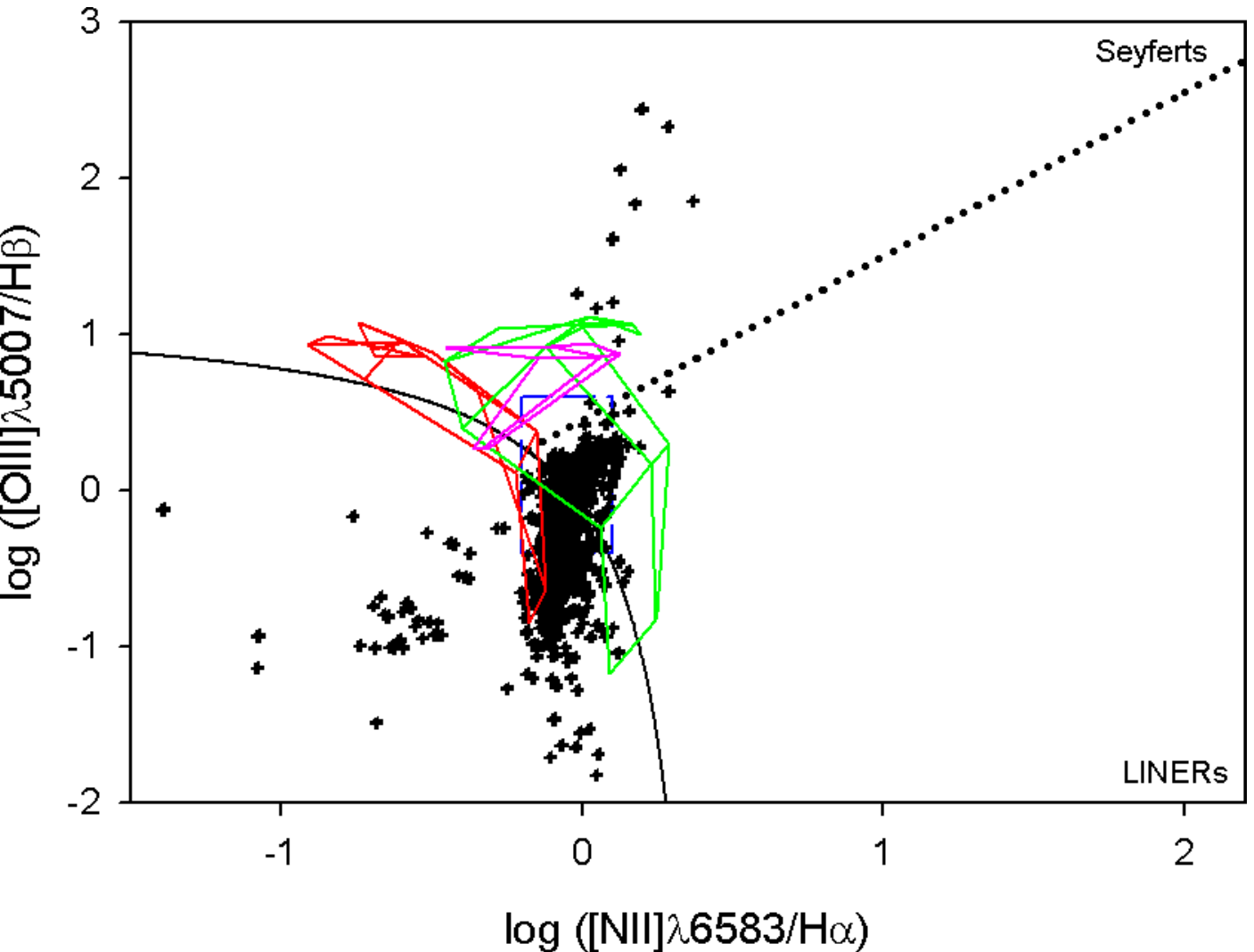}}\quad
         \subfigure[PGC026269]{\includegraphics[scale=0.38]{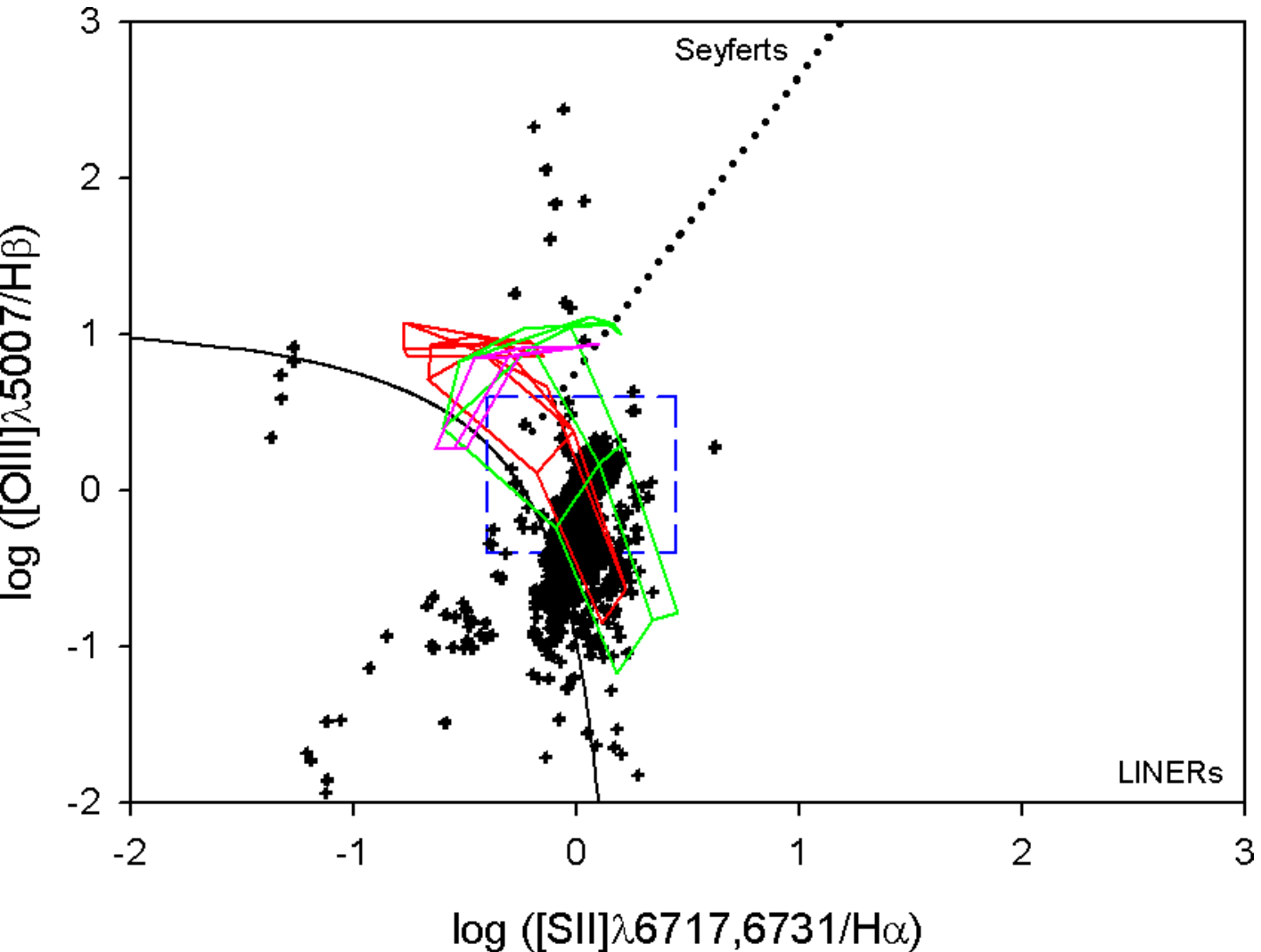}}\quad
         \subfigure[PGC026269]{\includegraphics[scale=0.38]{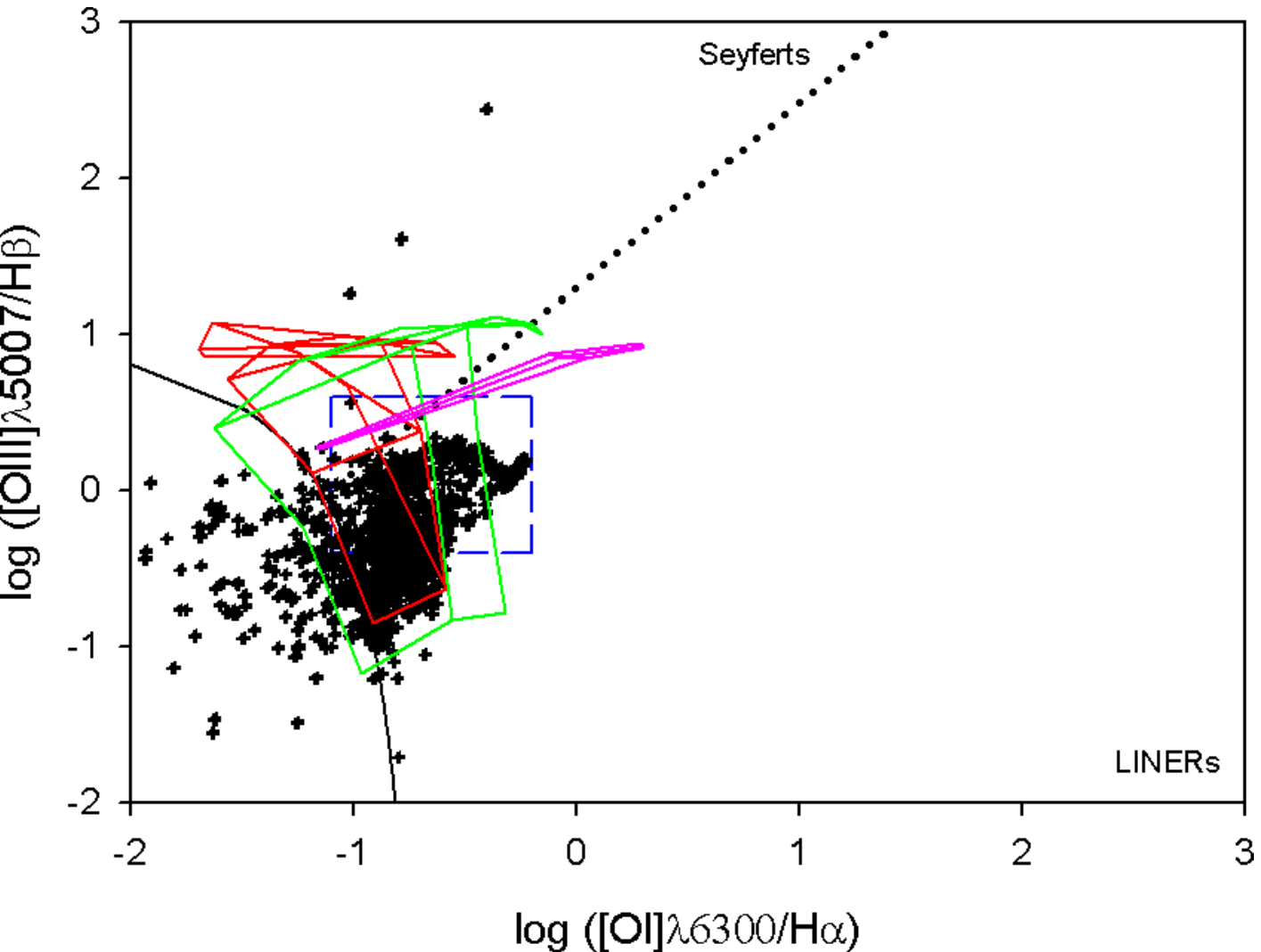}}}
   \mbox{\subfigure[PGC044257]{\includegraphics[scale=0.38]{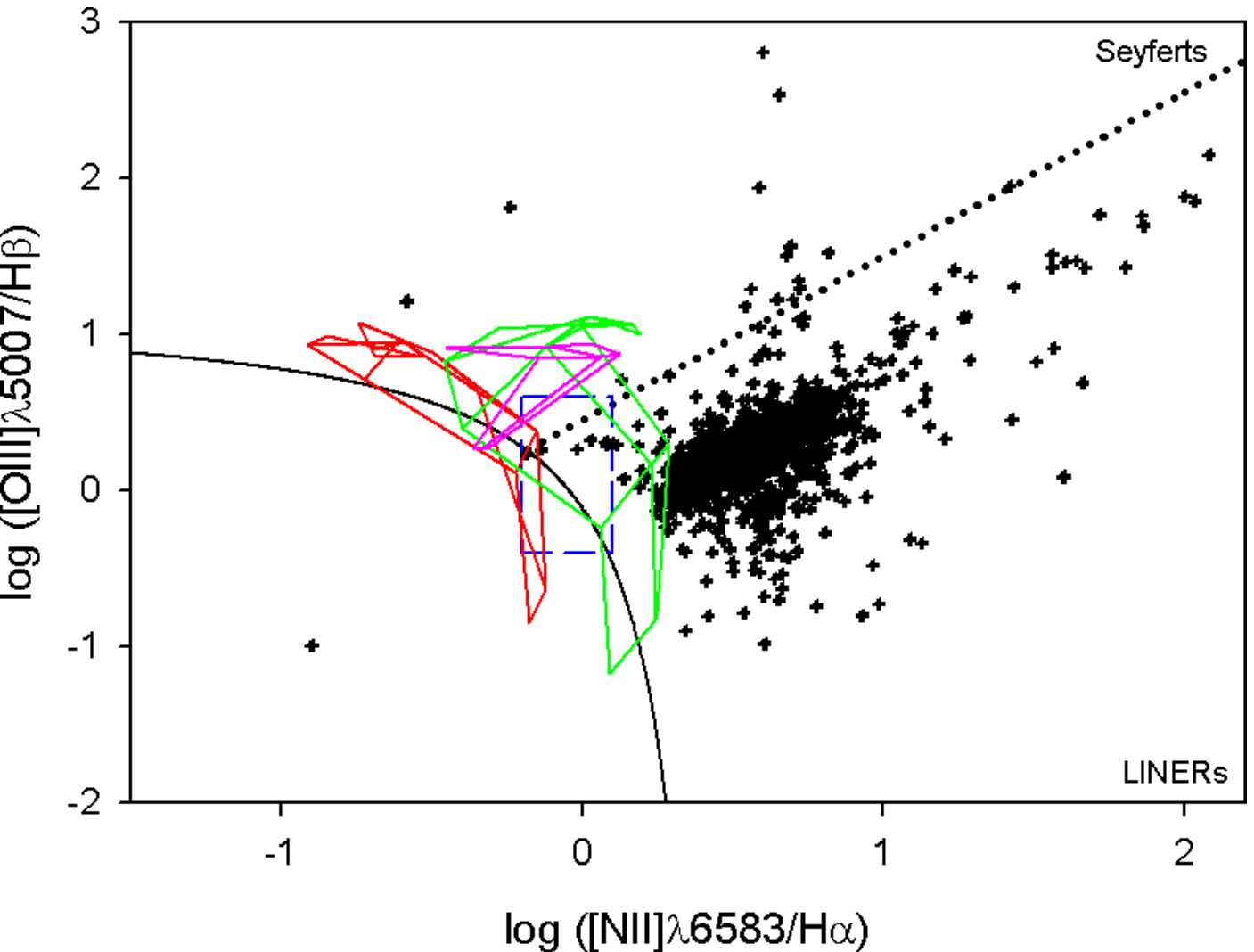}}\quad
         \subfigure[PGC044257]{\includegraphics[scale=0.38]{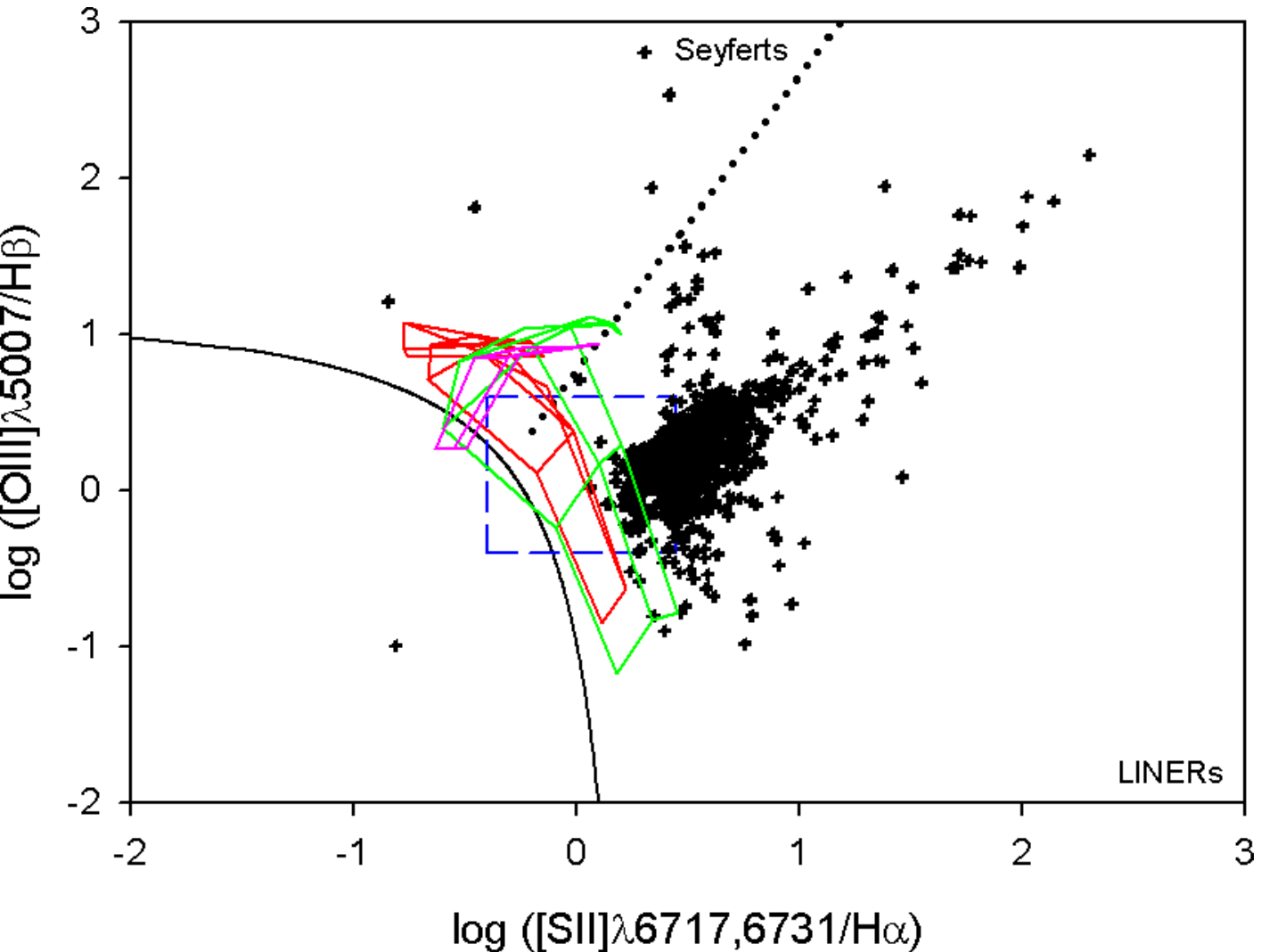}}\quad
         \subfigure[PGC044257]{\includegraphics[scale=0.38]{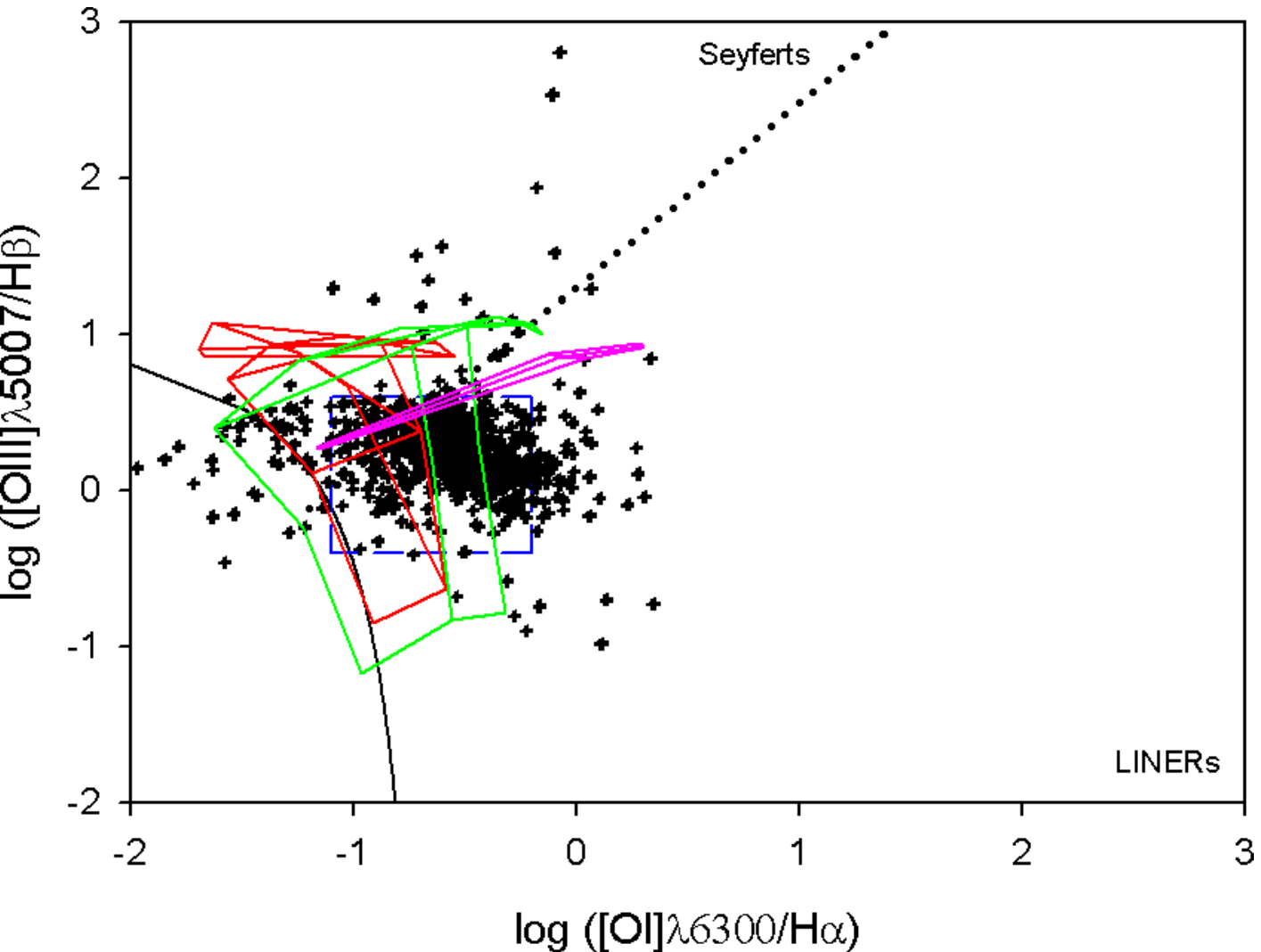}}}
\mbox{\subfigure[UGC09799]{\includegraphics[scale=0.38]{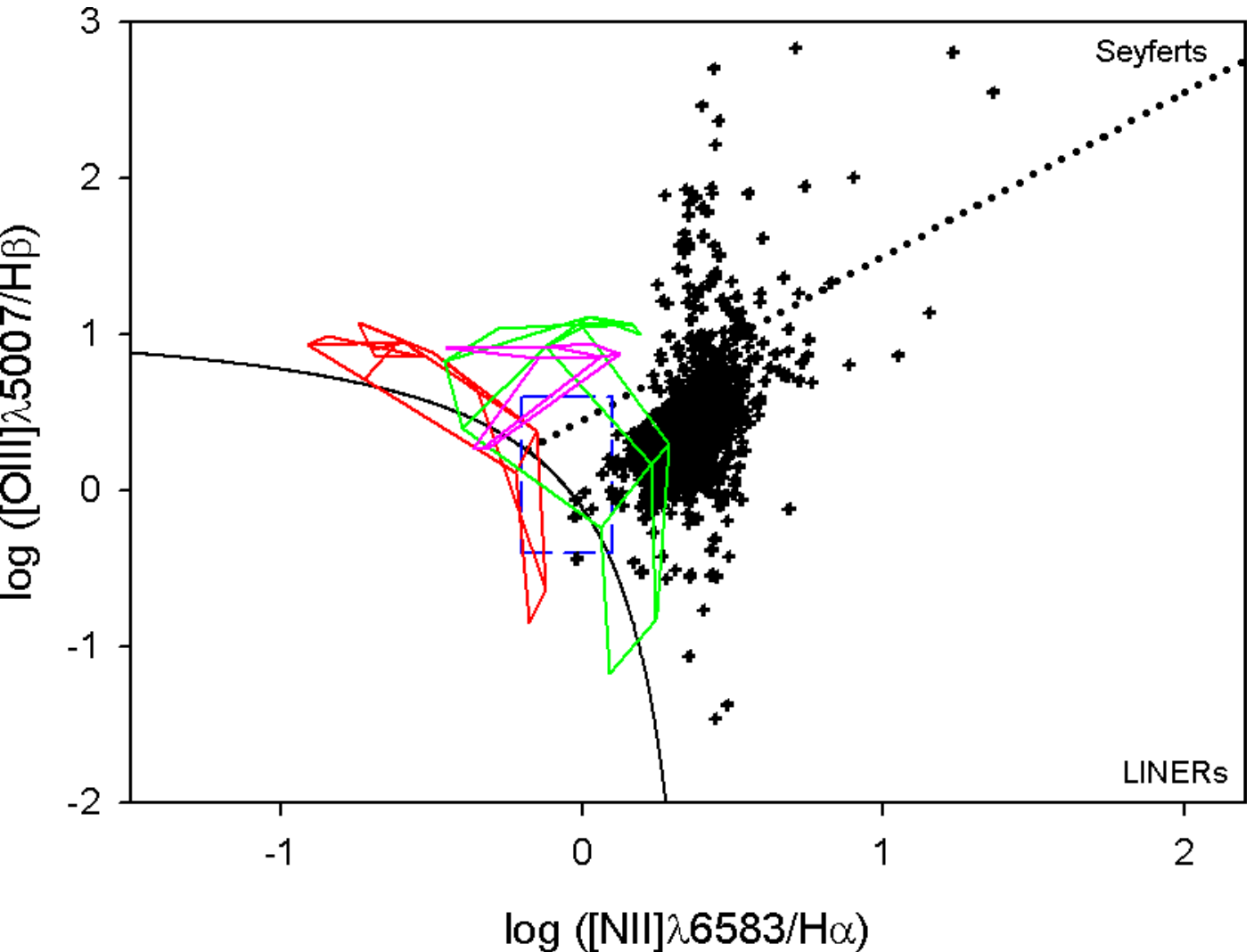}}\quad
         \subfigure[UGC09799]{\includegraphics[scale=0.38]{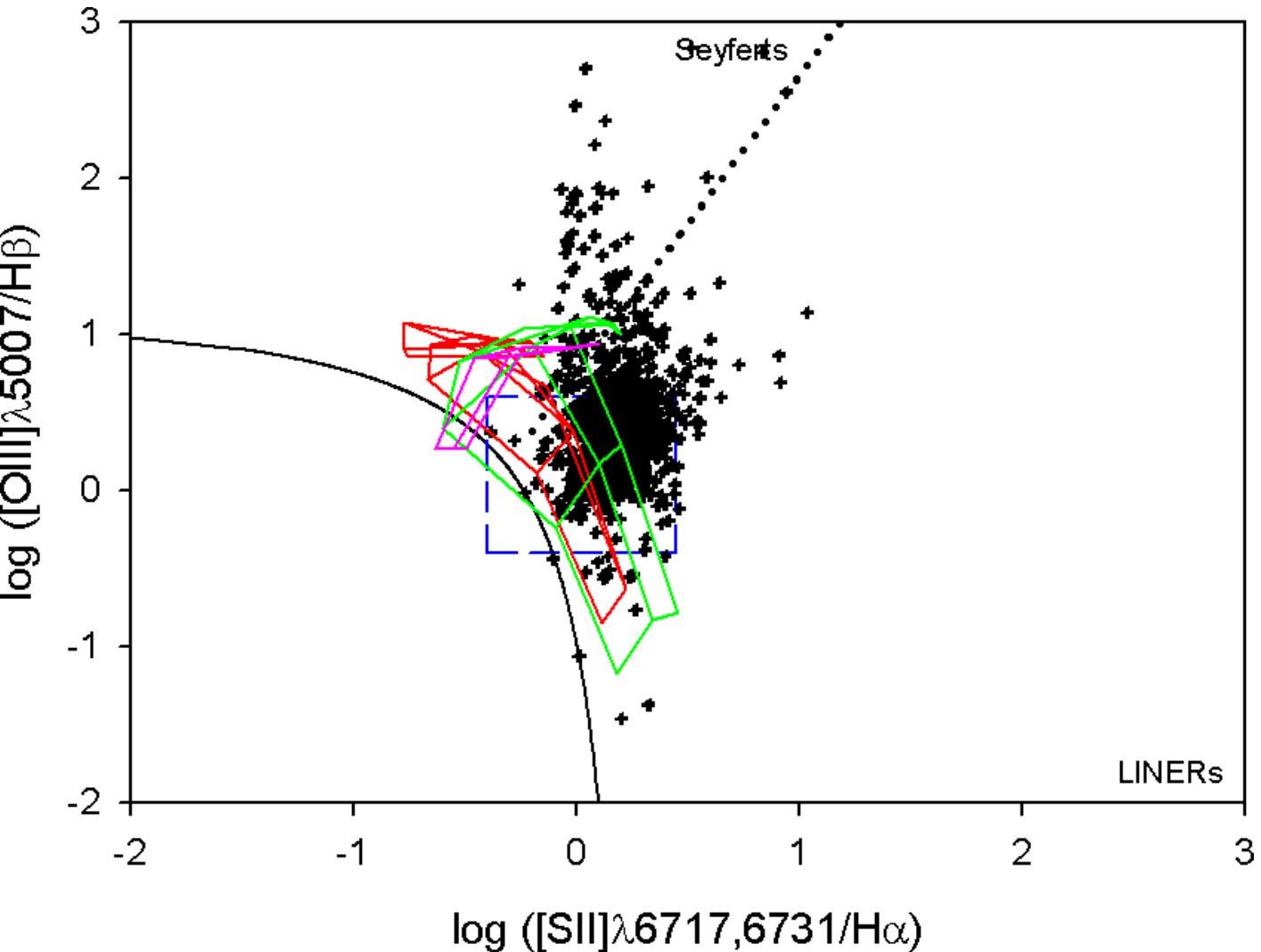}}\quad
         \subfigure[UGC09799]{\includegraphics[scale=0.38]{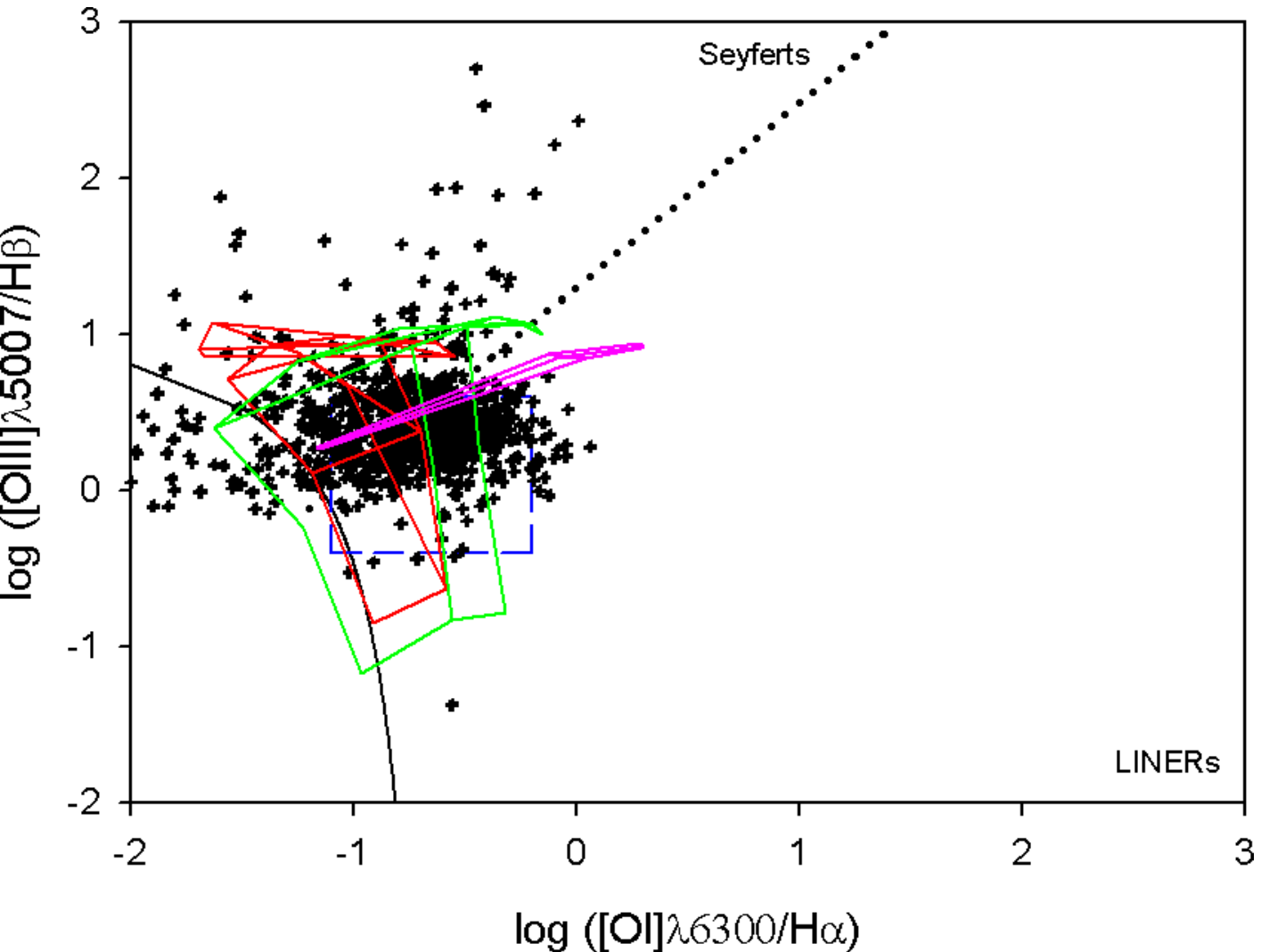}}}
\caption{Diagnostic diagrams for all four galaxies. From left to right: log ([OIII]$\lambda$5007/H$\beta$) vs. log ([NII]$\lambda$6584/H$\alpha$), log ([OIII]$\lambda$5007/H$\beta$) vs. log ([SII]$\lambda\lambda$6731,6717/H$\alpha$) and log ([OIII]$\lambda$5007/H$\beta$) vs. log ([OI]$\lambda$6300/H$\alpha$). The black solid curve is the theoretical maximum starburst model from Kewley et al.\ (2001), devised to isolate objects whose emission line ratios can be accounted for by the photoionisation by massive stars (below and to the left of the curve) from those where some other source of ionisation is required. The black-dashed curves in the [SII]$\lambda\lambda$6731,6717/H$\alpha$ and [OI]$\lambda$6300/H$\alpha$ diagrams represent the Seyfert-LINER dividing line from Kewley et al.\ (2006) and transposed to the [NII]$\lambda$6584/H$\alpha$ diagram by Schawinski et al.\ (2007). The predictions of different ionisation models for ionising the gas are overplotted in each diagram. The boxes show the predictions of photoionisation models by pAGB stars for Z = Z$_{\odot}$ and a burst age of 13 Gyr (Binette et al.\ 1994). The purple lines represent the shock grids of Allen et al.\ (2008) with solar metallicity and preshock magnetic fields B=1.0, 5.0 and 10 $\mu$G (left to right). The horizontal purple lines show models with increasing shock velocity V  = 100, 500 and, 1000 km s$^{-1}$, and the densities n$_{e}$ is 100 cm$^{-3}$. Grids of photoionization by an AGN (Groves et al.\ 2004) are indicated by green and red curves, with n$_{e}$ = 100 cm$^{-3}$ and a power-law spectral index of $\alpha$ = --2, --1.4 and --1.2 (from left to right). The models for Z = Z$_{\odot}$ (red) and Z = 2Z$_{\odot}$ (green), and the horizontal lines trace the ionisation parameter $\log$ U, which increases with the [OIII]$\lambda$5007/H$\beta$ ratio from $\log$ U = --3.6, --3.0, --2.0, --1.0, 0.0.}
\label{fig:BPTs} 
\end{figure*}

We investigate this further by plotting the other BPT diagrams, on a spaxel-by-spaxel basis, for all four galaxies (shown in Figure \ref{fig:BPTs}): [OIII]$\lambda$5007/H$\beta$ vs [NII]$\lambda$6584/H$\alpha$, [OIII]$\lambda$5007/H$\beta$ vs [SII]$\lambda\lambda$6717,6731/H$\alpha$, and [OIII]$\lambda$5007/H$\beta$ vs [OI]$\lambda$6300/H$\alpha$. One BCG PGC026269 show several HII pixels in all three BPT diagrams. Figure \ref{fig:MCG_BPT} show that these HII pixels occur throughout the centre of the galaxy.

On these plots, we compare our observations to the photoionisation models for pAGB stars with Z=Z$_{\odot}$ (Binette et al.\ 1994). These models are consistent with most of our observations. A model with Z=1/3 Z$_{\odot}$ will be shifted towards lower values on the x-axis of the three BPTs. The pAGB scenario has recently been revisited by Stas\'inska et al.\ (2008), whose extensive grid of photoionisation models (see their Figure 5) cover most of the regions occupied by our spatially resolved measurements.

Three other grids of ionisation models are overplotted on the BPT diagrams (Figure \ref{fig:BPTs}). The plotted AGN photoionisation models (Groves, Dopita $\&$ Sutherland 2004) have an electron density, n$_{e}$ = 100 cm$^{-3}$, metallicities of solar, Z=Z$_{\odot}$ (red grids), and twice solar (green grids), a range of ionisation parameter (--3.6 $< \log U < $ 0.0) and a power-law ionising spectrum with spectral index $\alpha$ = --2, --1.4, and --1.2. A harder ionising continuum, with $\alpha$ = --1.2, boosts [SII]$\lambda,\lambda$6717,6731 and [OI]$\lambda$6300 relative to H$\alpha$.

We also compared our results with shock models (Allen et al.\ 2008, purple grids). In Figure \ref{fig:BPTs}, we plot the grids with Z=Z$_{\odot}$, preshock densities of 100 cm$^{-3}$, shock velocities of 100, 500 and 1000 km s$^{-1}$, and preshock magnetic fields of B=1, 5 and 10 $\mu$G. Shock models with a range of magnetic field strengths (B=1,5 and 10 $\mu$G) match our observations. Interstellar magnetic fields of B $\sim$ 1 -- 10 $\mu$G are typical of what is observed in elliptical galaxies (Mathews $\&$ Brighenti 1997). Overall, shock models reproduce the majority of our data in the three emission-line ratio diagrams. The shock grids with lower metallicity (e.g. LMC and SMC metallicities) are not consistent with our measurements.\footnote{We downloaded the shock and AGN grids from the web page http://www.strw.leidenuniv.nl/$\sim$brent/itera.html.}

As shown in Table \ref{table:objects2} -- we have very weak as well as strong radio fluxes in our small sample. We therefore believe that we are not particularly prone to biases such as the fact that a priori choice of galaxies with strong radio fluxes will result in finding a sample where high [NII]/H$\alpha$ would be more common than in an optically, or H$\alpha$ selected sample only -- where starforming high H$\alpha$/[NII] sources might be more abundant.

\section{Conclusion}
\label{summary}
We present detailed integral field unit (IFU) observations of the central few kiloparsecs of the ionised nebulae surrounding four active 
CCGs in cooling flow clusters (Abell 0496, 0780, 1644 and 2052). Our sample consists of CCGs with H$\alpha$ filaments. We observed the detailed optical 
emission-line (and simultaneous absorption line) data over a broad wavelength range to probe the dominant ionisation processes, excitation sources, morphology and 
kinematics of the hot gas (as well as the morphology and kinematics of the stars). Two of the four sources have not been observed with IFU data before (Abell 0780 and 1644), and for the other two sources we observed with significantly improved integration times (and number of lines) than previous studies (Hatch et al.\ 2007, Edwards et al.\ 2009). This will help form a complete view of the different phases (hot gas and stars) and how they interact in the processes of star formation and 
feedback detected in central galaxies in cooling flow clusters, as well as the influence of the host cluster. 

The total extinction maps are presented in Figure \ref{fig:MCG_extinct} and shows extinction which agrees well with previously derived long-slit values (where available).
From long-slit spectra, Crawford et al.\ (1999) derived the total extinction of PGC044257 as 0.46 to 0.63 mag. This agrees with the extinction we derived for the very centre of the galaxy in Figure \ref{fig:MCG_extinct}, but on average our spatially resolved extinction is slightly lower (0.195 mag). Crawford et al.\ (1999) derived an integrated internal extinction of $E(B-V)_{internal}$ of 0.22 mag for the centre of UGC09799. This corresponds very well to what we derived and plotted in Figure \ref{fig:MCG_extinct}, although we find that some regions show much higher internal extinction (on average 0.42 mag).

We derive a range of different kinematic properties, given the small sample size. For Abell 0496 and 0780, we find that the stars and gas are kinematically decoupled, and in the case of Abell 1644 we find that these components are aligned. For Abell 2052, we find that the gaseous components show rotation even though no rotation is apparent in the stellar components. To the degree that our spatial resolution reveals, it appears that all the optical forbidden and hydrogen recombination lines originate in the same gas for all the galaxies.

All galaxies show important LINER emission, but that at least one has significant Seyfert emission areas, and at least one other has significant HII like emission line ratios for many pixels (consistent with the long-slit observations of McDonald et al.\ (2012)). We also show that the hardness of the ionising continuum does not decrease with galactocentric distance (except for PGC044257 which show an interesting core separation of the emission in the very centre of the galaxy in Figure \ref{fig:UGC_BPT}a). The radial profiles of diagnostic line ratios, [OIII]$\lambda$5007/H$\beta$ and [NII]$\lambda$6584/H$\alpha$, show that they are roughly constant with radius for three of the four galaxies (all except PGC044257). This indicates that the dominant ionising source is not confined to the nuclear region in the two objects and that the ionised gas properties are homogeneous in the emission-line regions across each galaxy.

Overall, it remains difficult to dissentangle the dominant photoionisation mechanisms, even with more line measurements. The AGN photoionisation models (with higher metallicity) are the best able to reproduce our spatially-resolved line ratios the in all of the three BPT diagrams simultaneously of most objects, even though shock models and pAGB stars can not be conclusively eliminated. We also do not see extended [OI]$\lambda$6300 emission, following the morphology of the strong emission lines (e.g.\ Farage et al.\ 2010), therefore, it is unlikely that shock excitation is the dominant ionising source in the galaxies of this limited sample. In addition, multiwavelength observations (as discussed in the previous section) favour the AGN photoionisation mechanism, especially in the case of PGC026269 and UGC09799.

\section*{Acknowledgments}
SIL is financially supported by the South African National Research Foundation. We thank James Turner, Bryan Miller, and Michele Cappellari for providing helpful scripts, as well as Marc Sarzi for helpful discussions. Based on observations obtained on the Gemini South telescope. The Gemini Observatory is operated by the Association of Universities for Research in Astronomy, Inc., under cooperative agreement with the NSF on behalf of the Gemini Partnership: the National Science Foundation (USA), the Science and Technology Facilities Council (UK), the National Research Council (Canada), CONICYT (Chile), the Australian Research Council (Australia), CNPq (Brazil) and CONICET (Argentina). This research has made use of the NASA/IPAC Extragalactic Database (NED) which is operated by the Jet Propulsion Laboratory, California Institute of Technology.

\appendix

\section{Stellar kinematics comparison}
\label{stellar_kinematics_comparison}

\begin{figure*}
   \centering
   \includegraphics[scale=0.6]{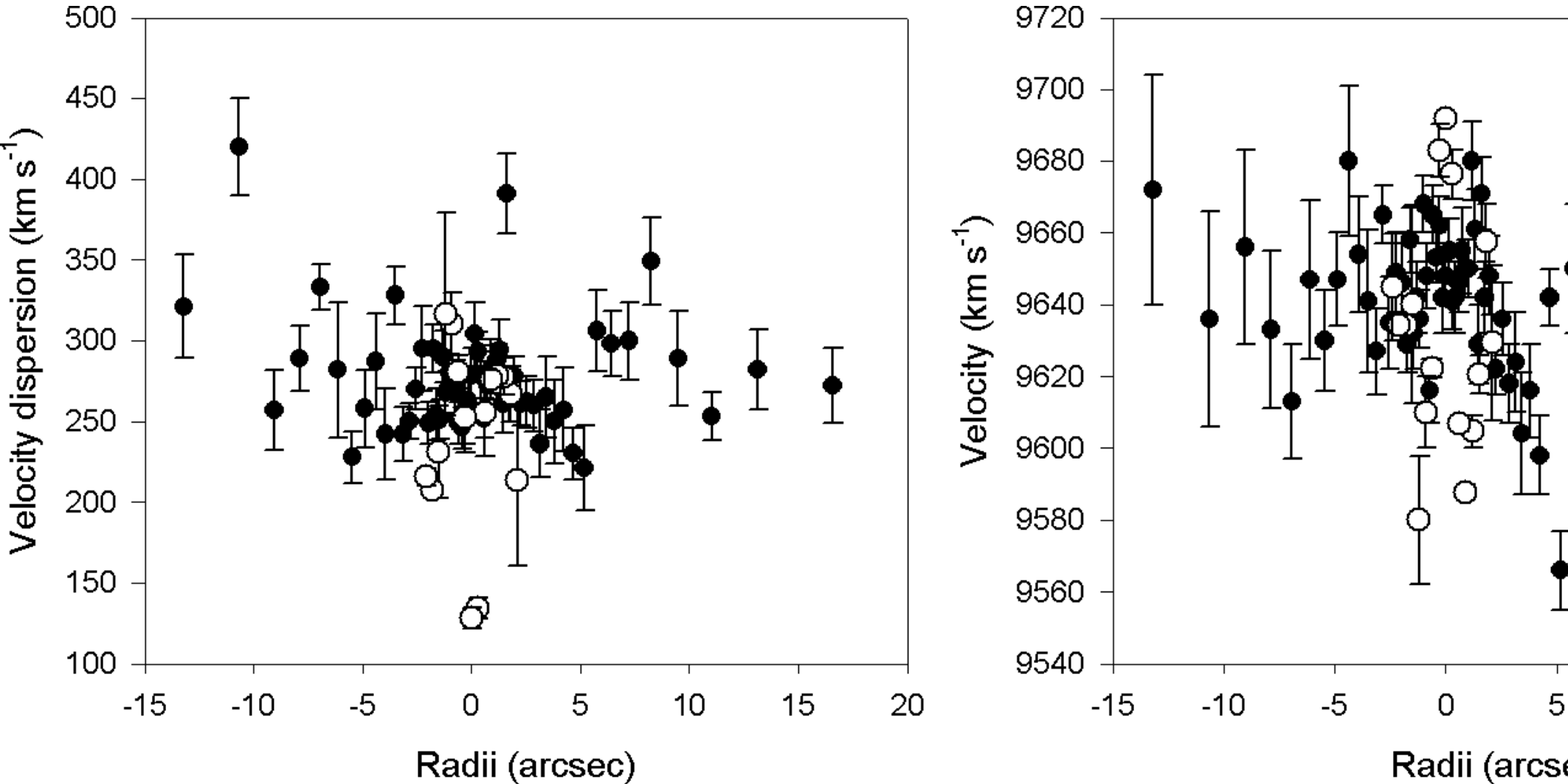}
   \caption{MCG-02-12-039: Kinematic data (velocity and velocity dispersion) extracted from the IFU data (empty symbols), compared with long-slit data along the same axis from Paper 1.}
   \label{fig:MCG_kinematics}
\end{figure*}

\begin{figure*}
   \centering
   \includegraphics[scale=0.6]{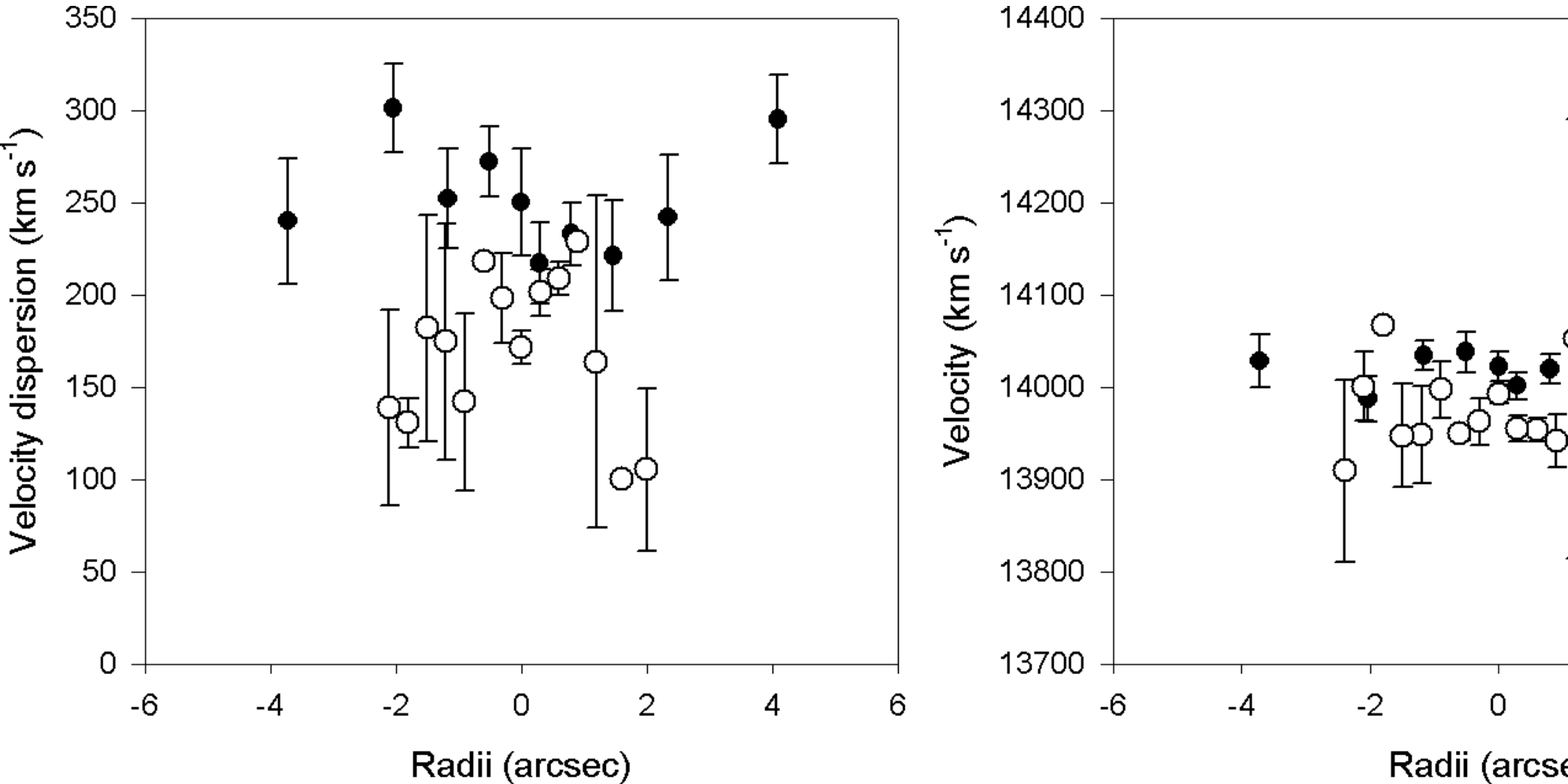}
   \caption{PGC026269: Kinematic data (velocity and velocity dispersion) extracted from the IFU data (empty symbols), compared with long-slit data along the same axis from Paper 1.}
   \label{fig:PGC026_kinematics}
\end{figure*}

\begin{figure*}
   \centering
   \includegraphics[scale=0.6]{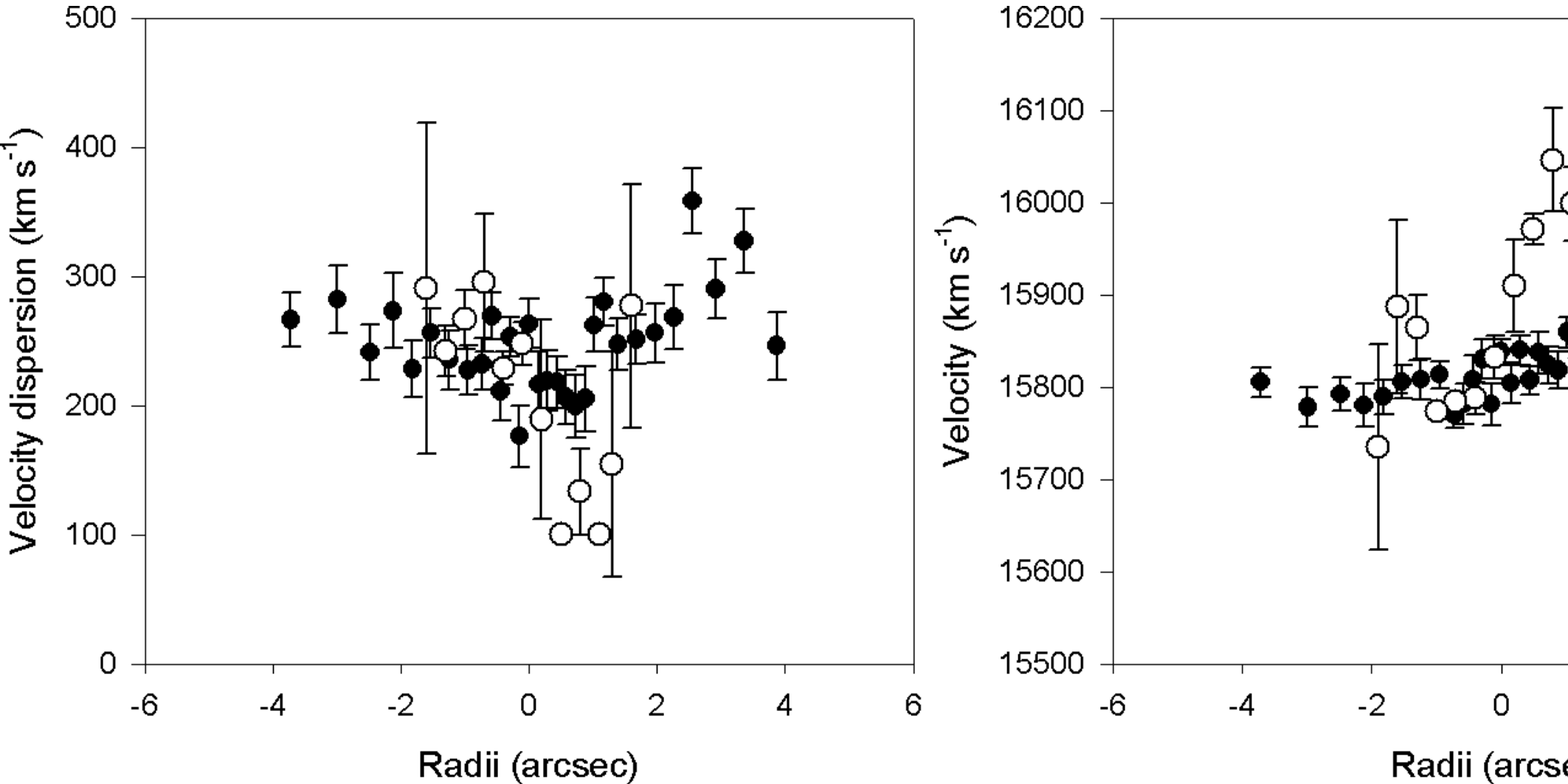}
   \caption{PGC044257: Kinematic data (velocity and velocity dispersion) extracted from the IFU data (empty symbols), compared with long-slit data along the same axis from Paper 1.}
   \label{fig:PGC044_kinematics}
\end{figure*}

For comparison purposes, the kinematics were extracted from the IFU image along the same slit position as the long-slit data in Paper 1. This comparison is showed in Figures \ref{fig:MCG_kinematics} to \ref{fig:PGC044_kinematics}, and the data points compare satisfactory. We emphasize that the long-slit data were binned to a constant (higher) S/N, where the IFU data are three binned spaxels (0.3'') along the same alignment.

\section{Line maps}

\begin{figure*}
\mbox{\subfigure[MCG-02-12-039: H$\alpha$]{\includegraphics[scale=0.4, trim=1mm 1mm 50mm 1mm, clip]{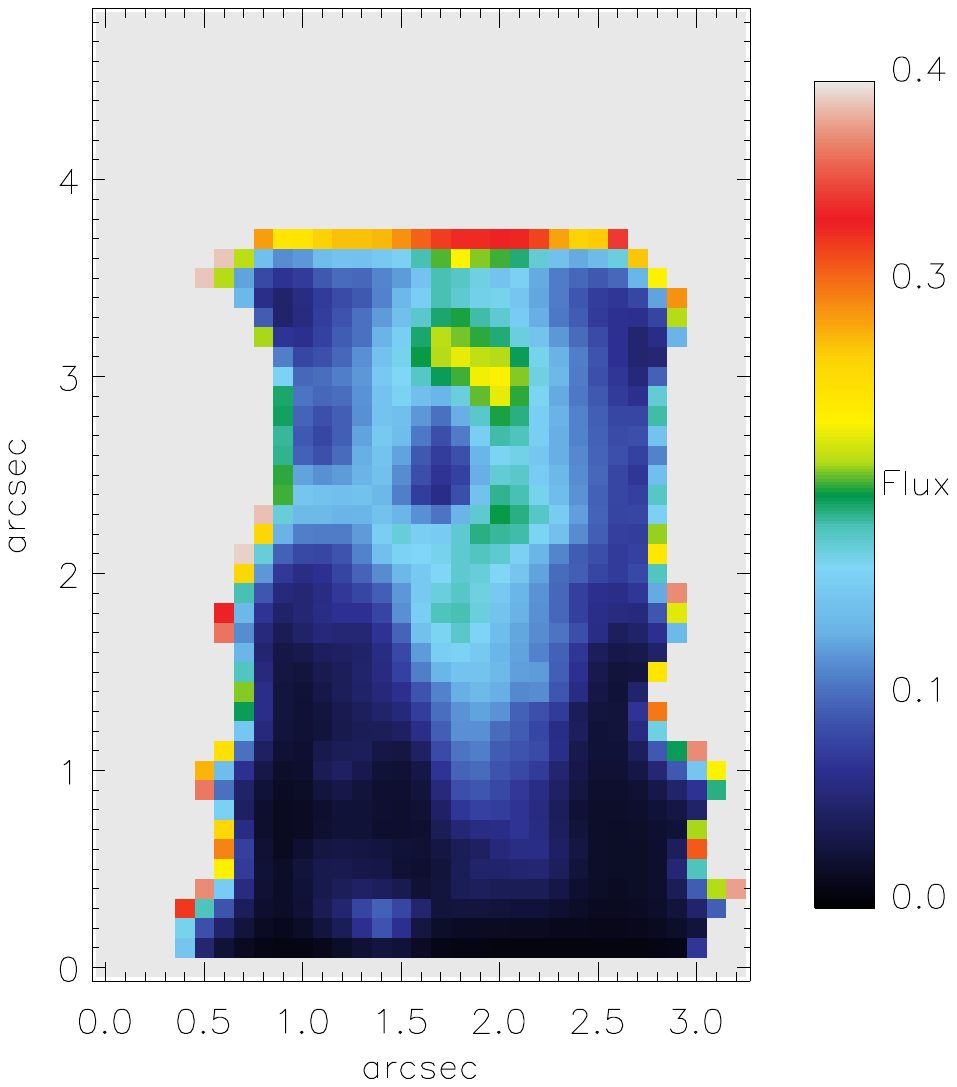}}\quad
\subfigure[MCG-02-12-039: {[NII]}$\lambda$6583/H$\alpha$]{\includegraphics[scale=0.4, trim=1mm 1mm 50mm 1mm, clip]{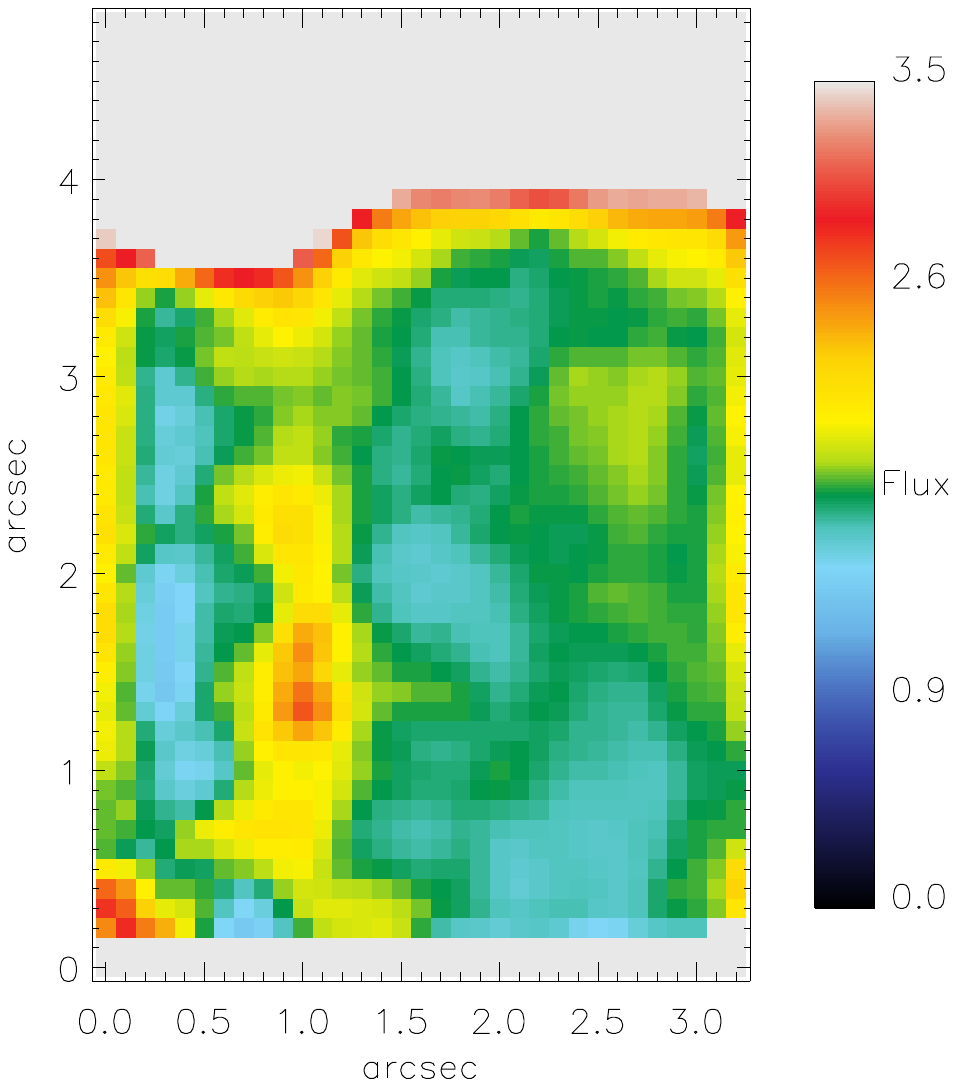}}
\subfigure[MCG-02-12-039: {[OIII]}$\lambda$5007/H$\beta$]{\includegraphics[scale=0.4, trim=1mm 1mm 50mm 1mm, clip]{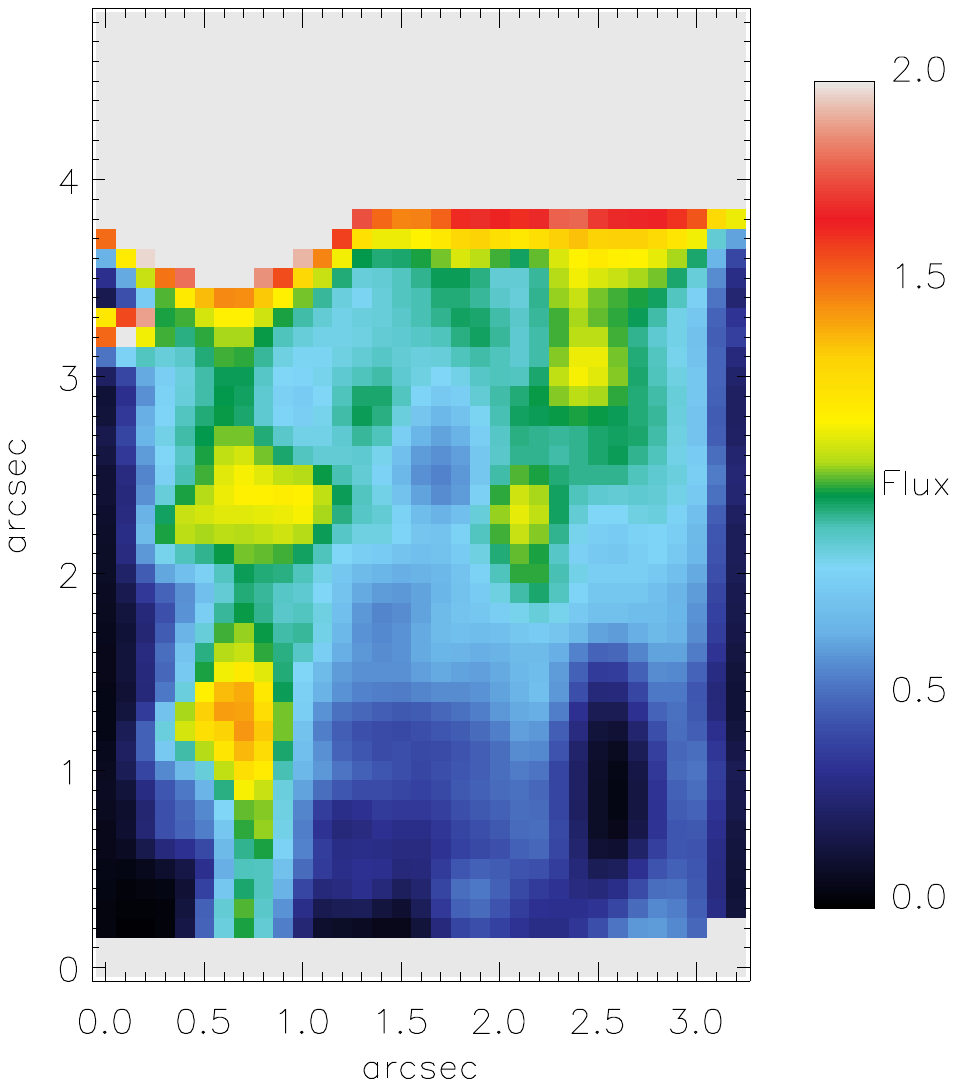}}}
\mbox{\subfigure[PGC026269: H$\alpha$]{\includegraphics[scale=0.4, trim=1mm 1mm 50mm 1mm, clip]{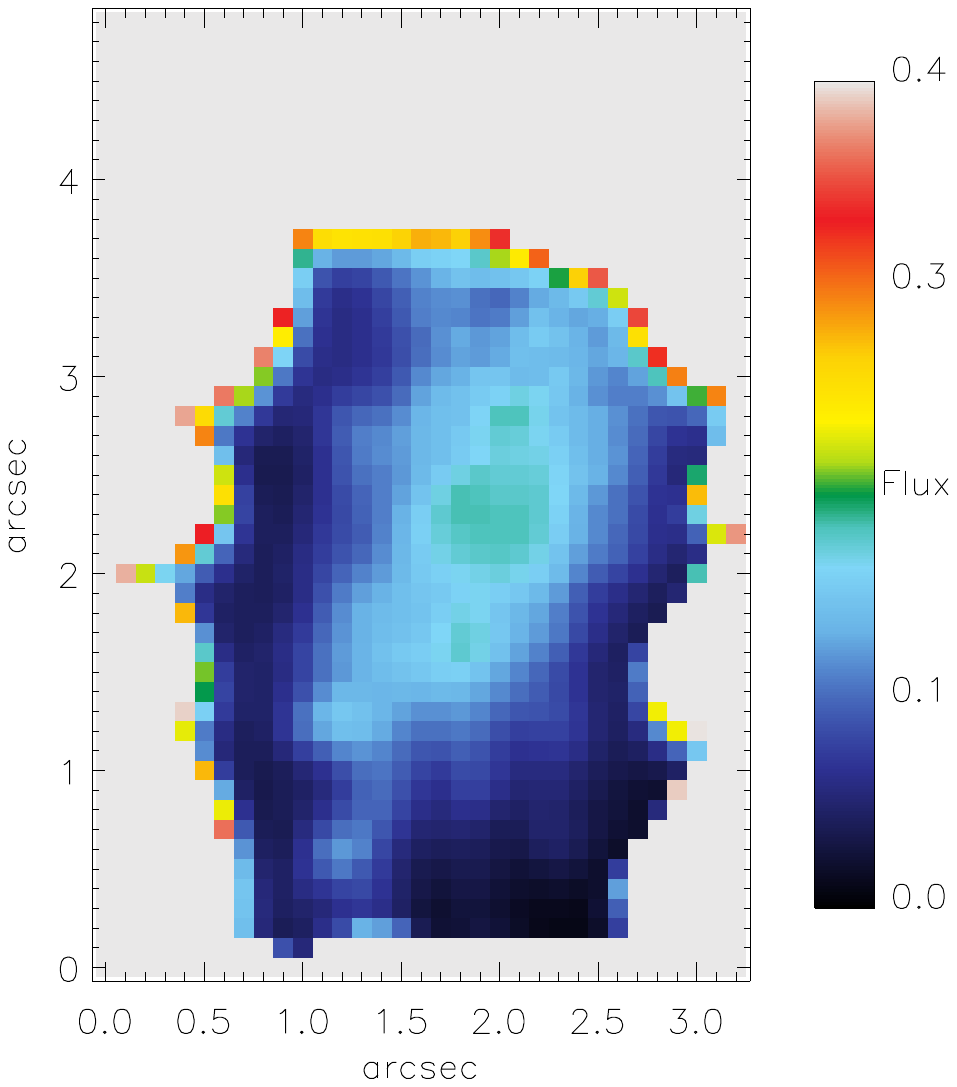}}\quad
\subfigure[PGC026269: {[NII]}$\lambda$6583/H$\alpha$]{\includegraphics[scale=0.4, trim=1mm 1mm 50mm 1mm, clip]{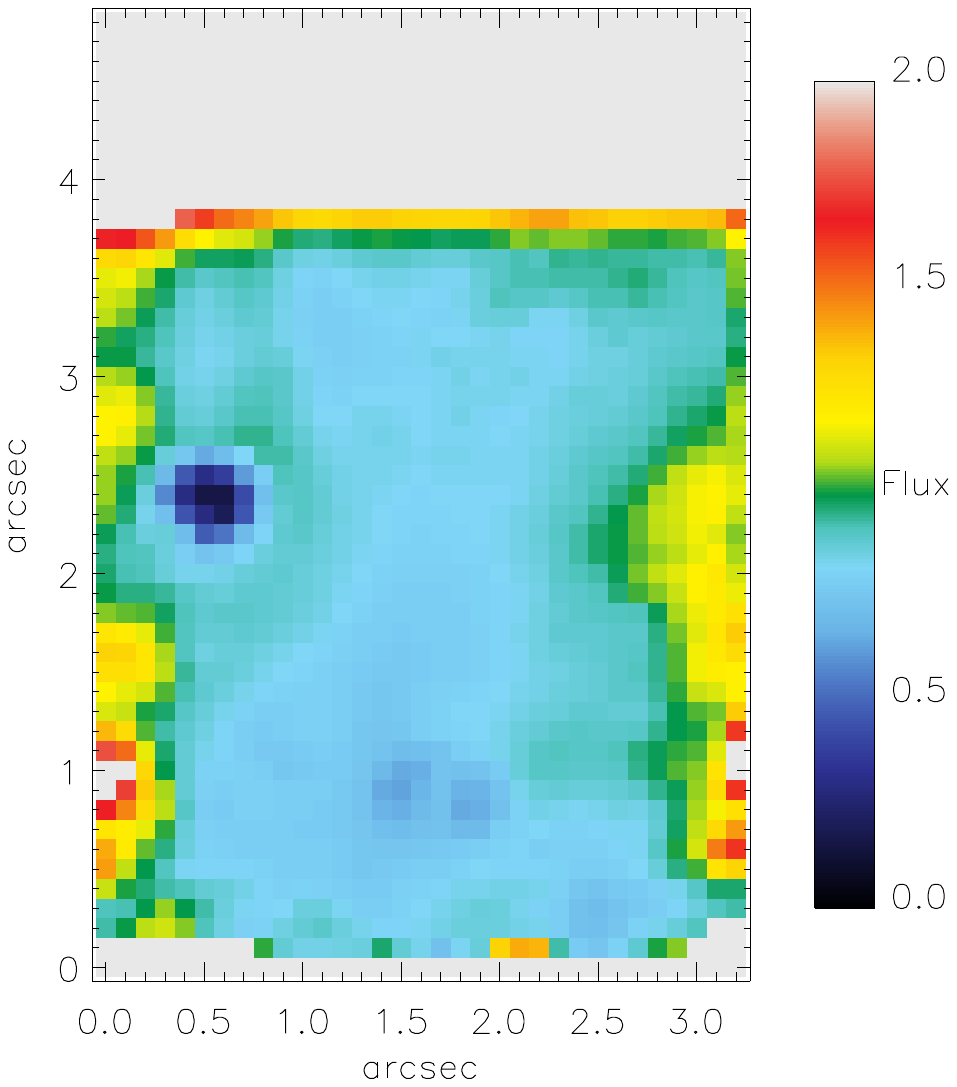}}
\subfigure[PGC026269: {[OIII]}$\lambda$5007/H$\beta$]{\includegraphics[scale=0.4, trim=1mm 1mm 50mm 1mm, clip]{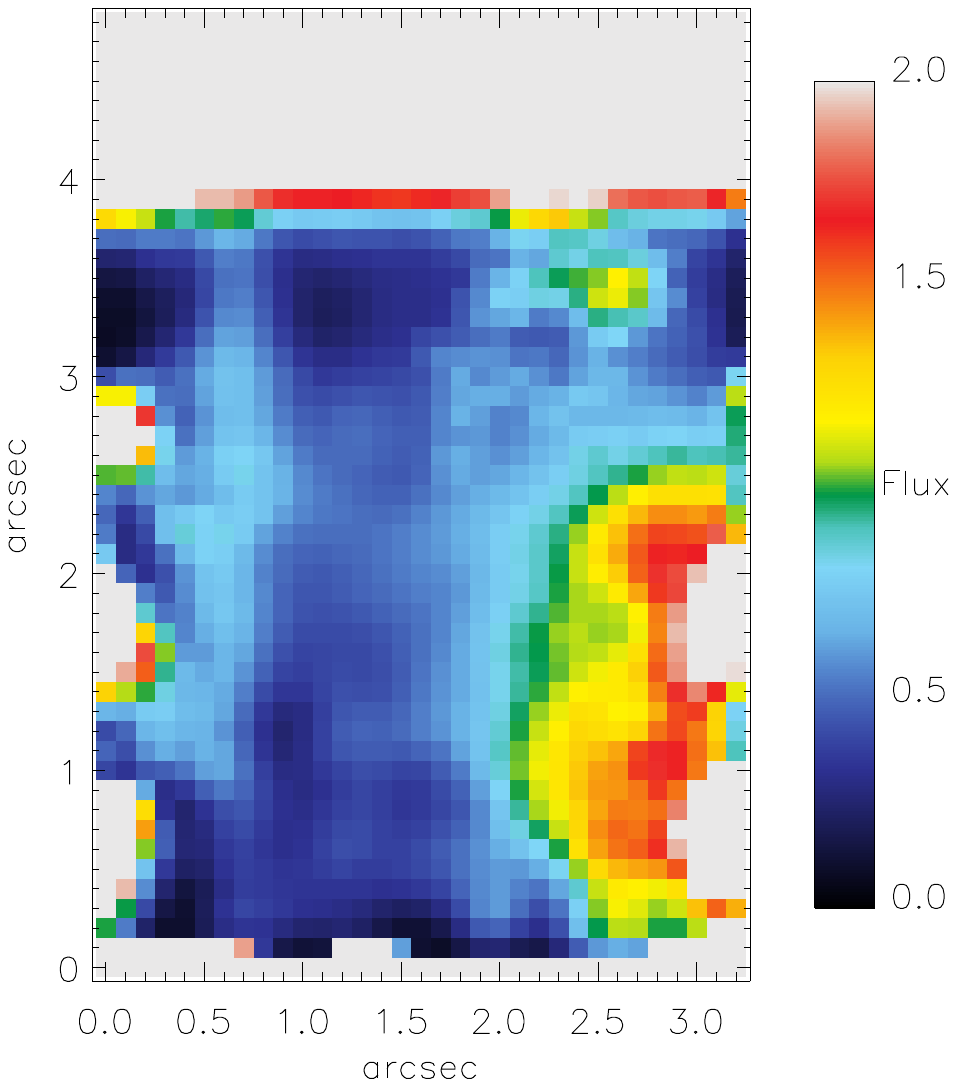}}}
\mbox{\subfigure[PGC044257: H$\alpha$]{\includegraphics[scale=0.4, trim=1mm 1mm 50mm 1mm, clip]{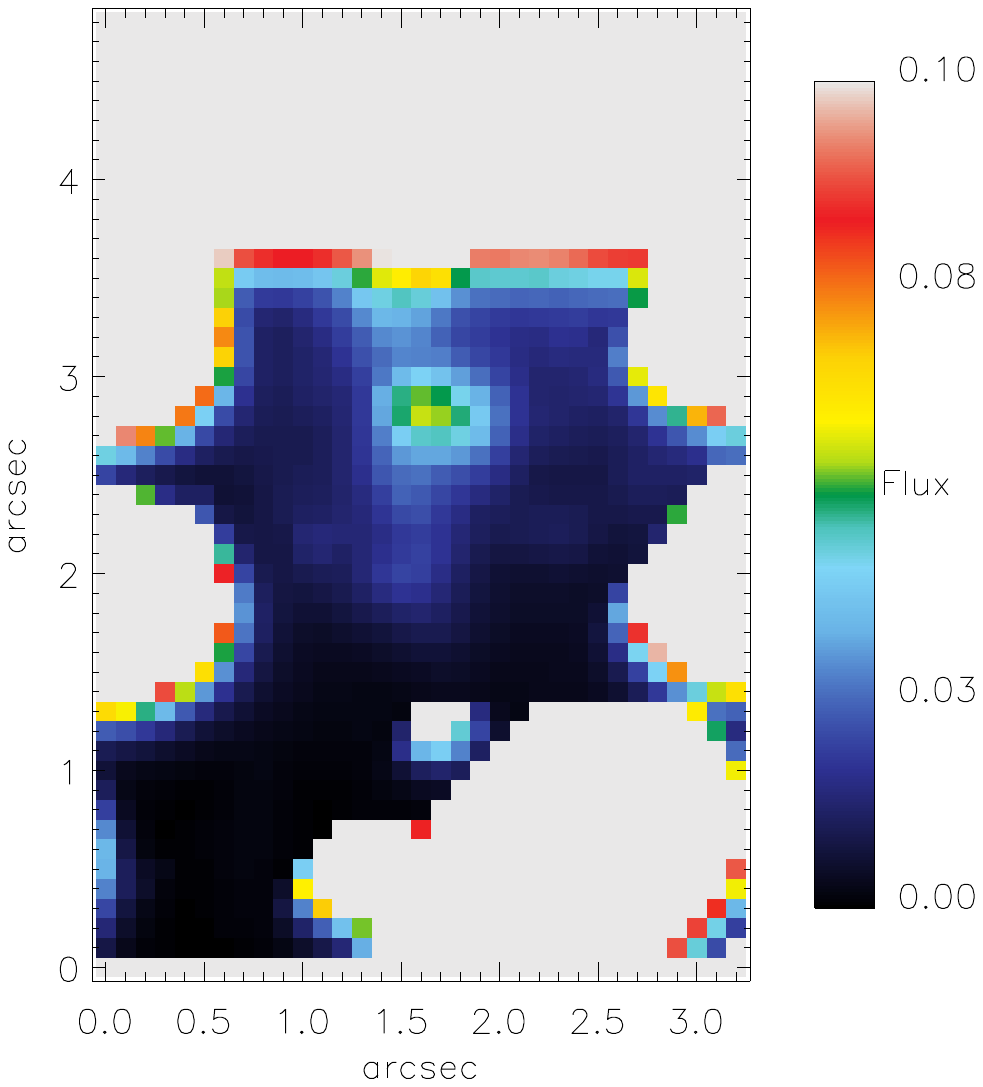}}\quad
\subfigure[PGC044257: {[NII]}$\lambda$6583/H$\alpha$]{\includegraphics[scale=0.4, trim=1mm 1mm 50mm 1mm, clip]{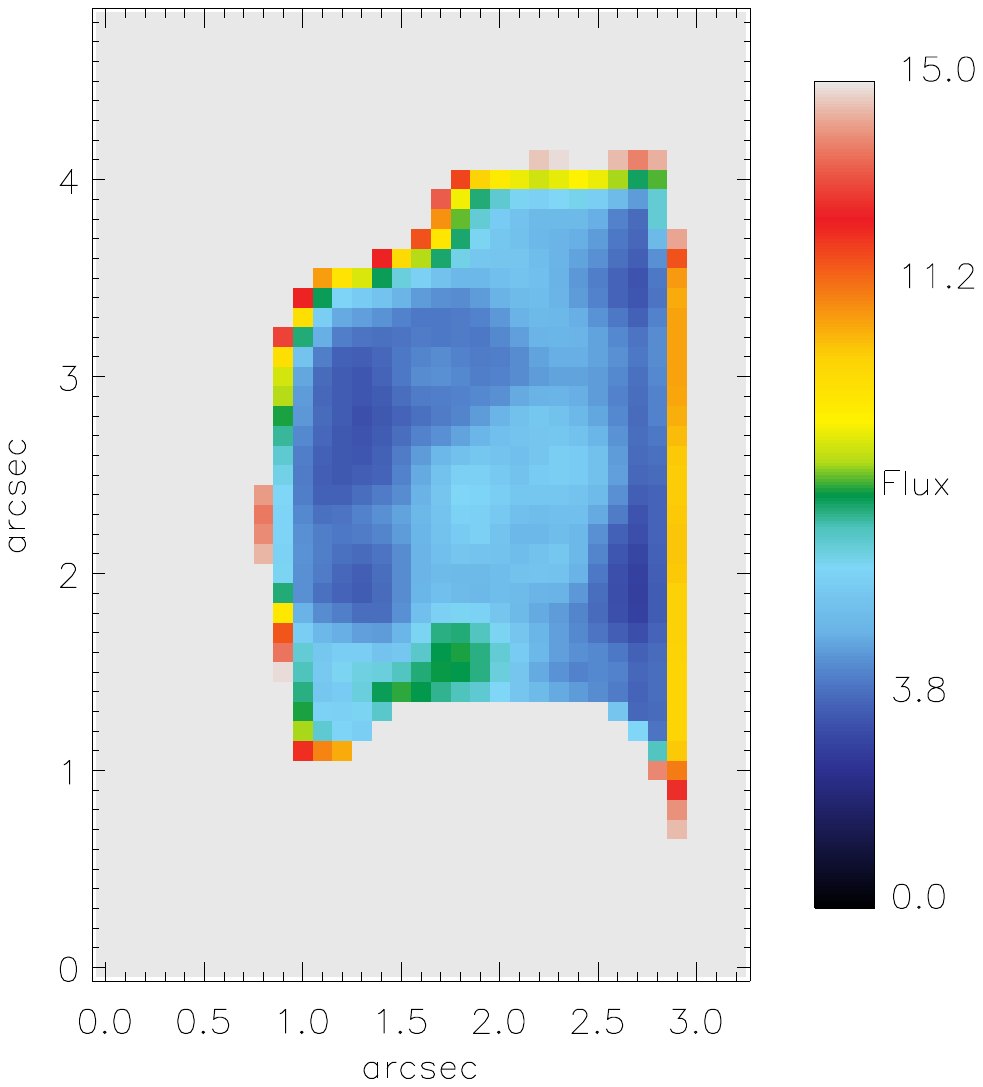}}
\subfigure[PGC044257: {[OIII]}$\lambda$5007/H$\beta$]{\includegraphics[scale=0.4, trim=1mm 1mm 50mm 1mm, clip]{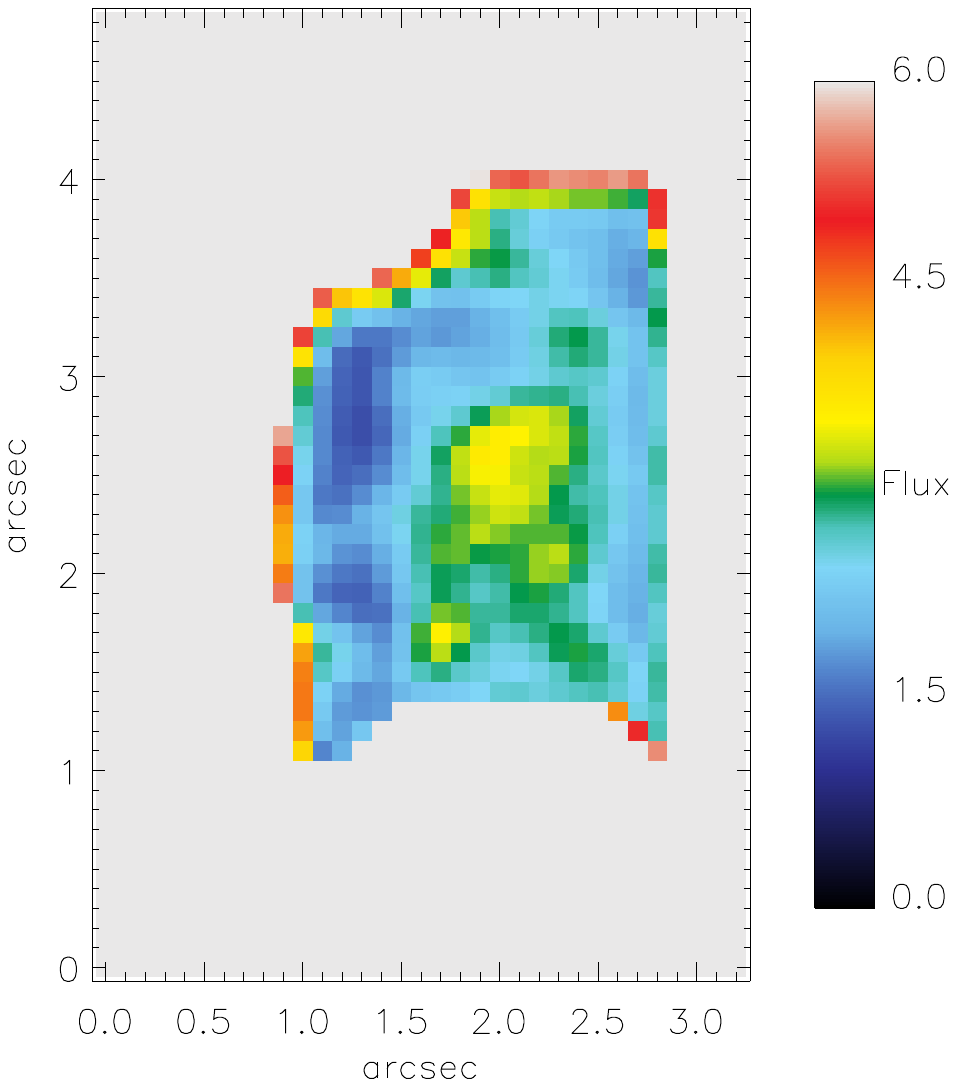}}}
\mbox{\subfigure[UGC09799: H$\alpha$]{\includegraphics[scale=0.4, trim=1mm 1mm 50mm 1mm, clip]{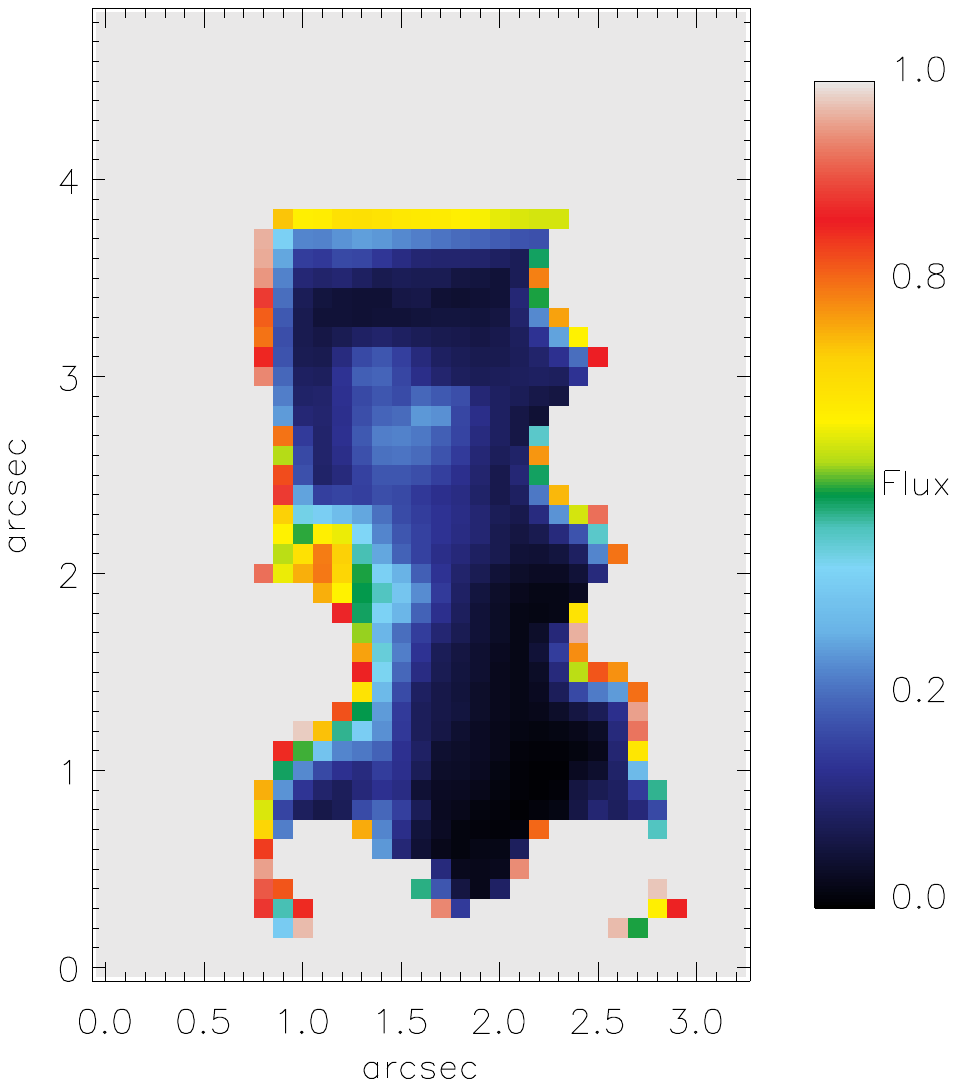}}\quad
\subfigure[UGC09799: {[NII]}$\lambda$6583/H$\alpha$]{\includegraphics[scale=0.4, trim=1mm 1mm 50mm 1mm, clip]{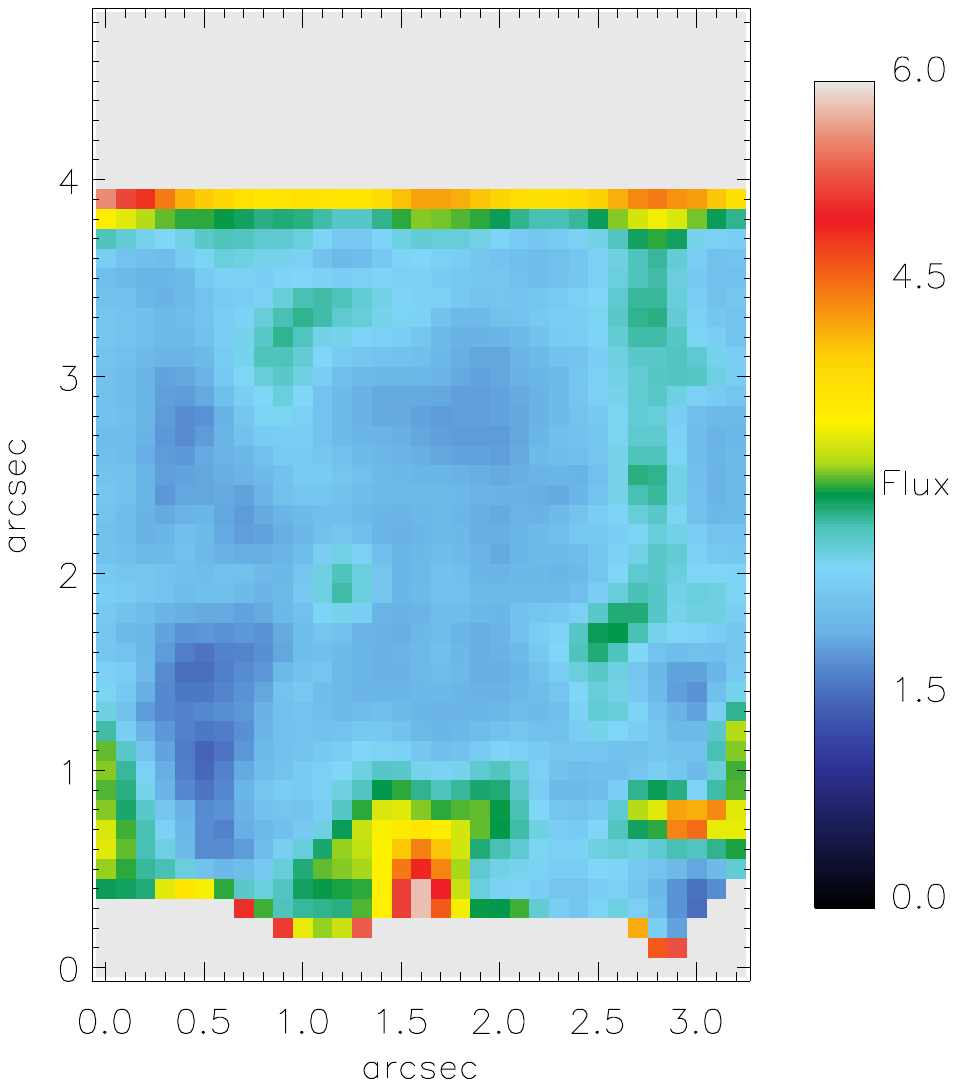}}
\subfigure[UGC09799: {[OIII]}$\lambda$5007/H$\beta$]{\includegraphics[scale=0.4, trim=1mm 1mm 50mm 1mm, clip]{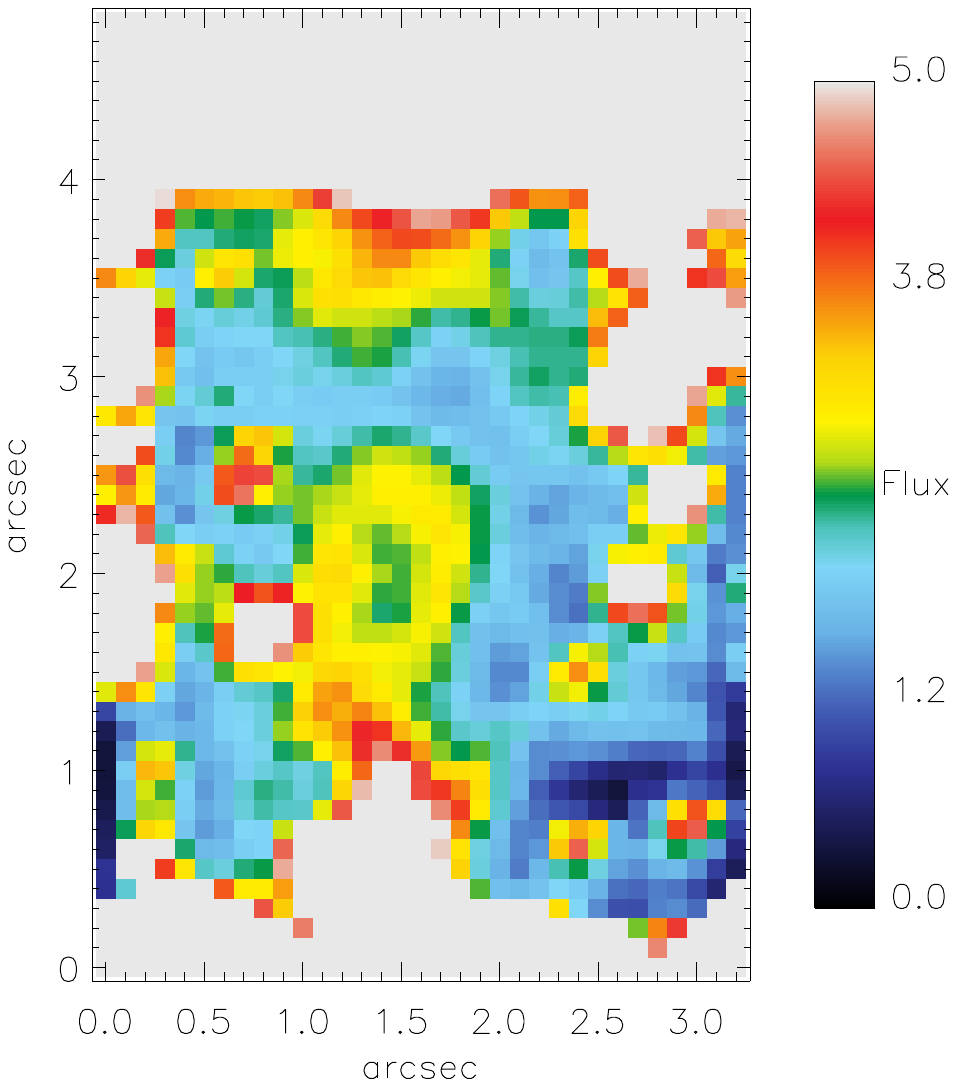}}}
\caption{Absorption-extracted, dereddened H$\alpha$, [NII]$\lambda$6583/H$\alpha$, and [OIII]$\lambda$5007/H$\beta$ emission maps in units of 10$^{-15}$ erg cm$^{-2}$ s$^{-1}$.}
\label{Haflux} 
\end{figure*}

\bsp

\label{lastpage}


\begin{thebibliography}{99}
\bibitem{All2} Allen M.G., Groves B.A., Dopita M.A., Sutherland R.S., Kewley L.J., 2008, ApJS, 178, 20
\bibitem{Anni} Annibali F., Bressan A., Rampazzo R., Zeilinger W.W., Vega O., Panuzzo P., 2010, A$\&$A, 519, 40
\bibitem{Atha} Athanassoula E., Garijo A., Garc\'ia-G\'omez C., 2001, MNRAS, 321, 353
\bibitem{Bald} Baldwin J.A., Phillips M.M., Terlevich R., 1981, PASP, 93, 5
\bibitem{Bild} Bildfell C., Hoekstra H., Babul A., Mahdavi A., 2008, MNRAS, 389, 1637
\bibitem{Bin} Binette L., Raga A.C., Calvet N., Canto J., 1990, PASP, 102, 723
\bibitem{Bin2} Binette L., Magris C.G., Stasi\'nska G., Bruzual A.G., 1994, A$\&$A, 292, 13
\bibitem{Blan} Blanton E.L., Sarazin C.L., McNamara B.R., 2003, ApJ, 585, 227
\bibitem{Boh} Bohringer H., et al., 2001, A$\&$A, 365, 181
\bibitem{Brou} Brough S., Tran K.-V., Sharp R.G., Von der Linden A., Couch W.J., 2011, MNRAS, 414L, 80
\bibitem{Cal} Calzetti D., Armus L., Bohlin R.C., Kinney A.L., Koornneef J., Storchi-Bergmann T., 2000, ApJ, 533, 682
\bibitem{Capp} Cappellari M., Emsellem E., 2004, PASP, 116, 138
\bibitem{Car} Cardelli J.A., Clayton G.C., Mathis J.S., 1989, ApJ, 345, 245
\bibitem{Card1} Cardiel N., Gorgas J., Arag\'{o}n-Salamanca A., 1998, MNRAS, 298, 977
\bibitem{Craw} Crawford C.S., Allen S.W., Ebeling H., Edge A.C., Fabian A.C., 1999, MNRAS, 306, 857
\bibitem{Cra1} Crawford C.S., Hatch N.A., Fabian A.C., Sanders J.S., 2005, MNRAS, 363, 216
\bibitem{DeLu} De Lucia G., Blaizot J., 2007, MNRAS, 375, 2
\bibitem{Dop} Dopita M.A., Sutherland R.S., 1995, ApJ, 455, 468
\bibitem{Edg} Edge A.C., Stewart G.C., Fabian A.C., 1992, MNRAS, 258, 177
\bibitem{Edwa} Edwards L.O.V., Hudson M.J., Balogh M.L., Smith R.J., 2007, MNRAS, 379, 100
\bibitem{Edw} Edwards L.O.V., Robert C., Moll\'a M., McGee S.L., 2009, MNRAS, 396, 1953
\bibitem{Fab} Fabian A.C., Nulsen P.E.J., Steward G.C., Ku W.H.-M., Malin D.F., Mushotzky R.F., 1981, MNRAS, 196, 35
\bibitem{Far} Farage C.L., McGregor P.J., Dopita M.A., Bicknell G.V., 2010, ApJ, 724, 267
\bibitem{Fer} Ferland G.J., Fabian A.C., Hatch N.A., Johnstone R.M., Porter R.L., van Hoof P.A.M., Williams R.J.R., 2009, MNRAS, 392, 1475
\bibitem{Fish} Fisher D., Illingworth G., Franx M., 1995, ApJ, 438, 539
\bibitem{Gas} Gaskell C.M., Ferland G.J., 1984, PASP, 96, 393
\bibitem{Gers} Gerssen J., Allington-Smith J., Miller B.W., Turner J.E.H., Walker A., 2006, MNRAS, 365, 29
\bibitem{Grove} Groves B.A., Dopita M.A., Sutherland R.S., 2004, ApJS, 153, 9 
\bibitem{Ham} Hamer S.L., Edge A.C., Swinbank A.M., Wilman R.J., Russell H.R., Fabian A.C., Sanders J.S., Salom\'{e} P., 2012, MNRAS, 421, 3409
\bibitem{Hatc} Hatch N.A., Crawford C.S., Fabian A.C., 2007, MNRAS, 380, 33
\bibitem{Hic} Hicks A.K., Mushotzky R., 2005, ApJ, 635, 9
\bibitem{Hu} Hu E.M., Cowie L.L., Wang Z., 1985, ApJS, 59, 447
\bibitem{Jon} Johnson R.E., Markevitch M., Wegner G.A., Jones C., Forman W.R., 2010, ApJ, 710, 1776
\bibitem{Jord} Jord\'an A., C\^ot\'e P., West M.J., Marzke R.O., Minniti D., Rejkuba M., 2004, AJ, 127, 24
\bibitem{Kauf} Kauffmann G., et al., 2003, MNRAS, 341, 33
\bibitem{Kew} Kewley L.J., Dopita M.A., Sutherland R.S., Heisler C.A., Trevena J., 2001, ApJ, 556, 121
\bibitem{Kew2} Kewley L.J., Groves B., Kauffmann G., Heckman T., 2006, MNRAS, 372, 961
\bibitem{Lain} Laine S., Van der Marel R.P., Lauer T.R., Postman M., O'Dea C.P., Owen F.N., 2003, AJ, 125, 478
\bibitem{Lou1} Loubser S.I., Sansom A.E., S\'{a}nchez-Bl\'{a}zquez P., Soechting I.K., Bromage G., 2008, MNRAS, 391, 1009 (Paper 1)
\bibitem{Lou2} Loubser S.I., S\'{a}nchez-Bl\'{a}zquez P., Sansom A.E., Soechting I.K., 2009, MNRAS, 398, 133 
\bibitem{Lou5} Loubser S.I., S\'{a}nchez-Bl\'{a}zquez P., 2012, MNRAS, 425, 841
\bibitem{Mar} Markovi\'c T., Owen F.N., Eilik J.A., 2004, Proceedings of The Riddle of Cooling Flows in Galaxies and Clusters of Galaxies, Eds. T. Reiprich, J. Kempner, and N. Soker.
\bibitem{Mat} Mathews W.G., Brighenti F., 1997, in ASP Conference Series ``The Nature of Elliptical Galaxies'', Vol 116, 396
\bibitem{McC} McCarthy I.G., et al., 2010, MNRAS, 406, 822
\bibitem{Mac} McDonald M., Veilleux S., Rupke D.S.N., Mushotzky R., 2010, ApJ, 721, 1262
\bibitem{Mac1} McDonald M., Veilleux S., Rupke D.S.N., Mushotzky R., Reynolds C., 2011, ApJ, 734, 95
\bibitem{Mc2} McDonald M., Veilleux S., Rupke D.S.N., 2012, ApJ, 746, 153
\bibitem{MacN} McNamara B.R., O'Connell R.W., 1992, ApJ, 393, 579
\bibitem{Mac2} McNamara B.R., Wise M., Sarazin C.L., Jannuzi B.T., Elston R., 1996, ApJL, 466, 9
\bibitem{McNa} McNamara B.R. et al., 2006, ApJ, 648, 164
\bibitem{ODea} O'Dea C.P. et al., 2008, ApJ, 681, 1035
\bibitem{Ost} Osterbrock D.E., 1989, Astrophysics of gaseous nebulae and active galactic nuclei, University Science Books, Mill Valley CA
\bibitem{Ost2} Osterbrock D.E., Ferland G.J., 2006, Astrophysics of gaseous nebulae and active galactic nuclei, University Science Books, Sausalito CA
\bibitem{Pere} Peres C.B., Fabian A.C., Edge A.C., Allen S.W., Johnstone R.M., White D.A., 1998, MNRAS, 298, 416
\bibitem{Pete} Peterson J.R., Kahn S.M., Paerels F.B.S., Kaastra K.S., Tamura T., Bleeker J.A.M., Ferrigno C., Jernigan J.G., 2003, ApJ, 590, 207
\bibitem{Pipi} Pipino A., Kaviraj S., Bildfell C., Hoekstra H., Babul A., Silk J., 2009, MNRAS, 395, 462
\bibitem{Rome} Romeo A.D., Napolitano N.R., Covone G., Sommer-Larsen J., Antonuccio-Delogu V., Capacciolo M., 2008, MNRAS, 389, 13
\bibitem{Salo} Salom\'{e} P., Combes F., 2003, A$\&$A, 412, 657 
\bibitem{San4} S\'{a}nchez-Bl\'{a}zquez P. et al., 2006, MNRAS, 371, 703
\bibitem{Sarz} Sarzi M. et al., 2006, MNRAS, 366, 1151
\bibitem{Sar1} Sarzi M., et al., 2010, MNRAS, 402, 2187
\bibitem{Scha} Schawinsky K., et al., 2007, MNRAS, 382, 1415
\bibitem{Sch} Schlegel D.J., Finkbeiner D.P., Davis M., 1998, ApJ, 500, 525
\bibitem{Sta1} Stasi\'nska G., Cid Fernandes R., Mateus A., Sodr\'e L.Jr., Asari N.V., 2006, MNRAS, 371, 972
\bibitem{Sta} Stasi\'nska G., Vale Asari N., Cid Fernandes R., Gomez J.M., Schlickmann M., Mateus A., Schoenel W., Sodr\'e L.Jr., 2008, MNRAS, 391, 29
\bibitem{Tam} Tamura T., Bleeker J.A.M., Kaastra J.S., Ferrigno C., Molendi S., 2001, A$\&$A, 379, 107
\bibitem{van} van Dokkum P.G., 2001, PASP, 113, 1420
\bibitem{Ven} Venturi T., Dallacasa D., Stefanachi F., 2004, A$\&$A, 422, 515
\bibitem{Voi1} Voit G.M., Donahue M., 1997, ApJ, 486, 242
\bibitem{Voit} Voit G.M., Cavagnolo K.M., Donahue M., Rafferty D.A., McNamara B.R., Nulsen P.E.J., 2008, ApJ, 681, 5
\bibitem{vond} Von der Linden A., Best P.N., Kauffmann G., White S.D.M., 2007, MNRAS, 379, 867
\bibitem{Whit} White D.A., Jones C., Forman W., 1997, MNRAS, 292, 419
\bibitem{Wil} Wilman R.J., Edge A.C., Swinbank A.M., 2006, MNRAS, 371, 93
\bibitem{Wise} Wise M.W., McNamara B.R., Nulsen P.E.J., Houck J.C., David L.P., 2007, ApJ, 659, 1153
\end{thebibliography}
\end{document}